\documentclass[10pt,dvips,a4paper]{report}
\usepackage[dvips]{graphicx}
\usepackage{fancyhdr}
\usepackage{amssymb}
\usepackage[english]{babel}

\def\beq{\begin{equation}}
\def\eeq{\end{equation}}
\def\beqy{\begin{eqnarray}}
\def\eeqy{\end{eqnarray}}

\def\bfig{\begin{figure}[htb]}
\def\efig{\end{figure}}

\def\btb{\begin{table}[ht]}
\def\etb{\end{table}}

\def\btab{\begin{tabular}}
\def\etab{\end{tabular}}

\def\bc{\begin{center}}
\def\ec{\end{center}}

\newcommand{\lparder}[3]{\stackrel{\leftarrow}{\frac{\partial^{#3} #1 }{\partial #2^{#3}}}} 
\newcommand{\rparder}[3]{\stackrel{\rightarrow}{\frac{\partial^{#3} #1 }{\partial #2^{#3}}}}

\begin{document}
\thispagestyle{empty}

\vskip 0.2truein
\begin{centering}
{\large Universit\`a degli Studi di Perugia - Dipartimento di Fisica}\\
\end{centering}

\vskip06\bigskipamount
\begin{centering}
{\huge \bf Strongly Coupled Lattice}\\[\bigskipamount]
{\huge \bf Gauge Theories}\\[\bigskipamount]
{\huge \bf and}\\[\bigskipamount]
{\huge \bf Antiferromagnetic Spin Systems}\\[06\bigskipamount]
{\Large Tesi presentata da}\\[\bigskipamount]
{\Large  Federico Berruto}\\[\bigskipamount]
{\large per il titolo di}\\[\bigskipamount]
{\Large  Dottore di Ricerca}\\
\vskip 0.5truein
\end{centering}

\begin{figure}[h]
\vspace{4.0cm}
\includegraphics{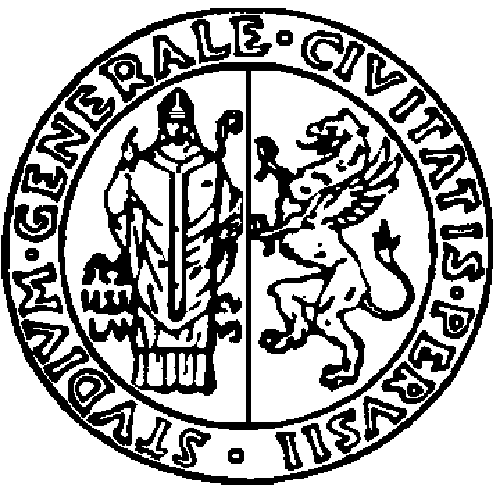}
\end{figure}

\vskip6.0\bigskipamount

\vskip3.0\bigskipamount
\begin{centering}
\large Dottorato di Ricerca, X Ciclo - Dicembre 1998\\
\end{centering}
\vglue-1.5\bigskipamount
\newpage
\thispagestyle{empty}
\pagenumbering{roman}
\thispagestyle{plain}

\vskip 0.2truein
\begin{centering}
{\Large \bf Strongly Coupled Lattice Gauge Theories and}\\[\medskipamount]
{\Large \bf Antiferromagnetic Spin Systems}\\[\bigskipamount]
{\large  Federico Berruto}\\
\vskip 0.01truein
{\large Dipartimento di Fisica, Universit\`a di Perugia}\\[\bigskipamount]
{\Large \bf Abstract}\\[\bigskipamount]
\end{centering}
\noindent
In this thesis, I use the strong coupling expansion to investigate the multiflavor lattice Schwinger models in the hamiltonian 
formalism using staggered fermions. In particular, I am interested in analysing the mapping of these gauge theories onto 
quantum spin-$1/2$ antiferromagnetic Heisenberg chains. Exploting this mapping, the chiral symmetry breaking patterns are 
studied and the spectra are computed. The extrapolation to zero lattice spacing of the results compare favorably 
qualitatively and quantitatively with the weak coupling studies of the gauge models in the continuum.

I studied the one-flavor lattice Schwinger model, in the strong coupling limit, showing that it is equivalent to an 
Ising model with long-range Coulomb interaction.  Even though the continuous chiral symmetry is explicitly broken by the 
staggered  fermions, a discrete chiral symmetry remains and appears in the lattice theory as a translation by one lattice site 
which forbids fermion mass and must be broken in order for the lattice model to exhibit the features of the spectrum of the continuum theory. 
The effects of the anomaly in the continuum appear on the lattice through spontaneous symmetry breaking: indeed I show that the ground state 
spontaneously breaks translations by one lattice site. The dicrete symmetries are carefully defined on the lattice. 
The masses of the low lying bosonic excitations as well as the chiral condensate of the model are computed up to the fourth order in the 
strong coupling expansion. Very good agreement between lattice calculations and continuum values is found.

The two-flavor lattice Schwinger model is analysed using the same computational scheme adopted in the one-flavor case. 
The problem of finding the low lying excitations is showed to be equivalent to solving the spin-$1/2$ antiferromagnetic Heisenberg chain. 
In fact the Heisenberg Hamiltonian is the effective Hamiltonian of the gauge model at the second order in the strong coupling expansion. 
The ground state of the spin model represents the vacuum of the gauge theory in the strong coupling limit. 
Two kinds of excitations can be created from the 
ground state; one kind involves only spin flipping and has lower energy since no electric flux is created, the other involves 
fermion transport besides spin flipping and thus has higher energy. The massless excitations of the gauge model are the spinons of the 
antiferromagnetic chain. The excitation masses can be expressed in terms of spin-spin correlators evaluated 
on the ground state. Either the massless or the massive excitations have identical 
$P$- and $G$-parity to the low lying excitations of the continuum two flavor gauge model. 
Good agreement between lattice and continuum results is found. 
The ground state is translationally invariant: both the 
isoscalar or the isovector chiral condensates are zero to every order in the strong coupling expansion as it happens in the continuum theory as a 
consequence of the Coleman theorem. In addition the vacuum 
expectation value of the umklapp operator is non zero as it happens for the corresponding operator in the continuum and this is the only relic 
of the chiral anomaly on the lattice.

The ${\cal N}$-flavor lattice Schwinger models are analysed generalizing the ${\cal N}=2$ case and at the second order in the 
strong coupling expansion are effectively described by spin-$1/2$  $SU({\cal N})$ antiferromagnetic Heisenberg chains. 
The generalization is quite straightforward, even if the theory is much different depending on if ${\cal N}$ is even or odd.

\begin{centering}
{\large Tesi presentata per il titolo di Dottore di Ricerca}\\
\end{centering}
\thispagestyle{plain}
\vskip 10.0truein
\noindent
{\Large \bf Preface}
\vskip 0.1truein 
\noindent
This thesis analyses the strong coupling limit of the multiflavor 
lattice Schwinger models and their relationship with generalized quantum 
spin-$1/2$  $SU({\cal N})$ antiferromagnetic Heisenberg chains. 
My original contributions are summaryzed in chapters 4, 5 and 6. In chapter 4 
the issues of chiral symmetry breaking and the mass spectrum are studied for the one-flavor 
lattice Schwinger model. 
Part of this work appears in the pubblication
\begin{itemize}
\item F. Berruto, G. Grignani, G. W. Semenoff and P. Sodano, Phys. Rev. {\bf D57}, 5070 (1998).
\end{itemize}

In chapter 5 the two-flavor lattice Schwinger model and its mapping onto the quantum spin-$1/2$ antiferromagnetic Heisenberg chain are 
analysed in detail. At variance with the one-flavor theory, 
the chiral symmetry breaking is now explicit and the spectrum of the gauge model is derived starting from the 
ground state of the Heisenberg model. Many of the results concerning the analysis of this model can be found in
\begin{itemize}
\item F. Berruto, G. Grignani, G. W. Semenoff and P. Sodano, Phys. Rev. {\bf D59}, 034504 (1999).
\item F. Berruto, G. Grignani, G. W. Semenoff and P. Sodano, hep-th/9901142.
\end{itemize}

Chapter 6 is devoted to a generalization to the ${\cal N}$-flavor lattice Schwinger models and their mapping onto 
quantum spin-$1/2$ $SU({\cal N})$ antiferromagnetic Heisenberg chains.
\thispagestyle{plain}
\vskip 10.0truein
\noindent
{\Large \bf Acknowledgements}
\vskip 0.1truein 
\noindent
I am very grateful to Prof. P. Sodano for stimulating my research interests with enlightening discussions, for the huge amount of time he dedicated to 
our research project and for providing me with the chance to attend several meetings dealing with the topics addressed in my thesis. I thank 
Dr. G. Grignani for his constant guidance when problems arose, for the many suggestions and for the many pleasant working hours spent 
together. I thank also Prof. G. W. Semenoff for many discussions and for the kind hospitality extended to me at the University of British Columbia. 
Prof. R. B. Laughlin is acknowledged for a very interesting discussion concerning low dimensional quantum antiferromagnets and 
quantum disorder in spin chains.     
\tableofcontents
\chapter{Introduction}
\pagenumbering{arabic}
Quantum field theory and condensed matter physics had an amazing development in this century and 
each one borrowed from the other new ideas. 
For instance the concept of spontaneous symmetry breaking was originally introduced to study ferromagnetism and 
the Higgs  phenomenon appears studing the Meissner effect in the theory of superconductivity~\cite{b0}. 
Mathematical analogies between the two disciplines were pointed out by Wilson~\cite{b1} who joined the concepts of renormalization, 
typical of particle physics, with the concepts of universality and scaling, originally introduced to describe second 
order phase transitions. Later Wilson~\cite{b2} introduced a lattice formulation of gauge theories. 
Once an euclidean field theory is defined on the lattice it becomes a statistical mechanics problem. In general a quantum 
theory in $d$ dimensions is equivalent to a classical statistical mechanical system in $d+1$ dimensions. 
The phase diagram of a statistical system determines its physical properties. Landau introduced the general description of all phase transitions 
as changes of symmetry~\cite{b3}. Fluctuations grow in a substance as its critical point is approached and they 
interact very strongly with each other so that the system is strongly correlated. The order of a phase transition is 
crucial: only if the phase transition is continuous, it is possible to obtain from a lattice field theory a relativistic continuous theory. 
The lattice approach to gauge theories provides a powerful tool to study non-perturbative properties. 

One of the most important properties of non-Abelian gauge theories, such as Quantum Chromodynamics ($QCD$), is asymptotic freedom~\cite{b4}. At short 
distances or high momenta quarks are weakly interacting and it is possible to perform a perturbative expansion. At large distances 
or small momenta the coupling constant increases indefinitely confining quarks. Confinement is an observed property of the strong 
interactions and it is an unproven, but widely believed feature of most non-Abelian gauge theories in four and lower space-time dimensions. The theoretical 
mechanism behind confinement is difficult to analyze since it escapes the weak coupling analysis: it is intrinsecally a strong coupling 
and non-perturbative phenomenon.

Wilson~\cite{b2} proposed on the lattice a mechanism for quark confinement. Gauge theories have been 
analysed in the strong coupling limit on a $(3+1)$-dimensional space-time lattice. In the weak coupling limit the gauge field behaves like 
a normal free massless field and quarks are unbound. In the strong coupling limit the gauge field is massive and the quarks are bound. There should be 
a confinement-deconfinement phase transition at  some intermediate value of the coupling constant. 

In the continuum, closely related to confinement there is dynamical chiral symmetry breaking, which creates the $\pi$ mesons 
and keep their masses light. 
Chiral symmetry~\cite{b6} is only an approximate symmetry of particle physics since the up and down quarks are light but not massless. 
Otherwise the pion, which is the Goldstone boson arising from the symmetry breaking would be strictly massless rather than being just 
light. The theory behind chiral symmetry breaking is only partially understood because this phenomenon occurs at strong 
coupling where the standard tools of perturbation theory are not reliable. There are conjectures~\cite{b7b} that if one could adjust the charge 
of the electron $e$ in ordinary $QED$, there would be a critical value of $e$ at which a phase transition would take place and 
the electron would become very massive. A transition to heavy electrons would break the approximate chiral symmetry of $QED$ which is 
there because electrons are very light. 

The strong coupling limit~\cite{b7} of lattice gauge theories though far from the scaling regime is often used to study the qualitative properties 
of a gauge model. 
Two important features of the spectrum of non-abelian gauge theories appear there. 
The strong coupling limit exhibits confinement~\cite{b2}, which is related in a rather natural 
way to the gauge symmetry and compactness of the non-abelian gauge group. 
Moreover, it is straightforward to show that strongly coupled gauge theories exhibit dynamical chiral symmetry breaking. 
Wilson fermions~\cite{b8} explicitly break chiral symmetry but for what concerns staggered fermions~\cite{b9},
 even if the continuous chiral symmetry is broken explicitly, a 
discrete axial symmetry survives the lattice regularization and appears in the lattice theory as a translation by one spacing. 
Several quantitative investigations of gauge theories using strong coupling techniques have been performed and there have been attempts to 
compute the mass spectrum of realistic models such as $QCD$~\cite{b10}. 
The strong coupling expansion is an expansion in the inverse of the gauge coupling constant. Due to asymptotic freedom, the continuum limit is found 
where the coupling constant is small, so that one should not expect a priori that the strong coupling expansion gives accurate quantitative 
information about the continuum gauge theory. However, the expansion does have a finite radius of convergence~\cite{b11}; this indicates that its properties 
are shared by the model for a finite range of the coupling parameter and certain informations can be obtained by analytic continuation to 
regions outside the convergence radius. 

It has been recognised for some time that the strong coupling limit of lattice gauge theories with dynamical fermions is related to 
certain quantum spin systems. The relationship between gauge and spin systems is particularly evident in the hamiltonian formalism~\cite{b13} and already appeared in 
some of the earliest analysis of chiral symmetry breaking in the strong coupling limit~\cite{b14}. In particular mesons emerge as spin waves when spontaneous 
chiral symmetry breaking takes place. It was noticed that there are several formal similarities between some condensed 
matter systems with lattice fermions $-$ and in particular certain antiferromagnetic spin systems $-$ and lattice gauge theories 
in their strong coupling limit~\cite{b15}. The quantum spin-$1/2$ Heisenberg antiferromagnet was recognised to be equivalent to the strong 
coupling limit of a $U(1)$ lattice gauge theory~\cite{b16} and can also be written as a kind of $SU(2)$ gauge theory~\cite{b17}. 
Furthermore, staggered fermions resemble ordinary lattice fermions used in tight binding models in condensed matter physics at half-filling, $i.e.$ 
when there is one electron per site, and placed in a background $U(1)$ magnetic field $\pi$ (mod $2\pi$) $-$ $1/2$ of a flux quantum $-$ 
through every plaquette of the lattice. In $(2+1)$-dimensions the parallel between lattice fermions in condensed matter and 
gauge theories was already recognized in the first work~\cite{b18} on the Azbel-Wannier-Hofstaeder problem and then has been discussed in the context of the so called 
flux phases of the Hubbard model. It is actually true for all $d\geq 2+1$~\cite{b19}, $i.e.$ the staggered lattice fermion approximation of the 
relativistic Dirac field with the minimal number of flavors is identical to a condensed matter fermion problem with simple nearest neighbor hopping 
in a background magnetic field which has $1/2$ flux quantum per plaquette. 
In the condensed matter context, the magnetic flux can be produced by a condensate, as in the flux phases of the Heisenberg and Hubbard models~\cite{b15}. 
A $1/2$ flux quantum per plaquette for ordinary lattice spacing is yet an experimentally inaccessible flux density. Nonetheless it could be achieved 
in analog experiments where macroscopic arrays of Josephson junctions, for example, take the place of atoms at lattice sites and their ground state in 
two dimensions is expected to be a flux phase~\cite{b20}. 

The analogy  between Dirac and tight-binding condensed matter fermions was exploited to find an exact mapping of the strong coupling limit 
of a large class of lattice gauge theories with dynamical staggered fermions onto certain quantum antiferromagnets, both the conventional 
Heisenberg antiferromagnets with spin $S$ and generalized antiferromagnets with spins taking values in Lie algebras other than $SU(2)$~\cite{b21}. The 
precise structure of the resulting antiferromagnet depends on the number of colours 
in the gauge group $-$ since $S={\cal N}_{c}/2$ $-$ and also on the number of fermion flavors. 
There are four different cases. $(a)$ An $U({\cal N}_{c})$ gauge theory with an even number ${\cal N}$ of lattice flavors of staggered fermions is 
an $SU({\cal N})$ antiferromagnet in the strong coupling limit. The number of colours ${\cal N}_{c}$ 
determines the representation of the antiferromagnet as being the one represented by the Young tableau with ${\cal N}_{c}$ columns and 
${\cal N}/2$ rows. $(b)$ An $SU({\cal N}_{c})$ gauge theory with ${\cal N}_{c}\geq 3$ is an $U({\cal N})$ antiferromagnet. The representation of the 
$U(1)$ subgroup of $U({\cal N})$ is not constrained as it was in the $(a)$ case where the local $U(1)$ charge had to vanish. The representation of 
the $SU({\cal N})$ subgroup of $U({\cal N})$ can now take on any representation on a given site, subject only to the constraint that the average 
fermion density is ${\cal N}_{c} {\cal N}/2$ and the number of fermions on each site is given by the number of boxes in the corresponding 
Young tableau. $(c)$ The $SU(2)$ lattice gauge theories are special. In the strong coupling limit additional terms compared to $SU({\cal N}_{c}\geq 3)$ 
appear and one obtains for ${\cal N}$ lattice flavors antiferromagnets associated with the simplectic group $Sp(2{\cal N})$ and for ${\cal N}=1$ a 
spin-$1/2$ Heisenberg antiferromagnet. $(d)$ A gauge theory with an odd number ${\cal N}$of lattice flavors of staggered fermions in the strong 
coupling limit is an $SU({\cal N})$ antiferromagnet where the size of representations on neighboring sites differ. One has to divide the sites 
in two sublattices such that the nearest  neighbors of all the sites of one sublattice are in the other sublattice; if this classification in 
two sublattices is possible, the lattice is called bipartite. On the bipartite lattice one has the representation of $SU({\cal N})$ with 
Young tableau with $({\cal N}+1)/2$ rows and ${\cal N}_{c}$ columns on one sublattice and with $({\cal N}-1)/2$ rows and ${\cal N}_{c}$ columns 
on the other sublattice. There exist analogs in spin systems of strong interactions behavior, particularly for what concerns confinement 
and chiral symmetry. 
Laughlin~\cite{b22} recently provided an appealing interpretation of confinement in quantum antiferromagnets. The physical idea is that the phase diagram of 
antiferromagnets consists of competing ordered phases regulated by a nearby quantum critical point. 
Exactly at the critical point the low lying elementary excitations of the magnet are gauge fields and particles with fractional quantum numbers 
analogous to the spinon and holon excitations found in spin chains. Away from the critical point, even if very close, the ``fundamental 
constituents" bind at low energy scales to make the familiar collective modes of ordered states which one renormalizes. 
The existence of ``fundamental constituents" in conventional materials and models may be indirectly inferred by high-energy spectroscopy 
and inconsistences in sum rules exactly the way the existence of quarks is inferred  in particle physics. There exist physically 
identifiable objects behaving like $U(1)$ quarks out of which the elementary excitations are built. In Laughlin's scenario quark-like 
objects and gauge fields would be liberated at the critical point and would become the true elementary excitations. 

In quantum antiferromagnets, the appearence of N\'eel order is the analog of chiral symmetry breaking. Exploiting the mapping existing 
between gauge and spin models, rigorous results about antiferromagnets can be used to prove spontaneous broken chiral symmetry in certain 
strong coupling gauge theories. Whenever the ground state of the antiferromagnet is not translationally invariant, the discrete chiral 
symmetry is spontaneously broken. The possibility of studing in detail the patterns of chiral symmetry breaking on the lattice is very 
important. In particular, an interesting question is how the axial anomaly~\cite{b23} appears in the lattice regularization. 
Anomalies occur when all of the symmetries of a classical field theory cannot be realized simultaneously in the corresponding 
quantum field theory. The classical example is the Adler-Bell-Jackiw anomaly in a vector like gauge theory where gauge invariance and 
global axial symmetry are incompatible. This is a manifestation of the fact that it is impossible to find a regularization of 
ultraviolet divergences preserving both axial and gauge invariance. 
If gauge invariance is preserved, the observable manifestation of the anomaly is the failure of the conservation laws for axial vector currents 
and the resulting abscence of their consequences in the spectrum of the quantized theory. Nielsen and Ninomiya~\cite{b24} proved that fermion theories on 
a lattice have an equal number of species of left- and right-handed Weyl particles in the continuum limit, thus preventing an axial anomaly. 
Lattice field theories are manifestly gauge invariant by construction and therefore, to produce continuum theories, they must find a way to 
violate axial current conservation. 

In the continuum an interesting mechanism of confinement in Quantum Electrodynamics ($QED$) with massless fermions in $(1+1)$-dimensions 
was discovered by Schwinger~\cite{b5} which demonstrated that the ``photon" acquires a mass and the spectrum of the model does not exhibit any  
free electron asymptotycal state. The ``photon" mass is proportional to the electromagnetic coupling constant if charges are totally 
screened by the vacuum polarization. After this investigation $QED$ in $(1+1)$-dimensions with massless fermions has been called the 
Schwinger model. Moreover Schwinger speculated that in four dimensions the photon mass would be zero for any 
electromagnetic charge $e$ less than $e_{c}$; for $e$ greater than $e_{c}$ the photon mass would be non-zero and vary with $e$. At 
the critical coupling constant $e_{c}$ a change of phase from zero to non-zero photon mass should take place.

In this thesis it is my purpose to investigate the one-flavor and the multiflavor lattice Schwinger models in the hamiltonian formalism using 
staggered fermions. Exploiting these toy models I will clarify the mechanisms of chiral symmetry breaking on the lattice and the vestiges of the chiral anomaly 
of the continuum theory. 
Moreover the spectra of these models will be computed in the strong coupling limit. Extrapolating the spectra to the zero lattice spacing limit 
by means of Pad\'e approximants~\cite{b26}, I find that the lattice compare favorably $-$ both qualitatively and quantitatively 
$-$ with the weak coupling continuum 
studies of the multiflavor Schwinger models. The success of the procedure is not completely surprising since the continuum theory of free 
massive mesons can be described in terms of quarks and antiquarks pairs bound by gauge strings and it is precisely this idea which is 
embodied in lattice methods. 
The ${\cal N}$-flavor lattice Schwinger models in the strong coupling limit are effectively described by generalized quantum $SU({\cal N})$ 
spin-$1/2$ Heisenberg antiferromagnetic Hamiltonians. 
Spinons are the fundamental spin-$1/2$ excitations of quantum antiferromagnets. In the same way as the Schwinger models spectra do not exhibit 
free quarks (electrons) but just massive bosons, the ``photons" or better mesons, also the Heisenberg models excitations contain an even number of 
spinons and so carry integer spin. One original result of my work has been to show explicitly how spinons appear in lattice gauge theories.

In chapter 2 I shall review the lattice approach to gauge theories. In particular boson, fermion and gauge fields are defined on the lattice. The 
``fermion doubling" problem is analysed and Wilson and staggered fermions are introduced. The hamiltonian formulation of gauge theories and the strong 
coupling limit are described. Furthermore the realization of the chiral anomaly on the lattice is discussed and the Nielsen and Ninomiya no-go theorem 
is demonstrated. 

Chapter 3 is devoted to a detailed analysis of quantum antiferromagnetic spin chains with special emphasis to the topics useful 
to understand the connection between Schwinger and Heisenberg models. First I demonstrate that spin wave theory fails in one dimension due to strong 
infrared divergences. Then I introduce the Haldane-Shastry model where it is particularly easy to see the spinons and analyse the spectrum. 
The Bethe ansatz solution of the spin-$1/2$ antiferromagnetic Heisenberg Hamiltonian is then reviewed in detail. 
I study the complete spectrum of finite size Heisenberg antiferromagnetic chains of 4 and 6 sites and I verify that the thermodynamic limit 
solution provided in \cite{b27} is already very well reproduced in the finite size chains. Moreover, I write down explicitly the ground state of 4, 6 and 8 sites chains. 
Spin-spin correlators relevant for evaluating the mass spectrum of the two-flavor lattice Schwinger model are calculated. 
In the last section, generalized Heisenberg antiferromagnets with symmetry group $SU({\cal N})$ are introduced.

In chapter 4 the one-flavor lattice Schwinger model is studied. The discrete chiral symmetry on the lattice is broken explicitly 
and this mechanism reproduces the effects that in the continuum are generated by the axial anomaly. The discrete symmetries of the model are analysed in 
detail and the vacuum of the model is found to be the ground state of the antiferromagnetic Ising spin chain with long range Coulomb interaction. 
The spectrum of the continuum theory is very well reproduced when the lattice results are extrapolated using the Pad\'e approximants. 

In chapter 5 I analyse the two-flavor lattice Schwinger model. In the strong coupling limit the gauge model is effectively described by  the quantum spin-$1/2$ 
antiferromagnetic Heisenberg model. 
The vacuum of the gauge theory in the strong coupling limit is the ground state of the spin model: a quantum disordered state very different from the classical 
N\'eel state. It is not possible to write down explicitly this ground state in the thermodynamic limit, but its energy can be computed using the Bethe ansatz 
as explained in chapter 3. 
The massless excitations of the gauge model are the spinons of the Heisenberg antiferromagnet, while the massive excitations are created by acting 
on the ground state with pertinent operators that have the right quantum numbers of the continuum theory and generate charge transport. 
The excitation masses of the gauge model are completely expressed in terms of vacuum expectation values (VEV's) of spin-spin correlation functions. 
The chiral symmetry breaking pattern is explicit in this case, due to the staggered fermions coupled by the gauge field. The isoscalar and 
isovector chiral condensates, that in the continuum are zero due to the Coleman theorem, are zero also on the lattice. 
The vacuum expectation value of the umklapp operator is non-zero due to the explicit breaking of $U_{A}(1)$; this can be 
viewed as a manifestation of the anomaly on the lattice.
 
In chapter 6 the analysis of chapter 5 is generalized to the Schwinger models with $SU({\cal N})$  flavor group. In the strong 
coupling limit the gauge models are equivalent to generalized $SU({\cal N})$ quantum Heisenberg antiferromagnets. 
Even if most of the analysis is a straightforward generalization of the ${\cal N}=2$ case, some surprising difference arises between the ${\cal N}$
 even and odd models. 
Some concluding remarks and a short summary of results is provided. 

The appendix aims at elucidating some features of Pad\'e approximants relevant for a better understanding of chapters 4 and 5.
\chapter{Lattice gauge theories}

It is my purpose to review in this chapter the lattice approach to gauge theories. 
I do not pretend to be exhaustive, but I want to illustrate the topics used in my approach to the lattice Schwinger model. 
The reader interested in knowing more about lattice gauge theories could usefully look up Refs.\cite{b7,b28}.
  
Quantum field theory is so far the most appropriate scheme for describing the strong, electromagnetic and weak interactions 
between elementary particles. It has been known for long time that electromagnetic interactions are described by a quantum gauge field theory, 
but the fundamental role played by the principle of gauge invariance in the construction of a theory for the strong and weak interactions has 
been recognized only much later. 

$QCD$, the theory of strong interactions, is a gauge theory based on the 
unbroken non-abelian $SU(3)$ group. The group $SU(3)$ has eight generators and so there are eight ``massless" 
gluons carrying a colour charge which mediate the strong 
interactions between the fundamental constituents of matter, the quarks. $QCD$ is an asymptotically free theory~\cite{b4}. 
Asymptotic freedom allowed to carry out quantitative perturbative calculations of observables which are 
sensitive to the short distance structure of $QCD$. Quarks have never been seen free in nature and only colour neutral baryons are observed: 
quarks are confined. 
A demonstration that $QCD$ accounts for quarks confinement can only come from a non-perturbative treatment of the theory; 
confinement is indeed a consequence of the dynamics at large distances where perturbation theory breaks down.

Wilson~\cite{b2} showed \footnote{Independently, gauge theories were discussed on a lattice by Wegner~\cite{b29}, who elevated the global up $\leftrightarrow$ down 
symmetry of the ordinary Ising to a local symmetry and in an unpublished work, which deals mostly with Abelian gauge theories, by Polyakov in 1974.}
that in the strong coupling limit $QCD$ confines quarks, however this is not a justified approximation when studying the continuum limit. 
There are numerical simulations which strongly suggest that $QCD$ accounts for quark confinement~\cite{b28}. Moreover, there are other 
fascinating questions that one would like to answer: does $QCD$ account for the observed spectrum? Are there other particles allowed by $QCD$ but not 
yet experimentally observed? Does $QCD$ account for chiral symmetry spontaneous breaking? 
In order to answer the above questions it is necessary a non perturbative  treatment of $QCD$ which at the moment is only provided by the lattice formulation. 

The lattice formulation of $QCD$ by Wilson opened the way to the study of non-perturbative phenomena using numerical methods.
His purpose was to study the long distance properties of $QCD$ in a format where the short distance properties which lead to ultraviolet 
divergences are regulated by the lattice cutoff. The space-time discretization provides a natural cut-off scheme as wave-lengths shorter than twice 
the lattice spacing, $a$, have no meaning and the momentum domain is restricted to a region bounded by $\pi / a$. The lattice provides 
then a non-perturbative regularization of ultraviolet divergences. Once a lattice field theory has been formulated, the original field theory problem 
becomes one of statistical mechanics. With a finite lattice there are a finite number 
of variables. It is then possible to study various physical interesting quantities: energy spectrum, correlation functions, critical 
exponents etc., for example by computer simulations based on the Monte Carlo method~\cite{b28}. 
By now lattice gauge theories have become a branch of particle physics in its own right, and their intimate connection to statistical mechanics 
raised the interest of elementary particle phisicists as well as of condensed matter physicists. 

Since the beginning of lattice gauge theory, chiral symmetries have been perplexing. The issues revolve around anomalies and fermion doubling. 
For vector-like theories such as $QCD$ the problems are largely resolved. The Wilson approach~\cite{b8} breaks chiral symmetry explicitly by adding a symmetry 
breaking term to give all the doublers a mass which becomes infinite with the cut-off scale. The Kogut and Susskind fermions~\cite{b9}, also called 
staggered fermions, eliminate the doubling by reducing the Brillouin zone, $i.e.$, by doubling the effective lattice spacing. 
The situation is much more clouded for the full Standard Model. Here chiral symmetry plays a fundamental role, with neutrinos maximally violating parity. 
Nielsen and Ninomiya demonstrated a no-go theorem~\cite{b24} whose most important consequence is that the weak interaction cannot be put on the lattice. 
For a general class of fermion theories on a lattice in the hamiltonian formalism an equal number of species of left- and right-handed Weyl particles, 
$i.e.$ neutrinos, necessary appears in the continuum limit. They also proved that this disease is not peculiar to the lattice theory, but some trouble 
appears universally in all regularization schemes, for instance dimensional regularization. 
In any lattice theory of chirally invariant fermions with locality there is an equal number of production and annihilation of Weyl fermions. Thus 
there is no net production, so that the axial charges are conserved. An analogy or a simulation exists between the Weyl fermion theory and 
gapless semiconductors, where two energy bands have point-like degeneracies. In section (\ref{lgt4}) I shall discuss thoroughly this no-go theorem. 
At the moment no reliable lattice computation for the 
Standard Model does exist.

In section (\ref{lgt2}) I discuss the differences arising between boson and fermion fields on the lattice and 
I analyse the fermion doubling, an obstacle to put fermions on a lattice. Gauge fields on the lattice are discussed.

Section (\ref{lgt3}) is devoted to introduce staggered fermions, a particular type of lattice fermions which partially avoid the fermion doubling problem. 
Moreover, hamiltonian formulation of lattice gauge theories is introduced.

Section (\ref{strongc}) explains the strong coupling approach to lattice gauge theories in the hamiltonian formalism. 

In section (\ref{lgt4}) the Nielsen and Ninomiya no-go theorem is reviewed and the chiral symmetry and chiral anomaly on the lattice are analysed 
and the implications and analogs of these topics in condensed matter theory are discussed.

\section{Bosons and fermions on the lattice}
\label{lgt2}
Any field (scalar, fermionic or gauge) can be defined on a lattice, but the result of putting a bosonic or a fermionic field on a lattice is 
distinctly different. While the procedure for a bosonic field is straightforward, placing the Dirac equation on a space-time lattice presents 
some surprisingly difficult problem. 

In this section I shall first illustrate the way to put the scalar field on the lattice. Then I shall discuss the so 
called doubling problem~\cite{b7} of lattice fermions. A prescription to cure it is discussed, the Wilson fermions~\cite{b8}, 
while section (\ref{lgt3}) is devoted to 
discuss the Kogut and Susskind fermions or staggered fermions~\cite{b9}, which have been used in my approach to the lattice Schwinger model. 
The link variable nature of gauge fields on the lattice is explained. 

\subsection{Bose fields on the lattice: scalar fields}

\begin{figure}[htb]
\begin{center}
\setlength{\unitlength}{0.240900pt}
\ifx\plotpoint\undefined\newsavebox{\plotpoint}\fi
\sbox{\plotpoint}{\rule[-0.200pt]{0.400pt}{0.400pt}}%
\begin{picture}(1000,900)(300,0)
\font\gnuplot=cmr10 at 10pt
\gnuplot
\sbox{\plotpoint}{\rule[-0.200pt]{0.400pt}{0.400pt}}%
\put(176.0,113.0){\rule[-0.200pt]{303.534pt}{0.400pt}}
\put(806.0,113.0){\rule[-0.200pt]{0.400pt}{173.207pt}}
\put(176.0,113.0){\rule[-0.200pt]{0.400pt}{4.818pt}}
\put(176,68){\makebox(0,0){${\rm -\pi/a}$}}
\put(176.0,812.0){\rule[-0.200pt]{0.400pt}{4.818pt}}
\put(1436.0,113.0){\rule[-0.200pt]{0.400pt}{4.818pt}}
\put(1436,68){\makebox(0,0){${\rm \pi/a}$}}
\put(1436.0,812.0){\rule[-0.200pt]{0.400pt}{4.818pt}}
\put(1400,158){\makebox(0,0){$k$}}
\put(806,877){\makebox(0,0){$S(k)-m^2$}}
\put(176,832){\usebox{\plotpoint}}
\multiput(176.58,827.88)(0.493,-1.131){23}{\rule{0.119pt}{0.992pt}}
\multiput(175.17,829.94)(13.000,-26.940){2}{\rule{0.400pt}{0.496pt}}
\multiput(189.58,798.71)(0.492,-1.186){21}{\rule{0.119pt}{1.033pt}}
\multiput(188.17,800.86)(12.000,-25.855){2}{\rule{0.400pt}{0.517pt}}
\multiput(201.58,771.01)(0.493,-1.091){23}{\rule{0.119pt}{0.962pt}}
\multiput(200.17,773.00)(13.000,-26.004){2}{\rule{0.400pt}{0.481pt}}
\multiput(214.58,743.14)(0.493,-1.052){23}{\rule{0.119pt}{0.931pt}}
\multiput(213.17,745.07)(13.000,-25.068){2}{\rule{0.400pt}{0.465pt}}
\multiput(227.58,716.26)(0.493,-1.012){23}{\rule{0.119pt}{0.900pt}}
\multiput(226.17,718.13)(13.000,-24.132){2}{\rule{0.400pt}{0.450pt}}
\multiput(240.58,689.99)(0.492,-1.099){21}{\rule{0.119pt}{0.967pt}}
\multiput(239.17,691.99)(12.000,-23.994){2}{\rule{0.400pt}{0.483pt}}
\multiput(252.58,664.39)(0.493,-0.972){23}{\rule{0.119pt}{0.869pt}}
\multiput(251.17,666.20)(13.000,-23.196){2}{\rule{0.400pt}{0.435pt}}
\multiput(265.58,639.39)(0.493,-0.972){23}{\rule{0.119pt}{0.869pt}}
\multiput(264.17,641.20)(13.000,-23.196){2}{\rule{0.400pt}{0.435pt}}
\multiput(278.58,614.52)(0.493,-0.933){23}{\rule{0.119pt}{0.838pt}}
\multiput(277.17,616.26)(13.000,-22.260){2}{\rule{0.400pt}{0.419pt}}
\multiput(291.58,590.40)(0.492,-0.970){21}{\rule{0.119pt}{0.867pt}}
\multiput(290.17,592.20)(12.000,-21.201){2}{\rule{0.400pt}{0.433pt}}
\multiput(303.58,567.65)(0.493,-0.893){23}{\rule{0.119pt}{0.808pt}}
\multiput(302.17,569.32)(13.000,-21.324){2}{\rule{0.400pt}{0.404pt}}
\multiput(316.58,544.77)(0.493,-0.853){23}{\rule{0.119pt}{0.777pt}}
\multiput(315.17,546.39)(13.000,-20.387){2}{\rule{0.400pt}{0.388pt}}
\multiput(329.58,522.54)(0.492,-0.927){21}{\rule{0.119pt}{0.833pt}}
\multiput(328.17,524.27)(12.000,-20.270){2}{\rule{0.400pt}{0.417pt}}
\multiput(341.58,500.90)(0.493,-0.814){23}{\rule{0.119pt}{0.746pt}}
\multiput(340.17,502.45)(13.000,-19.451){2}{\rule{0.400pt}{0.373pt}}
\multiput(354.58,479.90)(0.493,-0.814){23}{\rule{0.119pt}{0.746pt}}
\multiput(353.17,481.45)(13.000,-19.451){2}{\rule{0.400pt}{0.373pt}}
\multiput(367.58,459.03)(0.493,-0.774){23}{\rule{0.119pt}{0.715pt}}
\multiput(366.17,460.52)(13.000,-18.515){2}{\rule{0.400pt}{0.358pt}}
\multiput(380.58,438.96)(0.492,-0.798){21}{\rule{0.119pt}{0.733pt}}
\multiput(379.17,440.48)(12.000,-17.478){2}{\rule{0.400pt}{0.367pt}}
\multiput(392.58,420.16)(0.493,-0.734){23}{\rule{0.119pt}{0.685pt}}
\multiput(391.17,421.58)(13.000,-17.579){2}{\rule{0.400pt}{0.342pt}}
\multiput(405.58,401.29)(0.493,-0.695){23}{\rule{0.119pt}{0.654pt}}
\multiput(404.17,402.64)(13.000,-16.643){2}{\rule{0.400pt}{0.327pt}}
\multiput(418.58,383.29)(0.493,-0.695){23}{\rule{0.119pt}{0.654pt}}
\multiput(417.17,384.64)(13.000,-16.643){2}{\rule{0.400pt}{0.327pt}}
\multiput(431.58,365.23)(0.492,-0.712){21}{\rule{0.119pt}{0.667pt}}
\multiput(430.17,366.62)(12.000,-15.616){2}{\rule{0.400pt}{0.333pt}}
\multiput(443.58,348.54)(0.493,-0.616){23}{\rule{0.119pt}{0.592pt}}
\multiput(442.17,349.77)(13.000,-14.771){2}{\rule{0.400pt}{0.296pt}}
\multiput(456.58,332.54)(0.493,-0.616){23}{\rule{0.119pt}{0.592pt}}
\multiput(455.17,333.77)(13.000,-14.771){2}{\rule{0.400pt}{0.296pt}}
\multiput(469.58,316.51)(0.492,-0.625){21}{\rule{0.119pt}{0.600pt}}
\multiput(468.17,317.75)(12.000,-13.755){2}{\rule{0.400pt}{0.300pt}}
\multiput(481.58,301.67)(0.493,-0.576){23}{\rule{0.119pt}{0.562pt}}
\multiput(480.17,302.83)(13.000,-13.834){2}{\rule{0.400pt}{0.281pt}}
\multiput(494.58,286.80)(0.493,-0.536){23}{\rule{0.119pt}{0.531pt}}
\multiput(493.17,287.90)(13.000,-12.898){2}{\rule{0.400pt}{0.265pt}}
\multiput(507.58,272.80)(0.493,-0.536){23}{\rule{0.119pt}{0.531pt}}
\multiput(506.17,273.90)(13.000,-12.898){2}{\rule{0.400pt}{0.265pt}}
\multiput(520.00,259.92)(0.496,-0.492){21}{\rule{0.500pt}{0.119pt}}
\multiput(520.00,260.17)(10.962,-12.000){2}{\rule{0.250pt}{0.400pt}}
\multiput(532.00,247.92)(0.497,-0.493){23}{\rule{0.500pt}{0.119pt}}
\multiput(532.00,248.17)(11.962,-13.000){2}{\rule{0.250pt}{0.400pt}}
\multiput(545.00,234.92)(0.590,-0.492){19}{\rule{0.573pt}{0.118pt}}
\multiput(545.00,235.17)(11.811,-11.000){2}{\rule{0.286pt}{0.400pt}}
\multiput(558.00,223.92)(0.539,-0.492){21}{\rule{0.533pt}{0.119pt}}
\multiput(558.00,224.17)(11.893,-12.000){2}{\rule{0.267pt}{0.400pt}}
\multiput(571.00,211.92)(0.600,-0.491){17}{\rule{0.580pt}{0.118pt}}
\multiput(571.00,212.17)(10.796,-10.000){2}{\rule{0.290pt}{0.400pt}}
\multiput(583.00,201.92)(0.652,-0.491){17}{\rule{0.620pt}{0.118pt}}
\multiput(583.00,202.17)(11.713,-10.000){2}{\rule{0.310pt}{0.400pt}}
\multiput(596.00,191.92)(0.652,-0.491){17}{\rule{0.620pt}{0.118pt}}
\multiput(596.00,192.17)(11.713,-10.000){2}{\rule{0.310pt}{0.400pt}}
\multiput(609.00,181.93)(0.758,-0.488){13}{\rule{0.700pt}{0.117pt}}
\multiput(609.00,182.17)(10.547,-8.000){2}{\rule{0.350pt}{0.400pt}}
\multiput(621.00,173.93)(0.728,-0.489){15}{\rule{0.678pt}{0.118pt}}
\multiput(621.00,174.17)(11.593,-9.000){2}{\rule{0.339pt}{0.400pt}}
\multiput(634.00,164.93)(0.950,-0.485){11}{\rule{0.843pt}{0.117pt}}
\multiput(634.00,165.17)(11.251,-7.000){2}{\rule{0.421pt}{0.400pt}}
\multiput(647.00,157.93)(0.950,-0.485){11}{\rule{0.843pt}{0.117pt}}
\multiput(647.00,158.17)(11.251,-7.000){2}{\rule{0.421pt}{0.400pt}}
\multiput(660.00,150.93)(0.874,-0.485){11}{\rule{0.786pt}{0.117pt}}
\multiput(660.00,151.17)(10.369,-7.000){2}{\rule{0.393pt}{0.400pt}}
\multiput(672.00,143.93)(1.123,-0.482){9}{\rule{0.967pt}{0.116pt}}
\multiput(672.00,144.17)(10.994,-6.000){2}{\rule{0.483pt}{0.400pt}}
\multiput(685.00,137.93)(1.378,-0.477){7}{\rule{1.140pt}{0.115pt}}
\multiput(685.00,138.17)(10.634,-5.000){2}{\rule{0.570pt}{0.400pt}}
\multiput(698.00,132.93)(1.378,-0.477){7}{\rule{1.140pt}{0.115pt}}
\multiput(698.00,133.17)(10.634,-5.000){2}{\rule{0.570pt}{0.400pt}}
\multiput(711.00,127.94)(1.651,-0.468){5}{\rule{1.300pt}{0.113pt}}
\multiput(711.00,128.17)(9.302,-4.000){2}{\rule{0.650pt}{0.400pt}}
\multiput(723.00,123.95)(2.695,-0.447){3}{\rule{1.833pt}{0.108pt}}
\multiput(723.00,124.17)(9.195,-3.000){2}{\rule{0.917pt}{0.400pt}}
\multiput(736.00,120.95)(2.695,-0.447){3}{\rule{1.833pt}{0.108pt}}
\multiput(736.00,121.17)(9.195,-3.000){2}{\rule{0.917pt}{0.400pt}}
\put(749,117.17){\rule{2.500pt}{0.400pt}}
\multiput(749.00,118.17)(6.811,-2.000){2}{\rule{1.250pt}{0.400pt}}
\put(761,115.17){\rule{2.700pt}{0.400pt}}
\multiput(761.00,116.17)(7.396,-2.000){2}{\rule{1.350pt}{0.400pt}}
\put(774,113.67){\rule{3.132pt}{0.400pt}}
\multiput(774.00,114.17)(6.500,-1.000){2}{\rule{1.566pt}{0.400pt}}
\put(787,112.67){\rule{3.132pt}{0.400pt}}
\multiput(787.00,113.17)(6.500,-1.000){2}{\rule{1.566pt}{0.400pt}}
\put(812,112.67){\rule{3.132pt}{0.400pt}}
\multiput(812.00,112.17)(6.500,1.000){2}{\rule{1.566pt}{0.400pt}}
\put(825,113.67){\rule{3.132pt}{0.400pt}}
\multiput(825.00,113.17)(6.500,1.000){2}{\rule{1.566pt}{0.400pt}}
\put(838,115.17){\rule{2.700pt}{0.400pt}}
\multiput(838.00,114.17)(7.396,2.000){2}{\rule{1.350pt}{0.400pt}}
\put(851,117.17){\rule{2.500pt}{0.400pt}}
\multiput(851.00,116.17)(6.811,2.000){2}{\rule{1.250pt}{0.400pt}}
\multiput(863.00,119.61)(2.695,0.447){3}{\rule{1.833pt}{0.108pt}}
\multiput(863.00,118.17)(9.195,3.000){2}{\rule{0.917pt}{0.400pt}}
\multiput(876.00,122.61)(2.695,0.447){3}{\rule{1.833pt}{0.108pt}}
\multiput(876.00,121.17)(9.195,3.000){2}{\rule{0.917pt}{0.400pt}}
\multiput(889.00,125.60)(1.651,0.468){5}{\rule{1.300pt}{0.113pt}}
\multiput(889.00,124.17)(9.302,4.000){2}{\rule{0.650pt}{0.400pt}}
\multiput(901.00,129.59)(1.378,0.477){7}{\rule{1.140pt}{0.115pt}}
\multiput(901.00,128.17)(10.634,5.000){2}{\rule{0.570pt}{0.400pt}}
\multiput(914.00,134.59)(1.378,0.477){7}{\rule{1.140pt}{0.115pt}}
\multiput(914.00,133.17)(10.634,5.000){2}{\rule{0.570pt}{0.400pt}}
\multiput(927.00,139.59)(1.123,0.482){9}{\rule{0.967pt}{0.116pt}}
\multiput(927.00,138.17)(10.994,6.000){2}{\rule{0.483pt}{0.400pt}}
\multiput(940.00,145.59)(0.874,0.485){11}{\rule{0.786pt}{0.117pt}}
\multiput(940.00,144.17)(10.369,7.000){2}{\rule{0.393pt}{0.400pt}}
\multiput(952.00,152.59)(0.950,0.485){11}{\rule{0.843pt}{0.117pt}}
\multiput(952.00,151.17)(11.251,7.000){2}{\rule{0.421pt}{0.400pt}}
\multiput(965.00,159.59)(0.950,0.485){11}{\rule{0.843pt}{0.117pt}}
\multiput(965.00,158.17)(11.251,7.000){2}{\rule{0.421pt}{0.400pt}}
\multiput(978.00,166.59)(0.728,0.489){15}{\rule{0.678pt}{0.118pt}}
\multiput(978.00,165.17)(11.593,9.000){2}{\rule{0.339pt}{0.400pt}}
\multiput(991.00,175.59)(0.758,0.488){13}{\rule{0.700pt}{0.117pt}}
\multiput(991.00,174.17)(10.547,8.000){2}{\rule{0.350pt}{0.400pt}}
\multiput(1003.00,183.58)(0.652,0.491){17}{\rule{0.620pt}{0.118pt}}
\multiput(1003.00,182.17)(11.713,10.000){2}{\rule{0.310pt}{0.400pt}}
\multiput(1016.00,193.58)(0.652,0.491){17}{\rule{0.620pt}{0.118pt}}
\multiput(1016.00,192.17)(11.713,10.000){2}{\rule{0.310pt}{0.400pt}}
\multiput(1029.00,203.58)(0.600,0.491){17}{\rule{0.580pt}{0.118pt}}
\multiput(1029.00,202.17)(10.796,10.000){2}{\rule{0.290pt}{0.400pt}}
\multiput(1041.00,213.58)(0.539,0.492){21}{\rule{0.533pt}{0.119pt}}
\multiput(1041.00,212.17)(11.893,12.000){2}{\rule{0.267pt}{0.400pt}}
\multiput(1054.00,225.58)(0.590,0.492){19}{\rule{0.573pt}{0.118pt}}
\multiput(1054.00,224.17)(11.811,11.000){2}{\rule{0.286pt}{0.400pt}}
\multiput(1067.00,236.58)(0.497,0.493){23}{\rule{0.500pt}{0.119pt}}
\multiput(1067.00,235.17)(11.962,13.000){2}{\rule{0.250pt}{0.400pt}}
\multiput(1080.00,249.58)(0.496,0.492){21}{\rule{0.500pt}{0.119pt}}
\multiput(1080.00,248.17)(10.962,12.000){2}{\rule{0.250pt}{0.400pt}}
\multiput(1092.58,261.00)(0.493,0.536){23}{\rule{0.119pt}{0.531pt}}
\multiput(1091.17,261.00)(13.000,12.898){2}{\rule{0.400pt}{0.265pt}}
\multiput(1105.58,275.00)(0.493,0.536){23}{\rule{0.119pt}{0.531pt}}
\multiput(1104.17,275.00)(13.000,12.898){2}{\rule{0.400pt}{0.265pt}}
\multiput(1118.58,289.00)(0.493,0.576){23}{\rule{0.119pt}{0.562pt}}
\multiput(1117.17,289.00)(13.000,13.834){2}{\rule{0.400pt}{0.281pt}}
\multiput(1131.58,304.00)(0.492,0.625){21}{\rule{0.119pt}{0.600pt}}
\multiput(1130.17,304.00)(12.000,13.755){2}{\rule{0.400pt}{0.300pt}}
\multiput(1143.58,319.00)(0.493,0.616){23}{\rule{0.119pt}{0.592pt}}
\multiput(1142.17,319.00)(13.000,14.771){2}{\rule{0.400pt}{0.296pt}}
\multiput(1156.58,335.00)(0.493,0.616){23}{\rule{0.119pt}{0.592pt}}
\multiput(1155.17,335.00)(13.000,14.771){2}{\rule{0.400pt}{0.296pt}}
\multiput(1169.58,351.00)(0.492,0.712){21}{\rule{0.119pt}{0.667pt}}
\multiput(1168.17,351.00)(12.000,15.616){2}{\rule{0.400pt}{0.333pt}}
\multiput(1181.58,368.00)(0.493,0.695){23}{\rule{0.119pt}{0.654pt}}
\multiput(1180.17,368.00)(13.000,16.643){2}{\rule{0.400pt}{0.327pt}}
\multiput(1194.58,386.00)(0.493,0.695){23}{\rule{0.119pt}{0.654pt}}
\multiput(1193.17,386.00)(13.000,16.643){2}{\rule{0.400pt}{0.327pt}}
\multiput(1207.58,404.00)(0.493,0.734){23}{\rule{0.119pt}{0.685pt}}
\multiput(1206.17,404.00)(13.000,17.579){2}{\rule{0.400pt}{0.342pt}}
\multiput(1220.58,423.00)(0.492,0.798){21}{\rule{0.119pt}{0.733pt}}
\multiput(1219.17,423.00)(12.000,17.478){2}{\rule{0.400pt}{0.367pt}}
\multiput(1232.58,442.00)(0.493,0.774){23}{\rule{0.119pt}{0.715pt}}
\multiput(1231.17,442.00)(13.000,18.515){2}{\rule{0.400pt}{0.358pt}}
\multiput(1245.58,462.00)(0.493,0.814){23}{\rule{0.119pt}{0.746pt}}
\multiput(1244.17,462.00)(13.000,19.451){2}{\rule{0.400pt}{0.373pt}}
\multiput(1258.58,483.00)(0.493,0.814){23}{\rule{0.119pt}{0.746pt}}
\multiput(1257.17,483.00)(13.000,19.451){2}{\rule{0.400pt}{0.373pt}}
\multiput(1271.58,504.00)(0.492,0.927){21}{\rule{0.119pt}{0.833pt}}
\multiput(1270.17,504.00)(12.000,20.270){2}{\rule{0.400pt}{0.417pt}}
\multiput(1283.58,526.00)(0.493,0.853){23}{\rule{0.119pt}{0.777pt}}
\multiput(1282.17,526.00)(13.000,20.387){2}{\rule{0.400pt}{0.388pt}}
\multiput(1296.58,548.00)(0.493,0.893){23}{\rule{0.119pt}{0.808pt}}
\multiput(1295.17,548.00)(13.000,21.324){2}{\rule{0.400pt}{0.404pt}}
\multiput(1309.58,571.00)(0.492,0.970){21}{\rule{0.119pt}{0.867pt}}
\multiput(1308.17,571.00)(12.000,21.201){2}{\rule{0.400pt}{0.433pt}}
\multiput(1321.58,594.00)(0.493,0.933){23}{\rule{0.119pt}{0.838pt}}
\multiput(1320.17,594.00)(13.000,22.260){2}{\rule{0.400pt}{0.419pt}}
\multiput(1334.58,618.00)(0.493,0.972){23}{\rule{0.119pt}{0.869pt}}
\multiput(1333.17,618.00)(13.000,23.196){2}{\rule{0.400pt}{0.435pt}}
\multiput(1347.58,643.00)(0.493,0.972){23}{\rule{0.119pt}{0.869pt}}
\multiput(1346.17,643.00)(13.000,23.196){2}{\rule{0.400pt}{0.435pt}}
\multiput(1360.58,668.00)(0.492,1.099){21}{\rule{0.119pt}{0.967pt}}
\multiput(1359.17,668.00)(12.000,23.994){2}{\rule{0.400pt}{0.483pt}}
\multiput(1372.58,694.00)(0.493,1.012){23}{\rule{0.119pt}{0.900pt}}
\multiput(1371.17,694.00)(13.000,24.132){2}{\rule{0.400pt}{0.450pt}}
\multiput(1385.58,720.00)(0.493,1.052){23}{\rule{0.119pt}{0.931pt}}
\multiput(1384.17,720.00)(13.000,25.068){2}{\rule{0.400pt}{0.465pt}}
\multiput(1398.58,747.00)(0.493,1.091){23}{\rule{0.119pt}{0.962pt}}
\multiput(1397.17,747.00)(13.000,26.004){2}{\rule{0.400pt}{0.481pt}}
\multiput(1411.58,775.00)(0.492,1.186){21}{\rule{0.119pt}{1.033pt}}
\multiput(1410.17,775.00)(12.000,25.855){2}{\rule{0.400pt}{0.517pt}}
\multiput(1423.58,803.00)(0.493,1.131){23}{\rule{0.119pt}{0.992pt}}
\multiput(1422.17,803.00)(13.000,26.940){2}{\rule{0.400pt}{0.496pt}}
\put(800.0,113.0){\rule[-0.200pt]{2.891pt}{0.400pt}}
\put(176,401){\rule{1pt}{1pt}}
\put(189,401){\rule{1pt}{1pt}}
\put(201,401){\rule{1pt}{1pt}}
\put(214,400){\rule{1pt}{1pt}}
\put(227,398){\rule{1pt}{1pt}}
\put(240,396){\rule{1pt}{1pt}}
\put(252,393){\rule{1pt}{1pt}}
\put(265,389){\rule{1pt}{1pt}}
\put(278,385){\rule{1pt}{1pt}}
\put(291,381){\rule{1pt}{1pt}}
\put(303,376){\rule{1pt}{1pt}}
\put(316,370){\rule{1pt}{1pt}}
\put(329,364){\rule{1pt}{1pt}}
\put(341,357){\rule{1pt}{1pt}}
\put(354,350){\rule{1pt}{1pt}}
\put(367,343){\rule{1pt}{1pt}}
\put(380,335){\rule{1pt}{1pt}}
\put(392,327){\rule{1pt}{1pt}}
\put(405,319){\rule{1pt}{1pt}}
\put(418,311){\rule{1pt}{1pt}}
\put(431,302){\rule{1pt}{1pt}}
\put(443,293){\rule{1pt}{1pt}}
\put(456,284){\rule{1pt}{1pt}}
\put(469,275){\rule{1pt}{1pt}}
\put(481,266){\rule{1pt}{1pt}}
\put(494,256){\rule{1pt}{1pt}}
\put(507,247){\rule{1pt}{1pt}}
\put(520,238){\rule{1pt}{1pt}}
\put(532,229){\rule{1pt}{1pt}}
\put(545,220){\rule{1pt}{1pt}}
\put(558,211){\rule{1pt}{1pt}}
\put(571,202){\rule{1pt}{1pt}}
\put(583,194){\rule{1pt}{1pt}}
\put(596,186){\rule{1pt}{1pt}}
\put(609,178){\rule{1pt}{1pt}}
\put(621,170){\rule{1pt}{1pt}}
\put(634,163){\rule{1pt}{1pt}}
\put(647,156){\rule{1pt}{1pt}}
\put(660,150){\rule{1pt}{1pt}}
\put(672,144){\rule{1pt}{1pt}}
\put(685,139){\rule{1pt}{1pt}}
\put(698,134){\rule{1pt}{1pt}}
\put(711,129){\rule{1pt}{1pt}}
\put(723,125){\rule{1pt}{1pt}}
\put(736,122){\rule{1pt}{1pt}}
\put(749,119){\rule{1pt}{1pt}}
\put(761,117){\rule{1pt}{1pt}}
\put(774,115){\rule{1pt}{1pt}}
\put(787,114){\rule{1pt}{1pt}}
\put(800,113){\rule{1pt}{1pt}}
\put(812,113){\rule{1pt}{1pt}}
\put(825,114){\rule{1pt}{1pt}}
\put(838,115){\rule{1pt}{1pt}}
\put(851,117){\rule{1pt}{1pt}}
\put(863,119){\rule{1pt}{1pt}}
\put(876,122){\rule{1pt}{1pt}}
\put(889,125){\rule{1pt}{1pt}}
\put(901,129){\rule{1pt}{1pt}}
\put(914,134){\rule{1pt}{1pt}}
\put(927,139){\rule{1pt}{1pt}}
\put(940,144){\rule{1pt}{1pt}}
\put(952,150){\rule{1pt}{1pt}}
\put(965,156){\rule{1pt}{1pt}}
\put(978,163){\rule{1pt}{1pt}}
\put(991,170){\rule{1pt}{1pt}}
\put(1003,178){\rule{1pt}{1pt}}
\put(1016,186){\rule{1pt}{1pt}}
\put(1029,194){\rule{1pt}{1pt}}
\put(1041,202){\rule{1pt}{1pt}}
\put(1054,211){\rule{1pt}{1pt}}
\put(1067,220){\rule{1pt}{1pt}}
\put(1080,229){\rule{1pt}{1pt}}
\put(1092,238){\rule{1pt}{1pt}}
\put(1105,247){\rule{1pt}{1pt}}
\put(1118,256){\rule{1pt}{1pt}}
\put(1131,266){\rule{1pt}{1pt}}
\put(1143,275){\rule{1pt}{1pt}}
\put(1156,284){\rule{1pt}{1pt}}
\put(1169,293){\rule{1pt}{1pt}}
\put(1181,302){\rule{1pt}{1pt}}
\put(1194,311){\rule{1pt}{1pt}}
\put(1207,319){\rule{1pt}{1pt}}
\put(1220,327){\rule{1pt}{1pt}}
\put(1232,335){\rule{1pt}{1pt}}
\put(1245,343){\rule{1pt}{1pt}}
\put(1258,350){\rule{1pt}{1pt}}
\put(1271,357){\rule{1pt}{1pt}}
\put(1283,364){\rule{1pt}{1pt}}
\put(1296,370){\rule{1pt}{1pt}}
\put(1309,376){\rule{1pt}{1pt}}
\put(1321,381){\rule{1pt}{1pt}}
\put(1334,385){\rule{1pt}{1pt}}
\put(1347,389){\rule{1pt}{1pt}}
\put(1360,393){\rule{1pt}{1pt}}
\put(1372,396){\rule{1pt}{1pt}}
\put(1385,398){\rule{1pt}{1pt}}
\put(1398,400){\rule{1pt}{1pt}}
\put(1411,401){\rule{1pt}{1pt}}
\put(1423,401){\rule{1pt}{1pt}}
\put(1436,401){\rule{1pt}{1pt}}
\end{picture}
\end{center}
\caption{The dispersion relation \protect $S(k)$ for a free scalar field. 
The solid line, \protect $k^2$, is for the continuum theory, the dotted line
for the latticized system.}
\label{fig4}
\end{figure}
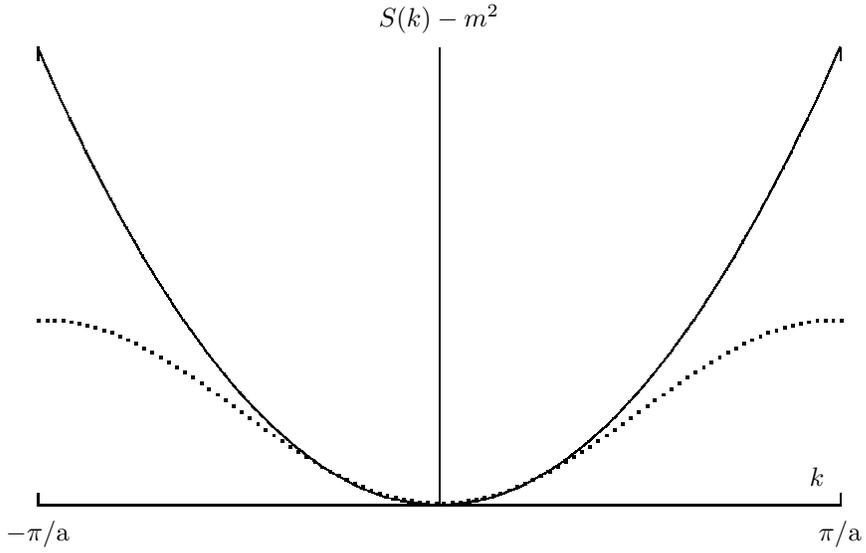

Let us study how a scalar field $\phi(x)$ can be defined on a lattice.
Consider the continuum field theory in Euclidean space 
\begin{equation}
S(\phi)=\int d^dx \left[\frac{1}{2}(\partial_\mu\phi)^2+V(\phi)\right]\ .
\end{equation}
where
\begin{equation}
V(\phi)=\frac{1}{2} m^2 \phi^2+\frac{\lambda}{4}\phi^4\ .
\end{equation}
A matter field is attributed to the lattice sites and it is natural to
approximate a continuous field by its values at the lattice sites
\begin{equation}
\phi(x)\longrightarrow \phi_x
\end{equation}
Clearly, in order for the lattice field $\phi_x$ to be a good
approximation of a continuous field configuration $\phi(x)$, 
the lattice spacing should be much smaller than the characteristic 
size of this configuration.
The derivative can be replaced by
\begin{equation}
\partial_\mu\phi\longrightarrow\frac{1}{a}\left(\phi_{x+\hat\mu}-\phi_x\right)
\end{equation}
where $\hat\mu$ is a $d$-vector of length $a$ in the direction of $\mu$.
The $d$ dimensional integration is replaced by a sum
\begin{equation}
\int d^dx\longrightarrow a^d\sum_x
\end{equation}
so that the scalar action on the lattice becomes
\begin{equation}
S(\phi)=\sum_x\left[\frac{a^{d-2}}{2}\sum_{\mu=1}^d
(\phi_{x+\hat\mu}-\phi_x)^2+a^d\left(\frac{m^2}{2}\phi^2_x+
\frac{\lambda}{4}\phi^4_x\right)\right]\quad .
\label{phil}
\end{equation}
It is instructive to go to momentum space to see the spectrum of the
free field theory. For this I take the Fourier transform
\begin{equation}
\phi_x=\int\frac{d^d k}{(2\pi)^d} e^{i k\cdot x}\phi(k)\ .
\label{phik}
\end{equation}
Since it is meaningless to consider wavelengths less than twice the
lattice spacing, the above integral is taken over only one ``Brillouin zone''
of the reciprocal lattice
\begin{equation}
-\frac{\pi}{a}\le k_\mu\le\frac{\pi}{a}\quad \hbox{for each $\mu$}
\end{equation}
where $k_\mu=k\cdot \hat{\mu}$. Substituting Eq.(\ref{phik}) into Eq.(\ref{phil})
the free field action can be written as
\begin{equation}
S_0(\phi)=\frac{1}{2}\int\frac{d^d
k}{(2\pi)^d}\left[\sum_\mu\frac{4}{a^{d-2}}\sin^2\left(\frac{a
k_\mu}{2}\right)+ m^2\right]\phi(-k)\phi(k)\ \ .
\end{equation}
Each mode contributes to the action in the momentum space a quantity
\begin{equation}
S(k)=m^2+\sum_x\frac{4}{a^{d-2}} \sin^2\left(\frac{a
k_\mu}{2}\right)
\end{equation}
rather than the standard $m^2+k^2$. Nevertheless these two expressions
have the same continuum limit as they coincide at the minimum value of
$k=0$ (see fig. (\ref{fig4})).

The lattice action Eq.(\ref{phil}) is quantized by using the Feynman
path-integral formalism, in which the expectation value of a product
of fields is given by
\begin{equation}
<0|\phi_{x_1}\phi_{x_2}\dots\phi_{x_l}|0>=\frac{1}{Z}\int\prod_x[d\phi_x]
\phi_{x_1}\phi_{x_2}\dots\phi_{x_l}\, e^{-S(\phi)}
\label{phicor}
\end{equation}
where
\begin{equation}
Z=\int\prod_x[d\phi_x]\, e^{-S(\phi)}\quad .
\label{phiz}
\end{equation}
The meaning of the integrals should be clear as one recalls that the
usual functional integrals are actually defined on a discretized
space-time lattice and an appropriate continuum limit is taken at the
end. If one rescales the field as
\begin{equation}
\phi'_x=\sqrt{\lambda}\,\phi_x\quad ,
\end{equation}
the lattice action scales as
\begin{equation}
S(\phi)=\frac{1}{\lambda}S'(\phi')
\end{equation}
with
\begin{equation}
S'(\phi')=\sum_x\left[\frac{a^{d-2}}{2}\sum_{\mu=1}^d
(\phi'_{x+\hat\mu}-\phi'_x)^2+a^d\left(\frac{m^2}{2}{\phi'}^2_x+
\frac{1}{4}{\phi'}^4_x\right)\right]
\label{100}
\end{equation}
$i.e.$ the coupling constant $\lambda$ has become an overall factor in
the action. In this way Eqs.(\ref{phicor}) and (\ref{phiz}) may be written as
\begin{eqnarray}
<0|\phi'_{x_1}\phi'_{x_2}&\dots&\phi'_{x_l}|0>=\frac{1}{Z}\int\prod_x[d\phi'_x]
\phi'_{x_1}\phi'_{x_2}\dots\phi'_{x_l}\, e^{-\frac{1}{\lambda}
S'(\phi')}\nonumber\\
Z&=&\int\prod_x[d\phi'_x]\, e^{-\frac{1}{\lambda}S'(\phi')}\quad .
\label{phizp}
\end{eqnarray}
Note that Eq.(\ref{phizp}) has the same structure of the partition
function in statistical mechanics, once the identification
\begin{equation}
\frac{1}{\lambda}\longrightarrow\beta\equiv\frac{1}{KT} 
\end{equation}
is made. The strong coupling expansion (in powers of $\lambda^{-1}$)
 corresponds to the high temperature expansion in statistical
mechanics.

\subsection{Fermi fields on the lattice}

\begin{figure}[htb]
\begin{center}
\setlength{\unitlength}{0.00041700in}%
\begingroup\makeatletter\ifx\SetFigFont\undefined
\def\x#1#2#3#4#5#6#7\relax{\def\x{#1#2#3#4#5#6}}%
\expandafter\x\fmtname xxxxxx\relax \def\y{splain}%
\ifx\x\y   
\gdef\SetFigFont#1#2#3{%
  \ifnum #1<17\tiny\else \ifnum #1<20\small\else
  \ifnum #1<24\normalsize\else \ifnum #1<29\large\else
  \ifnum #1<34\Large\else \ifnum #1<41\LARGE\else
     \huge\fi\fi\fi\fi\fi\fi
  \csname #3\endcsname}%
\else
\gdef\SetFigFont#1#2#3{\begingroup
  \count@#1\relax \ifnum 25<\count@\count@25\fi
  \def\x{\endgroup\@setsize\SetFigFont{#2pt}}%
  \expandafter\x
    \csname \romannumeral\the\count@ pt\expandafter\endcsname
    \csname @\romannumeral\the\count@ pt\endcsname
  \csname #3\endcsname}%
\fi
\fi\endgroup
\begin{picture}(10824,7224)(589,-6973)
\thicklines
\put(601,-3361){\makebox(13.3333,20.0000){\SetFigFont{10}{12}{rm}.}}
\put(601,-3361){\vector( 1, 0){10800}}
\put(3601,-5761){\line( 1, 1){4800}}
\put(6001,-6961){\vector( 0, 1){7200}}
\put(3601,-961){\line( 1,-1){4800}}
\put(2251,-1261){\makebox(0,0)[lb]{\smash{\SetFigFont{10}{12.0}{rm}$\gamma_{5}=-1$}}}
\put(5626,-61){\makebox(0,0)[lb]{\smash{\SetFigFont{10}{12.0}{rm}E}}}
\put(11251,-3661){\makebox(0,0)[lb]{\smash{\SetFigFont{10}{12.0}{rm}k}}}
\put(8401,-1261){\makebox(0,0)[lb]{\smash{\SetFigFont{10}{12.0}{rm}$\gamma_{5}=+1$}}}
\end{picture}
\end{center}
\caption{Spectrum of the free, continuum Dirac fermion in \protect $1+1$ dimensions.}
\label{f1}
\end{figure}
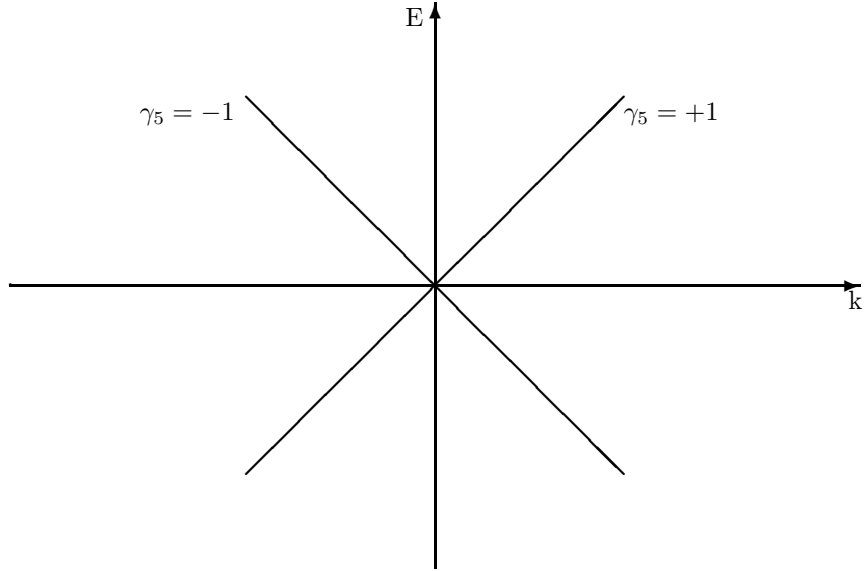

I shall now introduce the procedure to put Fermi fields on the lattice. A new phenomenon, the ``fermion doubling"~\cite{b7,b28}
 appears discretizing the Dirac equation, $i.e.$ additional fermionic species are generated. 
 Let us illustrate the fermion doubling problem starting from an example on a spatial lattice and continuum time in 
1+1 dimensions. The massless Dirac equation in 1+1 dimensions in the continuum reads
\begin{equation}
i\dot{\psi}=-i\alpha \partial_{x}\psi=-i\gamma_{5}\partial_{x}\psi
\label{dire}
\end{equation}
with  $\psi=\left(\begin{array}{c}
\psi_{1}\\
\psi_{2}
\end{array}\right)$
and $\alpha=\gamma_{5}=\sigma_{1}$.  
Choosing for the plane waves 
\begin{equation}
\psi_{\pm}(x,t)=e^{-i(kx-Et)} \chi_{\pm}
\end{equation}
with
\begin{equation}
\gamma_{5}\chi_{\pm}=\pm \chi_{\pm}
\end{equation}
the dispersion relation is 
\begin{equation}
E(k)=\pm k\quad,\quad -\infty<k<\infty
\end{equation}
and the excitations are left- and right-handed particles and antiparticles $-$ see fig.(\ref{f1}).

\begin{figure}[htb]
\begin{center}
\setlength{\unitlength}{0.240900pt}
\ifx\plotpoint\undefined\newsavebox{\plotpoint}\fi
\sbox{\plotpoint}{\rule[-0.200pt]{0.400pt}{0.400pt}}%
\begin{picture}(1500,900)(0,0)
\font\gnuplot=cmr10 at 10pt
\gnuplot
\sbox{\plotpoint}{\rule[-0.200pt]{0.400pt}{0.400pt}}%
\put(176.0,473.0){\rule[-0.200pt]{303.534pt}{0.400pt}}
\put(806.0,68.0){\rule[-0.200pt]{0.400pt}{194.888pt}}
\put(706,816){\makebox(0,0)[l]{E}}
\put(1436,574){\makebox(0,0)[l]{k}}
\put(491,735){\makebox(0,0)[l]{$\gamma_{5}=-1$}}
\put(1121,735){\makebox(0,0)[l]{$\gamma_{5}=+1$}}
\put(1436,371){\makebox(0,0)[l]{$\frac{\pi}{a}$}}
\put(176,371){\makebox(0,0)[l]{$\frac{\pi}{a}$}}
\put(176,472){\usebox{\plotpoint}}
\multiput(176.00,470.92)(0.539,-0.492){21}{\rule{0.533pt}{0.119pt}}
\multiput(176.00,471.17)(11.893,-12.000){2}{\rule{0.267pt}{0.400pt}}
\multiput(189.58,457.79)(0.492,-0.539){21}{\rule{0.119pt}{0.533pt}}
\multiput(188.17,458.89)(12.000,-11.893){2}{\rule{0.400pt}{0.267pt}}
\multiput(201.00,445.92)(0.497,-0.493){23}{\rule{0.500pt}{0.119pt}}
\multiput(201.00,446.17)(11.962,-13.000){2}{\rule{0.250pt}{0.400pt}}
\multiput(214.00,432.92)(0.539,-0.492){21}{\rule{0.533pt}{0.119pt}}
\multiput(214.00,433.17)(11.893,-12.000){2}{\rule{0.267pt}{0.400pt}}
\multiput(227.00,420.92)(0.497,-0.493){23}{\rule{0.500pt}{0.119pt}}
\multiput(227.00,421.17)(11.962,-13.000){2}{\rule{0.250pt}{0.400pt}}
\multiput(240.00,407.92)(0.496,-0.492){21}{\rule{0.500pt}{0.119pt}}
\multiput(240.00,408.17)(10.962,-12.000){2}{\rule{0.250pt}{0.400pt}}
\multiput(252.00,395.92)(0.590,-0.492){19}{\rule{0.573pt}{0.118pt}}
\multiput(252.00,396.17)(11.811,-11.000){2}{\rule{0.286pt}{0.400pt}}
\multiput(265.00,384.92)(0.539,-0.492){21}{\rule{0.533pt}{0.119pt}}
\multiput(265.00,385.17)(11.893,-12.000){2}{\rule{0.267pt}{0.400pt}}
\multiput(278.00,372.92)(0.590,-0.492){19}{\rule{0.573pt}{0.118pt}}
\multiput(278.00,373.17)(11.811,-11.000){2}{\rule{0.286pt}{0.400pt}}
\multiput(291.00,361.92)(0.600,-0.491){17}{\rule{0.580pt}{0.118pt}}
\multiput(291.00,362.17)(10.796,-10.000){2}{\rule{0.290pt}{0.400pt}}
\multiput(303.00,351.92)(0.590,-0.492){19}{\rule{0.573pt}{0.118pt}}
\multiput(303.00,352.17)(11.811,-11.000){2}{\rule{0.286pt}{0.400pt}}
\multiput(316.00,340.93)(0.728,-0.489){15}{\rule{0.678pt}{0.118pt}}
\multiput(316.00,341.17)(11.593,-9.000){2}{\rule{0.339pt}{0.400pt}}
\multiput(329.00,331.93)(0.669,-0.489){15}{\rule{0.633pt}{0.118pt}}
\multiput(329.00,332.17)(10.685,-9.000){2}{\rule{0.317pt}{0.400pt}}
\multiput(341.00,322.93)(0.824,-0.488){13}{\rule{0.750pt}{0.117pt}}
\multiput(341.00,323.17)(11.443,-8.000){2}{\rule{0.375pt}{0.400pt}}
\multiput(354.00,314.93)(0.824,-0.488){13}{\rule{0.750pt}{0.117pt}}
\multiput(354.00,315.17)(11.443,-8.000){2}{\rule{0.375pt}{0.400pt}}
\multiput(367.00,306.93)(0.950,-0.485){11}{\rule{0.843pt}{0.117pt}}
\multiput(367.00,307.17)(11.251,-7.000){2}{\rule{0.421pt}{0.400pt}}
\multiput(380.00,299.93)(0.874,-0.485){11}{\rule{0.786pt}{0.117pt}}
\multiput(380.00,300.17)(10.369,-7.000){2}{\rule{0.393pt}{0.400pt}}
\multiput(392.00,292.93)(1.378,-0.477){7}{\rule{1.140pt}{0.115pt}}
\multiput(392.00,293.17)(10.634,-5.000){2}{\rule{0.570pt}{0.400pt}}
\multiput(405.00,287.93)(1.378,-0.477){7}{\rule{1.140pt}{0.115pt}}
\multiput(405.00,288.17)(10.634,-5.000){2}{\rule{0.570pt}{0.400pt}}
\multiput(418.00,282.93)(1.378,-0.477){7}{\rule{1.140pt}{0.115pt}}
\multiput(418.00,283.17)(10.634,-5.000){2}{\rule{0.570pt}{0.400pt}}
\multiput(431.00,277.95)(2.472,-0.447){3}{\rule{1.700pt}{0.108pt}}
\multiput(431.00,278.17)(8.472,-3.000){2}{\rule{0.850pt}{0.400pt}}
\multiput(443.00,274.95)(2.695,-0.447){3}{\rule{1.833pt}{0.108pt}}
\multiput(443.00,275.17)(9.195,-3.000){2}{\rule{0.917pt}{0.400pt}}
\put(456,271.17){\rule{2.700pt}{0.400pt}}
\multiput(456.00,272.17)(7.396,-2.000){2}{\rule{1.350pt}{0.400pt}}
\put(469,269.67){\rule{2.891pt}{0.400pt}}
\multiput(469.00,270.17)(6.000,-1.000){2}{\rule{1.445pt}{0.400pt}}
\put(494,269.67){\rule{3.132pt}{0.400pt}}
\multiput(494.00,269.17)(6.500,1.000){2}{\rule{1.566pt}{0.400pt}}
\put(507,270.67){\rule{3.132pt}{0.400pt}}
\multiput(507.00,270.17)(6.500,1.000){2}{\rule{1.566pt}{0.400pt}}
\multiput(520.00,272.61)(2.472,0.447){3}{\rule{1.700pt}{0.108pt}}
\multiput(520.00,271.17)(8.472,3.000){2}{\rule{0.850pt}{0.400pt}}
\multiput(532.00,275.61)(2.695,0.447){3}{\rule{1.833pt}{0.108pt}}
\multiput(532.00,274.17)(9.195,3.000){2}{\rule{0.917pt}{0.400pt}}
\multiput(545.00,278.61)(2.695,0.447){3}{\rule{1.833pt}{0.108pt}}
\multiput(545.00,277.17)(9.195,3.000){2}{\rule{0.917pt}{0.400pt}}
\multiput(558.00,281.59)(1.378,0.477){7}{\rule{1.140pt}{0.115pt}}
\multiput(558.00,280.17)(10.634,5.000){2}{\rule{0.570pt}{0.400pt}}
\multiput(571.00,286.59)(1.267,0.477){7}{\rule{1.060pt}{0.115pt}}
\multiput(571.00,285.17)(9.800,5.000){2}{\rule{0.530pt}{0.400pt}}
\multiput(583.00,291.59)(1.123,0.482){9}{\rule{0.967pt}{0.116pt}}
\multiput(583.00,290.17)(10.994,6.000){2}{\rule{0.483pt}{0.400pt}}
\multiput(596.00,297.59)(0.950,0.485){11}{\rule{0.843pt}{0.117pt}}
\multiput(596.00,296.17)(11.251,7.000){2}{\rule{0.421pt}{0.400pt}}
\multiput(609.00,304.59)(0.758,0.488){13}{\rule{0.700pt}{0.117pt}}
\multiput(609.00,303.17)(10.547,8.000){2}{\rule{0.350pt}{0.400pt}}
\multiput(621.00,312.59)(0.824,0.488){13}{\rule{0.750pt}{0.117pt}}
\multiput(621.00,311.17)(11.443,8.000){2}{\rule{0.375pt}{0.400pt}}
\multiput(634.00,320.59)(0.824,0.488){13}{\rule{0.750pt}{0.117pt}}
\multiput(634.00,319.17)(11.443,8.000){2}{\rule{0.375pt}{0.400pt}}
\multiput(647.00,328.58)(0.652,0.491){17}{\rule{0.620pt}{0.118pt}}
\multiput(647.00,327.17)(11.713,10.000){2}{\rule{0.310pt}{0.400pt}}
\multiput(660.00,338.59)(0.669,0.489){15}{\rule{0.633pt}{0.118pt}}
\multiput(660.00,337.17)(10.685,9.000){2}{\rule{0.317pt}{0.400pt}}
\multiput(672.00,347.58)(0.590,0.492){19}{\rule{0.573pt}{0.118pt}}
\multiput(672.00,346.17)(11.811,11.000){2}{\rule{0.286pt}{0.400pt}}
\multiput(685.00,358.58)(0.590,0.492){19}{\rule{0.573pt}{0.118pt}}
\multiput(685.00,357.17)(11.811,11.000){2}{\rule{0.286pt}{0.400pt}}
\multiput(698.00,369.58)(0.590,0.492){19}{\rule{0.573pt}{0.118pt}}
\multiput(698.00,368.17)(11.811,11.000){2}{\rule{0.286pt}{0.400pt}}
\multiput(711.00,380.58)(0.543,0.492){19}{\rule{0.536pt}{0.118pt}}
\multiput(711.00,379.17)(10.887,11.000){2}{\rule{0.268pt}{0.400pt}}
\multiput(723.00,391.58)(0.539,0.492){21}{\rule{0.533pt}{0.119pt}}
\multiput(723.00,390.17)(11.893,12.000){2}{\rule{0.267pt}{0.400pt}}
\multiput(736.00,403.58)(0.497,0.493){23}{\rule{0.500pt}{0.119pt}}
\multiput(736.00,402.17)(11.962,13.000){2}{\rule{0.250pt}{0.400pt}}
\multiput(749.00,416.58)(0.496,0.492){21}{\rule{0.500pt}{0.119pt}}
\multiput(749.00,415.17)(10.962,12.000){2}{\rule{0.250pt}{0.400pt}}
\multiput(761.00,428.58)(0.497,0.493){23}{\rule{0.500pt}{0.119pt}}
\multiput(761.00,427.17)(11.962,13.000){2}{\rule{0.250pt}{0.400pt}}
\multiput(774.00,441.58)(0.539,0.492){21}{\rule{0.533pt}{0.119pt}}
\multiput(774.00,440.17)(11.893,12.000){2}{\rule{0.267pt}{0.400pt}}
\multiput(787.00,453.58)(0.497,0.493){23}{\rule{0.500pt}{0.119pt}}
\multiput(787.00,452.17)(11.962,13.000){2}{\rule{0.250pt}{0.400pt}}
\multiput(800.58,466.00)(0.492,0.539){21}{\rule{0.119pt}{0.533pt}}
\multiput(799.17,466.00)(12.000,11.893){2}{\rule{0.400pt}{0.267pt}}
\multiput(812.00,479.58)(0.497,0.493){23}{\rule{0.500pt}{0.119pt}}
\multiput(812.00,478.17)(11.962,13.000){2}{\rule{0.250pt}{0.400pt}}
\multiput(825.00,492.58)(0.539,0.492){21}{\rule{0.533pt}{0.119pt}}
\multiput(825.00,491.17)(11.893,12.000){2}{\rule{0.267pt}{0.400pt}}
\multiput(838.00,504.58)(0.497,0.493){23}{\rule{0.500pt}{0.119pt}}
\multiput(838.00,503.17)(11.962,13.000){2}{\rule{0.250pt}{0.400pt}}
\multiput(851.00,517.58)(0.496,0.492){21}{\rule{0.500pt}{0.119pt}}
\multiput(851.00,516.17)(10.962,12.000){2}{\rule{0.250pt}{0.400pt}}
\multiput(863.00,529.58)(0.497,0.493){23}{\rule{0.500pt}{0.119pt}}
\multiput(863.00,528.17)(11.962,13.000){2}{\rule{0.250pt}{0.400pt}}
\multiput(876.00,542.58)(0.539,0.492){21}{\rule{0.533pt}{0.119pt}}
\multiput(876.00,541.17)(11.893,12.000){2}{\rule{0.267pt}{0.400pt}}
\multiput(889.00,554.58)(0.543,0.492){19}{\rule{0.536pt}{0.118pt}}
\multiput(889.00,553.17)(10.887,11.000){2}{\rule{0.268pt}{0.400pt}}
\multiput(901.00,565.58)(0.590,0.492){19}{\rule{0.573pt}{0.118pt}}
\multiput(901.00,564.17)(11.811,11.000){2}{\rule{0.286pt}{0.400pt}}
\multiput(914.00,576.58)(0.590,0.492){19}{\rule{0.573pt}{0.118pt}}
\multiput(914.00,575.17)(11.811,11.000){2}{\rule{0.286pt}{0.400pt}}
\multiput(927.00,587.58)(0.590,0.492){19}{\rule{0.573pt}{0.118pt}}
\multiput(927.00,586.17)(11.811,11.000){2}{\rule{0.286pt}{0.400pt}}
\multiput(940.00,598.59)(0.669,0.489){15}{\rule{0.633pt}{0.118pt}}
\multiput(940.00,597.17)(10.685,9.000){2}{\rule{0.317pt}{0.400pt}}
\multiput(952.00,607.58)(0.652,0.491){17}{\rule{0.620pt}{0.118pt}}
\multiput(952.00,606.17)(11.713,10.000){2}{\rule{0.310pt}{0.400pt}}
\multiput(965.00,617.59)(0.824,0.488){13}{\rule{0.750pt}{0.117pt}}
\multiput(965.00,616.17)(11.443,8.000){2}{\rule{0.375pt}{0.400pt}}
\multiput(978.00,625.59)(0.824,0.488){13}{\rule{0.750pt}{0.117pt}}
\multiput(978.00,624.17)(11.443,8.000){2}{\rule{0.375pt}{0.400pt}}
\multiput(991.00,633.59)(0.758,0.488){13}{\rule{0.700pt}{0.117pt}}
\multiput(991.00,632.17)(10.547,8.000){2}{\rule{0.350pt}{0.400pt}}
\multiput(1003.00,641.59)(0.950,0.485){11}{\rule{0.843pt}{0.117pt}}
\multiput(1003.00,640.17)(11.251,7.000){2}{\rule{0.421pt}{0.400pt}}
\multiput(1016.00,648.59)(1.123,0.482){9}{\rule{0.967pt}{0.116pt}}
\multiput(1016.00,647.17)(10.994,6.000){2}{\rule{0.483pt}{0.400pt}}
\multiput(1029.00,654.59)(1.267,0.477){7}{\rule{1.060pt}{0.115pt}}
\multiput(1029.00,653.17)(9.800,5.000){2}{\rule{0.530pt}{0.400pt}}
\multiput(1041.00,659.59)(1.378,0.477){7}{\rule{1.140pt}{0.115pt}}
\multiput(1041.00,658.17)(10.634,5.000){2}{\rule{0.570pt}{0.400pt}}
\multiput(1054.00,664.61)(2.695,0.447){3}{\rule{1.833pt}{0.108pt}}
\multiput(1054.00,663.17)(9.195,3.000){2}{\rule{0.917pt}{0.400pt}}
\multiput(1067.00,667.61)(2.695,0.447){3}{\rule{1.833pt}{0.108pt}}
\multiput(1067.00,666.17)(9.195,3.000){2}{\rule{0.917pt}{0.400pt}}
\multiput(1080.00,670.61)(2.472,0.447){3}{\rule{1.700pt}{0.108pt}}
\multiput(1080.00,669.17)(8.472,3.000){2}{\rule{0.850pt}{0.400pt}}
\put(1092,672.67){\rule{3.132pt}{0.400pt}}
\multiput(1092.00,672.17)(6.500,1.000){2}{\rule{1.566pt}{0.400pt}}
\put(1105,673.67){\rule{3.132pt}{0.400pt}}
\multiput(1105.00,673.17)(6.500,1.000){2}{\rule{1.566pt}{0.400pt}}
\put(481.0,270.0){\rule[-0.200pt]{3.132pt}{0.400pt}}
\put(1131,673.67){\rule{2.891pt}{0.400pt}}
\multiput(1131.00,674.17)(6.000,-1.000){2}{\rule{1.445pt}{0.400pt}}
\put(1143,672.17){\rule{2.700pt}{0.400pt}}
\multiput(1143.00,673.17)(7.396,-2.000){2}{\rule{1.350pt}{0.400pt}}
\multiput(1156.00,670.95)(2.695,-0.447){3}{\rule{1.833pt}{0.108pt}}
\multiput(1156.00,671.17)(9.195,-3.000){2}{\rule{0.917pt}{0.400pt}}
\multiput(1169.00,667.95)(2.472,-0.447){3}{\rule{1.700pt}{0.108pt}}
\multiput(1169.00,668.17)(8.472,-3.000){2}{\rule{0.850pt}{0.400pt}}
\multiput(1181.00,664.93)(1.378,-0.477){7}{\rule{1.140pt}{0.115pt}}
\multiput(1181.00,665.17)(10.634,-5.000){2}{\rule{0.570pt}{0.400pt}}
\multiput(1194.00,659.93)(1.378,-0.477){7}{\rule{1.140pt}{0.115pt}}
\multiput(1194.00,660.17)(10.634,-5.000){2}{\rule{0.570pt}{0.400pt}}
\multiput(1207.00,654.93)(1.378,-0.477){7}{\rule{1.140pt}{0.115pt}}
\multiput(1207.00,655.17)(10.634,-5.000){2}{\rule{0.570pt}{0.400pt}}
\multiput(1220.00,649.93)(0.874,-0.485){11}{\rule{0.786pt}{0.117pt}}
\multiput(1220.00,650.17)(10.369,-7.000){2}{\rule{0.393pt}{0.400pt}}
\multiput(1232.00,642.93)(0.950,-0.485){11}{\rule{0.843pt}{0.117pt}}
\multiput(1232.00,643.17)(11.251,-7.000){2}{\rule{0.421pt}{0.400pt}}
\multiput(1245.00,635.93)(0.824,-0.488){13}{\rule{0.750pt}{0.117pt}}
\multiput(1245.00,636.17)(11.443,-8.000){2}{\rule{0.375pt}{0.400pt}}
\multiput(1258.00,627.93)(0.824,-0.488){13}{\rule{0.750pt}{0.117pt}}
\multiput(1258.00,628.17)(11.443,-8.000){2}{\rule{0.375pt}{0.400pt}}
\multiput(1271.00,619.93)(0.669,-0.489){15}{\rule{0.633pt}{0.118pt}}
\multiput(1271.00,620.17)(10.685,-9.000){2}{\rule{0.317pt}{0.400pt}}
\multiput(1283.00,610.93)(0.728,-0.489){15}{\rule{0.678pt}{0.118pt}}
\multiput(1283.00,611.17)(11.593,-9.000){2}{\rule{0.339pt}{0.400pt}}
\multiput(1296.00,601.92)(0.590,-0.492){19}{\rule{0.573pt}{0.118pt}}
\multiput(1296.00,602.17)(11.811,-11.000){2}{\rule{0.286pt}{0.400pt}}
\multiput(1309.00,590.92)(0.600,-0.491){17}{\rule{0.580pt}{0.118pt}}
\multiput(1309.00,591.17)(10.796,-10.000){2}{\rule{0.290pt}{0.400pt}}
\multiput(1321.00,580.92)(0.590,-0.492){19}{\rule{0.573pt}{0.118pt}}
\multiput(1321.00,581.17)(11.811,-11.000){2}{\rule{0.286pt}{0.400pt}}
\multiput(1334.00,569.92)(0.539,-0.492){21}{\rule{0.533pt}{0.119pt}}
\multiput(1334.00,570.17)(11.893,-12.000){2}{\rule{0.267pt}{0.400pt}}
\multiput(1347.00,557.92)(0.590,-0.492){19}{\rule{0.573pt}{0.118pt}}
\multiput(1347.00,558.17)(11.811,-11.000){2}{\rule{0.286pt}{0.400pt}}
\multiput(1360.00,546.92)(0.496,-0.492){21}{\rule{0.500pt}{0.119pt}}
\multiput(1360.00,547.17)(10.962,-12.000){2}{\rule{0.250pt}{0.400pt}}
\multiput(1372.00,534.92)(0.497,-0.493){23}{\rule{0.500pt}{0.119pt}}
\multiput(1372.00,535.17)(11.962,-13.000){2}{\rule{0.250pt}{0.400pt}}
\multiput(1385.00,521.92)(0.539,-0.492){21}{\rule{0.533pt}{0.119pt}}
\multiput(1385.00,522.17)(11.893,-12.000){2}{\rule{0.267pt}{0.400pt}}
\multiput(1398.00,509.92)(0.497,-0.493){23}{\rule{0.500pt}{0.119pt}}
\multiput(1398.00,510.17)(11.962,-13.000){2}{\rule{0.250pt}{0.400pt}}
\multiput(1411.58,495.79)(0.492,-0.539){21}{\rule{0.119pt}{0.533pt}}
\multiput(1410.17,496.89)(12.000,-11.893){2}{\rule{0.400pt}{0.267pt}}
\multiput(1423.00,483.92)(0.539,-0.492){21}{\rule{0.533pt}{0.119pt}}
\multiput(1423.00,484.17)(11.893,-12.000){2}{\rule{0.267pt}{0.400pt}}
\put(1118.0,675.0){\rule[-0.200pt]{3.132pt}{0.400pt}}
\put(176,473){\usebox{\plotpoint}}
\put(176.00,473.00){\usebox{\plotpoint}}
\put(191.08,487.25){\usebox{\plotpoint}}
\put(205.33,502.33){\usebox{\plotpoint}}
\put(220.24,516.76){\usebox{\plotpoint}}
\put(235.18,531.18){\usebox{\plotpoint}}
\put(249.85,545.85){\usebox{\plotpoint}}
\multiput(252,548)(15.844,13.407){0}{\usebox{\plotpoint}}
\put(265.51,559.47){\usebox{\plotpoint}}
\put(280.86,573.42){\usebox{\plotpoint}}
\put(296.74,586.79){\usebox{\plotpoint}}
\put(312.63,600.15){\usebox{\plotpoint}}
\multiput(316,603)(17.065,11.814){0}{\usebox{\plotpoint}}
\put(329.42,612.32){\usebox{\plotpoint}}
\put(346.35,624.29){\usebox{\plotpoint}}
\put(364.03,635.17){\usebox{\plotpoint}}
\multiput(367,637)(18.275,9.840){0}{\usebox{\plotpoint}}
\put(382.16,645.26){\usebox{\plotpoint}}
\put(400.74,654.36){\usebox{\plotpoint}}
\multiput(405,656)(19.372,7.451){0}{\usebox{\plotpoint}}
\put(420.11,661.81){\usebox{\plotpoint}}
\put(439.82,668.20){\usebox{\plotpoint}}
\multiput(443,669)(20.224,4.667){0}{\usebox{\plotpoint}}
\put(460.09,672.63){\usebox{\plotpoint}}
\put(480.70,674.97){\usebox{\plotpoint}}
\multiput(481,675)(20.756,0.000){0}{\usebox{\plotpoint}}
\put(501.43,674.43){\usebox{\plotpoint}}
\multiput(507,674)(20.694,-1.592){0}{\usebox{\plotpoint}}
\put(522.07,672.48){\usebox{\plotpoint}}
\put(542.25,667.64){\usebox{\plotpoint}}
\multiput(545,667)(20.224,-4.667){0}{\usebox{\plotpoint}}
\put(562.28,662.35){\usebox{\plotpoint}}
\put(581.54,654.61){\usebox{\plotpoint}}
\multiput(583,654)(18.845,-8.698){0}{\usebox{\plotpoint}}
\put(600.27,645.70){\usebox{\plotpoint}}
\put(618.02,634.99){\usebox{\plotpoint}}
\multiput(621,633)(17.677,-10.878){0}{\usebox{\plotpoint}}
\put(635.63,624.00){\usebox{\plotpoint}}
\put(652.87,612.49){\usebox{\plotpoint}}
\put(669.41,599.95){\usebox{\plotpoint}}
\multiput(672,598)(15.844,-13.407){0}{\usebox{\plotpoint}}
\put(685.37,586.69){\usebox{\plotpoint}}
\put(701.21,573.28){\usebox{\plotpoint}}
\put(716.85,559.64){\usebox{\plotpoint}}
\put(732.12,545.58){\usebox{\plotpoint}}
\put(746.94,531.06){\usebox{\plotpoint}}
\multiput(749,529)(14.676,-14.676){0}{\usebox{\plotpoint}}
\put(761.62,516.38){\usebox{\plotpoint}}
\put(776.39,501.80){\usebox{\plotpoint}}
\put(791.46,487.54){\usebox{\plotpoint}}
\put(805.89,472.62){\usebox{\plotpoint}}
\put(820.30,457.70){\usebox{\plotpoint}}
\put(835.37,443.43){\usebox{\plotpoint}}
\put(850.15,428.85){\usebox{\plotpoint}}
\multiput(851,428)(14.676,-14.676){0}{\usebox{\plotpoint}}
\put(864.82,414.18){\usebox{\plotpoint}}
\put(879.64,399.64){\usebox{\plotpoint}}
\put(894.91,385.58){\usebox{\plotpoint}}
\put(910.54,371.93){\usebox{\plotpoint}}
\put(926.38,358.52){\usebox{\plotpoint}}
\multiput(927,358)(15.844,-13.407){0}{\usebox{\plotpoint}}
\put(942.33,345.25){\usebox{\plotpoint}}
\put(958.87,332.71){\usebox{\plotpoint}}
\put(976.09,321.17){\usebox{\plotpoint}}
\multiput(978,320)(17.677,-10.878){0}{\usebox{\plotpoint}}
\put(993.70,310.20){\usebox{\plotpoint}}
\put(1011.44,299.46){\usebox{\plotpoint}}
\multiput(1016,297)(18.845,-8.698){0}{\usebox{\plotpoint}}
\put(1030.16,290.52){\usebox{\plotpoint}}
\put(1049.41,282.76){\usebox{\plotpoint}}
\multiput(1054,281)(20.224,-4.667){0}{\usebox{\plotpoint}}
\put(1069.43,277.44){\usebox{\plotpoint}}
\put(1089.62,272.60){\usebox{\plotpoint}}
\multiput(1092,272)(20.694,-1.592){0}{\usebox{\plotpoint}}
\put(1110.24,270.60){\usebox{\plotpoint}}
\put(1130.98,270.00){\usebox{\plotpoint}}
\multiput(1131,270)(20.684,1.724){0}{\usebox{\plotpoint}}
\put(1151.59,272.32){\usebox{\plotpoint}}
\multiput(1156,273)(20.224,4.667){0}{\usebox{\plotpoint}}
\put(1171.86,276.72){\usebox{\plotpoint}}
\put(1191.58,283.07){\usebox{\plotpoint}}
\multiput(1194,284)(19.372,7.451){0}{\usebox{\plotpoint}}
\put(1210.95,290.52){\usebox{\plotpoint}}
\put(1229.56,299.57){\usebox{\plotpoint}}
\multiput(1232,301)(18.275,9.840){0}{\usebox{\plotpoint}}
\put(1247.69,309.66){\usebox{\plotpoint}}
\put(1265.37,320.53){\usebox{\plotpoint}}
\put(1282.31,332.49){\usebox{\plotpoint}}
\multiput(1283,333)(17.065,11.814){0}{\usebox{\plotpoint}}
\put(1299.12,344.64){\usebox{\plotpoint}}
\put(1315.00,358.00){\usebox{\plotpoint}}
\put(1330.88,371.36){\usebox{\plotpoint}}
\put(1346.25,385.31){\usebox{\plotpoint}}
\multiput(1347,386)(15.844,13.407){0}{\usebox{\plotpoint}}
\put(1361.92,398.92){\usebox{\plotpoint}}
\put(1376.59,413.59){\usebox{\plotpoint}}
\put(1391.51,428.01){\usebox{\plotpoint}}
\put(1406.43,442.43){\usebox{\plotpoint}}
\put(1420.70,457.51){\usebox{\plotpoint}}
\put(1435.76,471.78){\usebox{\plotpoint}}
\put(1436,472){\usebox{\plotpoint}}
\end{picture}
\end{center}
\caption{Spectrum of the naively latticized Dirac fermion in \protect $1+1$ dimensions.}
\label{f2}
\end{figure}
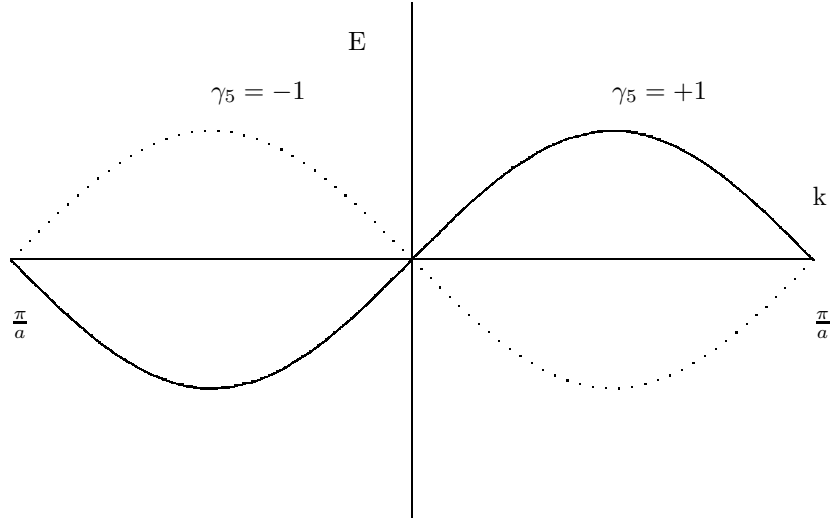

Let us consider the naively latticized Dirac equation. 
I place the spinor $\psi=\left(\begin{array}{c}
\psi_{1}\\
\psi_{2}
\end{array}\right)$ on each site of a spatial lattice and change the partial derivative $\partial_{x}$ with a finite difference
\begin{equation}
i\dot{\psi}(x)=-\frac{i}{2a}\gamma_{5}[\psi(x+1)-\psi(x-1)]\quad .
\end{equation}
The dispersion relation fig.(\ref{f2}) is now 
\begin{equation}
E(k)=\pm \frac{\sin (ka)}{a}\ .
\label{disp}
\end{equation}

This is a relativistic 
dispersion relation for $ka\ll 1$
\begin{equation}
E(k)\simeq \pm k+O(k^{3}a^{2})
\end{equation}
and for $ka=\pi -k'a$, with $k'a\ll 1$,
\begin{equation}
E(k)\simeq \mp k'+O(k'^{3}a^{2})\quad .
\end{equation}
There are two two-component Dirac particles in the continuum limit, 
and the total chiral charge of the fermions is zero. Consider, 
for example, the excitations associated with the field $\psi_{+}$
on the lattice. 
There are right-movers ($k\simeq 0$) and left-movers ($k\simeq \pi$), and since chirality (helicity for particles) is
 just velocity in $1+1$ dimensions this lattice field describes a pair of fermions with net chirality zero. 
 
The same phenomenon happens in 3+1 dimensions, even if it is more difficult to visualize it. The Dirac Hamiltonian 
for massive fermions now reads 
\begin{equation}
{\cal H}=-i\alpha_{l}\frac{\partial}{\partial x_{l}}+m \beta
\label{hd1}
\end{equation}
with
\begin{equation}
\alpha_{l}=\gamma_5 \gamma_l \quad,\quad \beta=\gamma ^{0}\quad .
\end{equation}
The field operator describing the multifermion system will be denoted by $\psi_{\alpha}(\vec{x})$ where $\alpha=0,1,2,3$ is the Dirac index 
and the three-component vector $\vec{x}$ denotes a space point. The Hamiltonian of the free fermion field is obtained from the single particle Hamiltonian 
Eq.(\ref{hd1}) as
\begin{equation}
H=\int d^{3}x \psi_{\alpha}^{\dagger}(\vec{x}){\cal H}_{\alpha \beta}\psi_{\beta}(\vec{x})
\label{hd2}
\end{equation}
or rewriting Eq.(\ref{hd2}) with a symmetrized derivative, one gets
\begin{equation}
H=\int d^3 x \psi_{\gamma}^{\dagger}(x)\left\{ m\beta+\frac{i}{2} (\lparder{}{x_{l}}{}-\rparder{}{x_{l}}{})
\alpha_{l} \right\}_{\gamma \delta}\psi_{\delta}(x)\quad .
\label{hd3}
\end{equation}
Let us now introduce a regular cubic lattice in space with lattice spacing $a$ and elementary cubes of volume $a^3$. On a finite lattice with total 
volume $V=(aL)^3$ the lattice points have integer indices $\vec{x}=\left\{x_{l}\quad ;\quad l=1,2,3\right\}$ satisfying $0\leq x_{l}\leq L-1$. The 
continuous field will be replaced by a discrete variable associated with elementary cubes, and the volume integral becomes a finite sum
\begin{equation}
\psi_{\alpha}(x)\rightarrow \psi_{\alpha \vec{x}}\quad,\quad \int d^3 x\rightarrow \sum_{x} a^3\quad .
\end{equation}
The discrete field variables are assumed to satisfy the anticommutation relations
\begin{equation}
\left\{\psi_{\alpha \vec{x}},\psi_{\beta \vec{y}}\right\}=\left\{\psi_{\alpha \vec{x}}^{\dagger},\psi_{\beta \vec{y}}^{\dagger}\right\}=0\quad,\quad
\left\{\psi_{\alpha \vec{x}}^{\dagger},\psi_{\beta \vec{y}}\right\}=\frac{1}{a^3}\delta_{\alpha \beta}\delta_{\vec{x} \vec{y}}\quad 
\end{equation}
and the continuous derivatives are discretized as
\begin{equation}
\frac{\partial}{\partial x_{l}}\psi_{\alpha}(\vec{x})\rightarrow \frac{1}{a}(\psi_{\alpha \vec{x}+\vec{l}}-\psi_{\alpha \vec{x}})
\end{equation}
where $\vec{l}$ denotes a unit vector on the lattice in direction $\vec{l}$. Applying these  rules in the Hamiltonian Eq.(\ref{hd3}) one obtains the 
naive discretized form
\begin{equation}
H=\sum_{\vec{x}} a^3 \left\{ \psi_{\gamma \vec{x}}^{\dagger} m\beta _{\gamma \delta}\psi_{\delta \vec{x}}+\frac{i}{2a}\sum_{l=1}^{3}
[\psi_{ \gamma \vec{x}+\vec{l}}^{\dagger}(\alpha_{l})_{\gamma \delta}\psi_{\delta \vec{x}}- 
\psi_{\gamma \vec{x}}^{\dagger}(\alpha_{l})_{\gamma \delta}\psi_{\delta \vec{x}+\vec{l}}] \right\}\quad .
\label{hd4}
\end{equation}
The free field Hamiltonian describes a collection of harmonic fermionic oscillators in momentum space. Assuming periodic 
boundary conditions in all three orthogonal directions of the $L^{3}$ cube, the allowed momentum components are discrete points in the 
Brillouin zone, namely 
\begin{equation}
\vec{p}\equiv \left\{p_{l}=\frac{2\pi }{L}n_{l}\quad;\quad l=1,2,3\right\}\quad,\quad 0\leq n_{l}\leq L-1\quad .
\end{equation}
Due to periodicity, instead of $0\leq p_{l} < 2\pi$, the Brillouin zone can also be equivalently defined in the interval $-\pi <p_{l}\leq \pi$.
Fourier transformed field components read
\begin{eqnarray}
\psi_{\alpha \vec{p}}&=&\sum_{\vec{x}}a^3 e^{-ip_{l}x_{l}}\psi_{\alpha \vec{x}}\\
\psi_{\alpha\vec{x}}&=&\frac{1}{(aL)^3}\sum_{\vec{p}}e^{ip_{l}x_{l}}\psi_{\alpha \vec{p}}\quad .
\end{eqnarray}
In momentum space the Hamiltonian (\ref{hd4}) reads
\begin{equation}
H=\frac{1}{(aL)^{3}}\sum_{\vec{p}}\psi_{\gamma \vec{p}}^{\dagger}\left\{m\beta +
\frac{1}{a}\alpha_{l} \sin p_{l} \right\}_{\gamma \delta}\psi_{\delta \vec{p}}\quad .
\label{hd5}
\end{equation}
The matrix in curly brackets in Eq.(\ref{hd5}) has eigenvalues
\begin{equation}
E_{\vec{p}}=\pm \sqrt{m^2 +\frac{1}{a^2 }\sum_{l=1}^{3}\sin ^{2}p_{l}}\quad .
\label{lspe}
\end{equation}
Eq.(\ref{lspe}) gives the possible energies of the free naive lattice fermions. In the continuum limit $a\rightarrow 0$, Eq.(\ref{lspe}) 
approaches the usual dispersion relation for free fermions,  since the coordinate vector is $a\vec{x}$ and correspondingly, 
the momentum vector is $\vec{k}=\frac{1}{a}\vec{p}$. For fixed momentum $\vec{k}$ and $a\rightarrow 0$ one has 
\begin{equation}
E_{\vec{p}}^{2}=m^2+\vec{k}^{2}+O(a^2)\quad .
\end{equation}
This is as expected, but still the naive lattice Hamiltonian Eq.(\ref{hd4}) is not entirely satisfactory. Due to the fact that 
$\sin (p+\pi)=-\sin p $ one has
\begin{equation}
E_{\vec{p}}^{2}=E_{\vec{p}+\vec{\pi}}^{2}
\label{eppi}
\end{equation}
where $\vec{\pi }$ is one of the eight vectors
\begin{eqnarray}
\vec{\pi}= \{ \quad (0,0,0),(\pi,0,0),(0,\pi,0),(0,0,\pi)&,&\nonumber\\
(\pi,\pi,0),(\pi,0,\pi),(0,\pi,\pi),(\pi,\pi,\pi)& & \}\quad .
\end{eqnarray}
These are the eight corners of the part of the Brillouin zone satisfying  $0\leq p_{l}\leq \pi$ $(l=1,2,3)$. The consequence of 
Eq.(\ref{eppi}) is that in this naive lattice formulation there are eight fermion states per field component. This phenomena is called fermion 
doubling although a  more appropriate name would be fermion octupling. The basic reason for fermion doubling is that the Dirac equation is of first 
order. 
In the case of free fermions one could, perhaps, tolerate this unwanted proliferation of degrees of freedom, but in an interacting theory the extra fermions 
influence the physical content in a non-trivial way, because the additional states can be pair-produced through the interactions of the 
fermion field. For example, even if the external particles in a process are the states at the zero $(0,0,0)$ corner of the Brillouin zone, the states 
of the other zones appear in virtual loops. 

In order to cure the disease, Wilson introduced an additional second order term in the lattice 
Hamiltonian   
\begin{eqnarray}
H&=&\sum_{\vec{x}}a^3 \{\psi_{\gamma \vec{x}}^{\dagger}(m+\frac{3r}{a})\beta_{\gamma \delta}\psi_{\delta \vec{x}} \nonumber\\
&+&\frac{i}{2a}\sum_{l=1}^{3}[\psi_{\gamma \vec{x}+\vec{l}}^{\dagger}(\alpha_{l})_{\gamma \delta}\psi_{\delta \vec{x}}-
\psi_{\gamma \vec{x}}^{\dagger}(\alpha_{l})_{\gamma \delta}\psi_{\delta \vec{x}+\vec{l}}]\nonumber\\
&-&\frac{r}{2a}\sum_{l=1}^{3}[\psi_{\gamma \vec{x}+\vec{l}}^{\dagger}(\beta_{l})_{\gamma \delta}\psi_{\delta \vec{x}}+
\psi_{\gamma \vec{x}}^{\dagger}(\beta_{l})_{\gamma \delta}\psi_{\delta \vec{x}+\vec{l}}] \}\quad .
\label{wh}
\end{eqnarray}
The Wilson parameter $r$ is assumed to be in the interval $0<r\leq 1$. In momentum space, instead of Eq.(\ref{hd5}) one has now  
\begin{eqnarray}
H&=&\frac{1}{(aL)^3}\sum_{\vec{p}}\psi_{\gamma \vec{p}}^{\dagger}\{ m \beta +\frac{r}{a}\beta \sum_{l=1}^{3}(1-\cos p_{l}) \nonumber\\
&+&\frac{1}{a}\alpha_{l}\sin p_{l}\}_{\gamma \delta}\psi_{\delta \vec{p}}
\end{eqnarray}
and the correspondingly energy eigenvalues are 
\begin{equation}
E_{\vec{p}}=\pm \sqrt{[m+\frac{r}{a}\sum_{l=1}^{3}(1-\cos p_{l})]^{2}+\sum_{l=1}^{3}\frac{1}{a^{2}}\sin ^{2}p_{l}}\quad .
\end{equation}
In the continuum limit $a\rightarrow 0$ the mass m is replaced by $m+\frac{2r}{a}n_{\pi}$, where $n_{\pi}$ is the number of momentum components 
equal to $\pi$. Therefore, the states with $n_{\pi} \neq 0$ become infinitely heavy. The only physical fermion state with finite energy is at the 
zero corner of the Brillouin zone. In section (\ref{lgt3}) I shall provide a detailed description of a different method to cure the doubling 
disease, $i.e.$ the staggered fermions which are the closest possible analog to the lattice fermions in condensed matter physics.

\subsection{Gauge fields on the lattice}

Let us now discuss how to put gauge fields on the lattice and briefly review how one arrives at the gauge invariant action in the continuum 
with the simple example of QED. The starting point is the action of the free Dirac field
\begin{equation}
S_{F}=\int d^4 x \overline{\psi}(x)(i \gamma^{\mu}\partial_{\mu}-m)\psi(x)\quad .
\label{sf}
\end{equation}
The action is invariant under the transformation
\begin{eqnarray}
\psi(x)&\longrightarrow& e^{i\Lambda} \psi(x) \\
\overline{\psi}(x)&\longrightarrow& \overline{\psi}(x) e^{-i\Lambda} 
\end{eqnarray}
with $\Lambda$ independent of x ($i.e.$ a global transformation). 

The next step is to require to the action to be also invariant under local $U(1)$ transformations with $\psi$ ($\overline{\psi}$) transforming independently 
at different space-time points. This is accomplished by introducing a four-vector potential $A_{\mu}(x)$ and replacing the ordinary four 
derivative $\partial_{\mu}$ by the covariant derivative ${\cal D}_{\mu}$ defined by
\begin{equation}
{\cal D}_{\mu}=\partial_{\mu}+ieA_{\mu}\quad .
\label{covd}
\end{equation}
The resulting new action 
\begin{equation}
S_{F}=\int d^4 x \overline{\psi}(x)(i\gamma_{\mu}{\cal D}^{\mu}-m)\psi(x)
\label{sfc}
\end{equation}
is then invariant under the local transformations 
\begin{eqnarray} 
\psi(x)&\longrightarrow& e^{i\Lambda(x)}\psi(x) \\
\overline{\psi}(x)&\longrightarrow& \overline{\psi}(x) e^{-i\Lambda(x)}\\
A_{\mu}(x)&\longrightarrow&e^{i\Lambda(x)}(A_{\mu}(x)-\frac{i}{e}\partial_{\mu})e^{-i\Lambda(x)}\quad .
\label{tga}
\end{eqnarray}
The crucial property which ensures the gauge invariance of Eq.(\ref{sfc}) is that, while $A_{\mu}$ transforms inhomogeneously, the transformation 
law for the covariant derivative Eq.(\ref{covd}) is homogeneous
\begin{equation}
{\cal D}_{\mu}\longrightarrow e^{i\Lambda(x)}{\cal D}_{\mu}e^{-i\Lambda(x)}\quad .
\end{equation}
By introducing a four-vector field $A_{\mu}$ gauge invariance of the action (\ref{sfc}) is ensured; one now must introduce a kinetic term which allows 
$A_{\mu}$ to propagate
\begin{equation}
S_{G}=-\frac{1}{4}\int d^{4}x F_{\mu \nu}F^{\mu \nu}
\label{ga}
\end{equation}
$F_{\mu \nu}$ is the gauge invariant field strenght tensor
\begin{equation}
F_{\mu \nu}=\partial_{\mu}A_{\nu}-\partial_{\nu}A_{\mu}\quad .
\end{equation}
The gauge invariant action describing the dynamics of the fermionic and gauge field is 
\begin{equation}
S_{QED}=S_{F}+S_{G}\quad .
\end{equation}

Let us start our approach to lattice QED by considering the lattice action of a free Dirac field. To parallel as closely as possible the steps in the 
continuum formulation, I shall consider Wilson fermions where every lattice site is occupied by all Dirac components $\psi_{\alpha}$. 
The results presented in the following are valid also for staggered fermions. 
The Hamiltonian Eq.(\ref{wh}) is invariant under the global transformations
\begin{eqnarray}
\psi_{\alpha x}&\longrightarrow& e^{i\Lambda}\psi_{\alpha x}\\
\overline{\psi}_{\alpha x}&\longrightarrow& \overline{\psi}_{\alpha x}e^{-i\Lambda}
\end{eqnarray}    
where $e^{i\Lambda}$ is an element of the $U(1)$ group. The next step is in requiring the theory to be invariant under local $U(1)$ 
transformations, with the group element $e^{i\Lambda_{x}}$ depending on the lattice site. Because of the non diagonal structure of the 
second and third term in the Hamiltonian Eq.(\ref{wh}) (whose origin is the derivative in the continuum formulation) one is forced to 
introduce new degrees of freedom. Since the group elements $e^{i\Lambda_{x}}$ do not act on the Dirac indices, it is sufficient 
for the following argument to focus our attention on a typical bilinear term
\begin{equation}
\overline{\psi}_{\vec{x}} \psi_{\vec{x}+\vec{l}}\quad .
\end{equation}
In the continuum it is well known how such bilinear terms should be modified in order to arrive at a gauge invariant expression. 
Since the bilinear $\overline{\psi}(\vec{x})\psi(\vec{y})$  transforms under gauge transformations
\begin{equation}
\overline{\psi}(\vec{x})\psi(\vec{y}) \longrightarrow \overline{\psi}(\vec{x})  e^{-i\Lambda(\vec{x})} e^{i\Lambda(\vec{y})}\psi(\vec{y})
\label{bilin}
\end{equation}
one must include a factor depending on the gauge potential which compensates the Eq.(\ref{bilin}) gauge variation. 
This factor, called Schwinger line integral is given by
\begin{equation}
U(\vec{x},\vec{y})=e^{ie\int _{\vec{x}}^{\vec{y}}dz_{\mu}A^{\mu}(\vec{z})}
\label{sli}
\end{equation}
and the line integral is carried out along a path connecting the point $\vec{x}$ with the point $\vec{y}$. $U(\vec{x},\vec{y})$ is an element of 
the $U(1)$ group. Under a gauge transformation Eq.(\ref{tga}), Eq.(\ref{sli}) transforms as
\begin{equation}
U(\vec{x},\vec{y})\longrightarrow e^{i\Lambda(\vec{x})}U(\vec{x},\vec{y})e^{-i\Lambda(\vec{y})}\quad .
\end{equation}
From these considerations, one concludes that the following bilinear expression in the fermion fields $\psi$ and $\overline{\psi}$ is gauge 
invariant
\begin{equation}
\overline{\psi}(\vec{x}) U(\vec{x},\vec{y}) \psi(\vec{y})=\overline{\psi}(\vec{x})e^{ie\int _{\vec{x}}^{\vec{y}}dz_{\mu}A^{\mu}(\vec{z})} 
\psi(y)\quad .
\label{pap}
\end{equation}
Let us suppose now that $\vec{y}=\vec{x}+\vec{\epsilon}$. 
From Eq.(\ref{pap}) one concludes that the bilinears should be defined as
\begin{eqnarray}
\overline{\psi}(\vec{x})\psi(\vec{x}+\vec{\epsilon})&\longrightarrow& \overline{\psi}(\vec{x})U(\vec{x},\vec{x}+\vec{\epsilon})
\psi(\vec{x}+\vec{\epsilon})\\
\overline{\psi}(\vec{x}+\vec{\epsilon})\psi(\vec{x})&\longrightarrow& \overline{\psi}(\vec{x}+\vec{\epsilon})U^{\dagger}(\vec{x},\vec{x}+\vec{\epsilon})
\psi(\vec{x})
\end{eqnarray}
with
\begin{equation}
U(\vec{x},\vec{x}+\vec{\epsilon})=e^{ie ~\vec{\epsilon}\cdot \vec{A}(\vec{x})}
\end{equation}
and $\vec{\epsilon} \cdot \vec{A}=\sum_{\mu}\epsilon_{\mu}A^{\mu}$. 

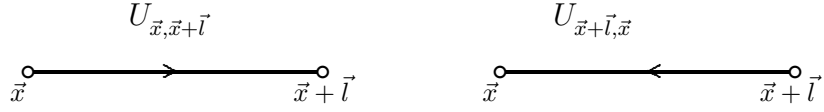
\begin{figure}[htb]
\begin{center}
\setlength{\unitlength}{0.00041700in}%
\begingroup\makeatletter\ifx\SetFigFont\undefined
\def\x#1#2#3#4#5#6#7\relax{\def\x{#1#2#3#4#5#6}}%
\expandafter\x\fmtname xxxxxx\relax \def\y{splain}%
\ifx\x\y   
\gdef\SetFigFont#1#2#3{%
  \ifnum #1<17\tiny\else \ifnum #1<20\small\else
  \ifnum #1<24\normalsize\else \ifnum #1<29\large\else
  \ifnum #1<34\Large\else \ifnum #1<41\LARGE\else
     \huge\fi\fi\fi\fi\fi\fi
  \csname #3\endcsname}%
\else
\gdef\SetFigFont#1#2#3{\begingroup
  \count@#1\relax \ifnum 25<\count@\count@25\fi
  \def\x{\endgroup\@setsize\SetFigFont{#2pt}}%
  \expandafter\x
    \csname \romannumeral\the\count@ pt\expandafter\endcsname
    \csname @\romannumeral\the\count@ pt\endcsname
  \csname #3\endcsname}%
\fi
\fi\endgroup
\begin{picture}(10058,1317)(901,-4390)
\thicklines
\put(4876,-3961){\circle{150}}
\put(7126,-3961){\circle{150}}
\put(10876,-3961){\circle{150}}
\put(1201,-3961){\line( 1, 0){3600}}
\put(7201,-3961){\line( 1, 0){3600}}
\multiput(3001,-3961)(-15.00000,7.50000){11}{\makebox(13.3333,20.0000){\SetFigFont{7}{8.4}{rm}.}}
\multiput(3001,-3961)(-15.00000,-7.50000){11}{\makebox(13.3333,20.0000){\SetFigFont{7}{8.4}{rm}.}}
\put(1126,-3961){\circle{150}}
\multiput(9001,-3961)(15.00000,7.50000){11}{\makebox(13.3333,20.0000){\SetFigFont{7}{8.4}{rm}.}}
\put(10426,-4336){\makebox(0,0)[lb]{\smash{\SetFigFont{10}{12.0}{rm}$\vec{x}+\vec{l}$}}}
\multiput(9001,-3961)(15.00000,-7.50000){11}{\makebox(13.3333,20.0000){\SetFigFont{7}{8.4}{rm}.}}
\put(901,-4336){\makebox(0,0)[lb]{\smash{\SetFigFont{10}{12.0}{rm}$\vec{x}$}}}
\put(6901,-4336){\makebox(0,0)[lb]{\smash{\SetFigFont{10}{12.0}{rm}$\vec{x}$}}}
\put(2401,-3361){\makebox(0,0)[lb]{\smash{\SetFigFont{12}{14.4}{rm}$U_{\vec{x},\vec{x}+\vec{l}}$}}}
\put(7801,-3361){\makebox(0,0)[lb]{\smash{\SetFigFont{12}{14.4}{rm}$U_{\vec{x}+\vec{l},\vec{x}}$}}}
\put(4501,-4336){\makebox(0,0)[lb]{\smash{\SetFigFont{10}{12.0}{rm}$\vec{x}+\vec{l}$}}}
\end{picture}
\end{center}
\caption{Link variables.}
\label{lvd}
\end{figure}

To arrive at a gauge-invariant expression for the fermionic action on the lattice, one should 
make the following substitution in Eq.(\ref{wh})
\begin{eqnarray}
\overline{\psi}_{\vec{x}+\vec{l}}(\gamma_{0}\gamma_{l}+ir)\psi_{\vec{x}}&\longrightarrow&
\overline{\psi}_{\vec{x}+\vec{l}}(\gamma_{0}\gamma_{l}+ir)U_{\vec{x}+\vec{l},\vec{x}}\psi_{\vec{x}} 
\label{bl1}\\
\overline{\psi}_{\vec{x}}(\gamma_{0}\gamma_{l}-ir)\psi_{\vec{x}+\vec{l}}&\longrightarrow&
\overline{\psi}_{\vec{x}}(\gamma_{0}\gamma_{l}-ir)U_{\vec{x},\vec{x}+\vec{l}}\psi_{\vec{x}+\vec{l}}
\label{bl2} 
\end{eqnarray}
where 
\begin{equation}
U_{\vec{x}+\vec{l},\vec{x}}=U^{\dagger}_{\vec{x},\vec{x}+\vec{l}}
\end{equation}
and $U_{\vec{x},\vec{x}+\vec{l}}$ is an element of the $U(1)$ gauge group and can therefore be written
\begin{equation}
U_{\vec{x},\vec{x}+\vec{l}}=e^{i\Phi_{\vec{x}}^{l}}
\end{equation}
where $\Phi_{\vec{x}}^{l}$ is restricted to the compact domain $[0,2\pi]$. The right hand side of Eqs.(\ref{bl1},\ref{bl2}) 
are now invariant under the following set of local transformations
\begin{eqnarray}
\psi_{\vec{x}}&\longrightarrow& e^{i\Lambda_{\vec{x}}}\psi_{\vec{x}}\\
\overline{\psi}_{\vec{x}}&\longrightarrow& \overline{\psi}_{\vec{x}}e^{-i\Lambda_{\vec{x}}} \\
U_{\vec{x},\vec{x}+\vec{l}}&\longrightarrow& e^{i\Lambda_{\vec{x}}}U_{\vec{x},\vec{x}+\vec{l}}e^{-i\Lambda_{\vec{x}+\vec{l}}}
\label{ugi1}\\ 
U_{\vec{x}+\vec{l},\vec{x}}&\longrightarrow& e^{i\Lambda_{\vec{x}+\vec{l}}}U_{\vec{x}+\vec{l},\vec{x}}e^{-i\Lambda_{\vec{x}}}\quad .
\label{ugi2} 
\end{eqnarray}
At variance with the matter fields discussed before, the group element $U_{\vec{x},\vec{x}+\vec{l}}$ lives on the links connecting two neighbouring 
lattice sites; hence I shall refer to them as link variables. The link variables have a direction and one can introduce a graphical representation 
- fig.(\ref{lvd}). 

The Hamiltonian Eq.(\ref{wh}) with the Wilson fermions can be written in the following gauge invariant form
\begin{eqnarray}
H &=& \sum_{\vec{x}} a^3 \{ \quad (m+\frac{3r}{a})\overline{\psi}_{\alpha \vec{x}}\psi_{\alpha \vec{x}} \nonumber\\
&+& \frac{i}{2a}\sum_{l=1}^{3}[ \quad \overline{\psi}_{\alpha \vec{x}+\vec{l}}(\gamma_{0}\gamma_{l}+ir)_{\alpha \beta} U_{\vec{x}+\vec{l},\vec{x}}
\psi_{\beta \vec{x}}\nonumber\\
&-& \overline{\psi}_{\alpha \vec{x} }(\gamma_{0}\gamma_{l}-ir)_{\alpha \beta} U_{\vec{x},\vec{x}+\vec{l}}\psi_{\beta \vec{x}+\vec{l}} ] \}\quad .
\label{whgi}
\end{eqnarray}

By requiring that $U_{\vec{x},\vec{x}+\vec{l}}$ and $U_{\vec{x}+\vec{l},\vec{x}}$ transform according to Eqs.(\ref{ugi1},\ref{ugi2}),
 one naturally implements $U(1)$-gauge invariance in the Hamiltonian and, a posteriori, one can state that the link variables are elements of the 
 $U(1)$ gauge group due to the requirement that their gauge transforms must also be elements of $U(1)$. 
  
Taking the continuum limit $a\rightarrow 0$ of Eq.(\ref{whgi}) one should get the usual continuum Hamiltonian Eq.(\ref{hd3}). This 
 requirement is fulfilled by establishing a relation between the link variables and the vector potential $A_{\mu}(\vec{x})$. 
The vector potential $A_{\mu}(\vec{x})$ at the lattice site $\vec{x}$ is real valued and carries a Lorentz index. 
The same happens for $\Phi_{\vec{x}}^{\mu}$ which parametrizes the link variable $U_{\vec{x},\vec{x}+\vec{\mu}}$. The difference is 
that $\Phi_{\vec{x}}^{\mu}$ takes only values in the interval $[0,2\pi]$, while the values taken by the vector potential $A_{\mu}(\vec{x})$ in the 
continuum theory extends over the entire real line. 

Since $A_{\mu}$ carries the dimension of an inverse length, while $\Phi_{\mu}$ is dimensionless one may try the ansatz
\begin{equation}
\Phi_{\mu}(\vec{x})=caA_{\mu}(\vec{x})
\end{equation}
where $a$ is the lattice spacing and c is a constant to be determined. In the continuum limit $a\rightarrow 0$ the range of $A_{\mu}$ will be 
infinite. 

By scaling m,$\psi$ and $\overline{\psi}$ as $m\rightarrow am$, $\psi\rightarrow a^{\frac{3}{2}}\psi$ and 
$\overline{\psi}\rightarrow a^{\frac{3}{2}}\overline{\psi}$ and replacing $U_{\vec{x},\vec{x}+\vec{\mu}}$ for small values of $a$ with
\begin{equation}
U_{\vec{x},\vec{x}+\vec{\mu}}\simeq 1+icaA_{\mu}(\vec{x})\quad .
\end{equation}
Eq.(\ref{whgi}) reduces to the continuum Eq.(\ref{sfc}) if one sets $c=e$. From here on I use the notation 
\begin{equation}
U_{\mu}(\vec{x})=U_{\vec{x},\vec{x}+\vec{\mu}}=e^{ieaA_{\mu}(\vec{x})}\quad .
\end{equation}
With this identification it is now an easy matter to verify that for $a\rightarrow 0$ $U_{\mu}(\vec{x})$ transforms as follows 
under gauge transformations
\begin{equation}
U_{\mu}(\vec{x})\rightarrow e^{i\Lambda_{\vec{x}}}U_{\mu}(\vec{x})e^{-i\Lambda_{\vec{x}+\vec{\mu}}}=e^{ieaA_{\mu}^{G}(\vec{x})}
\label{lk}
\end{equation}
where $A_{\mu}^{G}(\vec{x})$ comes from the discretized version of Eq.(\ref{tga}).

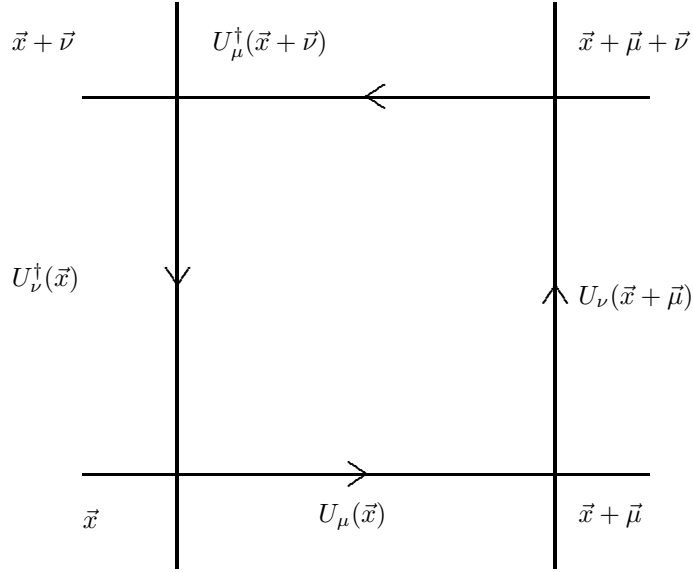
\begin{figure}[htb]
\begin{center}
\setlength{\unitlength}{0.00041700in}%
\begingroup\makeatletter\ifx\SetFigFont\undefined
\def\x#1#2#3#4#5#6#7\relax{\def\x{#1#2#3#4#5#6}}%
\expandafter\x\fmtname xxxxxx\relax \def\y{splain}%
\ifx\x\y   
\gdef\SetFigFont#1#2#3{%
  \ifnum #1<17\tiny\else \ifnum #1<20\small\else
  \ifnum #1<24\normalsize\else \ifnum #1<29\large\else
  \ifnum #1<34\Large\else \ifnum #1<41\LARGE\else
     \huge\fi\fi\fi\fi\fi\fi
  \csname #3\endcsname}%
\else
\gdef\SetFigFont#1#2#3{\begingroup
  \count@#1\relax \ifnum 25<\count@\count@25\fi
  \def\x{\endgroup\@setsize\SetFigFont{#2pt}}%
  \expandafter\x
    \csname \romannumeral\the\count@ pt\expandafter\endcsname
    \csname @\romannumeral\the\count@ pt\endcsname
  \csname #3\endcsname}%
\fi
\fi\endgroup
\begin{picture}(8112,7224)(1501,-7573)
\thicklines
\put(3601,-361){\line( 0,-1){7200}}
\put(8401,-361){\line( 0,-1){7200}}
\put(2401,-6361){\line( 1, 0){7200}}
\multiput(6001,-6361)(-14.06250,9.37500){17}{\makebox(13.3333,20.0000){\SetFigFont{7}{8.4}{rm}.}}
\multiput(6001,-6361)(-14.06250,-9.37500){17}{\makebox(13.3333,20.0000){\SetFigFont{7}{8.4}{rm}.}}
\multiput(8401,-3961)(-9.37500,-14.06250){17}{\makebox(13.3333,20.0000){\SetFigFont{7}{8.4}{rm}.}}
\multiput(8401,-3961)(9.37500,-14.06250){17}{\makebox(13.3333,20.0000){\SetFigFont{7}{8.4}{rm}.}}
\multiput(6001,-1561)(14.06250,9.37500){17}{\makebox(13.3333,20.0000){\SetFigFont{7}{8.4}{rm}.}}
\multiput(6001,-1561)(14.06250,-9.37500){17}{\makebox(13.3333,20.0000){\SetFigFont{7}{8.4}{rm}.}}
\put(2401,-1561){\line( 1, 0){7200}}
\multiput(3601,-3961)(-9.37500,14.06250){17}{\makebox(13.3333,20.0000){\SetFigFont{7}{8.4}{rm}.}}
\put(1501,-3961){\makebox(0,0)[lb]{\smash{\SetFigFont{10}{12.0}{rm}$U_{\nu}^{\dag}(\vec{x})$}}}
\multiput(3601,-3961)(9.37500,14.06250){17}{\makebox(13.3333,20.0000){\SetFigFont{7}{8.4}{rm}.}}
\put(2401,-7036){\makebox(0,0)[lb]{\smash{\SetFigFont{10}{12.0}{rm}$\vec{x}$}}}
\put(8701,-6961){\makebox(0,0)[lb]{\smash{\SetFigFont{10}{12.0}{rm}$\vec{x}+\vec{\mu}$}}}
\put(8701,-961){\makebox(0,0)[lb]{\smash{\SetFigFont{10}{12.0}{rm}$\vec{x}+\vec{\mu}+\vec{\nu}$}}}
\put(5401,-6961){\makebox(0,0)[lb]{\smash{\SetFigFont{10}{12.0}{rm}$U_{\mu}(\vec{x})$}}}
\put(8701,-4186){\makebox(0,0)[lb]{\smash{\SetFigFont{10}{12.0}{rm}$U_{\nu}(\vec{x}+\vec{\mu})$}}}
\put(1501,-961){\makebox(0,0)[lb]{\smash{\SetFigFont{10}{12.0}{rm}$\vec{x}+\vec{\nu}$ }}}
\put(4051,-961){\makebox(0,0)[lb]{\smash{\SetFigFont{10}{12.0}{rm}$U_{\mu}^{\dag}(\vec{x}+\vec{\nu})$}}}
\end{picture}
\end{center}
\caption{A plaquette.}
\label{plaquette}
\end{figure}

To complete the construction of lattice QED one has to construct the lattice version of the kinetic term Eq.(\ref{ga}) for gauge fields which should be 
strictly gauge invariant and a functional of link variables only. Such gauge-invariant functionals are easily constructed by taking the product of 
link variables around closed loops on the euclidean space-time lattice. Moreover, because of the local structure of the integrand in 
Eq.(\ref{ga}) one should focus his attention on the smallest possible loops on the euclidean space time lattice. Hence one is led to consider the 
product of link variables around an elementary plaquette as shown in fig.(\ref{plaquette}).
One then defines
\begin{equation}
U_{\mu \nu}(\vec{x})=U_{\mu}(\vec{x})U_{\nu}(\vec{x}+\vec{\mu})U_{\mu}^{\dagger}(\vec{x}+\vec{\nu})U_{\nu}^{\dagger}(\vec{x})
\label{plaq}
\end{equation}
where I have path ordered the link variables. The path ordering is irrelevant in the Abelian case, but becomes important in $QCD$.
Inserting Eq.(\ref{lk}) into Eq.(\ref{plaq}) one gets
\begin{equation}
U_{\mu \nu}(\vec{x})=e^{iea^{2}F_{\mu \nu}(\vec{x})}
\label{efmn}
\end{equation}
where $F_{\mu \nu}(\vec{x})$ is a discretized version of the continuum field strength tensor
\begin{equation}
F_{\mu \nu}(\vec{x})=\frac{1}{a}\left[(A_{\nu}(\vec{x}+\vec{\mu})-A_{\nu}(\vec{x}))-(A_{\mu}(\vec{x}+\vec{\nu})-A_{\mu}(\vec{x}))\right]\quad .
\end{equation}
From Eq.(\ref{efmn}), in the limit of small lattice spacing $a$, one has
\begin{equation}
\frac{1}{e^2}\sum_{\vec{x}}\sum_{\mu<\nu}\left[1-\frac{1}{2}(U_{\mu\nu}(\vec{x})+U_{\mu\nu}^{\dagger}(\vec{x}) )\right]\simeq 
\frac{1}{4}\sum_{\vec{x},\mu,\nu}a^4 F_{\mu \nu}(\vec{x})F^{\mu \nu}(\vec{x})
\end{equation}
where the sum appearing on the left hand side extends over the contributions coming from all distinct plaquettes on the lattice. 
The lattice action~\cite{b2} for the gauge potential in a compact form reads
\begin{equation}
S_{G}(U)=\frac{1}{e^{2}}\sum_{P}\left\{ 1-\frac{1}{2} (U_{P}+U^{\dagger}_{P}) \right\}
\end{equation}
where $U_{P}$, the plaquette variable, stands for the product of link variables around the boundary of the plaquette $P$, taken in 
a counterclockwise direction.  

\section{Staggered fermions}
\label{lgt3}
In what follows I shall introduce the staggered fermion formalism~\cite{b9}
that I used to study the lattice Schwinger models.
In this formalism one eliminates the unwanted fermion modes by reducing the Brillouin zone, $i.e.$ by doubling the effective lattice spacing. 
In the case of a $d$-dimensional lattice, one subdivides it into elementary $d$-dimensional hypercubes of unit length. At each site within a given hypercube 
one places a different degree of freedom and repeats this structure periodically throughout the lattice. Since there are $2^{d}$ sites within a hypercube, 
but only $2^{\frac{d}{2}}$ components of a Dirac field, one needs $2^{\frac{d}{2}}$ different Dirac fields to reduce the Brillouin zone by a factor of $\frac{1}{2}$. 

In the continuum, the massless Dirac spinors are invariant under the chiral rotations
\begin{equation}
\psi(x)\longrightarrow e^{i\gamma_{5}\theta}\psi(x)\quad ,\quad \overline{\psi}(x)\longrightarrow \overline{\psi}(x) e^{-i\gamma_{5}\theta}
\label{crot}
\end{equation}
The continuum axial symmetry (\ref{crot}) is broken explicitly by the staggered fermions, but a discrete axial symmetry remains. It is a chiral rotation 
of $\theta=\pi/2$ and corresponds to the continuum transformation
\begin{equation} 
\psi(x)\longrightarrow \gamma_{5}\psi(x)\quad ,\quad \overline{\psi}(x)\longrightarrow -\overline{\psi}(x) \gamma_{5}
\label{cdrot}
\end{equation}
and appears on the lattice as a translation by one site.  

Let us start to see how staggered fermions work in $(1+1)$-dimensions. 
The main idea is to reduce the number of degrees of freedom by using a single component Fermi field $\psi_{x} $ 
on each site of the lattice
\begin{equation}
\{\psi_{x},\psi_{y}\}=0\quad,\quad \{\psi^{\dag}_{x},\psi_{y}\}=\delta_{xy}
\end{equation}
so that the lattice Dirac Hamiltonian reads
\begin{equation}
H_{D}=-\frac{it}{2a}\sum_{x}(\psi_{x+1}^{\dag}\psi_{x}-\psi_{x}^{\dag}\psi_{x+1})
\label{dirl}
\end{equation}
and the equations of motion are
\begin{equation}
\dot{\psi}_{x}=-i[H_{D},\psi_{x}]=\frac{t}{2a}[\psi_{x+1}-\psi_{x-1}]\quad .
\end{equation}
If one decomposes the lattice into 
even and odd sublattices 
(characterized by $x$ even and odd), one can identify a single two component 
Dirac spinor by associating the upper component 
with even sites and the lower component with odd sites (or vice versa). 
In this way the equations of 
motion become
\begin{equation}
\dot{\psi_{1}}(x)=\frac{1}{2a}[\psi_{2}(x+1)-\psi_{2}(x-1)]
\label{direla}
\end{equation}
\begin{equation}
\dot{\psi_{2}}(x)=\frac{1}{2a}[\psi_{1}(x+1)-\psi_{1}(x-1)]\ .
\label{direlb}
\end{equation}
Comparing Eq.(\ref{direla}) and Eq.(\ref{direlb}) with the
the continuum Dirac equation,
one immediately sees that the staggered fermion 
formalism avoids the species doubling problem by doubling the lattice spacing and so 
 halving the size of the Brillouin zone.

Let us now discuss the symmetries of the 
lattice Dirac Hamiltonian Eq.(\ref{dirl}). 
It is invariant under translations of the spatial lattice 
by an even number of sites. 
Translation by two lattice spacings is the ordinary continuum translation, 
whose generator is 
\begin{equation}
p=-i\int dx \psi^{\dag}\partial _{x} \psi =-i\int dx(\psi_{1}^{\dag}\partial_{x}\psi_{1}+\psi_{2}^{\dag}\partial_{x}\psi_{2}) 
\end{equation}
and it does not mix upper and lower spinor components. 
Consequently, one expects the lattice generator not to mix the two sublattices
\begin{equation}
p=-i\sum_{x}(\psi_{x+2}^{\dag}\psi_{x}+\psi_{x}^{\dag}\psi_{x+2})\quad .
\end{equation}

The Hamiltonian Eq.(\ref{dirl}) is also invariant under translations by an odd number of sites. The lattice generator of this symmetry is
\begin{equation}
Q_{5}=\sum_{x}(\psi_{x}^{\dag}\psi_{x+1}+\psi_{x+1}^{\dag}\psi_{x})
\end{equation}
which in the continuum would be $ Q_{5}=\int dx \psi^{\dag}\gamma_{5}\psi  $. 
The matrix $\gamma_5$ applied to the bispinor interchanges upper and lower component, this on the lattice corresponds to
translation by one lattice site.

I illustrate now staggered fermions on a $(d-1)$-dimensional hypercubic
lattice and continuum time ( hamiltonian formalism) which are obtained by spin-diagonalization
\cite{b30} of the naively
latticized Dirac Hamiltonian
\begin{equation}
H_f=-{i\over2}\sum_{\vec{x},j}\left(\psi_\sigma^{\dagger}(\vec{x}+\vec{j})\alpha^j_{\sigma\sigma'}\psi_{\sigma'}(\vec{x})-
\psi_\sigma^{\dagger}(\vec{x})\alpha_{\sigma\sigma'}^j
\psi_{\sigma'}(\vec{x}+\vec{j})\right)
\label{2.1}
\end{equation}
where $\psi_{\sigma}(x)$ are the fermion field operators,
$\sigma=1,2,\ldots 2^{[d/2]}$ is the spin index,
$\vec{x}\equiv(x_1,x_2\ldots x_{(d-1)})=\sum_{j=1}^{d-1}x_j\vec{j}$ (the
$x_j$ are integers) refers to lattice sites, $\hat j$ are unit
vectors and $\alpha^j$ are the $2^{[d/2]}\times2^{[d/2]}$ Hermitean Dirac
matrices - here $[d/2]$ is the largest integer less than or equal to $d/2$. 
Hamiltonian (\ref{2.1}) describes a
fermion hopping problem in a U($2^{[d/2]}$) background gauge field
given by the unitary\footnote{Dirac's $\alpha$-matrices are
Hermitean, $$\alpha^{i\dagger}=\alpha^i~~,$$ 
and, due to their
anticommutator algebra, $$\bigl\{
\alpha^i,\alpha^j\bigr\}=2\delta^{ij}~~ ,$$
they are also unitary.}
matrices $\alpha^j$.  The crucial observation which allows spin
diagonalization is that this background field has only U(1) curvature,
i.e. if one considers the product around any plaquette 
\begin{equation}
\alpha^j\alpha^k(\alpha^j)^{\dagger}(\alpha^k)^{\dagger}=-1~~.
\end{equation}
Hence the only information carried by the $\alpha$-matrices which is
invariant under a space dependent change of phase of the fermion
fields is that a product of $\alpha$'s around any plaquette is $-1$.
This allows diagonalization of the $\alpha$'s using a gauge
transformation such as 
\begin{equation}
\psi(\vec{x})\rightarrow
(\alpha^1)^{x_1}(\alpha^2)^{x_2}\dots(\alpha^{(d-1)})^{x_{(d-1)}}\psi(\vec{x})
\end{equation}
resulting in the Hamiltonian 
\begin{equation}
H_f=-{i\over2}\sum_{\vec{x},j}(-1)^{\sum_{p=1}^{j-1}x_p}
\left(\psi^{\dagger}(\vec{x}+\vec{j})\psi(\vec{x})-\psi^{\dagger}(\vec{x})\psi(\vec{x}+\vec{j})\right)
\label{2.2}
\end{equation}
which describes $2^{[d/2]}$ species of lattice fermions with
background U(1) magnetic flux $\pi$ through every plaquette of the
lattice. This flux is contained in the background U(1) field
$U^0[\vec{x},\vec{j}]=(-1)^{\sum_{p=1}^{j-1} x_p}$.  Each species of fermion must have the
same spectrum as the original one given by the Dirac Hamiltonian
(\ref{2.1}).  This allows reduction of the fermion multiplicity by a factor
of $2^{[d/2]}$ (by dropping the fermion spin index $\sigma$).  The
result resembles a condensed matter hopping problem with a single
species of fermion where there is a background magnetic field $\pi$
per plaquette.

Discrete Chiral symmetry interchanges the even ($\sum_{p=1}^{d-1} x_p$=even) and odd
($\sum_{p=1}^{d-1}x_p=$odd) sublattices.  The substitutions 
\begin{equation}
\psi(\vec{x})\rightarrow (-1)^{\sum_{p=j+1}^{d-1}x_p}\psi(\vec{x}+\vec{j})
\label{2.3}
\end{equation}
for $j=1,\ldots,(d-1)$ leave the Hamiltonian in Eq.(\ref{2.2}) invariant.
A candidate for Dirac mass operator, which changes sign under the
transformations in Eq.(\ref{2.3}), is the staggered charge density
operator
\begin{equation}
\mu=m\sum_{\vec{x}} (-1)^{\sum_{p=1}^{d-1}x_p}\psi^{\dagger}(\vec{x})\psi(\vec{x})~~.
\label{2.4}
\end{equation} 

To obtain the continuum limit and the number of fermion species, one first divides the lattice into $2^{d-1}$ sublattices according to
whether the components of their coordinates are even or odd.  For
example, when $(d-1)=3$, I label 8 fermion species as 
\begin{eqnarray}
\psi({\rm
even},{\rm even},{\rm even})\equiv\psi_1~~,~~\psi({\rm even},{\rm
odd},{\rm odd})\equiv\psi_2
\nonumber\\
\psi({\rm odd},{\rm even},{\rm
odd})\equiv\psi_3~~,~~\psi({\rm odd},{\rm odd},{\rm
even})\equiv\psi_4
\nonumber\\
\psi({\rm even},{\rm even},{\rm
odd})\equiv\psi_5~~,~~\psi({\rm even},{\rm odd},{\rm
even})\equiv\psi_6
\nonumber\\
\psi({\rm odd},{\rm even},{\rm
even})\equiv\psi_7~~,~~\psi({\rm odd},{\rm odd},{\rm
odd})\equiv\psi_8~~.  
\end{eqnarray}
Then, if one adds the mass operator (\ref{2.4}), in momentum space the
Hamiltonian (\ref{2.2}) has the form
\begin{equation}
H_f=\int_{\Omega_B}d^3k~\psi^{\dagger}(\vec{k})\left(A^i\sin k_i +B
m\right)\psi(\vec{k})
\label{2.5}
\end{equation}
with 
\begin{equation}
\psi^\dagger(\vec{k})\equiv(\psi^\dagger_1(\vec{k}),
\ldots,\psi^\dagger_8(\vec{k}))
\end{equation}
and $8\times8$ matrices 
\begin{eqnarray}
A^i=\left(\matrix{0&\alpha^i\cr
\alpha^i&0\cr}\right)~~,~~
B=\left(\matrix{1&0\cr0&-1\cr}\right) \nonumber \\
\alpha^1=\left(\matrix{0&1\cr 1&0\cr}\right)
~~,~~
\alpha^2=\left(\matrix{\sigma^1&0\cr 0&-\sigma^1\cr}\right)
~~,~~
\alpha^3=\left(\matrix{\sigma^3&0\cr 0&-\sigma^3\cr}\right)
\label{2.6}
\end{eqnarray} 
with $\sigma^i$ the Pauli matrices; I used the Fourier
transform 
\begin{equation}
\psi(\vec{x})=\int_{\Omega_B}{d^3k\over(2\pi)^{3/2}}e^{-i\vec{k}\cdot \vec{x}}\psi(\vec{k})
\end{equation}
where $\Omega_B=\{\vec{k}=(k_1,k_2,k_3), -\pi/2<k_i\leq\pi/2\}$ is the
Brillouin zone of the (even,even,even) sublattice. The fermion
spectrum is 
\begin{equation}
\omega(\vec{k})=\sqrt{\sum_{i=1}^{3}\sin^2k_i+m^2},
\end{equation}
and only the region $k_i\sim0$ is relevant to the continuum limit.
I have normalized $\psi(k)$ so that 
\begin{equation}
\left\{\psi(x),\psi^{\dagger}(y)\right\}=\delta(x-y)
~~,~~
\left\{\psi(k),\psi^{\dagger}(l)\right\}=\delta^3(k-l)~~.
\label{2.7}
\end{equation} 
If one defines $\beta=\left(\matrix{\sigma^2&0\cr0&-\sigma^2}\right)$
and the unitary matrix 
\begin{equation}
M={1\over2}\left(\matrix{1-\beta&1+\beta\cr1+\beta&1-\beta\cr}\right)
\end{equation}
and 
\begin{equation}
\psi=M\psi'
\end{equation}
with 
\begin{equation}
\psi'=\left(\matrix{\psi_a\cr\psi_b}\right),
\end{equation}
the Hamiltonian is 
\begin{eqnarray}
H_f=\int_{\Omega_B}d^3k~\left(\psi_a^{\dagger},\psi_b^{\dagger}\right)
\left(\matrix{\alpha^i\sin k_i-\beta 
m&0\nonumber\\
0&\alpha^i\sin k_i+\beta
m\cr}\right)\left(\matrix{\psi_a\cr\psi_b}\right)~~.
\label{2.8}
\end{eqnarray} 
the low momentum limit, $\sin k_i\sim k_i$, with fermion density 1/2
per site so that the Fermi level is at the intersection point of the
positive and negative energy bands, one obtains 2 continuum Dirac
fermions.  Furthermore, the staggered charge operator gives a Dirac
mass of differing sign for the two species.

If one considers ${\cal N}{\cal N}_C$ lattice species in $d$
dimensions, i.e. fermions $\psi^\alpha_A$ with $\alpha=1,\ldots,{\cal N}$
and $A=1,\ldots,{\cal N}_C$ a lattice flavor and color index, respectively, 
this yields ${\cal N}{\cal N}_C2^{(d-1)}/2^{[d/2]}$ continuum species of Dirac
fermions where the lattice fermion density is ${\cal N}{\cal N}_C/2$ per site.
For the Dirac mass operator one may choose 
\begin{equation}
\mu = \sum_{\vec{x}} (-1)^{\sum_{p=1}^{d-1}x_p}
\psi_A^{\alpha\dagger}(\vec{x})m_{\alpha\beta} \psi^\beta_B (\vec{x})~~,
\label{2.9}
\end{equation}
with $m_{\alpha\beta}=diag(m_1,\ldots,m_{{\cal N}})$, and this gives the
fermion spectrum 
\begin{equation}\omega(k)=\sqrt{\sum_{i=1}^{d-1}\sin^2k_i +
m_\alpha}
\end{equation}
where $-\pi/2\leq k_i<\pi/2$ and $\alpha=1,\ldots,{\cal N}$.

Here, I count the number of flavors of fermions obtained in the
continuum limit by noting that Eq.(\ref{2.2}) describes a
$2^{d-1}$-component fermion.  In $d$ dimensions the Dirac matrices are
$[d/2]$ dimensional, therefore the continuum limit of Eq.(\ref{2.2})
describes $2^{d-1}/2^{[d/2]}$ species of Dirac fermions.
\begin{equation}
\matrix{ d&{\rm ~no.~of~flavors}&{\rm ~no.~of~spinor~components}\cr 
2&1&2\cr 3&2&2\cr 4&2&4\cr} \nonumber
\end{equation}

\section{Strong coupling lattice gauge theory}
\label{strongc}

This section is devoted to a detailed analysis of the strong coupling limit of hamiltonian lattice gauge theories. 
I shall give a general introduction to the subject by considering a gauge theory with colour group $U({\cal N}_{C})$ or 
$SU({\cal N}_{C})$ and $SU({\cal N})$-flavor groups. 
The Hamiltonian formalism is particularly suitable for illustrating the relationship between several gauge and spin systems 
and was already exploited in some of the earliest studies of chiral symmetry breaking in the strong coupling limit~\cite{b15}. 
The strong coupling limit, though far from the scaling regime is often used to study qualitative properties related to confinement 
and chiral symmetry breaking. In chapter 4, 5 and 6 I shall study the one-flavor and multiflavor lattice Schwinger models. The one-flavor 
model will be mapped in the strong coupling limit onto an antiferromagnetic Ising model with long range Coulomb interaction. The multiflavor Schwinger 
models are effectively described by spin-$1/2$ antiferromagnetic $SU({\cal N})$ Heisenberg models.

\subsection{ Hamiltonian formulation of lattice gauge theory}

Let us consider an hamiltonian lattice gauge theory~\cite{b13} where 
space is a $(d-1)$-dimensional hypercubic
lattice with oriented plaquettes $[x,i,j]$ with
corners $x,x+\hat i,x+\hat i+\hat j,x+\hat j$ and orientation $\hat
i\times\hat j$.  The gauge field $U[x,i]$ associated with the link
$[x,i]$ is a group element in the fundamental representation $[{\cal N}_C]$
of $SU({\cal N}_C)$ and, if the color group is $U({\cal N}_C)$, also carries a
representation of $U(1)$.  It has the property
$U[x,-i]=U^{\dagger}[x-\hat i,i]$. The electric field operator
$E^a[x,i]$ associated with link $[x,i]$ has the Lie algebra
\begin{equation}
\bigl[ E^a[x,i], E^b[y,j]\bigr]=if^{abc}E^c[x,i]\delta\left([x,i]-[y,j]\right)
\label{3.1}
\end{equation}
and 
\begin{equation}
E[x,-i]=-U^{\dagger}[x-\hat i,i]E[x-\hat i,i]U[x-\hat i,i]
\end{equation}
where 
\begin{equation}
E[x,i]\equiv E^a[x,i]T^a
\label{3.2}
\end{equation} 
with $T^a=(T^a)^\dagger$, $a=0\dots,{\cal N}_C^2-1$ the generators of the
Lie algebra of $U({\cal N}_C)$ obeying 
\begin{equation}
[T^a,T^b]=if^{abc}T^c,
\end{equation}
$T^0={\bf 1}$ the ${\cal N}_C\times {\cal N}_C$ unit matrix representing $U(1)$ and
$T^a$, $a\neq0$, in the representation $[{\cal N}_C]$ of $SU({\cal N}_C)$.  It
generates the left-action of the Lie algebra on $U[x,i]$, i.e.,
\begin{eqnarray}
\bigl[ E^a[x,i], U[y,j]\bigr]= 
-T^aU[x,i]\delta\left([x,i]-[y,j]\right)~~.
\label{3.3}
\end{eqnarray}

The Hamiltonian for lattice gauge theory with staggered fermions and
discrete chiral symmetry is
\begin{eqnarray} 
H=\sum_{[x,i],a}{g^2\over2}E^a[x,i]^2&+&
\left(\sum_{[x,i,j]}{1\over2 g^2}{\rm
Tr}(U[x,i]U[x+\hat i,j]U^\dagger[x+\hat j,i]U^\dagger[x,j])+{\rm h.c} 
\right) \nonumber\\
&+&\left(
\sum_{[x,i]}t_{[x,i]}\psi^{\alpha\dagger}_A(x+\hat i)U_{AB}[x,i]
\psi_B^\alpha(x) +{\rm h.c.}\right)
\label{3.4}
\end{eqnarray} 
where ${\rm Tr}(\cdot)$ is the ${\cal N}_C\times {\cal N}_C$ matrix trace and where
\begin{equation}
t_{[x,i]}=t{i\over2}(-)^{\sum_{p=1}^i x_p},\quad t_{[x,-i]}=t_{[x-\hat
i,i]}^*
\end{equation} 
and the generators of static gauge transformations are 
\begin{equation}
{\cal G}^a(x)=\sum_{i=-(d-1)}^{(d-1)} E^a[x,i]
+\psi_A^{\alpha\dagger}(x)T^a_{AB}\psi^\alpha_B(x)
\label{3.5}
\end{equation} 
with $a=1,\dots,{\cal N}_C^2-1$.  If the gauge group is $U({\cal N}_C)$ rather
than $SU({\cal N}_C)$ there is also the U(1) generator 
\begin{equation} 
{\cal G}^0(x)=\sum_{i=1}^{(d-1)}\left( E^0[x,i]-E^0[x-\hat i,i]\right)
+{1\over2}\bigl[\psi_A^{\alpha\dagger}(x),\psi_A^\alpha(x)\bigr]
\label{3.6}
\end{equation} 
with the fermionic U(1) charge operator ordered so that, like $E$, it
changes sign under charge conjugation.  (Note that there is no ordering
ambiguity for the ${\cal G}^a$, $a\neq 0$.)  These generators obey the
Lie algebra

\begin{equation}
[ {\cal G}^a(x) , {\cal G}^b(y)]= if^{abc}
{\cal G}^c(x)\delta(x-y)~~. 
\label{3.7}
\end{equation} 
The Hamiltonian is gauge
invariant, 
\begin{equation}
[{\cal G}^a(x),H]=0~~.
\label{3.8}
\end{equation} 
The dynamical problem of lattice gauge theory is to find the
eigenstates of the Hamiltonian operator Eq.(\ref{3.4}) which are also
gauge invariant, i.e. which obey the physical state condition

\begin{equation}
 {\cal G}^a(x) \vert \psi_{\rm
phys} >~=~0 ~~.\label{3.9}
\end{equation}

\subsection{ Strong coupling}
\label{SC}

In order to perform a strong-coupling expansion~\cite{b28}, let us rewrite Eq.(\ref{3.4}) 
$$H=H_0+H_1+H_2$$ where $H_0=\sum{g^2\over2}E^2$, $H_1=\sum
(t\psi^{\dagger}U\psi+{\rm h.c.})$ and $H_2=\sum{1\over 2g^2}({\rm
Tr}UUUU+{\rm h.c.})$.  Each term is gauge invariant, $$[{\cal
G}^a(x),H_i]=0.$$ Therefore, if one finds a gauge invariant eigenstate
of $H_0$, perturbations in $H_1$ and $H_2$ remain gauge invariant.
$H_0$ is the sum of group manifold Laplacians for each link.  If
$\vert0>$ is a singlet of the algebra (\ref{3.1}), i.e.
$$E^a[x,i]\vert0>=0,$$ then $$H_0\vert0>=0$$ and 
\begin{equation}
H_0U[x,i]\vert0>={g^2\over2}C_2({\cal N}_C)U[x,i]\vert0>
\label{upe} 
\end{equation}
In Eq.(\ref{upe}) the Casimir operator is
$$C_2({\cal N}_C){\bf 1}=\sum_a T^aT^a$$ 
and $a$ runs either from 0 (for
$U({\cal N}_C))$ or 1 (for $SU({\cal N}_C)$) to ${\cal N}_C^2-1$.

Consider the empty vacuum which is a singlet of (\ref{3.1}) and which has no
fermions, $$\psi_A^\alpha(x)\vert0>=0~~,~~\forall x,A,\alpha.$$ It is
necessary to find the lowest energy ($E_0$= eigenvalue of $H_0$) 
eigenstates $\vert\Psi>$ of $H_0$
which are gauge invariant, $$ {\cal G}^a(x)\vert\Psi> = 0 \quad
\forall a,x, $$ with the constraint that the fermion states are
half-filled, $i.e.$ half of the fermion states are filled.

I shall first review the simplest case of $SU({\cal N}_C\geq 3)$ gauge
theory with 1 lattice flavor (${\cal N}=1$) and $2^d/2^{[d/2]}$ continuum
flavors~\cite{b10}.  The density is ${\cal N}_C/2$ fermions per
site.  The lowest energy eigenstates of $H_0$ with $E_0=0$ are the
states which are singlets of the electric field algebra (\ref{3.1}) and,
since they are gauge invariant they must also be color singlets, i.e.
singlets of the algebra (\ref{3.7}).  Then, they must also be singlets of
the algebra of the fermion currents
$\psi^{\dagger}(x)T^a\psi(x)$.  One can form a ``baryon" (=color
singlet) at a site by either leaving it unoccupied or putting ${\cal N}_C$
fermions with antisymmetrized singlet wave-function by applying the creation
operator 
\begin{equation}
S^{\dagger}(x)=\epsilon_{A_1\dots
A_{{\cal N}_C}}\psi^{\dagger}_{A_1}(x)\dots\psi^{\dagger}_{A_{{\cal N}_C}}(x)\label{3.10}
\end{equation} 
($\epsilon_{A_1,\ldots A_{{\cal N}_C}}$ is the usual antisymmetric
tensor).  

Fermi statistics allows at most one singlet per site.
Otherwise one can distribute the $N$/2 singlets arbitrarily ($N$ is the
total number of lattice sites).  Thus, there are $N!/((N/2)!)^2$
degenerate ground states with a typical state being $$\vert
\{\rho_x\}_x> =\prod_{{\rm occupied}\; x}S^{\dagger}(x)\vert0>$$ I label them by the eigenvalues $\rho_x$ of the local Fermion number
operators $$\rho(x)={1\over2}[\psi^{\dagger}(x),\psi(x)]=
\psi^{\dagger}(x)\psi(x)-{{\cal N}_C}/2.$$

All matrix elements in the vector space of degenerate vacua of the
first order Hamiltonian vanish, $$<\{\rho_x\}_x\vert
H_1\vert\{\rho_x'\}_x>=0$$ so second order perturbation theory must be considered and the matrix
with elements $$ <\{\rho_x\}_x\vert
H_2\vert\{\rho_x'\}_x>-<\{\rho_x\}_x\vert H_1{1\over
H_0-E_0}H_1\vert\{\rho_x'\}_x>. $$
must be diagonalized. 
The first term is zero and diagonalizing the second term is equivalent to diagonalizing the
four-fermion Hamiltonian 
\begin{equation}
H_{\rm eff}= -K\sum_{[x,i]}
\psi^{\dagger}_A(x+\hat i)\psi_B(x)\psi^{\dagger}_B(x)\psi_A(x+\hat i)
=K\sum_{[x,i]}\rho(x+\hat i)\rho(x)
\label{3.11}
\end{equation}
(up to an additive constant), with 
\begin{equation}
K={t^2\over
2g^2{\cal N}_CC_2({\cal N}_C)}>0~~,
\label{3.12}
\end{equation} 
in the space of pure fermion states where $\rho(x)$ has eigenvalues
$$\rho_x=\pm {\cal N}_C/2$$ and $$\sum_x\rho_x=0.$$ This is the
antiferromagnetic Ising model.

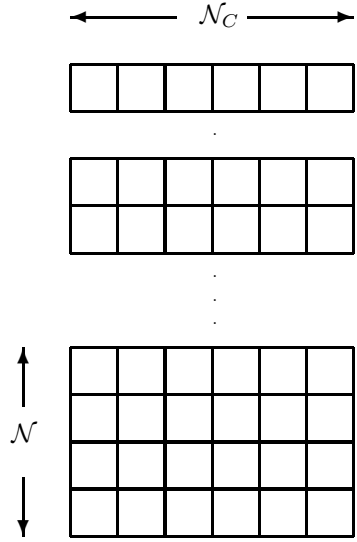
\begin{figure}[htb]
\begin{center}
\setlength{\unitlength}{0.00041700in}%
\begingroup\makeatletter\ifx\SetFigFont\undefined
\def\x#1#2#3#4#5#6#7\relax{\def\x{#1#2#3#4#5#6}}%
\expandafter\x\fmtname xxxxxx\relax \def\y{splain}%
\ifx\x\y   
\gdef\SetFigFont#1#2#3{%
  \ifnum #1<17\tiny\else \ifnum #1<20\small\else
  \ifnum #1<24\normalsize\else \ifnum #1<29\large\else
  \ifnum #1<34\Large\else \ifnum #1<41\LARGE\else
     \huge\fi\fi\fi\fi\fi\fi
  \csname #3\endcsname}%
\else
\gdef\SetFigFont#1#2#3{\begingroup
  \count@#1\relax \ifnum 25<\count@\count@25\fi
  \def\x{\endgroup\@setsize\SetFigFont{#2pt}}%
  \expandafter\x
    \csname \romannumeral\the\count@ pt\expandafter\endcsname
    \csname @\romannumeral\the\count@ pt\endcsname
  \csname #3\endcsname}%
\fi
\fi\endgroup
\begin{picture}(4362,6753)(4051,-6373)
\thicklines
\put(5401,-361){\line( 0,-1){600}}
\put(6001,-361){\line( 0,-1){600}}
\put(6601,-361){\line( 0,-1){600}}
\put(7201,-361){\line( 0,-1){600}}
\put(7801,-361){\line( 0,-1){600}}
\put(4801,-1561){\line( 1, 0){3600}}
\put(8401,-1561){\line( 0,-1){1200}}
\put(8401,-2761){\line(-1, 0){3600}}
\put(4801,-2761){\line( 0, 1){1200}}
\put(4801,-2161){\line( 1, 0){3600}}
\put(5401,-1561){\line( 0,-1){1200}}
\put(6001,-1561){\line( 0,-1){1200}}
\put(6601,-1561){\line( 0,-1){1200}}
\put(7201,-1561){\line( 0,-1){1200}}
\put(7801,-1561){\line( 0,-1){1200}}
\put(4801,-3961){\line( 1, 0){3600}}
\put(8401,-3961){\line( 0,-1){2400}}
\put(8401,-6361){\line(-1, 0){3600}}
\put(4801,-6361){\line( 0, 1){2400}}
\put(5401,-3961){\line( 0,-1){2400}}
\put(6001,-3961){\line( 0,-1){2400}}
\put(4801,-361){\line( 1, 0){3600}}
\put(8401,-361){\line( 0,-1){600}}
\put(8401,-961){\line(-1, 0){3600}}
\put(4801,-961){\line( 0, 1){600}}
\put(6601,-3961){\line( 0,-1){2400}}
\put(6451,164){\makebox(0,0)[lb]{\smash{\SetFigFont{10}{12.0}{rm}${\cal N}_{C}$}}}
\put(7201,-3961){\line( 0,-1){2400}}
\put(7801,-3961){\line( 0,-1){2400}}
\put(4801,-4561){\line( 1, 0){3600}}
\put(4801,-5161){\line( 1, 0){3600}}
\put(4801,-5761){\line( 1, 0){3600}}
\put(6151,239){\vector(-1, 0){1350}}
\put(7051,239){\vector( 1, 0){1350}}
\put(4201,-4711){\vector( 0, 1){750}}
\put(4201,-5611){\vector( 0,-1){750}}
\put(6601,-1261){\makebox(0,0)[lb]{\smash{\SetFigFont{6}{7.2}{rm}.}}}
\put(6601,-3061){\makebox(0,0)[lb]{\smash{\SetFigFont{6}{7.2}{rm}.}}}
\put(6601,-3361){\makebox(0,0)[lb]{\smash{\SetFigFont{6}{7.2}{rm}.}}}
\put(6601,-3661){\makebox(0,0)[lb]{\smash{\SetFigFont{6}{7.2}{rm}.}}}
\put(4051,-5161){\makebox(0,0)[lb]{\smash{\SetFigFont{10}{12.0}{rm}${\cal N}$}}}
\end{picture}
\end{center}
\caption{Young tableaux representations allowed at a given site.}
\label{y1}
\end{figure}

Now I consider arbitrary ${\cal N}$. Then the density is ${\cal N}_C{\cal N}/2$
fermions per site and the ``baryon" (= color singlet) creation operator
is 
\begin{equation} 
S^{\dagger}_{\alpha_1\dots\alpha_{{\cal N}_C}}(x)=\epsilon_{A_1\dots
A_{{\cal N}_C}}\psi_{A_1}^{\alpha_1\dagger}(x)
\dots\psi_{A_{{\cal N}_C}}^{\alpha_{{\cal N}_C}\dagger}(x)~~.\label{3.13D}
\end{equation}
Since this operator is symmetric in the lattice flavor indices
$\alpha_1,\dots,\alpha_{{\cal N}_C}$, it carries an irreducible
representation of $SU({\cal N})$ with Young Tableau with ${\cal N}_C$ columns and
1 row.  Fermi statistics allows at most ${\cal N}$ singlets on a given
site.  Thus, the allowed representations of the flavor $SU({\cal N})$ algebra at one site are the empty singlet and those with the Young
tableaux given in fig. (\ref{y1}) and which are
distinguished by the fermion numbers
\begin{equation}
\rho_x={\cal N}_C(2\nu
-{\cal N})/2,\quad
\nu=0,1,2,\ldots,{\cal N}
\label{3.14}
\end{equation}
respectively.
A ground state of $H_0$ is obtained by creating
$N {\cal N}/2$ color singlets.  The states are labeled by the local fermion
densities $\rho_x$ and the vector in the corresponding $SU({\cal N})$ representation at each site and are degenerate.

\begin{figure}[htb]
\begin{center}
\setlength{\unitlength}{0.00041700in}%
\begingroup\makeatletter\ifx\SetFigFont\undefined
\def\x#1#2#3#4#5#6#7\relax{\def\x{#1#2#3#4#5#6}}%
\expandafter\x\fmtname xxxxxx\relax \def\y{splain}%
\ifx\x\y   
\gdef\SetFigFont#1#2#3{%
  \ifnum #1<17\tiny\else \ifnum #1<20\small\else
  \ifnum #1<24\normalsize\else \ifnum #1<29\large\else
  \ifnum #1<34\Large\else \ifnum #1<41\LARGE\else
     \huge\fi\fi\fi\fi\fi\fi
  \csname #3\endcsname}%
\else
\gdef\SetFigFont#1#2#3{\begingroup
  \count@#1\relax \ifnum 25<\count@\count@25\fi
  \def\x{\endgroup\@setsize\SetFigFont{#2pt}}%
  \expandafter\x
    \csname \romannumeral\the\count@ pt\expandafter\endcsname
    \csname @\romannumeral\the\count@ pt\endcsname
  \csname #3\endcsname}%
\fi
\fi\endgroup
\begin{picture}(8112,5553)(1501,-5173)
\thicklines
\put(3601,-361){\line( 0,-1){4800}}
\put(4801,-361){\line( 0,-1){4800}}
\put(6001,-361){\line( 0,-1){4800}}
\put(7201,-361){\line( 0,-1){4800}}
\put(8401,-361){\line( 0,-1){4800}}
\put(2401,-1561){\line( 1, 0){7200}}
\put(2401,-361){\line( 1, 0){7200}}
\put(9601,-361){\line( 0,-1){4800}}
\put(9601,-5161){\line(-1, 0){7200}}
\put(2401,-5161){\line( 0, 1){4800}}
\put(2401,-2761){\line( 1, 0){7200}}
\put(5701,164){\makebox(0,0)[lb]{\smash{\SetFigFont{10}{12.0}{rm}${\cal N}_{C}$}}}
\put(2401,-3961){\line( 1, 0){7200}}
\put(5401,239){\vector(-1, 0){3000}}
\put(6601,239){\vector( 1, 0){3000}}
\put(1801,-2161){\vector( 0, 1){1800}}
\put(1801,-3361){\vector( 0,-1){1800}}
\put(1501,-2761){\makebox(0,0)[lb]{\smash{\SetFigFont{10}{12.0}{rm}$\frac{{\cal N}}{2}$}}}
\end{picture}
\end{center}
\caption{Young tableau with ${\cal N}_C$ columns and ${\cal N}/2$ rows.}
\label{y2}
\end{figure}
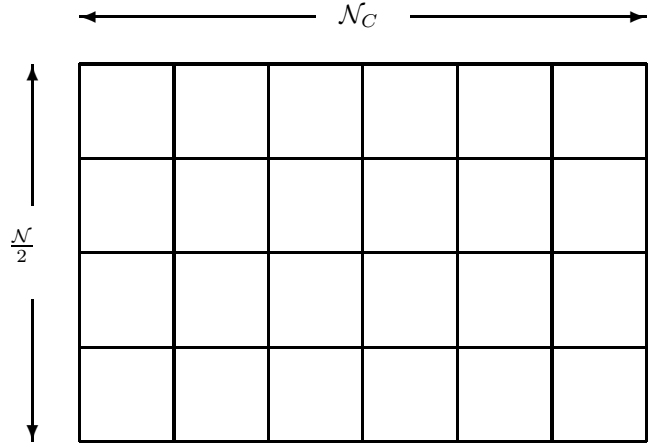

Again this degeneracy must be resolved by diagonalizing perturbations.
The first-order perturbation to the vacuum energy vanishes and
diagonalizing the second order perturbations is equivalent to
diagonalizing the effective Hamiltonian 
\begin{equation} H_{\rm
eff}=-K\sum_{[x,i]}\psi^{\alpha\dagger}_A(x+\hat i)
\psi^\alpha_B(x)\psi^{\beta\dagger}_B(x)\psi^\beta_A(x+\hat i)
=K\sum_{[x,i]}J^{\alpha\beta}(x+\hat i)J^{\beta\alpha}(x)
\label{3.15} 
\end{equation}
(up to an additive constant) where the operators
\begin{equation}
J^{\alpha\beta}(x)= {1\over
2}\left[\psi^{\alpha\dagger}_A(x),\psi^{\beta}_A(x)\right]
=\psi^{\alpha\dagger}_A(x)\psi^{\beta}_A(x)-
\delta^{\alpha\beta}{{\cal N}_{C}\over 2}=
(J^{\beta\alpha}(x))^\dagger\label{3.16}
\end{equation}
are generators of the Lie algebra of U(${\cal N}$),
\begin{equation}
\left[J^{\alpha\beta}(x),J^{\alpha'\beta'}(y)\right]=
\delta(x-y)\left( \delta^{\beta\alpha'}J^{\alpha\beta'}(x)- 
\delta^{\beta'\alpha}J^{\alpha'\beta}(x)\right)~~.\label{3.17}
\end{equation}
Thus, the effective Hamiltonian is that of a $U({\cal N})$ quantum
antiferromagnet with representations given in fig. (\ref{y2}) and with the
constraint $\sum_x\rho_x=0$, and I have shown that it is equivalent
to the strong coupling limit of $SU({\cal N}_C)$ lattice gauge theory with
${\cal N}$ lattice flavors of staggered fermions.

The SU(2) color group is peculiar and this can be traced back to the fact that for 2 colors the singlet creation
and annihilation operators involve only 2 fermion field operators, and
in second order perturbation theory terms are generated describing the
hopping of these singlets \cite{b21}. 
Due to this Eq.(\ref{3.15}) is not the right effective
Hamiltonian but 
\begin{eqnarray}
H_{\rm eff}&=&-{K}\sum_{[x,i]}\left(\psi^{\alpha\dagger}_A(x)
\psi^\alpha_B(x+\hat i)\psi^{\beta\dagger}_B(x+\hat i)\psi^\beta_A(x)-
\epsilon_{AC}\epsilon_{BD}\psi^{\alpha\dagger}_A(x+\hat i)
\psi^\alpha_B(x)\psi^{\beta\dagger}_C(x+\hat i)\psi^\beta_D(x)
\right)  
\quad \quad \quad \quad \quad \quad \quad \quad \nonumber\\
&=&{K}\sum_{[x,i]}\left(J^{\alpha\beta}(x)J^{\beta\alpha}(x+\hat i) +
S_+^{\alpha\beta}(x)S_-^{\beta\alpha}(x+\hat i) \right)
\label{3.18}
\end{eqnarray} 
(up to a constant), with
\begin{equation}
S_+^{\alpha\beta}(x)=\epsilon_{AB}\psi_A^{\alpha\dagger}(x)
\psi_B^{\beta\dagger}(x)~~, \label{3.19}
\end{equation}
\begin{equation}
S_-^{\alpha\beta}(x)=\epsilon_{AB}\psi_A^{\alpha}(x)
\psi_B^{\beta}(x)=(S_+^{\beta\alpha}(x))^\dagger~~, \label{3.20}
\end{equation}
and the
$J^{\alpha\beta}(x)$ as above. One can check that the operators
$J^{\alpha\beta}(x)$ and $S_\pm^{\alpha\beta}(x)$ obey the relations
of the Lie algebra of the symplectic group Sp($2N_L$), i.e. Eqs.
(\ref{3.17}) and 
\begin{equation}
\left[S_+^{\alpha\beta}(x), S_-^{\alpha'\beta'}(y)\right]
=
\delta(x-y)\left(\delta^{\beta\alpha'}J^{\alpha\beta'}(x) + 
\delta^{\alpha\beta'}J^{\beta\alpha'}(x) + 
\delta^{\alpha\alpha'}J^{\beta\beta'}(x) + 
\delta^{\beta\beta'}J^{\alpha\alpha'}(x)  \right)~~, 
\end{equation}
\begin{equation}
\left[J^{\alpha\beta}(x),
S_+^{\alpha'\beta'}(y)\right]=
\delta(x-y)\left(
\delta^{\beta\alpha'}S_+^{\alpha\beta'}(x) + 
\delta^{\beta\beta'}S_+^{\alpha\alpha'}(x)\right)~~,
\end{equation}
\begin{equation}
\left[J^{\alpha\beta}(x), S_-^{\alpha'\beta'}(y)\right]=
-\delta(x-y)\left(
\delta^{\alpha\alpha'}S_-^{\beta\beta'}(x) + 
\delta^{\alpha\beta'}S_+^{\beta\alpha'}(x)\right)~~,\label{3.21}
\end{equation}
hence (\ref{3.18}) is a Sp($2N_L$) antiferromagnet. Especially for ${\cal N}=1$
it is the spin $1/2$ Heisenberg antiferromagnet as originally found in~\cite{b21}.

If, instead of $SU({\cal N}_C)$ one had the gauge group U($N_C$), gauge
invariance would require that one imposes the extra constraint $${\cal
G}^0(x)\vert\Psi>=0 ~~,~~\forall x.$$  This can be fulfilled without
electric fields, i.e. with $\rho_x=0$ $\forall x$, only when ${\cal N}$ is
even.  Then the allowed $SU({\cal N}_C)$ representation at each site is
the one with the Young Tableau given in fig. (\ref{y2}),
and, with the constraint $$\sum_\alpha J^{\alpha\alpha}(x)=0~~,$$
$J^{\alpha\beta}(x)$ generate the $SU({\cal N})$ algebra in this
representation.  Thus, $U({\cal N}_C)$ gauge theory is equivalent to an
$SU({\cal N})$ quantum antiferromagnet in the representation fig. (\ref{y2}).  This
actually is true for all ${\cal N}_C\geq 1$ (${\cal N}_C=2$ is $not$ special here as
no hopping of color singlets is allowed for $U({\cal N}_C)$ color groups~\cite{b21}).

As a concrete example, let us consider the case ${\cal N}=2$ which gives 4
continuum flavors in $d=4$.  The $SU({\cal N}_C\leq 3)$ gauge theory is equivalent
to a $U(2)$ antiferromagnet.  Using the identity 
\begin{equation}
2\delta_{\beta\gamma}
\delta_{\alpha\epsilon}=
\vec\sigma_{\alpha\beta}\cdot\vec\sigma_{\gamma\epsilon}
+\delta_{\alpha\beta}\delta_{\gamma\epsilon}
\label{3.22}
\end{equation} 
(with $\vec\sigma$ the Pauli matrices) one can change basis in the U(2)
Lie algebra to find the effective strong coupling Hamiltonian 
\begin{equation}
H_{\rm eff}=\frac{K}{ 2}\sum_{[x,i]}\left(\psi^{\dagger}(x)
\vec\sigma\psi(x)\cdot\psi^{\dagger}(x+\hat i)
\vec\sigma\psi(x+\hat i)+\rho(x)\rho(x+\hat i)\right)
\label{3.23}
\end{equation}
where
\begin{equation} (\psi^{\dagger}\vec\sigma\psi\equiv
\psi^{\alpha\dagger}_A\vec\sigma_{\alpha\beta}\psi^\beta_A)
\end{equation}
and the representation at a given site has either $\rho_x=0$ and the
spin $S={\cal N}_C/2$ representation of $SU(2)$ or $\rho_x=\pm N_C$ and the
spin $S=0$.

If the gauge group is $U({\cal N}_C)$ then the density is constrained,
$\rho_x=0$ and each site is occupied by the $S={\cal N}_C/2$ representation
of $SU(2)$ and (\ref{3.23}) is the Hamiltonian of the quantum Heisenberg
antiferromagnet.

The most subtle case is that of a $U({\cal N}_C)$ lattice gauge theory with
odd ${\cal N}$.  There I first solve the problem of minimizing the
contribution of U(1) electric fields to the energy.  This is done by
solving the auxilliary problem of minimizing the energy function $$
{1\over2g^2}\sum_{[x,i]}\left(E^0[x,i]\right)^2 $$ with the constraint
\begin{equation}
\sum_i\left(E^0[x,i]-E^0[x-\hat i,i]\right)+ \rho(x) \sim 0 \label{3.24}
\end{equation} where $$\rho(x) =
\psi^{\alpha\dagger}_A(x)\psi^\alpha_A(x)-{{\cal N}_C{\cal N}/2}~~.$$ If ${\cal N}_C{\cal N}$ is
odd, the charge density $\rho(x)$ has only half-odd-integer
eigenvalues $\rho_x$.  Since they have no zero eigenvalue, the
constraint Eq.(\ref{3.24}) cannot be satisfied unless the $U(1)$ electric field
is non-zero.  This is also true of $SU({\cal N}_C)$ color singlet states,
since they always contain an integral multiple of ${\cal N}_C$ particles.
Then the operator $\psi^{\dagger}\psi$ has eigenvalue which is an
integer multiple of ${\cal N}_C$ and when ${\cal N}$ is odd, the charge density
$\rho(x)$ again has no nonzero eigenvalues. The configuration which
minimizes the $U(1)$ energy is the most symmetric one, $$ E^0[x,\pm
i]=\pm{{\cal N}_C\over 4(d-1)}(-1)^{\sum_{p=1}^{d-1} x_p} $$ with the
accompanying eigenvalues of the charge density $$
\rho_x= \pm {{\cal N}_C\over2}(-1)^{\sum_{p=1}^{d-1} x_p}~~.
$$ If, instead of $U(1)$ electric fields, one had color non-singlets, the
energy would be higher. 
In chapter 4 the one-flavor lattice Schwinger model is studied as a concrete example with ${\cal N}_c=1$ and ${\cal N}=1$.

\section{Chiral anomaly on the lattice}
\label{lgt4}
We shall see in section (\ref{nini}) that  under some mild assumption about the lattice action, there are always 
an equal number of left- and right-handed Weyl particles in the continuum limit of a lattice fermionic theory. 
In the continuum the axial or chiral charge $Q_{5}$ is defined as the integral of the chiral density of the current 
$j_{5}^{\mu}(x)=\overline{\psi}(x)\gamma^{\mu}\gamma_{5}\psi(x)$ 
\begin{equation}
Q_{5}=\int d^{3}x  j_{5}^{0}(x)=\int d^{3}x\psi^{\dagger}(\vec{x},t)\gamma_{5} \psi(\vec{x},t) 
\label{q5}
\end{equation}
According to Eq.(\ref{q5}) $Q_{5}$ is the number of right-handed fermions minus the number of left-handed fermions minus the number of 
right-handed antifermions plus the number of left-handed antifermions. 
In the case of massless fermions $Q_{5}$ is conserved at the classical level, not only in the case of free theories, but also  in the 
case of interacting theories like $QED$ or $\sigma$-models~\cite{b6}. 
At the quantum level the chiral symmetry is broken by the anomaly. For example in massless QED the divergence of the axial current is given by~\cite{b23} 
\begin{equation}
\partial _{\mu}j_{5}^{\mu}(x)=\frac{e^{2}}{16\pi^{2}}\epsilon_{\mu \nu \rho \sigma}F^{\mu \nu}(x)F^{\rho \sigma}(x)
\label{1anly}
\end{equation}   
There is no way to avoid the physical consequences of the anomaly; one might also define a conserved axial vector current, 
but it is not gauge invariant and the physical consequences are unchanged. 
The axial anomaly appears in perturbation theory as a consequence of the linear divergences in the triangle 
graph with an internal fermion loop, and either three axial vector current insertions or one axial vector and two vector current insertions. 
The one-loop quantum corrections to a chirally symmetric classical action are not chirally symmetric. It is impossible regularize 
a quantum field theory preserving both axial and gauge invariance. 
Since lattice gauge theories are manifestly gauge invariant, it is an interesting question how axial anomaly appears there. In general the fermion doublers 
do contribute to the anomaly. Since the doublers appear with opposite chiralities, one might conclude that the 
anomaly on the lattice is zero. However the question is more involved~\cite{b28}.

I am interested in analyzing how the anomaly manifests itself in the lattice Schwinger models with different flavor groups. 
I already illustrated that staggered fermions explicitly break the $U_{A}(1)$ symmetry but possess a discrete chiral symmetry 
corresponding to translations by one lattice spacing. As we shall see in chapters 4, 5 and 6 in the continuum the $U_{A}(1)$ symmetry is broken by the anomaly 
either in the one-flavor or in the multiflavor Schwinger models. In chapter 4 we shall see that on the lattice in the one-flavor case 
the discrete chiral symmetry is broken spontaneously. At variance, in the two-flavor model we shall see in chapter 5 
that the ground state is translationally invariant, $i.e.$ does not break the discrete chiral symmetry. The only relic of the continuum 
axial anomaly is the non-zero VEV of the umklapp operator. 
   
\subsection{The Nielsen-Ninomiya no-go theorem}
\label{nini}

Nielsen and Ninomiya~\cite{b24} demonstrated that fermion theories on a lattice have an equal number 
of species of left- and right-handed Weyl particles, $i.e.$ neutrinos, in the continuum limit. 
The consequence of this behavior is a no-go theorem for putting theories of weak interactions on a 
lattice. 
Moreover it is not possible in strong interaction models to solve the species doubling problem on a 
lattice in a chirally invariant way. The appearence of equally many right- and left-handed species of 
Weyl particles is an unavoidable consequence of a lattice theory under some mild assumptions. 

The hypothesis of this theorem are locality of the theory and exact conservation of discrete valued 
quantum numbers and charges assumed to have a density defined from a finite region. 

If one is interested in the strong interactions and wants to study a Dirac particle rather than a 
Weyl particle, there is no unsurmountable  problem because a Dirac particle can be thought of as a composite 
of two Weyl particles, a right-handed one and a left-handed one. 
However there is no way to construct chiral invariant lattice models for $QCD$. 

Let us consider the general class of lattice fermion theories for which the bilinear part of the action for the 
N-component complex fermion field $\psi(\vec{x})$ is of the form
\begin{equation}
S_{F}=-i\int dt \sum_{\vec{x}} \dot{\overline{\psi}}(\vec{x})\psi(\vec{x})-
\int dt \sum_{\vec{x},\vec{y}}\overline{\psi}(\vec{x})H(\vec{x}-\vec{y})\psi(\vec{y})
\label{ssff}
\end{equation}
$$
\psi =\left( \begin{array}{c}
\psi_{1}\\
\vdots\\
\psi_{N}
\end{array}
\right) 
$$
with H an Hamiltonian where interactions of quartic or higher degree in $\psi$ are neglected. 
In fact, one can define the number of species of Weyl fermions depending only on the bilinear part 
of the action, while the interactions do not change the dispersion relation but just cause scattering processes. 
The action $S_{F}$ should fulfill the following three conditions
\begin{itemize}

\item Locality of the interaction, $i.e.$ $H(\vec{x})\rightarrow 0$ fast enough as $|\vec{x}|\rightarrow \infty$ 
that its Fourier transform $\tilde{H}(\vec{p})$ is a smooth function. Thus the eigenvalues $\omega_{i}(\vec{p})$ 
$(i=1,\dots,N)$ are smooth except in the case that two eigenvalues $\omega_{i}(\vec{p})$ and 
$\omega_{i+1}(\vec{p})$ coincide
 
\item Translational invariance on the lattice

\item Hermiticity of the $N\times N$ matrix Hamiltonian H.

\end{itemize}
The assumptions made for the charges Q (lepton number, for instance) are
\begin{itemize}

\item Exact conservation of Q, even at the scales where the lattice cut-off is relevant

\item Q is locally defined $Q=\sum_{\vec{x}} \rho(\vec{x})$. The charge density $\rho(\vec{x})$ is a 
function of the field variables $\psi(\vec{y})$ related to $\vec{y}$ within a bounded distance from $\vec{x}$.

\item Q is quantized

\item Q is bilinear in the fermion field $\psi(\vec{x})$.

\end{itemize}
The weak interactions and the Standard Model exhibit right- and left-handed particles that do not have the same hypercharge, so that parity 
and charge conjugation are broken in weak interaction processes. 
This implies that if nature were indeed built on a lattice, it should be possible to realize Weyl fermions with different 
quantum numbers for left- and right-handed ones.

One has thus to understand where is the problem: one must give up the idea that nature is based on a fundamental lattice cut-off 
or some assumption of the no-go theorem, or weak interactions phenomenology, $i.e.$ the usual understanding of parity violation. 

Nielsen and Ninomiya gave two proofs of their no-go theorem, a very technical one based on algebraic topology and homotopy and 
a more intuitive topological proof. 
The proof is mainly supported by the fact that the momentum space of lattice theory is periodic, $i.e.$ there is the Brillouin 
zone 
\begin{equation}
-\frac{\pi}{a}\leq p_{i}\leq \frac{\pi}{a}
\end{equation}
where the end points $\pi/a$ and $-\pi/a$ are identified. From a topological point of view $p_{i}$ runs on a circle $S^{1}$, 
thus the momentum space makes up a hypertorus $S^{1}\otimes S^{1}\otimes S^{1}$.
 
\begin{figure}[htb]
\begin{center}
\setlength{\unitlength}{0.00041700in}%
\begingroup\makeatletter\ifx\SetFigFont\undefined
\def\x#1#2#3#4#5#6#7\relax{\def\x{#1#2#3#4#5#6}}%
\expandafter\x\fmtname xxxxxx\relax \def\y{splain}%
\ifx\x\y   
\gdef\SetFigFont#1#2#3{%
  \ifnum #1<17\tiny\else \ifnum #1<20\small\else
  \ifnum #1<24\normalsize\else \ifnum #1<29\large\else
  \ifnum #1<34\Large\else \ifnum #1<41\LARGE\else
     \huge\fi\fi\fi\fi\fi\fi
  \csname #3\endcsname}%
\else
\gdef\SetFigFont#1#2#3{\begingroup
  \count@#1\relax \ifnum 25<\count@\count@25\fi
  \def\x{\endgroup\@setsize\SetFigFont{#2pt}}%
  \expandafter\x
    \csname \romannumeral\the\count@ pt\expandafter\endcsname
    \csname @\romannumeral\the\count@ pt\endcsname
  \csname #3\endcsname}%
\fi
\fi\endgroup
\begin{picture}(9624,7224)(1189,-6973)
\thicklines
\put(1201,-3361){\vector( 1, 0){9600}}
\put(2401,239){\line( 0,-1){7200}}
\put(9601,239){\line( 0,-1){7200}}
\multiput(2401,-286)(15.00000,7.50000){11}{\makebox(13.3333,20.0000){\SetFigFont{7}{8.4}{rm}.}}
\multiput(2551,-211)(16.07143,5.35714){15}{\makebox(13.3333,20.0000){\SetFigFont{7}{8.4}{rm}.}}
\put(2776,-136){\line( 1, 0){150}}
\multiput(2926,-136)(15.00000,-7.50000){11}{\makebox(13.3333,20.0000){\SetFigFont{7}{8.4}{rm}.}}
\multiput(3076,-211)(12.50000,-12.50000){13}{\makebox(13.3333,20.0000){\SetFigFont{7}{8.4}{rm}.}}
\multiput(3226,-361)(9.37500,-14.06250){17}{\makebox(13.3333,20.0000){\SetFigFont{7}{8.4}{rm}.}}
\multiput(3376,-586)(4.16667,-16.66667){19}{\makebox(13.3333,20.0000){\SetFigFont{7}{8.4}{rm}.}}
\multiput(3451,-886)(7.50000,-15.00000){21}{\makebox(13.3333,20.0000){\SetFigFont{7}{8.4}{rm}.}}
\multiput(3601,-1186)(6.25000,-15.62500){25}{\makebox(13.3333,20.0000){\SetFigFont{7}{8.4}{rm}.}}
\put(3751,-1561){\line( 1,-4){132.353}}
\put(3901,-2086){\line( 3,-5){225}}
\multiput(4126,-2461)(4.16667,-16.66667){19}{\makebox(13.3333,20.0000){\SetFigFont{7}{8.4}{rm}.}}
\put(4201,-2761){\line( 1,-2){225}}
\multiput(4426,-3211)(6.25000,-15.62500){25}{\makebox(13.3333,20.0000){\SetFigFont{7}{8.4}{rm}.}}
\multiput(4576,-3586)(12.00000,-12.00000){26}{\makebox(13.3333,20.0000){\SetFigFont{7}{8.4}{rm}.}}
\put(4876,-3886){\line( 1, 0){150}}
\multiput(5026,-3886)(15.00000,7.50000){11}{\makebox(13.3333,20.0000){\SetFigFont{7}{8.4}{rm}.}}
\multiput(5176,-3811)(11.84211,11.84211){20}{\makebox(13.3333,20.0000){\SetFigFont{7}{8.4}{rm}.}}
\multiput(5401,-3586)(12.50000,12.50000){13}{\makebox(13.3333,20.0000){\SetFigFont{7}{8.4}{rm}.}}
\multiput(5551,-3436)(9.37500,14.06250){17}{\makebox(13.3333,20.0000){\SetFigFont{7}{8.4}{rm}.}}
\multiput(5701,-3211)(12.50000,12.50000){13}{\makebox(13.3333,20.0000){\SetFigFont{7}{8.4}{rm}.}}
\multiput(5851,-3061)(14.06250,9.37500){17}{\makebox(13.3333,20.0000){\SetFigFont{7}{8.4}{rm}.}}
\put(6076,-2911){\line( 1, 0){225}}
\multiput(6301,-2911)(12.50000,-12.50000){13}{\makebox(13.3333,20.0000){\SetFigFont{7}{8.4}{rm}.}}
\multiput(6451,-3061)(9.37500,-14.06250){17}{\makebox(13.3333,20.0000){\SetFigFont{7}{8.4}{rm}.}}
\multiput(6601,-3286)(7.50000,-15.00000){11}{\makebox(13.3333,20.0000){\SetFigFont{7}{8.4}{rm}.}}
\multiput(6676,-3436)(11.84211,-11.84211){20}{\makebox(13.3333,20.0000){\SetFigFont{7}{8.4}{rm}.}}
\multiput(6901,-3661)(16.66667,-4.16667){19}{\makebox(13.3333,20.0000){\SetFigFont{7}{8.4}{rm}.}}
\put(7201,-3736){\line( 5, 4){375}}
\multiput(7576,-3436)(10.22727,13.63636){23}{\makebox(13.3333,20.0000){\SetFigFont{7}{8.4}{rm}.}}
\multiput(7801,-3136)(6.25000,15.62500){25}{\makebox(13.3333,20.0000){\SetFigFont{7}{8.4}{rm}.}}
\put(7951,-2761){\line( 3, 5){225}}
\multiput(8176,-2386)(7.50000,15.00000){21}{\makebox(13.3333,20.0000){\SetFigFont{7}{8.4}{rm}.}}
\multiput(8326,-2086)(10.22727,13.63636){23}{\makebox(13.3333,20.0000){\SetFigFont{7}{8.4}{rm}.}}
\multiput(8551,-1786)(11.84211,11.84211){20}{\makebox(13.3333,20.0000){\SetFigFont{7}{8.4}{rm}.}}
\multiput(8776,-1561)(12.50000,12.50000){13}{\makebox(13.3333,20.0000){\SetFigFont{7}{8.4}{rm}.}}
\multiput(8926,-1411)(15.00000,7.50000){21}{\makebox(13.3333,20.0000){\SetFigFont{7}{8.4}{rm}.}}
\multiput(9226,-1261)(9.37500,14.06250){17}{\makebox(13.3333,20.0000){\SetFigFont{7}{8.4}{rm}.}}
\multiput(9376,-1036)(15.00000,7.50000){11}{\makebox(13.3333,20.0000){\SetFigFont{7}{8.4}{rm}.}}
\multiput(9526,-961)(7.50000,15.00000){11}{\makebox(13.3333,20.0000){\SetFigFont{7}{8.4}{rm}.}}
\put(6001,-6961){\vector( 0, 1){7200}}
\multiput(2401,-2161)(9.37500,-14.06250){17}{\makebox(13.3333,20.0000){\SetFigFont{7}{8.4}{rm}.}}
\multiput(2551,-2386)(7.50000,-15.00000){21}{\makebox(13.3333,20.0000){\SetFigFont{7}{8.4}{rm}.}}
\multiput(2701,-2686)(5.35714,-16.07143){15}{\makebox(13.3333,20.0000){\SetFigFont{7}{8.4}{rm}.}}
\put(2776,-2911){\line( 3,-5){225}}
\multiput(3001,-3286)(9.37500,-14.06250){17}{\makebox(13.3333,20.0000){\SetFigFont{7}{8.4}{rm}.}}
\multiput(3151,-3511)(10.22727,-13.63636){23}{\makebox(13.3333,20.0000){\SetFigFont{7}{8.4}{rm}.}}
\multiput(3376,-3811)(9.37500,-14.06250){17}{\makebox(13.3333,20.0000){\SetFigFont{7}{8.4}{rm}.}}
\multiput(3526,-4036)(6.25000,-15.62500){25}{\makebox(13.3333,20.0000){\SetFigFont{7}{8.4}{rm}.}}
\multiput(3676,-4411)(14.06250,-9.37500){17}{\makebox(13.3333,20.0000){\SetFigFont{7}{8.4}{rm}.}}
\multiput(3901,-4561)(11.84211,-11.84211){20}{\makebox(13.3333,20.0000){\SetFigFont{7}{8.4}{rm}.}}
\multiput(4126,-4786)(15.62500,-6.25000){25}{\makebox(13.3333,20.0000){\SetFigFont{7}{8.4}{rm}.}}
\multiput(4501,-4936)(16.07143,5.35714){15}{\makebox(13.3333,20.0000){\SetFigFont{7}{8.4}{rm}.}}
\put(4726,-4861){\line( 1, 0){525}}
\put(5251,-4861){\line( 1, 0){225}}
\multiput(5476,-4861)(15.00000,7.50000){21}{\makebox(13.3333,20.0000){\SetFigFont{7}{8.4}{rm}.}}
\multiput(5776,-4711)(17.04545,3.40909){23}{\makebox(13.3333,20.0000){\SetFigFont{7}{8.4}{rm}.}}
\put(6151,-4636){\line( 1, 0){225}}
\multiput(6376,-4636)(15.62500,6.25000){25}{\makebox(13.3333,20.0000){\SetFigFont{7}{8.4}{rm}.}}
\multiput(6751,-4486)(15.62500,6.25000){25}{\makebox(13.3333,20.0000){\SetFigFont{7}{8.4}{rm}.}}
\multiput(7126,-4336)(13.63636,10.22727){23}{\makebox(13.3333,20.0000){\SetFigFont{7}{8.4}{rm}.}}
\multiput(7426,-4111)(9.37500,14.06250){17}{\makebox(13.3333,20.0000){\SetFigFont{7}{8.4}{rm}.}}
\multiput(7576,-3886)(15.00000,7.50000){11}{\makebox(13.3333,20.0000){\SetFigFont{7}{8.4}{rm}.}}
\multiput(7726,-3811)(12.50000,12.50000){13}{\makebox(13.3333,20.0000){\SetFigFont{7}{8.4}{rm}.}}
\multiput(7876,-3661)(10.22727,13.63636){23}{\makebox(13.3333,20.0000){\SetFigFont{7}{8.4}{rm}.}}
\multiput(8101,-3361)(12.50000,12.50000){13}{\makebox(13.3333,20.0000){\SetFigFont{7}{8.4}{rm}.}}
\multiput(8251,-3211)(5.35714,16.07143){15}{\makebox(13.3333,20.0000){\SetFigFont{7}{8.4}{rm}.}}
\multiput(8326,-2986)(11.84211,11.84211){20}{\makebox(13.3333,20.0000){\SetFigFont{7}{8.4}{rm}.}}
\multiput(8551,-2761)(11.84211,11.84211){20}{\makebox(13.3333,20.0000){\SetFigFont{7}{8.4}{rm}.}}
\multiput(8776,-2536)(15.00000,7.50000){11}{\makebox(13.3333,20.0000){\SetFigFont{7}{8.4}{rm}.}}
\multiput(8926,-2461)(12.50000,12.50000){7}{\makebox(13.3333,20.0000){\SetFigFont{7}{8.4}{rm}.}}
\multiput(9001,-2386)(16.07143,5.35714){15}{\makebox(13.3333,20.0000){\SetFigFont{7}{8.4}{rm}.}}
\multiput(9226,-2311)(16.07143,-5.35714){15}{\makebox(13.3333,20.0000){\SetFigFont{7}{8.4}{rm}.}}
\multiput(9451,-2386)(7.50000,-15.00000){11}{\makebox(13.3333,20.0000){\SetFigFont{7}{8.4}{rm}.}}
\multiput(9526,-2536)(12.50000,-12.50000){7}{\makebox(13.3333,20.0000){\SetFigFont{7}{8.4}{rm}.}}
\put(1501,-3661){\makebox(0,0)[lb]{\smash{\SetFigFont{10}{12.0}{rm}$-\pi/a$}}}
\put(2401,-4786){\line( 2,-1){750}}
\put(3151,-5161){\line( 4,-1){529.412}}
\put(3676,-5311){\line( 6,-1){522.973}}
\put(4201,-5386){\line( 4,-1){900}}
\put(5101,-5611){\line( 1, 0){375}}
\put(5476,-5611){\line( 6, 1){522.973}}
\multiput(6001,-5536)(15.62500,6.25000){25}{\makebox(13.3333,20.0000){\SetFigFont{7}{8.4}{rm}.}}
\multiput(6376,-5386)(17.04545,3.40909){23}{\makebox(13.3333,20.0000){\SetFigFont{7}{8.4}{rm}.}}
\put(6751,-5311){\line( 4, 1){600}}
\put(7351,-5161){\line( 6, 1){595.946}}
\put(7951,-5086){\line( 4, 3){516}}
\put(8476,-4711){\line( 5, 3){375}}
\multiput(8851,-4486)(14.06250,9.37500){17}{\makebox(13.3333,20.0000){\SetFigFont{7}{8.4}{rm}.}}
\put(9076,-4336){\line( 3, 1){450}}
\multiput(9526,-4186)(12.50000,12.50000){7}{\makebox(13.3333,20.0000){\SetFigFont{7}{8.4}{rm}.}}
\put(10726,-3136){\makebox(0,0)[lb]{\smash{\SetFigFont{10}{12.0}{rm}p}}}
\put(6151,-61){\makebox(0,0)[lb]{\smash{\SetFigFont{10}{12.0}{rm}$\omega$}}}
\put(9676,-3661){\makebox(0,0)[lb]{\smash{\SetFigFont{10}{12.0}{rm}$+\pi/a$}}}
\end{picture}
\end{center}
\caption{Dispersion relations for $1+1$ dimensional lattice Weyl fermions. Each curve is closed since end points are identified. 
Each crossing point at $\omega =0$ is a Fermi ``surface" describing one species of Weyl fermions.}
\label{dis1}
\end{figure}
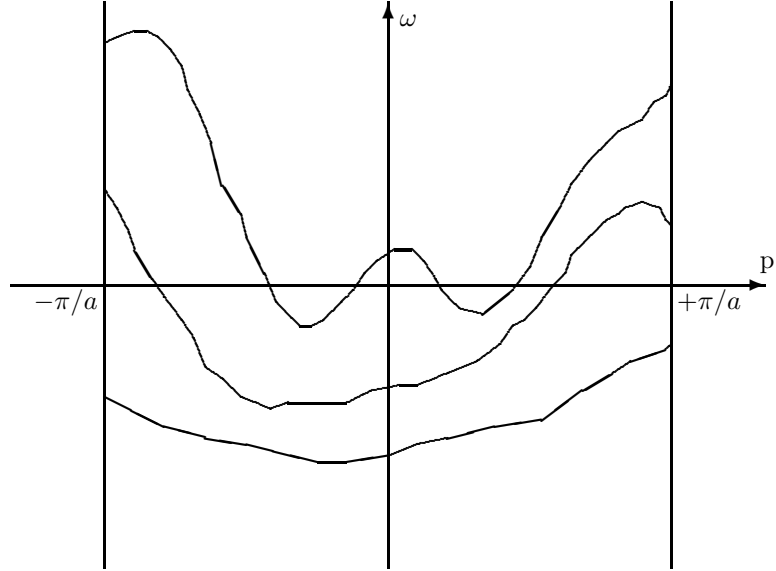

Let review the intuitive proof of the no-go theorem for a general class of $1+1$ dimensional lattice fermion theories, whose action 
is the $1+1$ dimensional version of the $3+1$ dimensional case Eq.(\ref{ssff}). One thus takes $\psi$ to be a complex 
N-component field in order to keep generality.
A generic $1+1$ dimensional Hamiltonian has non degenerate energy levels, in fact, in order to have just two-level 
degeneracy $\omega_{i}(p)=\omega_{i+1}(p)$ three parameters must be restricted in the Hamiltonian H(p), but there is only one p.
However the wave packet velocity
\begin{equation}
v_{i}=\frac{d\omega_{i}}{d p}|_{p=p_{f}}
\end{equation}
is non zero at the Fermi energy
\begin{equation}
\omega_{i}(p_{f})=0
\end{equation}
Low energy excitations are such that fermion states with p close to $p_{f}$ are excited and for particles relevant at low 
energies one has 
\begin{equation}
\omega_{i}(p)=(p-p_{f})\frac{d\omega_{i}}{dp}|_{p=p_{f}}+O((p-p_{f})^{2}) 
\end{equation}
The particle at the crossing point with $\omega_{i}=0$ line is called a right(left) mover when 
\begin{equation}
\frac{d\omega_{i}}{dp}|_{p=p_{f}}>0\quad (<0)
\end{equation}
Antiparticles and particles have the same velocity. 
Assuming locality $\omega_{i}(p)$ is analytic in p except for degeneracy points and a smooth curve 
$\omega_{i}(p)$ is defined in $\omega_{i}-p$ space. The curve is closed since on a lattice p is defined modulo $2\pi /a$ and thus runs on a circle $S_{1}$. 
Such curves must cross equally many times from $\omega_{i}(p)<0$ to 
$\omega_{i}(p)>0$ as they cross from $\omega_{i}(p)>0$ to $\omega_{i}(p)<0$. One can orient the curve such that it is along the 
increasing p direction. The curve going up through $\omega_{i}=0$ represents a right-mover and that going down a left-mover. 
From a topological point of view there must be the same number of upgoings and downgoings for the closed curve $-$ see fig. (\ref{dis1}). 
Therefore, there 
appear equally many right and left movers in the low energy regime. Let us introduce the shifted momentum
\begin{equation}
\tilde{p}=p-p_{f}
\end{equation}
The dispersion relation reads
\begin{equation}
\omega_{i}=\frac{d\omega_{i}}{dp}|_{\tilde{p}=0}\quad \tilde{p}
\end{equation}
This is a relativistically invariant dispersion relation for a massless mover along the right or left direction. 
This is true only when $\frac{d\omega_{i}}{dp}|_{p=p_{f}}$ is considered equal to the velocity of light. \\

Let us now turn our attention to the $3+1$ dimensional case. The eigenvalue equation for the Hamiltonian Eq.(\ref{ssff}) reads
\begin{equation}
H(\vec{p})\psi(\vec{p})=\omega_{i}(\vec{p})\psi(\vec{p})\quad i=1,\ldots,N
\label{H}
\end{equation}
with N ordered eigenvalues $\omega_{1}>\omega_{2}>\ldots >\omega_{N}$. When two energy levels, the i-th and (i+1)-th coincide, 
$i.e.$ are degenerate, at the momentum $\vec{p}_{deg}$, 
\begin{equation}
\omega_{i}(\vec{p}_{deg})=\omega_{i+1}(\vec{p}_{deg})=0
\end{equation}
one can expand the $N\times N$ matrix $H(p)$ near the degeneracy point $\vec{p}_{deg}$. The most general Hamiltonian $H^{(2)}(\vec{p})$ for the 
two component spinor $U(p)$ describing the i-th and (i+1)-th levels, is of the form
\begin{equation}
H^{(2)}(\vec{p})=\omega_{deg}(\vec{p}_{deg})+(\vec{p}-\vec{p}_{deg}) \vec{b}+(\vec{p}-\vec{p}_{deg})_{k}V^{k}_{\alpha}\sigma^{\alpha}+
O((\vec{p}-\vec{p}_{deg})^{2})
\end{equation}
with $\vec{b}$ and $\vec{V}$ constants. By introducing the shifted variables 
\begin{equation}
\tilde{\vec{p}}=\vec{p}-\vec{p}_{deg}\quad \tilde{\omega}=\omega -\omega_{deg}
\end{equation}
$H^{(2)}(\vec{p})$ reads
\begin{equation}
H^{(2)}(\vec{p})=\tilde{\vec{p}}_{k}V^{k}_{\alpha}\sigma^{\alpha}+\tilde{\vec{p}}\quad \vec{b}
\end{equation}
let us define further a new momentum
\begin{eqnarray}
P_{0}&=&\tilde{\omega}-(\vec{p}-\vec{p}_{deg})\vec{b}\\
P_{\alpha}&=&\pm \tilde{\vec{p}}_{k}V^{k}_{\alpha}
\end{eqnarray}
The sign $\pm$ depends on the sign of $det V$ which determines the relative handness of the $\tilde{\vec{p}}$ and $\vec{p}$ 
coordinates. Take conventionally $\tilde{\vec{p}}$ to be the right handed one. $H^{(2)}(\vec{p})$ becomes
\begin{equation}
H^{(2)}(\vec{p})=\vec{\sigma}\cdot \vec{P}
\end{equation}
and one obtains the right and left handed Weyl equation
\begin{equation}
\vec{P}\vec{\sigma}U(\vec{p})=\pm P_{0}U(\vec{p})
\label{psu}
\end{equation}
corresponding to the sign of $detV$. Wheter one has right- or left- handed Weyl fermions depends on the degeneracy point 
since the constant tensor $\vec{V}$ depends on $\vec{p}_{deg}$. 

One may call the degeneracy point a right or left handed one according to  Eq.(\ref{psu}). Each degeneracy point represents one species of Weyl 
fermions. If the Fermi energy surface lies at the degeneracy points of the i-th and (i+1)-th, each degeneracy 
point represents one species of Weyl particles. 

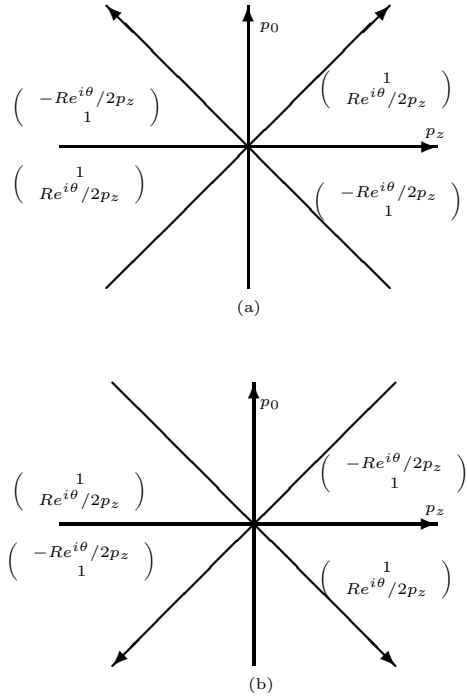
\begin{figure}[htb]
\begin{center}
\setlength{\unitlength}{0.00041700in}%
\begingroup\makeatletter\ifx\SetFigFont\undefined
\def\x#1#2#3#4#5#6#7\relax{\def\x{#1#2#3#4#5#6}}%
\expandafter\x\fmtname xxxxxx\relax \def\y{splain}%
\ifx\x\y   
\gdef\SetFigFont#1#2#3{%
  \ifnum #1<17\tiny\else \ifnum #1<20\small\else
  \ifnum #1<24\normalsize\else \ifnum #1<29\large\else
  \ifnum #1<34\Large\else \ifnum #1<41\LARGE\else
     \huge\fi\fi\fi\fi\fi\fi
  \csname #3\endcsname}%
\else
\gdef\SetFigFont#1#2#3{\begingroup
  \count@#1\relax \ifnum 25<\count@\count@25\fi
  \def\x{\endgroup\@setsize\SetFigFont{#2pt}}%
  \expandafter\x
    \csname \romannumeral\the\count@ pt\expandafter\endcsname
    \csname @\romannumeral\the\count@ pt\endcsname
  \csname #3\endcsname}%
\fi
\fi\endgroup
\begin{picture}(5487,8751)(2926,-8500)
\thicklines
\put(3601,-1561){\vector( 1, 0){4800}}
\put(4201,-3361){\vector( 1, 1){3600}}
\put(6076,-8161){\vector( 0, 1){3600}}
\put(3601,-6361){\vector( 1, 0){4800}}
\put(7876,-4561){\vector(-1,-1){3600}}
\put(4276,-4561){\vector( 1,-1){3600}}
\put(7801,-3361){\vector(-1, 1){3600}}
\put(6151,-61){\makebox(0,0)[lb]{\smash{\SetFigFont{6}{7.2}{rm}$p_0$}}}
\put(6151,-4861){\makebox(0,0)[lb]{\smash{\SetFigFont{6}{7.2}{rm}$p_0$}}}
\put(8251,-1411){\makebox(0,0)[lb]{\smash{\SetFigFont{6}{7.2}{rm}$p_z$}}}
\put(6001,-3361){\vector( 0, 1){3600}}
\put(8251,-6211){\makebox(0,0)[lb]{\smash{\SetFigFont{6}{7.2}{rm}$p_z$}}}
\put(2926,-6886){\makebox(0,0)[lb]{\smash{\SetFigFont{6}{7.2}{rm}$\left(\begin{array}{c}-Re^{i\theta}/2p_z\\1\end{array}\right)$}}}
\put(6901,-5761){\makebox(0,0)[lb]{\smash{\SetFigFont{6}{7.2}{rm}$\left(\begin{array}{c}-Re^{i\theta}/2p_z\\1\end{array}\right)$}}}
\put(5851,-3661){\makebox(0,0)[lb]{\smash{\SetFigFont{6}{7.2}{rm}(a)}}}
\put(6001,-8461){\makebox(0,0)[lb]{\smash{\SetFigFont{6}{7.2}{rm}(b)}}}
\put(6901,-886){\makebox(0,0)[lb]{\smash{\SetFigFont{6}{7.2}{rm}$\left(\begin{array}{c}1\\Re^{i\theta}/2p_z\end{array}\right)$}}}
\put(6826,-2311){\makebox(0,0)[lb]{\smash{\SetFigFont{6}{7.2}{rm}$\left(\begin{array}{c}-Re^{i\theta}/2p_z\\1\end{array}\right)$}}}
\put(6901,-7111){\makebox(0,0)[lb]{\smash{\SetFigFont{6}{7.2}{rm}$\left(\begin{array}{c}1\\Re^{i\theta}/2p_z\end{array}\right)$}}}
\put(3001,-1111){\makebox(0,0)[lb]{\smash{\SetFigFont{6}{7.2}{rm}$\left(\begin{array}{c}-Re^{i\theta}/2p_z\\1\end{array}\right)$}}}
\put(3001,-2086){\makebox(0,0)[lb]{\smash{\SetFigFont{6}{7.2}{rm}$\left(\begin{array}{c}1\\Re^{i\theta}/2p_z\end{array}\right)$}}}
\put(3001,-5986){\makebox(0,0)[lb]{\smash{\SetFigFont{6}{7.2}{rm}$\left(\begin{array}{c}1\\Re^{i\theta}/2p_z\end{array}\right)$}}}
\end{picture}
\end{center}
\caption{Around a degeneracy point all the curves of the type \protect $<a|\omega_{i}(\vec{p})>$ are oriented away on one level and inward on the other one.}
\label{orie}
\end{figure}

Let us consider curves in 4-dimensional $\omega-\vec{p}$ space or on the 3-dimensional dispersion relation surface defined by
\begin{equation}
\left\{ ( \vec{p},\omega_{i}(\vec{p}) ) \quad | \quad <a| \omega_{i} ( \vec{p} ) >=0 \right\}
\end{equation}
where
\begin{equation}
<a|\omega_{i}(\vec{p})>=a_{1}\psi_{1}^{(i)}+a_{2}\psi_{2}^{(i)}+\ldots+a_{N}\psi_{N}^{(i)}
\label{ao}
\end{equation}
The vector $|a>=(a_{1},a_{2},\ldots,a_{N})$ is an arbitrarily chosen constant in the complex N-dimensional vector space, 
which may be chosen to be  the basic vector corresponding to the field number one. Eq.(\ref{ao}) specifies a 1-dimensional curve 
in the generic case, in fact Eq.(\ref{ao}) fixes two variables since $<a|\omega_{i}(\vec{p}>$ is a complex number and it is a continuous 
and analytic complex-valued function of $\vec{p}$. 
The set of curves always passes through all the degeneracy points, as can be shown explicitly using the continuum 2-component Weyl 
equation near the degeneracy point $\vec{p}_{deg}$. Then one can construct a new state 
\begin{equation}
|\overline{\omega}_{i}(\vec{p})>=\alpha|\omega_{i}(\vec{p})>+\beta|\omega_{i+1}(\vec{p})>
\end{equation}
near $\vec{p}=\vec{p}_{deg}$. One can always choose $\alpha$ and $\beta$ such that 
\begin{equation}
<a|\overline{\omega}_{i}(\vec{p})>=0
\end{equation}
which is the condition of Eq.(\ref{ao}) for the curve. Due to the existence of the Brillouin zone and the fact that the curves 
continuously pass through all the degeneracy points between the energy levels $\omega_{i}$ and $\omega_{i+1}$, such a curve must be 
closed or consists of a number of closed curves.  If the curves are oriented  they must pass equally many 
times upward and downward between the energy levels $\omega_{i}$ and $\omega_{i+1}$, through all the degeneracy points. 
Let us now orient the curves and show that, when the curve crosses the level $\omega=0$ between the levels $\omega_{i}$ and 
$\omega_{i+1}$ upward (or downward), the crossing point, $i.e.$ the degeneracy point is a right- (or left-) handed degeneracy point. 
The eigenstate determined by Eq.(\ref{H}) is unique modulo a phase factor if one imposes the normalization condition 
$<\omega_{i}(\vec{p})|\omega_{i}(\vec{p})>=1$. 
One can choose the phase of $|\omega_{i}(\vec{p})>$ to be analytic in a simply connected region which should not include the degeneracy 
points. An orientation of the curves is assigned by means of the phase rotation of $<a|\omega_{i}(\vec{p})>$ on a small 
circle $S_{1}$ around the curve (\ref{ao}). The convention is that an increase of phase on $S_{1}$ should form a right-handed  screw together 
with the oriented curve when one takes a right-handed coordinate convention for $\vec{p}$. 
Let us consider for example the case the curve is along $p_{z}>0$ on the upper cone $p_{0}>0$ near the right handed degeneracy points. 
The two component field U satisfies the equation 
\begin{equation}
\vec{P}\vec{\sigma}U(\vec{p})=P_{0}U(\vec{p})
\label{ppssuu}
\end{equation}
One draws a circle $S_{1}$ with radius R around a point on the curve  at a distance d from the degeneracy point and let us 
take $R\ll d$. The point Q on $S_{1}$ is parametrized in the following way 
\begin{eqnarray}
p_{x}&=&R\cos \theta\\
p_{y}&=&R\sin \theta\\
p_{z}&=&d
\end{eqnarray}
The normalized eigenvector U of Eq.(\ref{ppssuu}) is solved in the limit $\frac{R}{d}\ll 1$
\begin{equation}
U=\left( \begin{array}{c}
1\\
\frac{R}{2d}e^{i\theta}
\end{array}
\right)\quad.
\end{equation}
Since one obtains 
\begin{equation}
U=\left( \begin{array}{c}
1\\
0
\end{array}
\right)\quad.
\end{equation} 
on the curve $R=0$, the vector $<a|$ must be $<a|=(0,1)$ so that on $S_{1}$
\begin{equation}
<a|\omega_{i}>=\frac{R}{2p_{z}}e^{i\theta}
\end{equation}
and according to the orientation convention the curve is oriented in the positive $p_{z}$ direction because it forms a 
right handed screw. Computing the case of $p_{0}>0$ and $p_{z}<0$, $p_{0}<0$ and $p_{z}<0$, $p_{0}<0$ and $p_{z}>0$ one obtains fig. (\ref{orie}a) 
for the right handed degeneracy point. The case of left handed degeneracy point is depicted in fig. (\ref{orie}b). 
As one can see in fig. (\ref{orie}), the curve continues through the degeneracy points conserving the orientation. 
The orientation determined in this way becomes the same all along the curve. The important fact is that the curve must be oriented 
away from the degeneracy point on the level sheet with right-handed particles ($i.e.$ with positive helicity), while the curve 
must be inward oriented on the level with left-handed particles. 
Finally the oriented curve is closed and thus must go equally many times upward and downward through all the degeneracy 
points $\omega_{i}(\vec{p}_{deg})=\omega_{i+1}(\vec{p}_{deg})=0$ between the energy levels $\omega_{i}(\vec{p})$ and $\omega_{i+1}(\vec{p})$. 
When the Dirac sea ($\omega_{i+1}$ level) is filled and the Fermi surface lies at the degeneracy point $\omega =0$, there appear 
equally many species of right- and left-handed Weyl fermions. This concludes the proof of this no-go theorem of Nielsen and Ninomiya.

The strong interaction lattice models for $QCD$ invented by Wilson, Susskind and Banks and Casher break chiral invariance in order 
to avoid the fermion doubling. 
Drell, Weinstein and Yankielowicz were able to construct a chiral invariant model for lattice $QCD$ but it breaks locality of the interaction. 
There is some proposal in the literature to avoid the no-go theorem, giving up some of its hypothesis, and to put the Standard Model on the 
lattice; but up to now no generally accepted scheme does exist.

Nielsen and Ninomiya pointed also out a similarity between the fermion system of lattice gauge theories and the electron system of crystals. 
More precisely they pointed out that there exists an effect analogous to the mechanism of the Adler-Bell-Jackiw (ABJ) axial anomaly in 
solid state physics. The main similarity between lattice fermions and ``real" electrons in crystals is that in both systems there is 
only the lattice translational invariance which implies the Brillouin zone existence. The crystal electrons are described 
by a one component non-relativistic Schr\"{o}dinger equation, but the energy eigenvalues form bands. 
It is possible to write the electron theory as a multicomponent lattice fermion theory with a matrix Hamiltonian. 
At points where two energy bands of the electrons make a contact in the energy-momentum dispersion law space, the two band wave functions of the 
electrons can be approximated by the Weyl equation, and so, according to the no-go theorem, such degeneracy points occur in pairs of the right-handed and left-handed type. 
There is no net production of electrons like the abscence of the net production of Weyl fermions in lattice fermions theories when parallel electric, E, 
and magnetic, H, fields are put on. This leads to the conservation of axial charge. 
The ABJ anomaly manifests itself by transferring electrons from the neighborhood of the left-handed degeneracy point to the 
right-handed one in energy-momentum space. 
In a parity non-invariant zero-gap semiconductor the effect analogous to the ABJ anomaly gives rise to a peculiar behavior of the 
conductivity of the electric current in the presence of the magnetic field.      

In chapter 5 I shall compute the VEV of the operator $F=\overline{\psi}_{L}^{(2)}\overline{\psi}_{L}^{(1)}\psi_{R}^{(1)}\psi_{R}^{(2)}$, describing an umklapp 
process in which two right-movers are annihilated and two left-movers are created. I shall show that VEV of the umklapp operator is non-zero 
due to the coupling of left- and right-movers by the gauge field. 
Right-handed fermions are annihilated and left-handed are created at a degeneracy point so that there is no net production of electrons. 
\chapter{Antiferromagnetic spin chains}
The title of this chapter deserves an entire book to be properly developed, and in fact some books, 
reviews and long articles about spin chains do already exist~\cite{b31}. 
Spin chains interested mathematical physicists for their exact solvability, field theorists for the possibility to test their methods 
by comparation with exact results and condensed matter physicists for the possibility to understand the spectra of real physical systems. 
It is my purpose to report a self-contained analysis of antiferromagnetic spin chains, useful to understand chapter 5 and 6 where I shall show 
that the multiflavor lattice Schwinger models in the strong coupling limit are effectively described by the 
spin-$1/2$ $SU({\cal N})$ antiferromagnetic Heisenberg chains. 

A common tool to study spin systems is spin-wave theory. Spin-wave theory was developed by Anderson, Kubo and others~\cite{b32} in the 1950s 
and it is a very powerful approach to quantum magnetic systems in dimensions greater than one, predicting long-range order and gapless Goldstone bosons. 
The situation remained clouded for magnetic  chains because it was known that long-range order could not occur in one dimension, but the Bethe ansatz 
predicts massless excitations. There is ``quasi-long-range-order" corresponding to a power law decay of spin-spin correlators and there are gapless excitations 
which are not true Goldstone bosons. 

Quantum spin chains have been extensively studied in the literature, starting from the seminal paper by Hans Bethe~\cite{b33} in 1931 for the spin-$1/2$ 
case, where he introduced an ansatz to write down the eigenfunctions of the Heisenberg Hamiltonian, describing a chain with periodic boundary conditions. 
Seven years later Hulthen~\cite{b34} was able to compute the ground state energy of the antiferromagnetic Heisenberg chain. We had to wait until 1984 
to know exactly the spectrum of this model, when Faddeev and Takhtadzhyan~\cite{b27} analyzed the model using the algebraic Bethe ansatz 
and showed that the only one-particle excitation is a doublet of spin-$1/2$ quantum 
excitations with gapless dispersion relation. This excitation is a kink rather than an ordinary particle and is called spinon. All the eigenstates of the 
antiferromagnetic Heisenberg Hamiltonian contain an even number of kinks, nevertheless the kinks are localizable objects and one can consider 
their scattering. There are no bound states of kinks in this model. 

The one dimensional spin systems are not only unusual because they are exactly solvable but also for their incompatibility with long range order. 
Actually solvability and absence of long range order are deeply related concepts; systems with spontaneously broken symmetries are more difficult to 
describe and resist analytical solutions. All isotropic antiferromagnetic quantum spin chains with short range interactions 
exhibit quantum spin-liquid ground states $-$ $i.e.$ states with short range antiferromagnetic correlations and no order. Moreover these 
chains have strange quantum elementary excitations above these ground states, that are not ordinary spin waves but are usually called spinons $-$ neutral 
spin-$1/2$ kinks. 

As R. B. Laughlin~\cite{b35} points out:``spinons are not ``like" anything familiar to most of us, but are instead an important and beautiful instance of 
fractional  quantization, the physical phenomenon in which particles carrying pieces of a fundamental quantum number, such as charge or spin, are created 
as a collective motion of many conventional particles obeying quantum mechanical laws. The fractional quantum number of the spinon is its spin. It is 
fractional because the particles out of which the magnetic states are constructed are spin flips, which carry spin 1."

One significant result of my research~\cite{b36,b36b} has been to show explicitly how spinons appear in lattice gauge theories. More precisely I showed that the massless 
excitations of the two-flavor Schwinger model coincide with the spinons of the spin-$1/2$ antiferromagnetic Heisenberg model.\\

Section (\ref{fsw}) shows the failure of spin wave theory in one dimension, due to the infrared divergencies. Spin-spin correlation functions are computed in 
the spin wave theory and are compared with exact expressions.
 
In section (\ref{hash}) I review the Haldane-Shastry model~\cite{b37}, a spin model discovered independently by Haldane and Shastry and in the context of one dimensional 
Bose gas by Sutherland. The Haldane-Shastry model is a realization of the spin-$1/2$ Heisenberg chain in which the quantum disordered 
ground state and spinon excitations are particularly easy to understand. 

Section (\ref{baa}) is devoted to review the Bethe ansatz solution of the antiferromagnetic Heisenberg chain. Moreover I shall compare the exact solution 
given in \cite{b27} with a study of finite size chains of 4,6, and 8 sites. An original result presented is the thermodynamic 
limit coefficient of the states with $N-2$ domain walls composing the ground state. The spin-spin correlation function 
$\sum_{x}<g.s.|\vec{S}_{x}\cdot \vec{S}_{x+2}|g.s.>$ originally computed by M. Takahashi~\cite{b38}, is derived in a slightly different way. The vacuum expectation 
value of the square of the vector $\vec{V}=\sum_{x}\vec{S}_{x}\wedge \vec{S}_{x+1}$ is computed in an approximate way. 

In section (\ref{sunc}) I introduce the spin-$1/2$ $SU({\cal N})$ antiferromagnetic Heisenberg chains. To provide an intuitive picture I study 
the $SU(3)$ two sites chain and I find the ground state. A short review of the literature on $SU({\cal N})$ antiferromagnetic Heisenberg chains 
is provided.

\section{Failure of spin wave theory in D=1}
\label{fsw}

In the limit of large spin $S\longrightarrow \infty$ the quantum nature of a magnetic system can be forgotten, and one can 
consider the spin $\vec{S}$ as a classical vector in the three dimensional space. In fact in this limit the spin commutator is much smaller than the square 
of the spin variables
\begin{equation}
\left[ S^{a},S^{b}\right]=i\epsilon^{abc}S^{c}=O(S)\ll O(S^{2})
\end{equation}
The classical ground states of the Heisenberg model are states with spins parallel for the ferromagnetic case 
\begin{center}
$$
\uparrow\uparrow\uparrow\uparrow\uparrow\uparrow\uparrow\uparrow\uparrow\uparrow\ldots
$$
\end{center}
or neighbouring spins 
antiparallel for the antiferromagnetic case
\begin{center}
$$
\uparrow\downarrow\uparrow\downarrow\uparrow\downarrow\uparrow\downarrow\uparrow\downarrow\ldots
$$
\end{center}
The classical ground state of the antiferromagnet is called N\'eel state. 
By introducing the raising and lowering spin operators
\begin{equation}
S^{\pm}_{x}=S_{x}^{1}\pm iS_{x}^{2}
\end{equation}
the antiferromagnetic ($J>0$) Heisenberg Hamiltonian reads
\begin{equation}
H_{J}=J\sum_{x=1}^{N}\left\{ S^{3}_{x}S_{x+1}^{3}+\frac{1}{2}(S^{+}_{x}S^{-}_{x+1}+S^{-}_{x}S^{+}_{x+1})\right\}
\label{hhej}
\end{equation}
The classical ferromagnetic ground state coincides with the quantum one, but the N\'eel state is not an eigenstate of $H_{J}$ due to the 
raising and lowering spin operators action. 

It is an interesting question to investigate whether or not long-range antiferromagnetic order 
occurs
\begin{equation}
<S^{3}_{x}S^{3}_{0}>=\pm m
\end{equation}
where the sign depends whether x belongs to the even or odd sublattice. 
The spin wave approach to this problem is to perturb away from the limit in which the N\'eel state is exact. 
This happens of course for a spin-S Heisenberg antiferromagnet in the limit $S\longrightarrow \infty$. 
In this limit one expects that the fluctuations of $S^{3}$ around $+S$ on one sublattice and $-S$ on the other sublattice are very small. 
It is customary to study the effects of these fluctuations using the Holstein-Primakov transformation, 
which consists in parametrizing the spin operators in terms of 
boson operators $a$ and $a^{\dagger}$ on the sublattice of even sites A and $b$ and $b^{\dagger}$ on the sublattice of odd sites B. 
On the sublattice A one writes 
\begin{eqnarray}
S^{3}&=&S-a^{\dagger}a
\label{hpa}\\
S^{-}&=&(2S)^{\frac{1}{2}}a^{\dagger}(1-\frac{a^{\dagger}a}{2S})^{\frac{1}{2}}
\label{hpb}
\end{eqnarray}
Eqs.(\ref{hpa},\ref{hpb}) for spin operators correctly reproduce the spin algebra. The state with maximal $S^{3}=S$ is a state with zero bosons. 
Applying $S^{-}$ on this state one creates a boson and so lowers $S^{3}$ by one. The state with $S^{3}=-S$ has $2S$ bosons 
and is annihilated by $S^{-}$. On sublattice B one writes
\begin{eqnarray}
S^{3}&=&-S+b^{\dagger}b
\label{2hpa}\\
S^{-}&=&(2S)^{\frac{1}{2}}(1-\frac{b^{\dagger}b}{2S})^{\frac{1}{2}}b
\label{2hpb}
\end{eqnarray}
Using the Holstein-Primakov representation given by Eqs.(\ref{hpa},\ref{hpb}) and Eqs.(\ref{2hpa},\ref{2hpb}) and studying 
the small fluctuations around the N\'eel state so that $a^{\dagger}a$ and $b^{\dagger}b$  are $O(1)\ll S$ in the large S limit, 
one may develop an expansion in $\frac{1}{S}$ by expanding the square roots in the definition of $S^{-}$ in Eqs.(\ref{hpb},\ref{2hpb}). 
To quadratic order the Hamiltonian (\ref{hhej}) reads
\begin{equation}
H=J\sum_{x=1}^{N}\left\{-S^{2}+S(a^{\dagger}_{x}a_{x}+b^{\dagger}_{x+1}b_{x+1}+a_{x}b_{x+1}+a_{x}^{\dagger}b_{x+1}^{\dagger})\right\}
\label{hhej2}
\end{equation}
where higher orders in $\frac{1}{S}$ have been dropped. 
In momentum space the Hamiltonian (\ref{hhej2}) reads 
\begin{equation}
H_{J}=2JS\sum_{k}\left\{\gamma_{k}(a_{k}b_{-k}+a_{k}^{\dagger}b_{-k}^{\dagger})+(a_{k}^{\dagger}a_{k}+b_{k}^{\dagger}b_{k})\right\}
\label{hhej3}
\end{equation}
where 
\begin{equation}
\gamma_{k}=\frac{e^{ik}+e^{-ik}}{2}=\cos k=\gamma_{-k}\quad .
\end{equation}
The Hamiltonian (\ref{hhej3}) describes free bosons. Let us take the Bogoliubov transformation 
\begin{equation}
C_{k}=U_{k}a_{k}-V_{k}b_{-k}^{\dagger}\quad D_{k}=U_{k}b_{k}-V_{k}a_{-k}^{\dagger}
\label{bogo}
\end{equation}
where, since $a$ and $b$ are bosonic operators, one has
\begin{equation}
|U|^{2}-|V|^{2}=1 \quad .
\label{bogo1}
\end{equation}
To put to zero the off diagonal terms inverting Eqs.(\ref{bogo}) one has
\begin{equation}
\gamma (U^{2}+V^{2})+2UV=0
\label{bogo2}
\end{equation}
Solving Eq.(\ref{bogo1}) and Eq.(\ref{bogo2}) one gets
\begin{eqnarray}
U_{k}&=&\frac{1}{2}\frac{1}{(1-\gamma_{k}^{2})^{\frac{1}{4}}}\left[ \sqrt{1-\gamma_{k}}+\sqrt{1+\gamma_{k}}\right] 
\label{uk}\\
V_{k}&=&\frac{1}{2}\frac{1}{(1-\gamma_{k}^{2})^{\frac{1}{4}}}\left[ \sqrt{1-\gamma_{k}}-\sqrt{1+\gamma_{k}}\right] \quad .
\label{vk}
\end{eqnarray}
Using Eqs.(\ref{bogo}) the Hamiltonian Eq.(\ref{hhej3}) reads 
\begin{equation}
H_{J}=2JS\sum_{k}(1-\gamma_{k}^{2})^{\frac{1}{2}}(C_{k}^{\dagger}C_{k}+D_{k}^{\dagger}D_{k})
\label{swh}
\end{equation}
The free excitations created by C and D are known as spin waves and correspond to the infinitesimal deviations of the spins away from the N\'eel state. 
The dispersion relation for spin waves is 
\begin{equation}
E(k)=2JS(1-\gamma_{k}^{2})^{\frac{1}{2}}
\end{equation}
that vanishes linearly at $k=0$
\begin{equation}
E(k)\approx v|k|
\end{equation}
where the effective ``light velocity" is 
\begin{equation}
v=2JS
\end{equation}
In a field theoretical language spin waves are gapless due to the Goldstone theorem. In fact there are two Goldstone modes C and 
D corresponding to the breaking of $SO(3)$ down to the $SO(2)$ rotations about the z axis. 
Low energy long-wavelength spin waves correspond to configurations where all regions are locally in some N\'eel ground state 
but the direction of the sublattice magnetization vector makes long-wavelength rotations. 
The two different modes C and D have spin $S^{3}=\pm 1$ and correspond to raising $S^{3}$ on one sublattice or lowering it on the other. 

Now one can compute the reduction in the sublattice magnetization due to quantum fluctuations 
\begin{equation}
<S^{3}(x)>=S-\int \frac{dk}{2\pi} \frac{1}{2}\left[ \frac{1}{\sqrt{1-\gamma_{k}^{2}}}-1\right]
\label{gsg}
\end{equation}
Quantum spins like quantum harmonic oscillators have zero point motion. In one dimension one finds that Eq.(\ref{gsg}) is divergent for small 
wavelengths 
\begin{equation}
\Delta<S^{3}(x)>= <S^{3}(x)> -S=-\int \frac{dk}{2\pi} \frac{1}{2k}=-\infty
\label{ds3}
\end{equation}
Eq.(\ref{ds3}) indicates that the N\'eel state is destabilized by quantum fluctuations in one dimension even in the large $S$ limit. 
The divergence of Eq.(\ref{ds3}) is in agreement with the Coleman and Mermin-Wagner theorems~\cite{b39} preventing the spontaneous breaking of 
continuous symmetries in (1+1)-dimensions due to infrared divergences connected with the Goldstone bosons. 

\subsection{Spin-spin correlators in spin wave theory} 

We have seen that the N\'eel state is destabilized by quantum fluctuations in D=1 even in the classical limit $S\rightarrow \infty$. 
The spin$-$spin correlation function to next order in $\frac{1}{S}$ is 
\begin{equation}
<\vec{S}_{0}\cdot \vec{S}_{r}>\approx \pm S^{2}\left[ 1-\frac{1}{\pi S}\log \frac{r}{a}+O(\frac{1}{S^{2}})\right]
\label{ccoo}
\end{equation}
Eq.(\ref{ccoo}) indicates that at large S the system is ordered at small distances while the disorder occurs at exponentially 
large scales
\begin{equation}
\xi \simeq e^{\pi S}
\label{xis}
\end{equation}
More precisely, one should distinguish the $r$ odd and even cases. To see this, it is instructive to compute the finite distance correlators from 
spin wave theory. If $r$ belongs to the odd sublattice, the spin-spin correlator reads
\begin{eqnarray}
<\vec{S}_{0}\cdot\vec{S}_{r}>&=&-S^{2}\left[1+\frac{1}{S}(1-\frac{1}{N}\sum_{k}\frac{1-\cos rk\cos k }{|\sin k|})+O(S^{-2})\right]\nonumber\\
&=&-S^{2}\left[1+\frac{1}{S}(1-I_{o}(r))+O(S^{-2})\right]
\label{sso}
\end{eqnarray}
while if $r$ belongs to the even sublattice one has
\begin{eqnarray}
<\vec{S}_{0}\cdot\vec{S}_{r}>&=&-S^{2}\left[1+\frac{1}{S}(1-\frac{1}{N}\sum_{k}\frac{1-\cos rk}{|\sin k|})+O(S^{-2})\right]\nonumber\\
&=&-S^{2}\left[1+\frac{1}{S}(1-I_{e}(r))+O(S^{-2})\right]\quad .
\label{sse}
\end{eqnarray}
In Eqs.(\ref{sso},\ref{sse})
\begin{eqnarray}
I_{o}(r)&=&\frac{1}{2\pi}\int_{0}^{2\pi}\frac{1-\cos rk \cos k}{|\sin k|} dk
\label{io}\\
I_{e}(r)&=&\frac{1}{2\pi}\int_{0}^{2\pi}\frac{1-\cos rk }{|\sin k|} dk
\label{ie}
\end{eqnarray}
and up to the distance $r=10$  one gets the values given in table (\ref{tabu})
\begin{table}[htbp]
\begin{center}
\caption{$I_{o}(r)$ and $I_{e}(r)$}\label{tabu}
\vspace{.1in}
\begin{tabular}{|c|c|}\hline
$I_{o}(r)$   &   $I_{e}(r)$ \rule{0in}{4ex}\\[2ex] \hline 
 
$I_{o}(1)=\frac{2}{\pi}$        &   $I_{e}(2)=\frac{4}{\pi}$   \rule{0in}{4ex}\\
  
$I_{o}(3)=\frac{14}{3\pi}$      &   $I_{e}(4)=\frac{16}{3\pi}$ \rule{0in}{4ex}\\

$I_{o}(5)=\frac{86}{15\pi}$     &   $I_{e}(6)=\frac{92}{15\pi}$  \rule{0in}{4ex}\\
 
$I_{o}(7)=\frac{674}{105\pi}$   &   $I_{e}(8)=\frac{704}{105\pi}$  \rule{0in}{4ex}\\
 
$I_{o}(9)=\frac{2182}{315\pi}$  &   $I_{e}(10)=\frac{2252}{315\pi}$\rule{0in}{4ex}\\[2ex] \hline
\end{tabular}
\end{center}
\end{table}
The spin-spin correlator $G(r)=<\vec{S}_{0}\cdot \vec{S}_{r}>$ are given in table (\ref{swtssc}).   
\begin{table}[htbp]
\begin{center}
\caption{Spin-wave theory spin-spin correlation functions}\label{swtssc}
\vspace{.1in}
\begin{tabular}{|rc|}\hline
$r$   &   $G(r)$ \rule{0in}{4ex}\\[2ex] \hline 
 
 1    & -0.4317 \rule{0in}{4ex}\\
  
 2    & 0.1817   \\

 3    & -0.0073  \\
 
 4    & 0.0756   \\
 
 5    & 0.1625   \\
 
 6    & 0.0119   \\
 
 7    & 0.2716   \\

 8    & -0.0335  \\
 
 9    & 0.3525   \\

 10   & -0.0689  \\[2ex] \hline
 \end{tabular}
 \end{center}
 \end{table}
It is interesting to compare table (\ref{swtssc}) with the numerical correlators given in table (\ref{30c}). While it is quite impressive the agreement 
of $G(r)$ for $r=1,2$ with the exact results, increasing the distance $r$ the agreement becomes very poor already for $r=3$, and starts to 
have no meaning at $r=5$, where one gets even a wrong sign, opposite to the right answer. 
I think that this is and interesting lesson that one can learn from the failure of spin-wave theory. While, in studying short 
distance correlations ($r=1,2$) one gets very good results in the spin-wave theory, quantum disorder prevails for larger distances and so a 
N\'eel state approximation cannot pick up the main features of the system. The expected scale for the appearence of quantum disorder for $S=\frac{1}{2}$ is, 
from Eq.(\ref{xis}), $\xi=e^{\frac{\pi}{2}}\approx 5$.

\section{The Haldane-Shastry spin model}
\label{hash}
The Haldane-Shastry spin model~\cite{b37} is a generalized spin-$1/2$ Heisenberg Hamiltonian with long range inverse-square exchange interaction 
between spins. The inverse-square interaction makes the model more easy to solve than the original Heisenberg model and the description of the 
excitation spectrum in terms of spin-$1/2$ spinons is particularly easy to understand. The model can be considered as a discretized version of 
the Calogero-Sutherland model~\cite{b40} describing a one-dimensional Bose gas with inverse square repulsions. 

This paragraph shortly reviews the approach to the Haldane-Shastry model given by Laughlin et al. in~\cite{b31}. My purpose is to stress some feature of 
one-dimensional antiferromagnetic chains that are more intuitive to understand in this model than in the Heisenberg model. 
The Haldane-Shastry Hamiltonian describes N spin-$1/2$ particles distributed equidistantly on a circle, coupled with an 
exchange interaction that decays proportional to the inverse square of the chord between the spins
\begin{equation}
H_{HS}=J(\frac{2\pi}{N})^{2} \sum_{\alpha<\beta=1}^{N}\frac{\vec{S}_{\alpha}\cdot \vec{S}_{\beta}}{|z_{\alpha}-z_{\beta}|^{2}}
\label{hahs}
\end{equation}
where each site is parametrized by the roots of the unity
\begin{equation}
z_{\alpha}^{N}-1=0
\end{equation}
and $\vec{S}_{\alpha}$ is the spin operator on site $\alpha$. The Hamiltonian Eq.(\ref{hahs}) is invariant, due to periodic boundary 
conditions, under translations generated by $\hat{T}$
\begin{equation}
\hat{T}\quad :\quad \vec{S}_{\alpha}\longrightarrow \vec{S}_{\alpha +1}
\end{equation}
and under global rotations in the spin space generated by
\begin{equation}
\vec{S}=\sum_{x=1}^{N}\vec{S}_{x}\quad .
\end{equation}
Of course,
\begin{equation}
\left[\vec{S},\hat{T}\right]=0\quad .
\end{equation}

Moreover the Hamiltonian Eq.(\ref{hahs}) is invariant under the special internal Yangian symmetry~\cite{b41}
\begin{equation}
\left[H_{HS},\vec{\Lambda}\right]=0
\end{equation}
where
\begin{equation}
\vec{\Lambda}=\frac{i}{2}\sum_{\alpha\neq \beta}(\frac{z_{\alpha}+z_{\beta}}{z_{\alpha}-z_{\beta}})(\vec{S}_{\alpha}\wedge \vec{S}_{\beta})
\label{yang}
\end{equation}
The ground state of this model has the same functional form as the fractional quantum Hall ground state~\cite{b42,b42bis}; it is a spin singlet 
in the case of N even and exhibits quantum disorder: $i.e.$ it has the same relation to the antiferromagnetically ordered N\'eel state that 
a quantum liquid has to a conventional crystal.
 
In the hard-core bosons picture, the spin $\uparrow$ on a site corresponds to the presence of a boson, the spin $\downarrow$ to the abscence of a boson. 
The ground state wavefunction is the complex coefficient assigned to each configuration with $N/2$ bosons. 
The ground state of the Hamiltonian (\ref{hahs}) reads
\begin{equation}
\psi_{g.s.}(z_{1},\ldots ,z_{\frac{N}{2}})=\prod_{\alpha < \beta}^{\frac{N}{2}}(z_{\alpha}-z_{\beta})^{2}\prod_{\alpha=1}^{\frac{N}{2}}z_{\alpha}
\label{psigs}
\end{equation}
and its energy is
\begin{equation}
E_{g.s.}=-J(\frac{\pi^{2}}{24}) (N+\frac{5}{N})
\end{equation}
where $z_{1},\ldots,z_{\frac{N}{2}}$ denote the localization of the $N/2$ bosons. 
It is easy to verify that the ground state wave function Eq.(\ref{psigs}) is real, and it is translationally invariant  with momentum 0 or $\pi$  
for $N/2$ even or odd. In fact the translated expression of one site of $\psi_{g.s.}$ is obtained by multiplying each of 
its arguments by $z=e^{i\frac{2\pi}{N}}$ and due to the fact $\psi_{g.s.}$ is an homogeneous polynomial of degree $(N/2)^{2}$  one gets
\begin{equation}
\psi_{g.s.}(z_{1}z,\ldots,z_{\frac{N}{2}}z)=e^{i\frac{N\pi}{2}}\psi_{g.s.}(z_{1},\ldots,z_{\frac{N}{2}})
\end{equation}
One can immediately verify that $\psi_{g.s.}$ is a spin singlet. Due to the fact that $z_{j}$ is constrained to take $N/2$ 
possible values, $i.e.$ that there are $N/2$ spins up (and $N/2$ down), $S^{z}\psi_{g.s.}=0$. Moreover it is the lowest weight state
\begin{equation}
S^{-}\psi_{g.s.}=0
\end{equation}
It is worth stressing that it is very peculiar of this model the possibility to write down in a closed form, so simple to read, 
the exact quantum ground state (\ref{psigs}). In general, it is very difficult to write down in an easy compact form the ground state 
of a quantum antiferromagnetic chain. In the case of the Heisenberg antiferromagnet this is not possible for a general N. 

$\psi_{g.s.}$ exhibits quantum disorder, $i.e.$ the spin-spin correlation function
\begin{equation}
<\vec{S}_{\alpha}\cdot\vec{S}_{\beta}>=\frac{<\psi_{g.s.}|\vec{S}_{\alpha}\cdot\vec{S}_{\beta}|\psi_{g.s.}>}{<\psi_{g.s.}|\psi_{g.s.}>}
\label{sasb}
\end{equation}
goes to zero as $|z_{\alpha}-z_{\beta}|\rightarrow \infty$ so that it is a spin-liquid state. 
However the fall-off of  Eq.(\ref{sasb}) is slow and this is typical of half-integer spin systems that exhibit gapless excitations~\cite{b43}. 
Different is the case of integer spin systems that are strongly-disordered spin liquids with exponential decay of correlations on a length scale 
$\xi$ and have an energy gap for excitations $\Delta=\frac{\hbar v}{\xi}$, where $v$ is the spin-wave velocity of a nearby ordered state.
 
The ground state Eq.(\ref{psigs}) is not degenerate, but the state
\begin{equation}
\psi_{g.s.}'(z_{1},\ldots,z_{\frac{N}{2}})=\prod_{\alpha<\beta=1}^{\frac{N}{2}}(z_{\alpha}-z_{\beta})^{2}
\left[1-\prod _{\alpha}^{\frac{N}{2}}z_{\alpha}^{2}\right]
\end{equation}
is almost degenerate with $\psi_{g.s.}$
\begin{equation} 
H_{HS}|\psi_{g.s.}'>=-J(\frac{\pi^{2}}{24})(N-\frac{7}{N})|\psi_{g.s.}'>
\end{equation}
and it has momentum $\pi$ greater than $\psi_{g.s.}$. The state $\psi_{g.s.}'$ can be thought as $\psi_{g.s.}$ plus a pair of spinons 
excited out in a singlet state of momentum $\pi$. 
When the number of sites N is odd, the ground state cannot be a spin singlet, but it is a spin-$1/2$ state, $i.e.$ 
it describes a spinon of momentum $\pm \pi/2$. 
The wave function 
\begin{equation}
\psi_{\alpha}(z_{1},\ldots,z_{\frac{N-1}{2}})=\prod_{\gamma=1}^{\frac{N-1}{2}}(z_{\alpha}-z_{\gamma})\prod_{\gamma<\delta=1}^{\frac{N-1}{2}}
(z_{\gamma} -z_{\delta})^{2}\prod_{\gamma=1}^{\frac{N-1}{2}}z_{\gamma}
\end{equation}
describes a $\downarrow$ spinon at site $\alpha$ and it can be interpreted as a $\downarrow$ spin on site $\alpha$ 
surrounded by an otherwise featureless singlet sea. 
The linear combination of the states given by
\begin{equation}
\psi_{m}(z_{1},\ldots,z_{M})=\sum_{\alpha=1}^{N}(z_{\alpha}^{*})^{m}\psi_{\alpha}(z_{1},\ldots,z_{M})
\label{psim}
\end{equation}
with $0\leq m \leq \frac{N-1}{2}$, is an exact eigenstate of $H_{HS}$ with eigenvalue
\begin{equation}
H_{HS}|\psi_{m}>=\left\{ -J(\frac{\pi ^{2}}{24})(n-\frac{1}{N})+\frac{J}{2}(\frac{2\pi}{N})^{2}m(\frac{N-1}{2}-m)\right\}|\psi_{m}>
\label{eshs}
\end{equation}
The state $|\psi_{m}>$ represents a $\downarrow$ spinon travelling with crystal momentum
\begin{equation}
q=\frac{\pi}{2}N-\frac{2\pi}{N}(m+\frac{1}{4})\quad (mod\quad 2\pi)
\end{equation}
due to the definition
\begin{equation}
\psi_{m}(z_{1}z,\ldots,z_{M}z)=e^{iq}\psi(z_{1},\ldots, z_{M})
\label{psimq}
\end{equation}
Rewriting Eq.(\ref{eshs}) as
\begin{equation}
H_{HS}|\psi_{m}>=\left\{ -J(\frac{\pi^{2}}{24})(N+\frac{5}{N}-\frac{3}{N^{2}})+E_{q}\right\}|\psi_{m}>
\end{equation}
one obtains the dispersion relation 
\begin{equation}
E_{q}=\frac{J}{2}\left[ (\frac{\pi}{2})^{2} -q^{2}\right]\quad(mod\quad \pi)
\end{equation}
The allowed momenta for the spinon are only half of the Brillouin zone (the inner $-\frac{\pi}{2}\leq q\leq \frac{\pi}{2}$ or the outer part
$-\pi \leq q\leq -\frac{\pi}{2}$ and $\frac{\pi}{2}\leq q\leq \pi$, depending on wheter $N-1$ is 
divisible by 4). 
The spinon dispersion relation is linear in $q$ close to $\pm \frac{\pi}{2}$ with velocity
\begin{equation}
v_{spinon}=\frac{\pi}{2}\frac{J}{\hbar}\frac{2\pi}{N}
\end{equation}
Spinons are ``relativistic" excitations travelling with velocity $v_{spinon}$ and they exist because quantum disorder exists 
whenever the total spin in the unit cell is half integral.
A spin liquid singlet ground state can exist only if the total number of spins is even. If N is odd the total spin cannot be less than $1/2$. 
Since in the thermodynamic $N\longrightarrow \infty$ there can be no difference between N even and odd, the spin system must have already had a neutral 
spin-$\frac{1}{2}$ spinon excitation at the beginning, $i.e.$ Eq.(\ref{psimq}) describes a spinon. 
In the case of N odd, the ground state is 4-times degenerate and is given by Eq.(\ref{psimq}) for $m=0$ and $\frac{N-1}{2}$ and their $\uparrow$ 
counterparts.
Even if more than one spinon is present in the chain, they keep their identity. 
In the thermodynamic limit spinons are non-interacting excitations, some kind of quantum kink-excitations. 
Spinons are neither bosons nor fermions but they are semions, $i.e.$ particles obeying $1/2$ statistics. 
The two-spinon wave function is given by 
\begin{equation}
\psi_{AB}=\prod_{j}(z_{j}-z_{A})(z_{j}-z_{B})\prod_{j<k}(z_{j}-z_{k})^{2}
\end{equation}
where $z_{A}$ and $z_{B}$ are not necessarly lattice sites, and it has the property 
\begin{equation}
\psi^{*}_{AB}(z_{1},\ldots,z_{\frac{N}{2}-1})=(z_{A}z_{B})^{1-\frac{N}{2}}\psi_{AB}(z_{1},\ldots,z_{\frac{N}{2}-1})\quad .
\end{equation}

The phase for adiabatic motion of spinon A in the presence of spinon B, called Berry phase, is 
\begin{equation}
\frac{1}{2<\psi_{AB}|\psi_{AB}>}\left[ 
<\psi_{AB}|z_{A}\frac{\partial}{\partial z_{A}}\psi_{AB}>+<z_{A}\frac{\partial}{\partial z_{A}}\psi_{AB}|\psi_{AB}>\right]=\frac{1}{2}(1-\frac{N}{2})
\end{equation}
By moving spinon A around the  loop to exchange it with B, one gets the phase
\begin{equation}
\Delta\Phi_{S}=\oint \frac{1}{2}(1-\frac{N}{2})\frac{dz_{\alpha}}{z_{\alpha}}=\pm \frac{\pi}{2}i\quad (mod\quad 2\pi)
\end{equation}
One would get in the case of bosons or fermions
\begin{eqnarray}
\Delta \Phi_{B}&=&0\nonumber\\
\Delta \Phi_{F}&=&\pi
\end{eqnarray}
Fractional statistics has nothing to do with the symmetry properties of $\psi_{AB}$ under the exchange of A with B, 
but it has to do with state-counting.
 
Last I want to say that the the Yangian symmetry operator Eq.(\ref{yang}) represents a scaled spin current. Acting 
on the propagating spinon Eq.(\ref{psim}) one gets
\begin{equation}
\Lambda^{z}|\psi_{m}>=\left\{ \frac{N-1}{4}-m\right\} |\psi_{m}>
\end{equation}
So the eigenvalue of $\Lambda^{z}$ is proportional to the spinon velocity
\begin{equation}
\frac{dE_{q}}{dq}=\frac{2\pi J}{N}\left\{\frac{N-1}{4}-m\right\}
\end{equation}
The action of $\Lambda^{z}$ on a multispinon state is the sum of all the spinon velocities. 

I analysed the model at half filling, $i.e.$ when every site is occupied by one particle. In the presence of doping the possible configurations on 
each site are spin up, spin down and empty site. When the filling is less than half, new quantum excitations called holons are created. 
Holons are charged quantum excitations of spin 0. New phenomena appear in presence of holons and it is necessary to add to $H_{HS}$ a term 
describing the holon dynamics. Since my thesis is concerned with the relationships between spin chains and field theories, the half
 filling constraint is required. The reader interested in knowing more about holons can look up Ref.~\cite{b44}.

\section{The Bethe Ansatz solution of the antiferromagnetic Heisenberg chain}
\label{baa}

In this paragraph the Bethe ansatz solution of the antiferromagnetic 
Heisenberg chain~\cite{b27} is discussed in detail. Moreover I shall compare the exact solution given by L.D. Faddeev and L.A. Takhtadzhyan in~\cite{b27} 
with a study of finite size chains of 4, 6 and 8 sites. I show that already these very small finite systems exhibit spectra 
that match very well with the thermodynamic limit solution. I suggest to the reader interested in the subject the references 
\cite{b31}.

The Bethe ansatz is a method of solution of a number of quantum field theory and statistical mechanics 
models in two space-time dimensions. This method was first suggested by Bethe~\cite{b33} in 1931 from which takes its name. 
Historically one can call this formulation the Coordinate Bethe ansatz to distinguish it from the modern formulation 
known as Algebraic Bethe ansatz. The eigenfunctions of some (1+1)-dimensional Hamiltonians can be constructed imposing 
periodic boundary conditions which lead to a system of equations for the permitted values of momenta. These are 
known as the Bethe equations which are also useful in the thermodynamic limit. The energy of the ground state may be 
calculated in this limit and its excitations can be investigated. 

I review here the method applyied to the study of 
the antiferromagnetic Heisenberg chain~\cite{b27}. In particular it is shown that there is only one excitation with spin-$1/2$ 
which is a kink: physical states have only an even number of kinks, therefore they always have an integer spin.
The one dimensional isotropic Heisenberg model describes a system of 
$N$ interacting spin-$\frac{1}{2}$ particles. 
The Hamiltonian of the model is
\begin{equation}
H_{J}=J\sum_{x=1}^{N}(\vec{S}_{x}\cdot \vec{S}_{x+1}-\frac{1}{4})\quad .
\label{heis}
\end{equation}
where $J>0$ ($J<0$ would describe a ferromagnet) and the spin operators 
have the following form
\begin{equation}
\vec{S}_{x}=1_{1}\otimes 1_{2}\otimes \ldots \otimes 
\frac{\vec{\sigma}_{x}}{2} 
\otimes \ldots \otimes 1_{N}\quad .
\end{equation}
They act nontrivially only on the Hilbert space of the $x^{th}$ site. 
Periodic boundary conditions are assumed. 

The Hamiltonian (\ref{heis}) is invariant under global rotations in the 
spin space, generated by
\begin{equation}
\vec{S}=\sum_{x=1}^{N}\vec{S}_{x}\quad .
\end{equation}
Due to the periodic boundary conditions, under translations generated by 
the operator $\hat{T}$, 
\begin{equation}
\hat{T}\quad : \vec{S}_{x} \longrightarrow \vec{S}_{x-1} 
\end{equation}
the Hamiltonian is invariant and $[\vec{S},\hat{T}]=0$.

In order to diagonalize $H_{J}$ it is convenient to use an eigenfunction basis of operators commuting with 
$H_{J}$, so obviously $\vec{S}^2$, $S^z$ and also $\hat{T}$.
Let us sketch the Coordinate Bethe ansatz technique. One has to introduce the ``false vacuum"
\begin{equation}
|\Omega>=\prod_{x=1}^{N}\otimes |\uparrow>_{x}
\end{equation}
with
\begin{eqnarray}
S_{x}^{+} |\uparrow>_{x}&=&0\\
S_{x}^{3} |\uparrow>_{x}&=&\frac{1}{2}|\uparrow>_{x}
\end{eqnarray}
where
\begin{equation}
S_{x}^{\pm}=S_{x}^{1}\pm S_{x}^{2}
\end{equation}
\begin{eqnarray}
S^{3}|\Omega>&=&\frac{N}{2}|\Omega>\\
\vec{S}^{2}|\Omega>&=&\frac{N}{2}(\frac{N}{2}+1)|\Omega>\\
\hat{T}|\Omega>&=&|\Omega>\quad .
\end{eqnarray}
All the other basis vectors have $S^{3}<\frac{N}{2}$ and one can get them by properly acting on $|\Omega>$ with 
the lowering operators $S_{x}^{-}$. Let us start from the generic state with $S^{3}=\frac{N}{2}-1$, with $N-1$ spins up and 
$M=1$ spin down
\begin{equation}
|M=1>=\sum_{x=1}^{N}\phi_{x}|x>\quad with \quad |x>=S_{x}^{-}|\Omega>
\end{equation}
where the coefficients $\phi_{x}$ must be such that $|M=1>$ is a translationally and rotationally invariant state
\begin{equation}
\hat{T}|M=1>=\sum_{x=1}^{N}\phi_{x}|x-1>=\sum_{x=1}^{N}\phi_{x+1}|x>=\mu \sum_{x=1}^{N}\phi_{x}|x>
\label{tm}
\end{equation}
and from Eq.(\ref{tm}) one gets
\begin{equation}
\phi_{x+1}=\mu \phi_{x}
\end{equation}
\begin{eqnarray}
\phi_{x+1}&=&\mu ^{x} \phi_{1}\quad \quad x\neq N \\
\phi_{1}&=&\mu \phi_{N}=\mu^{N} \phi_{1}\longrightarrow \mu^N =1\quad .
\end{eqnarray}
Setting $\phi_{1}=1$ one has $\phi_{x}=\mu ^{x-1}$. There are N possible values for $\phi_{x}$, 
corresponding to the N roots of the unity. One of these roots corresponds to a state with $S=\frac{N}{2}$ 
\begin{equation}
\sum_{x=1}^{N}|x>=S^{-}|\Omega>\quad\longrightarrow \quad \mu=1
\end{equation}
while the other $N-1$ roots have $S=\frac{N}{2}-1$.

The generic case with M spins down is more complicated. Let us consider the case $M=2$ to understand 
what happens 
\begin{equation}
|M=2>=\sum_{x_{1}<x_{2}=1}^{N}\phi(x_{1},x_{2})|x_{1},x_{2}>\quad with \quad |x_{1},x_{2}>=S_{x_{1}}^{-}S_{x_{2}}^{-}|\Omega>\quad .
\end{equation}      
By requiring the translational invariance of the state $|M=2>$ one has
\begin{equation}
\hat{T}|M=2>=T|M=2>
\end{equation}
and for $x_{2}<N$  one has
\begin{equation}
\phi(x_{1}+1,x_{2}+1)=T \phi(x_{1},x_{2})
\label{ee1}
\end{equation}
that would easily give
\begin{equation}
\phi(x_{1},x_{2})=\mu_{1}^{x_{1}-1} \mu_{2}^{x_{2}-1}\quad ,\quad T=\mu_{1} \mu_{2}
\label{trial}
\end{equation}
but due to the periodic boundary conditions one has to find coefficients $\phi(x_{1},x_{2})$ that satisfy not only Eq.(\ref{ee1}) 
but also 
\begin{equation}
\phi(1,x+1)=T\phi(x,N)\quad .
\label{ee2}
\end{equation}
Eq.(\ref{ee2}) is no more satisfied by (\ref{trial}). Bethe proposed the following ansatz for the coefficients $\phi(x_{1},x_{2})$
\begin{equation}
\phi(x_{1},x_{2})=A_{1,2}\mu_{1}^{x_{1}-1}\mu_{2}^{x_{2}-1}+A_{2,1}\mu_{2}^{x_{1}-1}\mu_{1}^{x_{2}-1}\quad .
\label{ba}
\end{equation}
Eq.(\ref{ee1},\ref{ee2}) are satisfied if the following equations hold
\begin{eqnarray}
A_{12}&=&A_{21}\mu_{1}^{N}
\label{cc1}\\
A_{21}&=&A_{12}\mu_{2}^{N}\quad .
\label{cc2}
\end{eqnarray}
By imposing to $|M=2>$ to be an highest weight state
\begin{equation}
S^{+}|M=2>=0
\end{equation}
taking into account Eq.(\ref{cc1},\ref{cc2}) and introducing the following change of variables
\begin{equation}
\mu_{\alpha}=\frac{\lambda_{\alpha}-\frac{i}{2}}{\lambda_{\alpha}-\frac{i}{2}} 
\end{equation}
one gets the so called Bethe ansatz equations
\begin{equation}
(\frac{\lambda_{\alpha}-\frac{i}{2}}{\lambda_{\alpha}+\frac{i}{2}})^{N}=-
\prod_{\beta=1}^{2}\frac{\lambda_{\alpha}-\lambda_{\beta}-i}{ \lambda_{\alpha}-\lambda_{\beta}+i}\quad .
\end{equation}
In the general case of M spins flipped one gets
\begin{equation}
(\frac{\lambda_{\alpha}-\frac{i}{2}}{\lambda_{\alpha}+\frac{i}{2}})^{N}=-
\prod_{\beta=1}^{M}\frac{\lambda_{\alpha}-\lambda_{\beta}-i}{ \lambda_{\alpha}-\lambda_{\beta}+i}\quad .
\label{baeq}
\end{equation}

The energy and the momentum of a given state with $M$ spins down can be expressed in terms of the parameters $\lambda_{\alpha}$
\begin{eqnarray}
E_{M}&=&\sum_{\alpha=1}^{M}\epsilon_{\alpha}=-\frac{J}{2}\sum_{\alpha=1}^{M}\frac{1}{\lambda_{\alpha}^{2}+\frac{1}{4}}\\
P_{M}&=&i\ln T=\sum_{\alpha=1}^{M}p_{\alpha}=i\sum_{\alpha=1}^{M}\ln \frac{\lambda_{\alpha}-\frac{i}{2}}{\lambda_{\alpha}+\frac{i}{2}}\quad .
\end{eqnarray}
Energy and momentum are thus additive as if there were M independent particles and the $\lambda_{\alpha}$ must satisfy the Bethe ansatz equations 
(\ref{baeq}) in order for $E_{M}$ and $P_{M}$ to be eigenvalues of the Hamiltonian and momentum operators. 

\begin{figure}[htb]
\begin{center}
\setlength{\unitlength}{0.00041700in}%
\begingroup\makeatletter\ifx\SetFigFont\undefined
\def\x#1#2#3#4#5#6#7\relax{\def\x{#1#2#3#4#5#6}}%
\expandafter\x\fmtname xxxxxx\relax \def\y{splain}%
\ifx\x\y   
\gdef\SetFigFont#1#2#3{%
  \ifnum #1<17\tiny\else \ifnum #1<20\small\else
  \ifnum #1<24\normalsize\else \ifnum #1<29\large\else
  \ifnum #1<34\Large\else \ifnum #1<41\LARGE\else
     \huge\fi\fi\fi\fi\fi\fi
  \csname #3\endcsname}%
\else
\gdef\SetFigFont#1#2#3{\begingroup
  \count@#1\relax \ifnum 25<\count@\count@25\fi
  \def\x{\endgroup\@setsize\SetFigFont{#2pt}}%
  \expandafter\x
    \csname \romannumeral\the\count@ pt\expandafter\endcsname
    \csname @\romannumeral\the\count@ pt\endcsname
  \csname #3\endcsname}%
\fi
\fi\endgroup
\begin{picture}(8712,7514)(2701,-7718)
\thicklines
\put(6001,-5161){\circle{300}}
\put(7201,-1561){\circle{300}}
\put(7201,-6361){\circle{300}}
\put(7201,-3961){\circle{300}}
\put(8401,-2761){\circle{300}}
\put(8401,-361){\circle{300}}
\put(8401,-5161){\circle{300}}
\put(8401,-7561){\circle{300}}
\put(6001,-2761){\circle{300}}
\put(4801,-3961){\circle{300}}
\put(3601,-7561){\vector( 0, 1){7200}}
\put(10801,-4561){\makebox(0,0)[lb]{\smash{\SetFigFont{10}{12.0}{rm}Re$\lambda$}}}
\put(3601,-3961){\vector( 1, 0){7800}}
\multiput(6001,-2986)(0.00000,-7.98817){254}{\line( 0,-1){  3.994}}
\multiput(7201,-1786)(0.00000,-7.98817){254}{\line( 0,-1){  3.994}}
\multiput(7201,-4186)(0.00000,-7.98817){254}{\line( 0,-1){  3.994}}
\multiput(8401,-511)(0.00000,-8.00000){263}{\line( 0,-1){  4.000}}
\multiput(8401,-2911)(0.00000,-8.00000){263}{\line( 0,-1){  4.000}}
\multiput(8401,-5386)(0.00000,-7.98817){254}{\line( 0,-1){  3.994}}
\put(2701,-511){\makebox(0,0)[lb]{\smash{\SetFigFont{10}{12.0}{rm}Im$\lambda$}}}
\end{picture}
\end{center}
\caption{Strings for \protect $L=0,\frac{1}{2},1,\frac{3}{2}$}
\label{strings}
\end{figure}
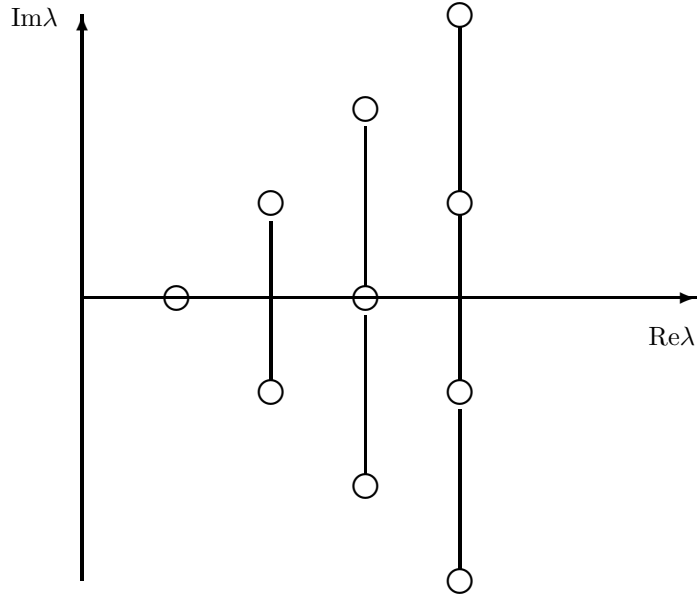

The solution of the antiferromagnetic Heisenberg chain is 
reduced to the solution of the system of the $M$ algebraic equations 
(\ref{baeq}). This, in general, is not an easy task.
It can be shown~\cite{b27}, however, that, in the thermodynamic limit 
$N\to\infty$, the complex parameters $\lambda$ have the form
\begin{equation}
\lambda_{l}=\lambda_{j,L}+il\quad,\quad l=-L,-L+1,\dots,L-1,L;
\label{complex}
\end{equation}
where $L$ is a non-negative integer or half-integer, $\lambda_{j,L}$
is the real part of the solution of (\ref{baeq}) and I shall 
define shortly the set of allowed values for the integer index $j$.
The $\lambda$'s that, for a given $\lambda_{j,L}$, are obtained varying $l$ 
between  $[-L,L]$ by integer steps, form a string of length $2L+1$,
see fig.(\ref{strings}). 
This arrangement of $\lambda$'s in the complex plane is called the 
``string hypothesis" \cite{b27}. 
In the following I shall verify 
that, even on finite size systems, the ``string hypothesis" 
is very well fulfilled. 

In a generic Bethe state with $M$ spins down, 
there are $M$ solutions to (\ref{baeq}), which can be grouped
according to the length of their strings.
Let us denote  by $\nu_{L}$ the number of strings of
length $2L+1$, $L=0,\frac{1}{2},\ldots$; strings of the same length
are obtained by changing the real parts, $\lambda_{j,L}$, of the $\lambda$'s 
in (\ref{complex}); as a consequence $\quad j=1,\ldots,\nu_{L}$. 
If one denotes the total number of strings by $q$ one has
\begin{equation}
q=\sum_{L}\nu _{L}\quad,\quad M=\sum_{L} (2L+1)\nu _{L}\quad .
\label{constraint}
\end{equation}

The set of integers $(M,q,\{\nu_{L}\})$ constrained 
by (\ref{constraint}), characterizes Bethe 
states up to the fixing of the $q$ numbers $\lambda_{j,L}$; this set 
is called the ``configuration''. 
Varying $M$, $q$ and $\nu_L$, one is able to construct all the
$2^N$ eigenstates of an Heisenberg antiferromagnetic chain of $N$ 
sites~\cite{b27}.
The energy and momentum of the Bethe's state, corresponding to a 
given configuration $-$ within exponential accuracy as
$N\rightarrow \infty$ $-$ consist of $q$ summands representing 
the energy and momentum of separate strings. 
For the parameters $\lambda_{j,L}$ of the given configuration, 
taking the logarithm of (\ref{baeq}) 
the following system of equations is obtained in the thermodynamic limit
\begin{equation}
2N\arctan \frac{\lambda_{j,L_{1}}}{L_{1}+\frac{1}{2}}=2\pi 
Q_{j,L_{1}}+\sum_{L_{2}}
\sum_{k=1}^{\nu_{L_{2}}}\Phi_{L_1 L_2}(\lambda_{j,L_{1}}-
\lambda_{k,L_{2}})\quad, 
\label{tbaeq}
\end{equation}
where
\begin{equation}  
\Phi_{L_1 L_2}(\lambda)=2\sum_{L=|L_1 -L_2|\neq 0}^{L_1 +L_2}
(\arctan \frac{\lambda}{L}+\arctan \frac{\lambda}{L+1})\quad .
\end{equation}

Integer and half integer numbers $Q_{j,L}$ parametrize the 
branches of the arcotangents and, consequently, the 
possible solutions of the system of Eqs.(\ref{tbaeq}). In ref.\cite{b27}
it was shown that the $Q_{j,L}$  are limited as
\begin{equation}
-Q_{L}^{max}\leq Q_{1,L}<Q_{2,L}<\ldots<Q_{\nu_{L},L}\leq Q_{L}^{max}\quad 
\label{vacan}
\end{equation}
with $Q_{L}^{max}$ given by
\begin{equation}
Q_{L}^{max}=\frac{N}{2}-\sum_{L'}J(L,L')\nu_{L'}-\frac{1}{2}
\label{qmax}
\end{equation}
and
\begin{equation}
 J(L_1 ,L_2)=\left\{\begin{array}{ll}
 2{\rm min}(L_1 ,L_2)+1 &\mbox{if $L_1 \neq L_2$}\\ 
 2L_1+\frac{1}{2} &\mbox{if $L_1 =L_2$}\quad .
\end{array}
\right.
\label{jll}
\end{equation}
The admissible values for the numbers $Q_{j,L}$ are called 
the ``vacancies'' and their number for every $L$ is denoted by $P_{L}$
\begin{equation}
P_{L}=2Q_{L}^{max}+1\quad .
\label{pelle}
\end{equation}

The main hypothesis formulated in~\cite{b27} is that to every 
admissible collection of $Q_{j,L}$ there corresponds a unique 
solution of the system of equations (\ref{tbaeq}). 
The solution always provides, in a multiplet, the state with the highest
value of the third spin component $S^3$. 

Let us now consider some simple example.
The simplest configuration has only strings of length 1,
$i.e.$ all the $\lambda$'s are real. 
The singlet associated to this configuration
\begin{equation}
M=q=\nu_0 =\frac{N}{2}\quad,\quad \nu_{L}=0\quad,\quad L>0\quad ,
\end{equation}
is the ground state.
The vacancies of the strings of length 1, $i.e.$ the 
admissible values of $Q_{j,0}$, due to 
eqs.(\ref{vacan},\ref{qmax},\ref{jll}), belong to the segment 
\begin{equation}
-\frac{N}{4}+\frac{1}{2}\leq Q_{j,0}\leq \frac{N}{4}-\frac{1}{2}\quad .
\label{igs}
\end{equation}
Therefore they are $N/2$.
All these vacancies must then be used to find the $N/2$ strings of length 1.
As a consequence this state is uniquely specified and no degeneracy 
is possible.

Next I consider the configuration that provides a singlet
with 1 string of length 2 and all the others of length 1:
\begin{equation}
M=\frac{N}{2}\quad,\quad q=\frac{N}{2}-1\quad,\quad \nu_{0}=
\frac{N}{2}-2 \quad,\quad \nu_{\frac{1}{2}}=1\quad,\quad \nu_{L}
=0\quad,\quad L>\frac{1}{2}\ .
\label{c1}
\end{equation}
For the strings of length 1 the 
number of vacancies is again ${N}/{2}$; 
for the string of length 2 there is one vacancy and the only 
admissible $Q_{j,1}$ equals 0. 
Thus, since the number of strings of length 1 is $\nu_{0}=
\frac{N}{2}-2$, there are two vacancies for which Eqs.(\ref{tbaeq}) 
have no solution; they are called ``holes'' and are denoted $Q_{1}^{(h)}$ 
and $Q_{2}^{(h)}$. 
This configuration is determined by two parameters: 
the positions of two ``holes" which vary independently in the interval 
(\ref{igs}). 

There is another state with only 2 holes:
the triplet corresponding to the configuration
\begin{equation}
M=q=\nu_{0}=\frac{N}{2}-1\quad,\quad \nu_{L}=0\quad,\quad L>0
\label{c2}
\end{equation}
The number of vacancies for the strings of length 1 equals $\frac{N}{2}+1$, 
while $\nu_{0}=\frac{N}{2}-1$.

The excitations determined by the configurations 
(\ref{c1},\ref{c2}) belong to the configuration class called in~\cite{b27}
$\cal{M}_{AF}$. The class $\cal{M}_{AF}$ is characterized as follows: 
the number of strings of length 1 in each configuration 
belonging to this class
differs by a finite quantity from ${N}/{2}$, $\nu_0=\frac{N}{2}-k_0$ 
where $k_0$
is a positive finite constant, so that 
the number of strings of length greater than 1 is finite. From (\ref{pelle}) 
one then has
\begin{eqnarray}
P_0&=&\frac{N}{2}+k_0-2\sum_{L>0}\nu_L\label{p0}\\
P_L&=&2k_0-2\sum_{L'>0}J(L,L')\nu_{L'}\ ,\quad L>0
\end{eqnarray}
so that
\begin{equation}
P_0\ge \frac{N}{2}\ ,\quad P_L<2 k_0\ ,\quad L>0\ .
\end{equation}
From (\ref{p0}) follows that the number of holes for the strings of length 1 
is always even and equals 2 only for the singlet and the triplet 
excitations discussed above.
One can imagine the class $\cal{M}_{AF}$ as a
``sea" of strings of length 1 with a finite number of 
strings of length greater than 1 immersed into it. 
It was proven in~\cite{b27} that, in the 
thermodynamic limit, the states belonging to 
$\cal{M}_{AF}$ have finite energy and momentum with 
respect to the antiferromagnetic vacuum, whereas 
each of the states which 
corresponds to a configuration not included in the class $\cal{M}_{AF}$ has 
an infinite energy relative to the antiferromagnetic vacuum.

Let us now sketch the computation of the thermodynamic limit 
ground state energy. Eqs.(\ref{tbaeq}) for the ground state have the form
\begin{equation}
\arctan 2\lambda_{j}=\frac{\pi Q_{j}}{N}+\frac{1}{N}
\sum_{k=1}^{{N}/{2}}\arctan (\lambda_{j} -\lambda_{k})\quad .
\label{gstbae}
\end{equation}
Taking the thermodynamic limit $N\rightarrow \infty$, one has 
\begin{equation}
\frac{Q_{j}}{N}\rightarrow x\quad,\quad -\frac{1}{4}\leq x\leq 
\frac{1}{4}\quad,\quad \lambda_{j}\rightarrow \lambda(x)\quad ,
\end{equation}
and Eqs.(\ref{gstbae}) can be rewritten in the form
\begin{equation}
\arctan 2\lambda(x)=\pi x+\int_{-\frac{1}{4}}^{\frac{1}{4}}
\arctan(\lambda(x)-\lambda(y)) dy\quad .
\label{2gsbae}
\end{equation}
Upon introducing the density of the numbers $\lambda(x)$ 
in the interval $d\lambda$ 
\begin{equation}
\rho(\lambda)=\frac{1}{\frac{d\lambda(x)}{dx}|_{x=x(\lambda)}}
\label{den}
\end{equation}
and differentiating Eqs.(\ref{2gsbae}), one gets
\begin{equation}
\rho(\lambda)=\frac{1}{2\pi}\int_{-\infty}^{\infty}
\frac{e^{-\frac{1}{2}|\xi|}}{1+e^{-|\xi|}}e^{-i\lambda |\xi|}
d\xi =\frac{1}{2\cosh\pi\lambda}\quad .
\label{den1}
\end{equation}
The density $\rho(\lambda)$ introduced in this way is normalized to $1/2$.
It is now easy to compute the energy and the momentum of the ground state 
\begin{equation}
E_{g.s.}=\sum_{\alpha =1}^{\frac{N}{2}}\epsilon_{\alpha}
=N\int_{-\infty}^{\infty}\epsilon(\lambda)\rho(\lambda)d\lambda=
-\frac{J N}{4}\int_{-\infty}^{\infty}d\lambda
\frac{1}{\left(\lambda^2+\frac{1}{4}
\right)\cosh\pi\lambda}=-JN\ln 2
\label{gse}
\end{equation}
\begin{equation}
P_{g.s.}=\sum_{\alpha=1}^{\frac{N}{2}}p_{\alpha}=N
\int_{-\infty}^{\infty}p(\lambda)
\rho(\lambda)d\lambda=-\frac{N}{2}
\int_{-\infty}^{\infty}d\lambda\frac{\pi}{\cosh\pi\lambda}=
\frac{N}{2}\pi\quad ({\rm mod}\quad 2\pi)\quad .
\label{gsm}
\end{equation}
According to Eq.(\ref{gsm}), $P_{g.s.}=0\ ({\rm mod} 2\pi)$ 
for $\frac{N}{2}$ even, and 
$P_{g.s.}=\pi\ ({\rm mod}\  2\pi)$ for $\frac{N}{2}$ odd. 
The ground state, as expected, is a singlet, in fact the spin $S$ 
is given by
\begin{equation}
S=\frac{N}{2}-\sum_{\alpha=1}^{N/2}1=\frac{N}{2}-
N\int_{-\infty}^{\infty}\rho(\lambda) d\lambda=0\quad .
\end{equation}

Let us analyze the triplet described by Eq.(\ref{c2}); 
Eqs.(\ref{tbaeq}) take the form
\begin{equation}
\arctan 2\lambda_{j}=\frac{\pi Q_{j}}{N}+\frac{1}{N}
\sum_{k=1}^{\frac{N}{2}-1}\arctan (\lambda_{j} -\lambda_{k})
\label{trbae}
\end{equation}
where now the numbers $Q_{j}$ lie in the segment 
$[-\frac{N}{4},\frac{N}{4}]$ and have two holes, $Q_1 ^{(h)}$ and 
$Q_2 ^{(h)}$ with $Q_1 ^{(h)} <Q_2 ^{(h)}$. Taking the 
thermodynamic limit one gets
\begin{equation}
\frac{Q_1 ^{(h)}}{N}\rightarrow x_1\quad,\quad 
\frac{Q_2 ^{(h)}}{N}\rightarrow x_2\quad,\quad \frac{Q_j}{N}
\rightarrow x+\frac{1}{N}(
\theta(x-x_1)+\theta(x-x_2))
\end{equation}
where $\theta (x)$ is the Heaviside function. Eqs.(\ref{trbae}) become 
\begin{equation}
\arctan 2\lambda(x)=\pi x+\frac{\pi}{N}(\theta(x-x_1)+\theta(x-x_2)) +
 \int_{-\frac{1}{4}}^{\frac{1}{4}}\arctan(\lambda(x)-\lambda(y)) dy\quad .
\label{2trbae}
\end{equation}
Eq.(\ref{2trbae}) gives, for this triplet, the density of 
$\lambda$, $\rho(\lambda)=\frac{d \lambda}{dx}$ 
\begin{equation}
\rho_{t}(\lambda)=\rho(\lambda)+\frac{1}{N}(\sigma (\lambda -
\lambda_1)-\sigma (\lambda -\lambda_2))
\end{equation}
where $\rho (\lambda)$ is given in Eq.(\ref{den1}) and  
\begin{equation}
\sigma(\lambda )=-\frac{1}{2\pi }\int_{-\infty}^{\infty}\frac{1}{1+e^{-|\xi|}}
e^{-i\lambda \xi} d\xi \quad .
\label{densig}
\end{equation}
$\lambda_1$ and $\lambda_2$ are the parameters of the holes, 
$\lambda_i=\lambda(x_{i})$, $i=1,2$.
The energy and the momentum of this state measured from the 
ground state are now easily computed
\begin{equation}
\epsilon_{T}(\lambda_1,\lambda_2)=
N\int_{-\infty}^{\infty}\epsilon(\lambda) 
(\rho_{t}(\lambda)-\rho(\lambda))d\lambda =
\epsilon(\lambda_1)+\epsilon(\lambda_2)
\label{tre}
\end{equation}
\begin{equation}
p_{T}(\lambda_1 ,\lambda_2)=N\int_{-\infty}^{\infty}p(\lambda) 
(\rho_{t}(\lambda)-\rho(\lambda))d\lambda=
p(\lambda_1)+p(\lambda_2)\quad (mod\quad 2\pi)
\end{equation}
where
\begin{equation}
\epsilon(\lambda)=\int_{-\infty}^{\infty}\epsilon(\mu)
\sigma(\lambda -\mu)d\mu=J\frac{\pi}{2\cosh \pi\lambda}
\label{et}
\end{equation}
\begin{equation}
p(\lambda)=\int_{-\infty}^{\infty}p(\mu)\sigma(\lambda -\mu)
d\mu=\arctan \sinh \pi \lambda -\frac{\pi}{2},\quad -\pi
\leq p(\lambda) \leq 0\quad .
\label{pt}
\end{equation}
From Eqs.(\ref{et},\ref{pt}) one gets
\begin{equation}
\epsilon=-\frac{J\pi}{2}\sin p\quad .
\label{disrel}
\end{equation}
The momentum $p_T(\lambda_1,\lambda_2)$ varies over 
the interval $[0,2\pi)$, when $\lambda_1$ and $\lambda_2$ run 
independently over the whole real axis.
The spin of this state can be computed by the formula
\begin{equation}
S=-\int_{-\infty}^{\infty}(\sigma(\lambda -\lambda_1)+
\sigma(\lambda -\lambda_2))d\lambda=1\quad .
\end{equation}

Let us finally analize the singlet excitation 
characterized by the configuration (\ref{c1}). 
Denoting by $\lambda_S$ 
the only number among the $\lambda_{j,{1}/{2}}$ 
which characterizes the string of length 2 and by $\lambda_{j}$ 
the numbers $\lambda_{j,0}$ 
for the strings of length 1, Eqs.(\ref{tbaeq}) read
\begin{eqnarray}
\arctan 2\lambda_{j}&=&\frac{\pi Q_j}{N}+\frac{1}{N}
\Phi(\lambda_j -\lambda_S)+\frac{1}{N}
\sum_{k=1}^{\frac{N}{2}-2}\arctan (\lambda_j -\lambda_k)\\
\arctan \lambda_S&=&\frac{1}{N}
\sum_{j=1}^{\frac{N}{2}-2}\Phi(\lambda_S -\lambda_j)
\end{eqnarray}
with
\begin{equation}
\Phi(\lambda)=\arctan 2\lambda+\arctan \frac{2}{3}\lambda\quad .
\end{equation}
The $\frac{N}{2}-2$ numbers $Q_{j}$ vary in the segment 
$[-\frac{N}{4}+\frac{1}{2},\frac{N}{4}-\frac{1}{2}]$; 
among them there are the two holes $Q_{1}^{(h)}$ and 
$Q_{2}^{(h)}$. Taking the thermodynamic limit one finds 
the density of $\lambda$'s for the singlet
\begin{equation}
\rho(\lambda_{S})=\rho(\lambda)+\frac{1}{N}
(\sigma(\lambda -\lambda_1)+\sigma(\lambda -\lambda_2)+
\omega(\lambda -\lambda_S))
\label{dens}
\end{equation}
where $\rho$ and $\sigma$ were given in Eqs.(\ref{den1}, 
\ref{densig}) and where 
\begin{equation}
\omega(\lambda)=-\frac{1}{2\pi}\int_{-\infty}^{\infty}
e^{-\frac{1}{2}|\xi|-i\lambda \xi} d\xi=
-\frac{2}{\pi(1+4\lambda^2)}\quad .
\end{equation}
In \cite{b27} it was demonstrated that the 
string parameter $\lambda_S$ is uniquely determined by 
the $\lambda$'s parametrizing the two holes
\begin{equation}
\lambda_S=\frac{\lambda_{1}^{(h)}+\lambda_{2}^{(h)}}{2}\quad .
\end{equation}
In \cite{b27} it was also proved the remarkable fact 
that the string of length 2 does not contribute to the 
energy and momentum of the excitation, 
so that the singlet and the triplet have the same dispersion relations
\begin{eqnarray}
\epsilon_S(\lambda_1,\lambda_2)&=&\epsilon_{T}(\lambda_1,\lambda_2)
=\epsilon(\lambda_1)+\epsilon(\lambda_2) \\
p_S(\lambda_1 ,\lambda_2)&=&p_T(\lambda_1 ,\lambda_2)=
p(\lambda_1)+p(\lambda_2)\quad ({\rm mod}\quad 2\pi)\quad .
\end{eqnarray}
The spin of this excitation is, of course, zero
\begin{equation}
S=-2-\int_{-\infty}^{\infty}(2\sigma(\lambda)+\omega(\lambda)) d\lambda=0
\end{equation}
The only difference between the state 
whose configuration is given in Eq.(\ref{c2}) and the state of
Eq.(\ref{c1}) is the spin. 

To summarize,
the finite energy 
excitations of the antiferromagnetic Heisenberg chain are only 
those belonging to the class ${\cal M_{AF}}$ and are described by
scattering states of an even number of quasiparticles or kinks. 
The momentum $p$ of these kinks runs over half the Brillouin 
zone $-\pi\le p\le 0$, the 
dispersion relation is $\epsilon(p)=\frac{J\pi}{2}\sin p$, Eq.(\ref{disrel}),
and the spin of a kink is $1/2$. The singlet and the triplet excitations
described above are the only states composed of two kinks, the spins of the 
kinks being parallel for the triplet and antiparallel for the singlet.
For vanishing total momentum all the states belonging to ${\cal M_{AF}}$
have the same energy of the ground state so that they are gapless excitations.
Since the eigenstates of $H_{J}$ always contain an even number of kinks,
the dispersion relation is determined by a set of two-parameters: the
momenta of the even number of kinks whose scattering provides the excitation. 
There are no bound states of kinks.

\subsection{Finite size antiferromagnetic Heisenberg chains}

Let us now turn to the computation of the spectrum of 
finite size quantum antiferromagnetic chains 
by exact diagonalization. We shall see that already 
for very small chains, the spectrum is well described  
by the Bethe ansatz solution in the thermodynamic limit. 
Furthermore, an intuitive picture 
of the ground state and of the lowest lying excitations
of the strongly coupled two-flavor 
lattice Schwinger model emerges, due to the mapping of the gauge model onto the spin chain $-$ see chapter 5. 
 
The states of an antiferromagnetic chain are 
classified according to the quantum numbers of spin, 
third spin component, energy and momentum $|S, S^3,E,p>$. 
For a 4 site chain the 
momenta allowed for the states are: $0,
\frac{\pi}{2},\frac{3\pi}{2}\ {\rm mod}\  2\pi $. The ground state is
\begin{equation}
|g.s.>=|0,0,-3J,0>=\frac{1}{\sqrt{12}}(2|\uparrow
\downarrow\uparrow\downarrow>+2|\downarrow\uparrow\downarrow\uparrow>-
|\uparrow\uparrow\downarrow \downarrow> -|\uparrow
\downarrow \downarrow\uparrow>-|\downarrow \downarrow\uparrow\uparrow>-
|\downarrow\uparrow\uparrow\downarrow>)\quad .
\label{gs4}
\end{equation}
This state is $P$-parity even.
In fact, by the definition of $P$-parity given in Eq.(\ref{par}),
the $P$-parity inverted state (\ref{gs4}) is
obtained by reverting the order of the spins in each vector $|\dots>$
appearing in (\ref{gs4}), e.g. $|\downarrow\downarrow\uparrow\uparrow>
\buildrel P\over\longrightarrow|\uparrow\uparrow\downarrow \downarrow>$.

The $\lambda$'s associated to the ground state (solution of the Bethe 
ansatz equations (\ref{baeq})) are 
$\lambda_1=-\frac{1}{2\sqrt{3}}$ and $\lambda_2=\frac{1}{2\sqrt{3}}$. 
There is also an excited singlet 
\begin{equation}
|0,0,-J,\pi>=\frac{1}{\sqrt{4}}(|\downarrow \downarrow
\uparrow\uparrow>-|\downarrow\uparrow\uparrow\downarrow>
-|\uparrow\downarrow \downarrow\uparrow>+ |\uparrow
\uparrow\downarrow \downarrow>)\quad .
\label{sex1}
\end{equation}
It is $P$-even, so that it is a $S^{P}=0^{+}$ excitation, with
the same quantum numbers (the isospin is replaced by the spin)
of the lowest lying singlet excitation of the
strongly coupled Schwinger model discussed by Coleman~\cite{b46}. 
The state (\ref{sex1}) also coincides with the excited 
singlet described by the configuration (\ref{c1}). It has only two 
complex $\lambda$'s which arrange 
themselves in a string approximately of length 2, 
$\lambda_1=-\lambda_2=i\sqrt{\frac{\sqrt{481}-17}{8}}$ and 
there are two holes with $Q_{1}^{(h)}=-\frac{1}{2}$ and 
$Q_{2}^{(h)}=\frac{1}{2}$. 

There are also three excited triplets, whose highest weight states are 
\begin{eqnarray}
|1,1,-J,\frac{\pi}{2}>&=&\frac{1}{\sqrt{4}}
(|\downarrow\uparrow\uparrow\uparrow>+i|\uparrow\downarrow\uparrow\uparrow>-
|\uparrow\uparrow\downarrow\uparrow>-i|\uparrow\uparrow\uparrow\downarrow>)\\
|1,1,-2J,\pi>&=&\frac{1}{\sqrt{4}}(|\downarrow
\uparrow\uparrow\uparrow>-|\uparrow\downarrow\uparrow\uparrow>+
|\uparrow\uparrow\downarrow\uparrow>-|\uparrow\uparrow\uparrow\downarrow>)
\label{tr63}\\
|1,1,-J,\frac{3\pi}{2}>&=&\frac{1}{\sqrt{4}}(|\downarrow
\uparrow\uparrow\uparrow>-i|\uparrow\downarrow\uparrow\uparrow>-
|\uparrow\uparrow\downarrow\uparrow>+i|\uparrow\uparrow\uparrow\downarrow>)\quad .
\label{6tri}
\end{eqnarray}
Among these, only the non-degenerate state with the lowest energy has a
well defined $P$-parity (\ref{tr63}). It is a $S^P=1^{-}$
like the lowest lying triplet of the two-flavor strongly coupled Schwinger
model. The degenerate states can be always combined
to form a $P$-odd state. 

We thus see that within the states in a given 
configuration there is always a representative state with well defined parity,
the others are degenerate and can be used to construct state of well defined 
energy and parity. Moreover the parity of the representative states
(with respect to the parity of the ground state) is
the same of the one of the lowest-lying
Schwinger model excitations in strong coupling.

\begin{figure}[htb]
\begin{center}
\setlength{\unitlength}{0.00041700in}%
\begingroup\makeatletter\ifx\SetFigFont\undefined
\def\x#1#2#3#4#5#6#7\relax{\def\x{#1#2#3#4#5#6}}%
\expandafter\x\fmtname xxxxxx\relax \def\y{splain}%
\ifx\x\y   
\gdef\SetFigFont#1#2#3{%
  \ifnum #1<17\tiny\else \ifnum #1<20\small\else
  \ifnum #1<24\normalsize\else \ifnum #1<29\large\else
  \ifnum #1<34\Large\else \ifnum #1<41\LARGE\else
     \huge\fi\fi\fi\fi\fi\fi
  \csname #3\endcsname}%
\else
\gdef\SetFigFont#1#2#3{\begingroup
  \count@#1\relax \ifnum 25<\count@\count@25\fi
  \def\x{\endgroup\@setsize\SetFigFont{#2pt}}%
  \expandafter\x
    \csname \romannumeral\the\count@ pt\expandafter\endcsname
    \csname @\romannumeral\the\count@ pt\endcsname
  \csname #3\endcsname}%
\fi
\fi\endgroup
\begin{picture}(10224,7224)(1189,-6973)
\thicklines
\put(6001,-3361){\circle{450}}
\put(7801,-2161){\circle{450}}
\put(6676,-5986){\circle{450}}
\put(2401,-6961){\vector( 0, 1){7200}}
\put(1201,-961){\vector( 1, 0){10200}}
\put(2401,-2161){\line( 1, 0){7200}}
\put(2401,-2161){\line( 1, 0){7200}}
\put(2401,-3361){\line( 1, 0){3600}}
\put(2401,-4561){\line( 1, 0){ 75}}
\put(2251,-2161){\line( 1, 0){150}}
\put(2251,-3361){\line( 1, 0){150}}
\put(2251,-4561){\line( 1, 0){150}}
\put(4201,-661){\line( 0,-1){1500}}
\put(6001,-661){\line( 0,-1){2700}}
\put(7801,-661){\line( 0,-1){1500}}
\put(9601,-661){\line( 0,-1){300}}
\multiput(2626,-961)(-8.91855,14.86426){14}{\makebox(13.3333,20.0000){\SetFigFont{7}{8.4}{rm}.}}
\put(2513,-766){\line(-1, 0){224}}
\multiput(2289,-766)(-8.91855,-14.86426){14}{\makebox(13.3333,20.0000){\SetFigFont{7}{8.4}{rm}.}}
\multiput(2176,-961)(8.91855,-14.86426){14}{\makebox(13.3333,20.0000){\SetFigFont{7}{8.4}{rm}.}}
\put(2289,-1156){\line( 1, 0){224}}
\multiput(2513,-1156)(8.91855,14.86426){14}{\makebox(13.3333,20.0000){\SetFigFont{7}{8.4}{rm}.}}
\multiput(9826,-961)(-8.91855,14.86426){14}{\makebox(13.3333,20.0000){\SetFigFont{7}{8.4}{rm}.}}
\put(9713,-766){\line(-1, 0){224}}
\multiput(9489,-766)(-8.91855,-14.86426){14}{\makebox(13.3333,20.0000){\SetFigFont{7}{8.4}{rm}.}}
\multiput(9376,-961)(8.91855,-14.86426){14}{\makebox(13.3333,20.0000){\SetFigFont{7}{8.4}{rm}.}}
\put(9489,-1156){\line( 1, 0){224}}
\multiput(9713,-1156)(8.91855,14.86426){14}{\makebox(13.3333,20.0000){\SetFigFont{7}{8.4}{rm}.}}
\put(4201,-2161){\circle{450}}
\put(5851,-2311){\framebox(300,300){}}
\put(1726,-2161){\makebox(0,0)[lb]{\smash{\SetFigFont{10}{12.0}{rm}-1}}}
\put(2326,-4711){\framebox(300,300){}}
\multiput(6901,-5311)(-8.91855,14.86426){14}{\makebox(13.3333,20.0000){\SetFigFont{7}{8.4}{rm}.}}
\put(6788,-5116){\line(-1, 0){224}}
\multiput(6564,-5116)(-8.91855,-14.86426){14}{\makebox(13.3333,20.0000){\SetFigFont{7}{8.4}{rm}.}}
\multiput(6451,-5311)(8.91855,-14.86426){14}{\makebox(13.3333,20.0000){\SetFigFont{7}{8.4}{rm}.}}
\put(6564,-5506){\line( 1, 0){224}}
\multiput(6788,-5506)(8.91855,14.86426){14}{\makebox(13.3333,20.0000){\SetFigFont{7}{8.4}{rm}.}}
\put(6601,-6736){\framebox(300,300){}}
\put(2401,-2161){\line( 1, 0){5400}}
\put(2401,-4561){\line( 1, 0){7200}}
\put(9601,-961){\line( 0,-1){3600}}
\put(9451,-4711){\framebox(300,300){}}
\put(11101,-1561){\makebox(0,0)[lb]{\smash{\SetFigFont{14}{16.8}{rm}p}}}
\put(1726,-3361){\makebox(0,0)[lb]{\smash{\SetFigFont{10}{12.0}{rm}-2}}}
\put(1726,-4561){\makebox(0,0)[lb]{\smash{\SetFigFont{10}{12.0}{rm}-3}}}
\put(7201,-5461){\makebox(0,0)[lb]{\smash{\SetFigFont{10}{12.0}{rm}QUINTET}}}
\put(7201,-6061){\makebox(0,0)[lb]{\smash{\SetFigFont{10}{12.0}{rm}TRIPLETS}}}
\put(7201,-6661){\makebox(0,0)[lb]{\smash{\SetFigFont{10}{12.0}{rm}SINGLETS}}}
\put(3976,-361){\makebox(0,0)[lb]{\smash{\SetFigFont{10}{12.0}{rm}$\pi /2$}}}
\put(5851,-361){\makebox(0,0)[lb]{\smash{\SetFigFont{10}{12.0}{rm}$\pi$}}}
\put(7576,-361){\makebox(0,0)[lb]{\smash{\SetFigFont{10}{12.0}{rm}$3\pi /2$}}}
\put(9451,-361){\makebox(0,0)[lb]{\smash{\SetFigFont{10}{12.0}{rm}$2\pi$}}}
\put(1201,-61){\makebox(0,0)[lb]{\smash{\SetFigFont{20}{24.0}{rm}E/J}}}
\end{picture}
\end{center}
\caption{Four sites chain spectrum}
\label{4spectrum}
\end{figure}
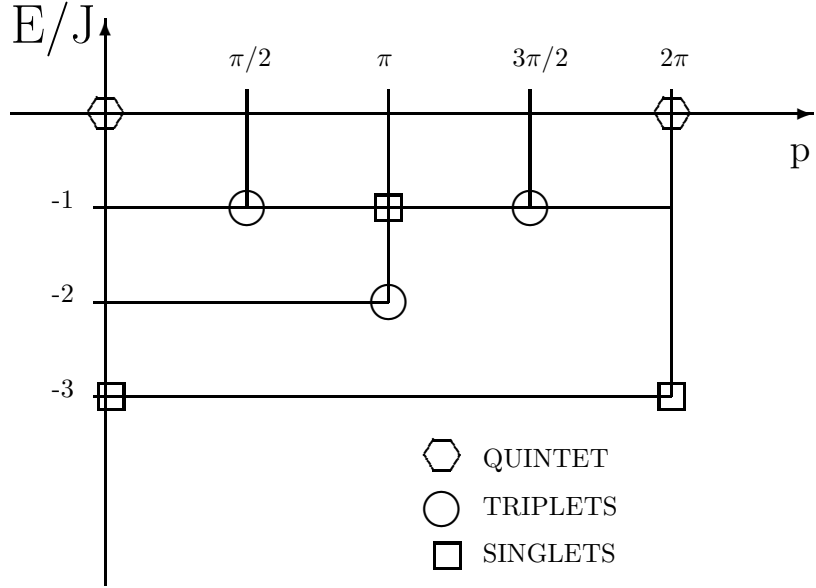

All the triplets in (\ref{6tri}) have one real $\lambda$ and two holes; 
they can be associated with the family of triplets (\ref{c2}).
 In table (\ref{tiqn}) I summarize the triplet $\lambda$'s and $Q^{(h)}$'s.
\begin{table}[htbp]
\begin{center}
\caption{Triplet internal quantum numbers }\label{tiqn}
\vspace{.1in}
\begin{tabular}{|lccc|}
\hline
TRIPLET&$\lambda$&$Q_{1}^{(h)}$&$Q_{2}^{(h)}$\rule{0in}{4ex}\\[2ex] \hline
$|1,1,-J,\frac{\pi}{2}>$& $\frac{1}{2} $&$-1$&$0$\rule{0in}{4ex}\\[2ex] \hline
$|1,1,-2J,\pi>$&$0$&$-1$&$1$\rule{0in}{4ex}\\[2ex] \hline
$|1,1,-J,\frac{3\pi}{2}>$&$-\frac{1}{2}$&$0$&$1$\rule{0in}{4ex}\\[2ex] \hline
\end{tabular}
\end{center}
\end{table}
The spectrum exhibits also a quintet, whose highest weight state is 
\begin{equation}
|2,2,0,0>=|\uparrow\uparrow\uparrow\uparrow>
\end{equation}
I report in fig.(\ref{4spectrum}) the spectrum of the 4 sites chain. 

\begin{figure}[htb]
\begin{center}
\setlength{\unitlength}{0.00050000in}%
\begingroup\makeatletter\ifx\SetFigFont\undefined
\def\x#1#2#3#4#5#6#7\relax{\def\x{#1#2#3#4#5#6}}%
\expandafter\x\fmtname xxxxxx\relax \def\y{splain}%
\ifx\x\y   
\gdef\SetFigFont#1#2#3{%
  \ifnum #1<17\tiny\else \ifnum #1<20\small\else
  \ifnum #1<24\normalsize\else \ifnum #1<29\large\else
  \ifnum #1<34\Large\else \ifnum #1<41\LARGE\else
     \huge\fi\fi\fi\fi\fi\fi
  \csname #3\endcsname}%
\else
\gdef\SetFigFont#1#2#3{\begingroup
  \count@#1\relax \ifnum 25<\count@\count@25\fi
  \def\x{\endgroup\@setsize\SetFigFont{#2pt}}%
  \expandafter\x
    \csname \romannumeral\the\count@ pt\expandafter\endcsname
    \csname @\romannumeral\the\count@ pt\endcsname
  \csname #3\endcsname}%
\fi
\fi\endgroup
\begin{picture}(12087,8499)(451,-8173)
\thicklines
\put(12001,-6661){\circle{450}}
\put(4801,-5461){\circle{450}}
\put(8401,-5461){\circle{450}}
\put(3001,-4861){\circle{450}}
\put(10201,-4861){\circle{450}}
\put(1201,-2761){\circle{450}}
\put(12001,-2761){\circle{450}}
\put(6601,-2161){\circle{450}}
\put(4801,-1786){\circle{450}}
\put(8401,-1786){\circle{450}}
\put(10726,-7186){\circle{450}}
\put(1201,-8161){\vector( 0, 1){8400}}
\put(3001,-61){\line( 0,-1){300}}
\put(4801,-61){\line( 0,-1){300}}
\put(6601,-61){\line( 0,-1){300}}
\put(8401,-61){\line( 0,-1){300}}
\put(10201,-61){\line( 0,-1){300}}
\put(12001,-61){\line( 0,-1){300}}
\put(976,-2161){\line( 1, 0){5625}}
\put(6601,-361){\line( 0,-1){1800}}
\put(976,-1261){\line( 1, 0){9225}}
\put(3001,-361){\line( 0,-1){900}}
\put(10201,-361){\line( 0,-1){900}}
\put(976,-1561){\line( 1, 0){5625}}
\put(4801,-361){\line( 0,-1){1425}}
\put(8401,-361){\line( 0,-1){1425}}
\put(976,-3061){\line( 1, 0){7425}}
\put(976,-2761){\line( 1, 0){225}}
\put(4801,-1786){\line( 0,-1){1275}}
\put(8401,-1786){\line( 0,-1){1275}}
\put(3001,-1261){\line( 0,-1){2700}}
\put(6601,-2161){\line( 0,-1){1800}}
\put(10201,-1261){\line( 0,-1){2700}}
\put(976,-3961){\line( 1, 0){9225}}
\put(976,-4861){\line( 1, 0){9225}}
\put(3001,-3961){\line( 0,-1){900}}
\put(10201,-3961){\line( 0,-1){900}}
\put(976,-5761){\line( 1, 0){225}}
\put(976,-5461){\line( 1, 0){7425}}
\put(8401,-3061){\line( 0,-1){2400}}
\put(4801,-3061){\line( 0,-1){2400}}
\put(976,-6661){\line( 1, 0){225}}
\put(976,-7861){\line( 1, 0){5625}}
\put(6601,-3961){\line( 0,-1){3900}}
\put(1201,-6661){\circle{450}}
\put(12001,-361){\line( 0,-1){6300}}
\put(1276,-7711){\makebox(0,0)[lb]{\smash{\SetFigFont{7}{8.4}{rm}$-(\sqrt{13}+5)/2$}}}
\put(1201,-6661){\line( 1, 0){10800}}
\put(1201,-5761){\line( 1, 0){10800}}
\put(1201,-2761){\line( 1, 0){10800}}
\put(6451,-8011){\framebox(300,300){}}
\put(6451,-1711){\framebox(300,300){}}
\put(1051,-5911){\framebox(300,300){}}
\put(11851,-5911){\framebox(300,300){}}
\put(2851,-4111){\framebox(300,300){}}
\put(10051,-4111){\framebox(300,300){}}
\multiput(6826,-3961)(-7.24632,12.07721){17}{\makebox(11.1111,16.6667){\SetFigFont{7}{8.4}{rm}.}}
\put(6713,-3766){\line(-1, 0){224}}
\multiput(6489,-3766)(-7.24632,-12.07721){17}{\makebox(11.1111,16.6667){\SetFigFont{7}{8.4}{rm}.}}
\multiput(6376,-3961)(7.24632,-12.07721){17}{\makebox(11.1111,16.6667){\SetFigFont{7}{8.4}{rm}.}}
\put(6489,-4156){\line( 1, 0){224}}
\multiput(6713,-4156)(7.24632,12.07721){17}{\makebox(11.1111,16.6667){\SetFigFont{7}{8.4}{rm}.}}
\multiput(5026,-3061)(-7.24632,12.07721){17}{\makebox(11.1111,16.6667){\SetFigFont{7}{8.4}{rm}.}}
\put(4913,-2866){\line(-1, 0){224}}
\multiput(4689,-2866)(-7.24632,-12.07721){17}{\makebox(11.1111,16.6667){\SetFigFont{7}{8.4}{rm}.}}
\multiput(4576,-3061)(7.24632,-12.07721){17}{\makebox(11.1111,16.6667){\SetFigFont{7}{8.4}{rm}.}}
\put(4689,-3256){\line( 1, 0){224}}
\multiput(4913,-3256)(7.24632,12.07721){17}{\makebox(11.1111,16.6667){\SetFigFont{7}{8.4}{rm}.}}
\multiput(8626,-3061)(-7.24632,12.07721){17}{\makebox(11.1111,16.6667){\SetFigFont{7}{8.4}{rm}.}}
\put(8513,-2866){\line(-1, 0){224}}
\multiput(8289,-2866)(-7.24632,-12.07721){17}{\makebox(11.1111,16.6667){\SetFigFont{7}{8.4}{rm}.}}
\multiput(8176,-3061)(7.24632,-12.07721){17}{\makebox(11.1111,16.6667){\SetFigFont{7}{8.4}{rm}.}}
\put(8289,-3256){\line( 1, 0){224}}
\multiput(8513,-3256)(7.24632,12.07721){17}{\makebox(11.1111,16.6667){\SetFigFont{7}{8.4}{rm}.}}
\multiput(3226,-1261)(-7.24632,12.07721){17}{\makebox(11.1111,16.6667){\SetFigFont{7}{8.4}{rm}.}}
\put(3113,-1066){\line(-1, 0){224}}
\multiput(2889,-1066)(-7.24632,-12.07721){17}{\makebox(11.1111,16.6667){\SetFigFont{7}{8.4}{rm}.}}
\multiput(2776,-1261)(7.24632,-12.07721){17}{\makebox(11.1111,16.6667){\SetFigFont{7}{8.4}{rm}.}}
\put(2889,-1456){\line( 1, 0){224}}
\multiput(3113,-1456)(7.24632,12.07721){17}{\makebox(11.1111,16.6667){\SetFigFont{7}{8.4}{rm}.}}
\multiput(10426,-1261)(-7.24632,12.07721){17}{\makebox(11.1111,16.6667){\SetFigFont{7}{8.4}{rm}.}}
\put(10313,-1066){\line(-1, 0){224}}
\multiput(10089,-1066)(-7.24632,-12.07721){17}{\makebox(11.1111,16.6667){\SetFigFont{7}{8.4}{rm}.}}
\multiput(9976,-1261)(7.24632,-12.07721){17}{\makebox(11.1111,16.6667){\SetFigFont{7}{8.4}{rm}.}}
\put(10089,-1456){\line( 1, 0){224}}
\multiput(10313,-1456)(7.24632,12.07721){17}{\makebox(11.1111,16.6667){\SetFigFont{7}{8.4}{rm}.}}
\put(1201,-61){\line(-2,-3){300}}
\put(901,-511){\line( 1, 0){600}}
\put(1501,-511){\line(-2, 3){300}}
\put(1201,-61){\line( 0, 1){  0}}
\put(12001,-136){\line(-2,-3){300}}
\put(11701,-586){\line( 1, 0){600}}
\put(12301,-586){\line(-2, 3){300}}
\put(12001,-136){\line( 0, 1){  0}}
\put(10576,-8011){\framebox(300,300){}}
\multiput(7951,-7861)(-7.24632,12.07721){17}{\makebox(11.1111,16.6667){\SetFigFont{7}{8.4}{rm}.}}
\put(7838,-7666){\line(-1, 0){224}}
\multiput(7614,-7666)(-7.24632,-12.07721){17}{\makebox(11.1111,16.6667){\SetFigFont{7}{8.4}{rm}.}}
\multiput(7501,-7861)(7.24632,-12.07721){17}{\makebox(11.1111,16.6667){\SetFigFont{7}{8.4}{rm}.}}
\put(7614,-8056){\line( 1, 0){224}}
\multiput(7838,-8056)(7.24632,12.07721){17}{\makebox(11.1111,16.6667){\SetFigFont{7}{8.4}{rm}.}}
\put(7726,-6961){\line(-2,-3){300}}
\put(7426,-7411){\line( 1, 0){600}}
\put(8026,-7411){\line(-2, 3){300}}
\put(7726,-6961){\line( 0, 1){  0}}
\put(601,-361){\vector( 1, 0){11925}}
\put(976,-1786){\line( 1, 0){7425}}
\put(2851, 14){\makebox(0,0)[lb]{\smash{\SetFigFont{12}{14.4}{rm}$\pi/3$}}}
\put(4651, 14){\makebox(0,0)[lb]{\smash{\SetFigFont{12}{14.4}{rm}$2\pi/3$}}}
\put(6526, 14){\makebox(0,0)[lb]{\smash{\SetFigFont{12}{14.4}{rm}$\pi$}}}
\put(8251, 14){\makebox(0,0)[lb]{\smash{\SetFigFont{12}{14.4}{rm}$4\pi/3$}}}
\put(10051, 14){\makebox(0,0)[lb]{\smash{\SetFigFont{12}{14.4}{rm}$5\pi/3$}}}
\put(11851, 89){\makebox(0,0)[lb]{\smash{\SetFigFont{12}{14.4}{rm}$2\pi$}}}
\put(451, 14){\makebox(0,0)[lb]{\smash{\SetFigFont{17}{20.4}{rm}E/J}}}
\put(12226,-1036){\makebox(0,0)[lb]{\smash{\SetFigFont{17}{20.4}{rm}p}}}
\put(8101,-7261){\makebox(0,0)[lb]{\smash{\SetFigFont{9}{10.8}{rm}SEVENTHPLET}}}
\put(8101,-7945){\makebox(0,0)[lb]{\smash{\SetFigFont{9}{10.8}{rm}QUINTETS}}}
\put(11026,-7261){\makebox(0,0)[lb]{\smash{\SetFigFont{9}{10.8}{rm}TRIPLETS}}}
\put(11026,-7945){\makebox(0,0)[lb]{\smash{\SetFigFont{9}{10.8}{rm}SINGLETS}}}
\put(1201,-1486){\makebox(0,0)[lb]{\smash{\SetFigFont{7}{8.4}{rm}$-(5-\sqrt{13})/2$}}}
\put(1276,-1711){\makebox(0,0)[lb]{\smash{\SetFigFont{7}{8.4}{rm}$-(7-\sqrt{17})/2$}}}
\put(1201,-2011){\makebox(0,0)[lb]{\smash{\SetFigFont{7}{8.4}{rm}$-1$}}}
\put(1276,-1186){\makebox(0,0)[lb]{\smash{\SetFigFont{7}{8.4}{rm}$-1/2$}}}
\put(1276,-2536){\makebox(0,0)[lb]{\smash{\SetFigFont{7}{8.4}{rm}$-(5-\sqrt{5})/2$}}}
\put(1501,-2986){\makebox(0,0)[lb]{\smash{\SetFigFont{7}{8.4}{rm}$-3/2$}}}
\put(1276,-3886){\makebox(0,0)[lb]{\smash{\SetFigFont{7}{8.4}{rm}$-2$}}}
\put(1276,-4786){\makebox(0,0)[lb]{\smash{\SetFigFont{7}{8.4}{rm}$-5/2$}}}
\put(1276,-5386){\makebox(0,0)[lb]{\smash{\SetFigFont{7}{8.4}{rm}$-(\sqrt{17}+7)/4$}}}
\put(1501,-5686){\makebox(0,0)[lb]{\smash{\SetFigFont{7}{8.4}{rm}$-3$}}}
\put(1501,-6511){\makebox(0,0)[lb]{\smash{\SetFigFont{7}{8.4}{rm}$-(\sqrt{5}+5)/2$}}}
\end{picture}
\end{center}
\caption{Six sites chain spectrum}
\label{sixspectrum}
\end{figure}
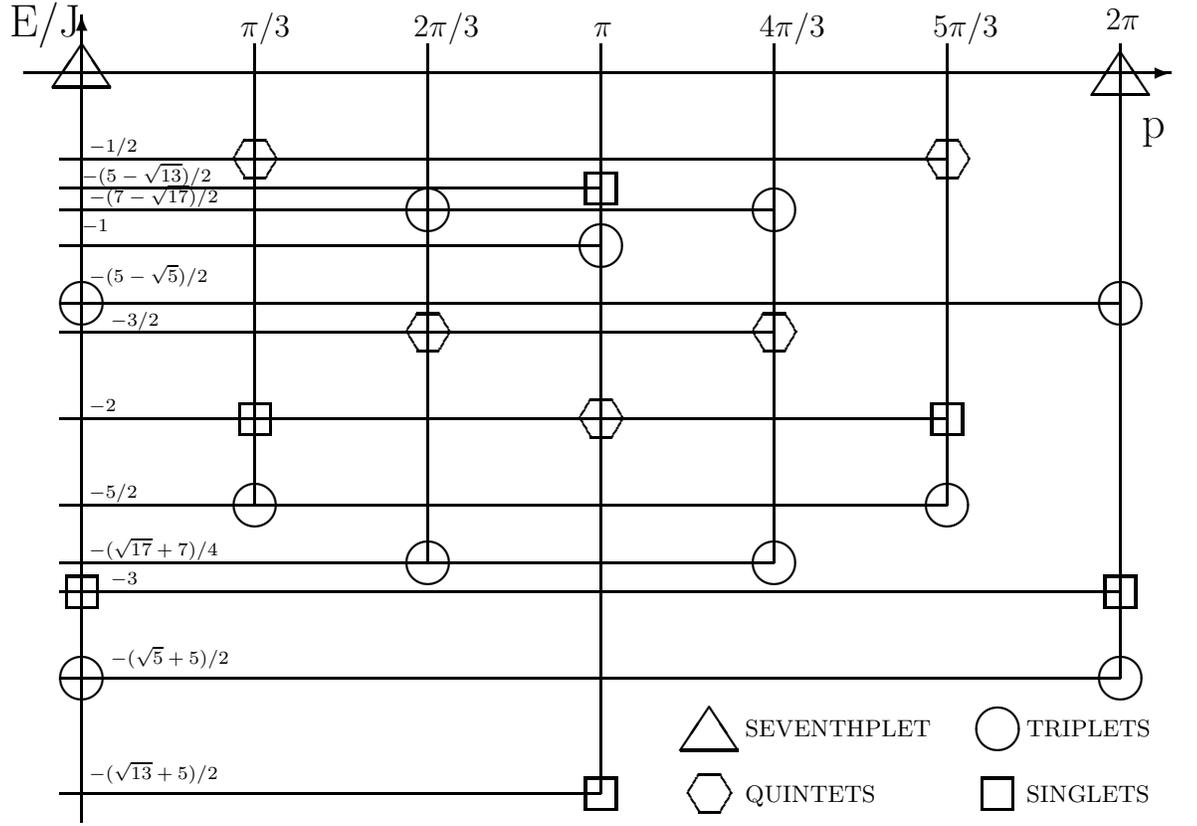

Let us analize the spectrum of the 6 site antiferromagnetic chain. 
The momenta allowed for 
the states are now $0,\frac{\pi}{3},\frac{2\pi}{3},\pi,\frac{4\pi}{3},
\frac{5\pi}{3}\ {\rm mod}\  2\pi$. 
The ground state is
\begin{eqnarray}
|g.s.>&=&|0,0,-\frac{J}{2}(5+\sqrt{13}),\pi>=
\frac{1}{\sqrt{26-6\sqrt{13}}}
\{|\downarrow\uparrow\downarrow\uparrow\downarrow\uparrow>-
|\uparrow\downarrow\uparrow\downarrow\uparrow\downarrow>\nonumber\\
&+&\frac{1-\sqrt{13}}{6}(|\uparrow\uparrow\downarrow
\uparrow\downarrow\downarrow>-|\uparrow\downarrow\uparrow
\downarrow\downarrow\uparrow>+
|\downarrow\uparrow\downarrow\downarrow\uparrow\uparrow>-|
\uparrow\downarrow\downarrow\uparrow\uparrow\downarrow>
+|\downarrow\downarrow\uparrow\uparrow\downarrow\uparrow>- |
\downarrow\uparrow\uparrow\downarrow\uparrow\downarrow>\nonumber\quad \quad \\
&-&|\downarrow\downarrow\uparrow\downarrow\uparrow\uparrow>+|
\downarrow\uparrow\downarrow\uparrow\uparrow\downarrow>-
|\uparrow\downarrow\uparrow\uparrow\downarrow\downarrow>+|
\downarrow\uparrow\uparrow\downarrow\downarrow\uparrow>
-|\uparrow\uparrow\downarrow\downarrow\uparrow\downarrow>+|
\uparrow\downarrow\downarrow\uparrow\downarrow\uparrow>)\nonumber\quad \quad \\
&+&\frac{4-\sqrt{13}}{3}(|\uparrow\uparrow\uparrow\downarrow
\downarrow\downarrow>-|\uparrow\uparrow\downarrow\downarrow\downarrow\uparrow>+
|\uparrow\downarrow\downarrow\downarrow\uparrow\uparrow>
-|\downarrow\downarrow\downarrow\uparrow\uparrow\uparrow>+|
\downarrow\downarrow\uparrow\uparrow\uparrow\downarrow>-
|\downarrow\uparrow\uparrow\uparrow\downarrow\downarrow>)\}
\quad \quad \quad .
\label{6gs}
\end{eqnarray}
This state is odd under $P$-parity.
The spectrum of the six sites chain is reported in 
fig.(\ref{sixspectrum}). There are 9 triplets in the spectrum. 
In~\cite{b45} 
it was already pointed out that the number of lowest 
lying triplets for a finite system with N sites is 
$N(N+2)/8$, so for $N=6$ there are 6 lowest lying triplet states. 
In order to identify these 6 states among the 9 that 
are exhibited by the spectrum of fig.(\ref{sixspectrum}), 
it is necessary to compute their $\lambda$'s and their $Q$'s.
In this way in fact, I can find out which are the triplets characterized by 
two holes and thus belonging to the triplet of type (\ref{c2}). In 
table (\ref{6tiqn}) I report the internal quantum numbers 
of the lowest lying triplets. The $Q^{(h)}$'s vary in the segment 
$[-\frac{3}{2},\frac{3}{2}]$.
The highest weight state of the triplet of zero 
momentum and energy $-(J/2)(5+\sqrt{5})$ reads
\begin{eqnarray}
|0,0,-\frac{J}{2}(5+\sqrt{5}),0>&=&\frac{1}{\sqrt{45-15\sqrt{5}}}
\{\frac{-3+\sqrt{5}}{2}(
|\downarrow\downarrow\uparrow\uparrow\uparrow\uparrow>+
|\downarrow\uparrow\uparrow\uparrow\uparrow\downarrow>+
|\uparrow\uparrow\uparrow\uparrow\downarrow\downarrow>+
|\uparrow\uparrow\uparrow\downarrow\downarrow\uparrow>\nonumber\\
&+&|\uparrow\uparrow\downarrow\downarrow\uparrow\uparrow>+
|\uparrow\downarrow\downarrow\uparrow\uparrow\uparrow>)\nonumber\\
&+&(|\downarrow\uparrow\downarrow\uparrow\uparrow\uparrow>+
|\uparrow\downarrow\uparrow\uparrow\uparrow\downarrow>+
|\downarrow\uparrow\uparrow\uparrow\downarrow\uparrow>+
|\uparrow\uparrow\uparrow\downarrow\uparrow\downarrow>+
|\uparrow\uparrow\downarrow\uparrow\downarrow\uparrow>+
|\uparrow\downarrow\uparrow\downarrow\uparrow\uparrow>)\nonumber\\
&+&(1-\sqrt{5})(|\downarrow\uparrow\uparrow\downarrow
\uparrow\uparrow>+|\uparrow\uparrow\downarrow\uparrow\uparrow\downarrow>+
|\uparrow\downarrow\uparrow\uparrow\downarrow\uparrow>)\}\quad\quad .
\label{tpo}
\end{eqnarray}
One can get the triplet of energy $-(J/2)(5-\sqrt{5})$ 
from (\ref{tpo}) by changing $\sqrt{5}\rightarrow -\sqrt{5}$.
As can be explicitly checked from (\ref{tpo}),
the two non-degenerate triplets of zero momentum are then $P$-parity even, 
namely they have opposite parity 
with respect to that of the  ground state, as it happens for the 
lowest lying triplet excitations of the two-flavor 
Schwinger model. 
For what concerns the degenerate triplets of momenta 
$\pi/3$ and $5\pi/3$ (or $2\pi/3$ and $4\pi/3$) they do not have definite 
$P$-parity, but it is always possible to take a 
linear combination of them with parity opposite to the ground state.  
\begin{table}[htbp]
\begin{center}
\caption{Triplet internal quantum numbers }\label{6tiqn}
\vspace{.1in}
\begin{tabular}{|lcccc|}
\hline
TRIPLET& $\lambda_1$ & $\lambda_2 $ & $Q_{1}^{(h)}$ & 
$Q_{2}^{(h)}$ \rule{0in}{4ex}\\[2ex] \hline

$|1,1,-\frac{5+\sqrt{5}}{2}J,0>$ & $-\sqrt{\frac{5-2\sqrt{5}}{20}}$ 
& $\sqrt{\frac{5-2\sqrt{5}}{20}}$ & $-\frac{1}{2}$ & $\frac{1}{2}$ 
\rule{0in}{4ex}\\[2ex] \hline

$|1,1,-\frac{5-\sqrt{5}}{2}J,0>$ & $-\sqrt{\frac{5+2\sqrt{5}}{20}}$ 
& $\sqrt{\frac{5+2\sqrt{5}}{20}}$ & $-\frac{3}{2}$ & $\frac{3}{2}$     
\rule{0in}{4ex}\\[2ex] \hline

$|1,1,-\frac{5}{2}J,\frac{\pi}{3}>$ & $-\frac{\sqrt{3}+\sqrt{\pi}}{8}$ 
& $-\frac{\sqrt{3}-\sqrt{\pi}}{8}$ & $-\frac{3}{2}$ & $\frac{1}{2}$     
\rule{0in}{4ex}\\[2ex] \hline

$|1,1,-\frac{7+\sqrt{17}}{4}J,\frac{2\pi}{3}>$ & $\frac{-2\sqrt{3}-
\sqrt{-2+2\sqrt{17}}}{2+2\sqrt{17}}$ & 
$\frac{-2\sqrt{3}+\sqrt{-2+2\sqrt{17}}}{2+2\sqrt{17}}$ 
& $-\frac{3}{2}$ & $-\frac{1}{2}$     
\rule{0in}{4ex}\\[2ex] \hline

$|1,1,-\frac{7+\sqrt{17}}{4}J,\frac{4\pi}{3}>$ 
& $\frac{2\sqrt{3}-\sqrt{-2+2\sqrt{17}}}{2+2\sqrt{17}}$ & 
$\frac{2\sqrt{3}+\sqrt{-2+2\sqrt{17}}}{2+2\sqrt{17}}$ 
& $\frac{1}{2}$ & $\frac{3}{2}$     
\rule{0in}{4ex}\\[2ex] \hline

$|1,1,-\frac{5}{2}J,\frac{5\pi}{3}>$ & $\frac{\sqrt{3}-
\sqrt{\pi}}{8}$ & $\frac{\sqrt{3}+\sqrt{\pi}}{8}$ 
& $-\frac{1}{2}$ & $\frac{3}{2}$\rule{0in}{4ex}\\[2ex] \hline
\end{tabular}
\end{center}
\end{table} 

The remaining three triplets in fig.(\ref{sixspectrum}) 
have no real $\lambda$'s and are characterized by a 
string of length 2 and four holes 
for $Q=-\frac{3}{2},-\frac{1}{2},\frac{1}{2},\frac{3}{2}$, $i.e.$ do 
not belong to the type (\ref{c2}). More precisely, two 
triplets have a string approximately of length 2, due to the finite size 
of the system, while the triplet of momentum $\pi$ has a 
string exactly of length 2. In table (\ref{26tiqn}) I summarize the quantum numbers of 
these triplets. 

\begin{table}[htbp]
\begin{center}
\caption{Four holes triplet internal quantum numbers }\label{26tiqn}
\vspace{.1in}
\begin{tabular}{|lcc|}
\hline
TRIPLET&$\lambda_1$&$\lambda_2$\rule{0in}{4ex}\\[2ex] \hline

$|1,1,-\frac{7-\sqrt{17}}{4}J,\frac{2\pi}{3}>$ 
& $\frac{2\sqrt{3}-i\sqrt{2+2\sqrt{17}}}{-2+2\sqrt{17}}$ 
& $\frac{2\sqrt{3}+i\sqrt{2+2\sqrt{17}}}{-2+2\sqrt{17}}$ 
\rule{0in}{4ex}\\[2ex] \hline

$|1,1,-J,\pi>$ & $-\frac{i}{2}$ & $\frac{i}{2}$ 
\rule{0in}{4ex}\\[2ex] \hline

$|1,1,-\frac{7-\sqrt{17}}{4}J,\frac{4\pi}{3}>$ 
& $\frac{2\sqrt{3}+i\sqrt{2+2\sqrt{17}}}{2-2\sqrt{17}}$ 
& $\frac{2\sqrt{3}-i\sqrt{2+2\sqrt{17}}}{2-2\sqrt{17}}$ 
\rule{0in}{4ex}\\[2ex]\hline
\end{tabular}
\end{center}
\end{table} 

In fig.(\ref{sixspectrum}) it is shown that the spectrum 
exhibits five singlet states. The lowest lying state
at momentum $\pi$ is the ground state. Then there are 
three excited singlets characterized by the 
configuration with two holes (\ref{c1}), $i.e.$ they 
have one real $\lambda$ and a string of length almost 2.
In table (\ref{esin}) I summarize their quantum numbers. 
Among these singlets, those which are
not degenerate, have   
$P$-parity equal to that of the
ground state (odd) as it happens in the two-flavor Schwinger 
model. The non-  singlet in fact reads
\begin{eqnarray}
|0,0,-3J,0>&=&\frac{1}{\sqrt{12}}\{|\uparrow\uparrow\downarrow
\uparrow\downarrow\downarrow>+|\uparrow\downarrow\uparrow
\downarrow\downarrow\uparrow>+
|\downarrow\uparrow\downarrow\downarrow\uparrow\uparrow>+
|\uparrow\downarrow\downarrow\uparrow\uparrow\downarrow>+
|\downarrow\downarrow\uparrow\uparrow\downarrow\uparrow>+
|\downarrow\uparrow\uparrow\downarrow\uparrow\downarrow>\nonumber\quad \quad\\
&-&|\downarrow\downarrow\uparrow\downarrow\uparrow\uparrow>-
|\downarrow\uparrow\downarrow\uparrow\uparrow\downarrow>-
|\uparrow\downarrow\uparrow\uparrow\downarrow\downarrow>-
|\downarrow\uparrow\uparrow\downarrow\downarrow\uparrow>-
|\uparrow\uparrow\downarrow\downarrow\uparrow\downarrow>-
|\uparrow\downarrow\downarrow\uparrow\downarrow\uparrow>\}\quad .
\label{spo}
\end{eqnarray}
The degenerate singlets are again not eigenstates of the $P$-parity, 
but it is always possible to take a linear combination of 
them with the a $P$-parity 
that coincides with that of the representative state (\ref{spo})
of the configuration.
\begin{table}[htbp]
\begin{center}
\caption{Singlet internal quantum numbers }\label{esin}
\vspace{.1in}
\begin{tabular}{|lcccc|}
\hline
SINGLET&$\lambda$ & $\lambda_S $ & $Q_1^{(h)}$ 
& $Q_2^{(h)}$\rule{0in}{4ex}\\[2ex] \hline

$|0,0,-3J,0>$ & 0 & 0 & $-1$ & 1 \rule{0in}{4ex}\\[2ex] \hline

$|0,0,-2J,\frac{\pi}{3}>$ & $-\frac{\sqrt{3}+2\sqrt{6}}{14}$ 
& $\frac{-2+3\sqrt{2}}{\sqrt{3}(4+\sqrt{2})}$ & $0$ 
& 1 \rule{0in}{4ex}\\[2ex] \hline

$|0,0,-2J,\frac{5\pi}{3}>$ & $\frac{\sqrt{3}+2\sqrt{6}}{14} $ 
&  $\frac{2-3\sqrt{2}}{\sqrt{3}(4+\sqrt{2})} $& $-1$ 
& 0 \rule{0in}{4ex}\\[2ex]\hline
\end{tabular}
\end{center}
\end{table}

The remaining singlet $|0,0,-\frac{5-\sqrt{13}}{2} J,\pi>$ 
it is not of the type (\ref{c1}). It is characterized by 
a string approximately of length 3 with $\lambda_{1,1}=
i\sqrt{\frac{5+2\sqrt{13}}{12}}$, $\lambda_{2,1}=0$ 
and $\lambda_{3,1}=-i\sqrt{\frac{5+2\sqrt{13}}{12}}$.

Even in finite systems very small like the 4 and 6 sites chains, 
the ``string hypothesis" is a very good approximation and 
it allows us to classify and distinguish among states with the same spin.   

The ground state of the antiferromagnetic Heisenberg 
chain with N sites is a linear combination of all the 
$\left( \begin{array}{c} N \\ 
\frac{N}{2} 
\end{array} \right)$ states with $\frac{N}{2}$ spins up and $\frac{N}{2}$ 
spins down. 
These states group themselves into sets with the same coefficient 
in the linear combination  according to the fact that 
the ground state is translationally invariant (with momentum 0 ($\pi$) 
for $\frac{N}{2}$ even (odd)), it 
is an eigenstate of $P$-parity and it is 
invariant under the exchange of up with down spins. 
The states belonging to the same set 
have the same number of domain walls, which ranges from 
$N$, for the two N\'eel states, to 2 for the states with $\frac{N}{2}$ 
adjacent spins up and $\frac{N}{2}$ adjacent spins down.  

The ground state of the 8 sites chain is 

\begin{equation}
|g.s.>= \frac{1}{ \sqrt{\cal{N}} }
(|\psi_{8} >+\alpha |\psi_6 ^{(1)} >+\beta |\psi_6^{(2)} >
+\gamma |\psi_4 ^{(1)}>+\delta |\psi_4 ^{(2)}>+\epsilon
|\psi_4 ^{(3)} >+\zeta |\psi_2>)
\label{8gs}
\end{equation} 
where 
\begin{eqnarray}
& &|\psi_{8}> = |\uparrow\downarrow\uparrow\downarrow\uparrow
\downarrow\uparrow\downarrow>+
|\downarrow\uparrow\downarrow\uparrow\downarrow\uparrow
\downarrow\uparrow>\\
& &|\psi_6 ^{(1)}> = |\uparrow\uparrow\downarrow\uparrow
\downarrow\uparrow\downarrow\downarrow>+
|\downarrow\downarrow\uparrow\downarrow\uparrow\downarrow
\uparrow\uparrow>+translated\quad states \\
& &|\psi_6^{(2)}>= |\uparrow\uparrow\downarrow\uparrow
\downarrow\downarrow\uparrow\downarrow>+
|\downarrow\downarrow\uparrow\downarrow\uparrow\uparrow
\downarrow\uparrow>+translated\quad states\\
& &|\psi_4 ^{(1)}>= |\uparrow\uparrow\downarrow\downarrow
\uparrow\uparrow\downarrow\downarrow>+translated\quad states\\
& &|\psi_4 ^{(2)}> = |\uparrow\uparrow\uparrow\downarrow
\downarrow\downarrow\uparrow\downarrow>+
|\downarrow\downarrow\downarrow \uparrow\uparrow\uparrow
\downarrow\uparrow>+translated\quad states \\
& &|\psi_4 ^{(3)}> = |\uparrow\uparrow\downarrow\downarrow
\downarrow\uparrow\uparrow\downarrow>+
|\downarrow\downarrow\uparrow\uparrow\uparrow\downarrow
\downarrow\uparrow>+translated\quad states\\
& &|\psi_2> = |\uparrow\uparrow\uparrow\uparrow\downarrow
\downarrow\downarrow\downarrow>+translated\quad states\ .
\label{8states}
\end{eqnarray}
By direct diagonalization one gets
\begin{eqnarray}
\alpha&=&-0.412773\\
\beta&=&0.344301\\
\gamma&=&0.226109\\
\delta&=&-0.087227\\
\epsilon&=&0.136945\\
\zeta&=&0.018754\\
\cal{N}&=&2+16\alpha^2+8\beta^2+4\gamma^2+16\delta^2
+16\epsilon^2+8\zeta^2=6.30356\quad .
\end{eqnarray}

The energy of the ground state is 
\begin{equation}
E_{g.s.}=-5.65109J\quad .
\label{egs8}
\end{equation}
Eq.(\ref{egs8}) differs only by $1.8\%$ from the thermodynamic 
limit expression $E_{g.s.}=-8\ln 2=-5.54518$. Moreover 
also the correlation function of distance 2 Eq.(\ref{corrd2}) 
computed for the 8 sites chain is $G(2)=0.1957N$, value which is $7\%$ 
higher than the exact answer Eq.(\ref{corrd2n}).

In the analysis of finite size systems I were able to find 
the coefficient $\beta$ of the first set of states containing
$N-2$ domain walls in 
the ground state. These states 
are obtained interchanging two adjacent spins in the N\'eel states. 
The $\beta$ is for a generic chain of $N$-sites
\begin{equation}
\beta=\frac{N+2E_{g.s.}}{N}=1-2\ln 2\quad .
\end{equation}    

\subsection{Spin-spin correlators}

The explicit computation of spin-spin correlations is far from being trivial since the correlator 
$G(r)=<g.s.|\vec{S}_{0}\cdot \vec{S}_{r}|g.s.>$ is not known for arbitrary lattice separations $r$.
For $r=2$ it was computed by M. Takahashi \cite{b38} 
in his perturbative analysis of the half filled Hubbard model in one dimension. For $r>2$ no exact numerical 
values of $G(r)$ are known. In \cite{b47} were given two representations of $G(r)$, while in \cite{b48,b49} the exact asymptotic 
($r\rightarrow \infty$) expression of $G(r)$ was derived. 

In order to explicitly compute the second order energies Eq.(\ref{senes}) and 
Eq.(\ref{senet}) one has to evaluate the correlation function
\begin{equation}
G(2)=\frac{1}{N}\sum_{x=1}^{N}<g.s.|\vec{S}_{x}\cdot \vec{S}_{x+2}|g.s>
\label{corrd2}
\end{equation}
which has been exactly computed in~\cite{b38} and is given by
\begin{equation}
G(2)=\frac{1}{4}(1-16\ln 2+9\zeta (3))=0.1820\quad . 
\label{corrd2n}
\end{equation}
In the following I shall show how the knowledge of this correlator 
allows one to compute explicitly the first three ``emptiness formation 
probabilities'', used in Ref.~\cite{b47} in the study of the Heisenberg
chain correlators, $G(r)$. The isotropy of the Heisenberg model implies that
\begin{equation}
\sum_{x=1}^{N}<g.s.|\vec{S}_{x}\cdot \vec{S}_{x+2}|g.s.>=3
\sum_{x=1}^{N}<g.s.|S^{3}_{x}\cdot S^3_{x+2}|g.s.>\quad .
\label{e1}
\end{equation}
Let us introduce the probability $P_3$ for finding three adjacent spins
in a given position in the Heisenberg antiferromagnetic vacuum. 
Taking advantage of the isotropy of the Heisenberg model 
ground state and of its
translational invariance, it is easy to see that the correlator (\ref{e1}) 
can be written in terms of the $P_3$'s as
\begin{equation}
\sum_{x=1}^{N}<g.s.|S^{3}_{x}\cdot S^3_{x+2}|g.s>=N~
\frac{1}{4}~2~(~P_3(\uparrow\uparrow\uparrow)+P_3(\uparrow\downarrow\uparrow)-
P_3(\uparrow\uparrow\downarrow)-P_3(\downarrow\uparrow\uparrow)~)\quad .
\label{e2}
\end{equation}
The factor 2 appears in (\ref{e2}) due again to the isotropy of the 
Heisenberg model: the probability of  
a configuration and of the configuration rotated by $\pi$ around the 
chain axis, are the same.  

In \cite{b47} the so called ``emptiness formation probability'' $P(x)$
was introduced.
\begin{equation}
P(x)=<g.s.|\prod_{j=1}^{x}P_{j}|g.s.>\quad ,
\label{e3}
\end{equation}
where 
\begin{equation}
P_{j}=\frac{1}{2}(\sigma_{j}^{3}+1)
\label{e4}
\end{equation}
and $\sigma_{j}^{3}$ is the Pauli matrix.
$P(x)$ determines the probability of finding $x$ adjacent spins up in the 
antiferromagnetic vacuum.
One gets
\begin{eqnarray}
P(\uparrow\uparrow\uparrow)&=&P(3)\\
P(\uparrow\downarrow\uparrow)&=&P(1)-2P(2)+P(3)\\
P(\downarrow\uparrow\uparrow)&=&P(\uparrow\uparrow\downarrow)=P(2)-P(3)
\end{eqnarray}
so that Eq.(\ref{corrd2}) reads
\begin{equation}
G(2)=2P(3)-2P(2)+\frac{1}{2}P(1)\quad .
\label{corrd22}
\end{equation}
Using the exact  value the correlator $G(2)$ computed in~\cite{b38}
from (\ref{corrd22}) and from the known values of $P(2)$
and $P(1)$ given in \cite{b47}
\begin{eqnarray}
P(1)&=&\frac{1}{2}\label{p1}\\
P(2)&=&\frac{1}{3}(1-\ln 2)\label{p2}\\
\label{p1p2}
\end{eqnarray}
one gets
\begin{eqnarray}
P(3)&=&\frac{1}{3}(1-7\ln 2)+\frac{9}{8} \zeta(3)\label{p3}\quad .
\label{efp3}
\end{eqnarray}
For the general emptiness formation probability $P(x)$ of the antiferromagnetic
Heisenberg chain,
an integral representation was given in~\cite{b47},
but, to my knowledge, the exact value of $P(3)$ (\ref{efp3}) was not
known.

I illustrate now the computation of a spin-spin correlator which appears in the mass spectrum of the two-flavor 
lattice Schwinger model
\begin{eqnarray}
& &<g.s.|\vec{V}\cdot \vec{V}|g.s.>=\sum_{x,y=1}^{N}(<g.s.|(\vec{S}_{x}\cdot \vec{S}_{y})(\vec{S}_{x+1}\cdot \vec{S}_{y+1})|g.s.>\nonumber\\
&-&<g.s.|(\vec{S}_{x}\cdot \vec{S}_{y+1})(\vec{S}_{x+1}\cdot \vec{S}_{y})|g.s.>)
-\sum_{x=1}^{N}<g.s.|\vec{S}_{x}\cdot \vec{S}_{x+1}|g.s.>\quad \quad .
\label{vv}
\end{eqnarray}
It is possible to extract a numerical value from Eq.(\ref{vv}) only within the random phase approximation \cite{b38,b50}. 
For this purpose it is first convenient 
to rewrite the unconstrained sum over the sites x and y as a sum where all the four spins involved in the VEV's lie on different sites,
\begin{eqnarray}
<g.s.|\vec{V}\cdot \vec{V}|g.s.>&=&\sum_{\stackrel{y\neq x}{ y\neq x\pm 1}}(<g.s.|(\vec{S}_{x}\cdot \vec{S}_{y})(\vec{S}_{x+1}\cdot \vec{S}_{y+1})|g.s.>
\nonumber\\
&-&<g.s.|(\vec{S}_{x}\cdot \vec{S}_{y+1})(\vec{S}_{x+1}\cdot \vec{S}_{y})|g.s.>)+\frac{3}{8}N\nonumber\\
&-&\frac{1}{2}\sum_{x=1}^{N}<g.s.|\vec{S}_{x}\cdot \vec{S}_{x+1}|g.s.>-\sum_{x=1}^{N}
<g.s.|\vec{S}_{x}\cdot \vec{S}_{x+2}|g.s.>\quad \quad 
\label{vvv}
\end{eqnarray}
and then factorize the four spin operators in Eq.(\ref{vvv}) as 
\begin{eqnarray}
<g.s.|\vec{V}\cdot \vec{V}|g.s.>&=&N\sum_{r=2}^{\infty}(<g.s.|\vec{S}_{0}\cdot \vec{S}_{r}|g.s.>^2 -<g.s.|\vec{S}_{0}\cdot \vec{S}_{r+1}|g.s.>
<g.s.|\vec{S}_{1}\cdot \vec{S}_{r}|g.s.>)\nonumber\\
&+&\frac{3}{8}N-\frac{1}{2}\sum_{x=1}^{N}<g.s.|\vec{S}_{x}\cdot \vec{S}_{x+1}|g.s.>-\sum_{x=1}^{N}<g.s.|\vec{S}_{x}\cdot \vec{S}_{x+2}|g.s.>\quad .
\label{rpavv}
\end{eqnarray}
Of course, Eq.(\ref{rpavv}) provides an answer larger than the exact result; 
terms such as $<(\ldots)(\ldots)>$ yield negative contributions 
which are eliminated once one factorizes them in the form $<(...)><(...)>$. 
This is easily checked also by direct computation on finite size systems. 

The spin-spin correlation functions G(r) are exactly known for $r=1,2$. For the spin-spin correlation functions $G(r)$ up to a distance of 
$r=30$ the results are reported in table (\ref{30c})  \cite{b48,b51}. 
\begin{table}[htbp]
\begin{center}
\caption{Spin-spin correlation functions}\label{30c}
\vspace{.1in}
\begin{tabular}{|rc|rc|}\hline
$r$   &   $G(r)$ & $r$   &  $G(r)$ \rule{0in}{4ex}\\[2ex] \hline 
 
 1    & -0.4431              & 16    &   0.0305 \rule{0in}{4ex}\\
  
 2    & 0.1821               & 17    &  -0.0296 \\

 3    & -0.1510              & 18    &   0.0274 \\
 
 4    & 0.1038               & 19    &  -0.0267 \\
 
 5    & -0.0925              & 20    &   0.0249 \\
 
 6    & 0.0731               & 21    &   -0.0242 \\
 
 7    & -0.0671              & 22    &   0.0228 \\

 8    & 0.0567              & 23    &   -0.0223 \\
 
 9    & -0.0532              & 24    &   0.0211 \\

 10   & 0.0465              & 25    &   -0.0206 \\
 
 11   &-0.0442              & 26    &   0.0196 \\
 
 12   & 0.0395              & 27    &   -0.0193 \\
 
 13   & -0.0379              & 28    &   0.0183 \\
 
 14   & 0.0344              & 29    &   -0.0181 \\
 
 15   & -0.0332              & 30    &   0.0172 \\[2ex] \hline
 \end{tabular}
 \end{center}
 \end{table}

 For $r>30$, one may write~\cite{b48}

 \begin{eqnarray}
 G(r)&=&\frac{3}{4}\sqrt{\frac{2}{\pi^3}}\frac{1}{r\sqrt{g(r)}}[ 1-\frac{3}{16}g(r)^2+\frac{156\zeta(3)-73}{384}g(r)^3+O(g(r)^4)-\nonumber \\
 & &\frac{0.4}{2r}((-1)^r +1+O(g(r))+O(\frac{1}{r^2}) ]
 \label{l1}
 \end{eqnarray}
 with $g(r)$ satisfying
 \begin{equation}
 g(r)=\frac{1}{C(r)}(1+\frac{1}{2}g(r)\ln (g(r)))
 \label{l2}
 \end{equation}
 and 
 \begin{equation}
 C(r)=\ln(2\sqrt{2\pi}e^{\gamma +1} r)\quad .
 \label{l3}
 \end{equation}
 Eq.(\ref{l2}) may be solved by iteration. To the lowest order in $\frac{1}{C}$ one finds
 \begin{equation}
 g(r)\approx \frac{1}{C(r)}-\frac{1}{C(r)^2}\ln C(r)\quad .
 \label{lg}
 \end{equation}
 Inserting (\ref{lg}) in Eq.(\ref{l1}) leads to 
\begin{equation}
G(r)\approx \sqrt{2}{\pi^3}\frac{1}{r}\sqrt{C(r)} [1+\frac{1}{4C(r)}\ln C(r)]+O(\frac{1}{C(r)^2})\quad .
\label{sslu}
\end{equation}
Inserting Eq.(\ref{sslu}) in (\ref{rpavv}), one finally gets  
\begin{equation}
<g.s.|\vec{V}\cdot \vec{V}|g.s.>=0.3816N
\label{vvapp}
\end{equation}

Last, I report the following exact three spin correlators that have been used in the determination of the mass spectrum of the two-flavor 
lattice Schwinger model
\begin{eqnarray}
<g.s.|\sum_{x=1}^{N}S^{3}_{x}S^{3}_{x+1}S^{3}_{x+2}|g.s.>&=&0\\
<g.s.|\sum_{x=1}^{N}S^{+}_{x}S^{-}_{x+1}S^{3}_{x+2}|g.s.>&=&0\\
<g.s.|\sum_{x=1}^{N}S^{-}_{x}S^{+}_{x+1}S^{3}_{x+2}|g.s.>&=&0\\
<g.s.|\sum_{x=1}^{N}S^{+}_{x}S^{3}_{x+1}S^{-}_{x+2}|g.s.>&=&0\\  
<g.s.|\sum_{x=1}^{N}S^{-}_{x}S^{3}_{x+1}S^{+}_{x+2}|g.s.>&=&0
\end{eqnarray}

So, on a spin singlet, not only the VEV of $\sum_{x=1}^{N}S^{3}_{x}$ is zero, but also every VEV with an odd number of $S^{3}$.

\section{$SU({\cal N})$ quantum antiferromagnetic chains}
\label{sunc}

It is my purpose to introduce spin-$1/2$ antiferromagnetic Heisenberg chains where  ``spins" are generators of the $SU({\cal N})$ 
group. In the limit ${\cal N}=2$ one has the usual antiferromagnetic Heisenberg chain discussed in the previous section. 
An $U({\cal N})$ spin-$1/2$ quantum antiferromagnetic chain is described by the Hamiltonian
\begin{equation}
H_{J}^{U({\cal N})}=J\sum_{x=1}^{N}S_{ab}(x)S_{ba}(x+1)
\label{hun}
\end{equation}
where $S_{ab}(x)$ $a,b=1,\ldots,{\cal N}$ are the generators of $U({\cal N})$ satisfying the Lie algebra
\begin{equation}
\left[S_{ab}(x),S_{cd}(y)\right]=(S_{ad}(x)\delta_{bc}- S_{cb}(x)\delta_{ad})\delta_{xy}
\end{equation}
and they can be conventionally represented by fermion bilinear operators
\begin{equation}
S_{ab}(x)=\psi_{ax}^{\dagger}\psi_{bx}-\frac{\delta_{ab}}{2}\quad .
\end{equation}
The representation of the algebra on each site is fixed by specifying the fermion number occupation
\begin{equation}
\rho(x)=\sum_{a=1}^{\cal N}S_{aa}(x)\quad .
\end{equation}
By fulfilling the global neutrality condition $\sum_{x=1}^{N}\rho(x)=0$, one may choose
\begin{equation}
\sum_{a=1}^{\cal N}\psi^{\dagger}_{ax}\psi_{ax}=\left\{\begin{array}{cc}
m&x\quad $even$\\
{\cal N}-m&x\quad $odd$
\end{array} \right.
\label{nnn}
\end{equation}
or viceversa the opposite choice for $x$ even-odd. 
Eq.(\ref{nnn}) restricts on each site the Hilbert space to a representation with Young tableau of $m$ rows for $x$ even and ${\cal N}-m$ rows for 
$x$ odd. 
For each site $x$ $\rho(x)$ is the generator of the $U(1)$ subgroup of $U({\cal N})$. 

Let us use the basis $T^{\alpha}=(T^{\alpha})^{*}$, 
$\alpha=1,\ldots,{\cal N}^{2}-1$, 
of the Lie algebra of $SU({\cal N})$ in the fundamental representation such that $tr(T^{\alpha}T^{\beta})=\delta ^{\alpha \beta}/2$ and 
$\left[T^{\alpha},T^{\beta}\right]=if^{\alpha \beta \gamma}T^{\gamma}$, where $f^{\alpha \beta \gamma}$ are the structure constants. By 
means of
\begin{equation}
T_{ab}^{\alpha} T_{cd}^{\alpha}=\frac{1}{2}\delta_{ad}\delta_{bc}-\frac{1}{2{\cal N}} \delta_{ab}\delta_{cd} 
\end{equation}
and redefining the group generators
\begin{equation}
S^{\alpha}_x=\psi_{ax}^{\dagger}T_{ab}^{\alpha}\psi_{bx}
\label{spisch}
\end{equation}
one can rewrite the Hamiltonian (\ref{hun}) as
\begin{equation}
H_{J}^{U({\cal N})}=J\sum_{x=1}^{N}\rho(x)\rho(x+1)+ H_{J}^{SU({\cal N})}
\label{hun2}
\end{equation}
where
\begin{equation}
H_{J}^{SU({\cal N})}=J\sum_{x=1}^{N}S^{\alpha}_x  S^{\alpha}_{x+1}
\label{sunn}
\end{equation}
is the Hamiltonian of an $SU({\cal N})$ quantum antiferromagnet. 
From Eq.(\ref{hun2}) it is clear that by fixing $\rho(x)$ $H_{J}^{U({\cal N})}$ is reduced to $H_{J}^{SU({\cal N})}$. 

The representations of the $SU({\cal N})$ spins relevant for the relationship of the spin Hamiltonians (\ref{sunn})  with 
the Schwinger models are different when ${\cal N}$ is even or odd. When ${\cal N}$ is even I shall consider the representation on each site 
with Young tableau of one column and ${\cal N}/2$ rows so that $\rho(x)=0$. When ${\cal N}$ is odd I shall consider on one sublattice 
the representation with Young tableau of one column and $({\cal N}+1)/2$ rows and of one column and $({\cal N}-1)/2$ rows on the other sublattice. 

As an explicit example let us consider the case ${\cal N}=3$. The generators $T^{\alpha}$ of the $SU(3)$ group in Eq.(\ref{spisch}) 
are the Gell Mann matrices. By denoting $u$,$d$ and $s$ the three flavors, the eight spin operators $S^{\alpha}_{x}$ read
\begin{eqnarray}
S_{x}^{1}&=&\psi_{ux}^{\dagger}\psi_{dx}+\psi_{dx}^{\dagger}\psi_{ux}\\  
S_{x}^{2}&=&-i\psi_{ux}^{\dagger}\psi_{dx}+i\psi_{dx}^{\dagger}\psi_{ux}\\ 
S_{x}^{3}&=&\psi_{ux}^{\dagger}\psi_{ux}-\psi_{dx}^{\dagger}\psi_{dx}\\
S_{x}^{4}&=&\psi_{ux}^{\dagger}\psi_{sx}+\psi_{sx}^{\dagger}\psi_{ux}\\  
S_{x}^{5}&=&-i\psi_{ux}^{\dagger}\psi_{sx}+i\psi_{sx}^{\dagger}\psi_{ux}\\     
S_{x}^{6}&=&\psi_{dx}^{\dagger}\psi_{sx}+\psi_{sx}^{\dagger}\psi_{dx}\\   
S_{x}^{7}&=&-i\psi_{dx}^{\dagger}\psi_{sx}+i\psi_{sx}^{\dagger}\psi_{dx}\\    
S_{x}^{8}&=&\frac{1}{\sqrt{3}}(\psi_{ux}^{\dagger}\psi_{ux}+\psi_{dx}^{\dagger}\psi_{dx}-2\psi_{sx}^{\dagger}\psi_{sx})
\end{eqnarray}
Let us find the ground state of the Hamiltonian (\ref{sunn}) for a chain of two sites. Taking the representations with one particle 
on site 1 and two particles on site 2, the ground state with energy $E_{g.s.}=-16J/3$ reads
\begin{equation}
|g.s.>=\frac{1}{\sqrt{3}}(|u\begin{array}{c} d\\ s \end{array}>-|d\begin{array}{c} u\\ s \end{array}>+|s\begin{array}{c} u\\ d \end{array}>)
\label{hsgs1}   
\end{equation}
The state (\ref{hsgs1}) is a singlet of $SU(3)$, $i.e.$ it is annihilated by the Casimir $\vec{S}^{2}=(\vec{S}_{1}+ \vec{S}_{2})^{2}=
\sum_{\alpha=1}^{8}\left[(S_{1}^{\alpha})^{2}+(S_{2}^{\alpha})^{2}+2S_{1}^{\alpha}S_{2}^{\alpha}\right]$. If one chooses the representation 
with two particles on site  1 and one particle on site 2, the ground state degenerate with (\ref{hsgs1}) is
\begin{equation}
|g.s.>'=\frac{1}{\sqrt{3}}(|\begin{array}{c} d\\ s \end{array}u>-|\begin{array}{c} u\\ s \end{array}d>+|\begin{array}{c} u\\ d \end{array}s>)
\label{hsgs2}   
\end{equation}
Diagonalizing the translation operator $\hat{T}$ with the Hamiltonian (\ref{sunn}), one gets two degenerate ground states
\begin{equation}
|G.S.>^{\pm}=\frac{|g.s.>\pm|g.s.>'}{\sqrt{2}}
\label{GSPM}
\end{equation}
In Eq.(\ref{GSPM}) the linear combination with the $+$ has momentum zero and the one with the $-$ has momentum $\pi$. 
By studying this very simple example, one can infer that the thermodynamic limit $N\rightarrow \infty$ analysis for 
a generic $SU({\cal N})$ chain is very involved.
 
Unfortunately, no analysis with a level of completeness such as the one given in~\cite{b27} for the $SU(2)$ case does exist 
for a generic symmetry group $SU({\cal N})$ with spins in the representation in which I am interested. In~\cite{b52} $SU({\cal N})$ antiferromagnetic models 
were solved for a particular spin representation such that the Hamiltonian (\ref{sunn}) becomes
\begin{equation}
H_{J}^{SU({\cal N})}=J\sum_{x=1}^{N}P_{x x+1}
\label{sunnp}
\end{equation}
with $P_{x x+1}$ the operator that permutes whatever objects occupy sites $x$ and $x+1$. The spectrum of Eq.(\ref{sunnp}) 
was shown to exhibit massless excitations. 

A large ${\cal N}$ expansion approach has been performed in~\cite{b53} for an $SU({\cal N})$ antiferromagnetic chain characterized 
by spins living in a representation with Young tableaux of one row on one sublattice and ${\cal N}-1$ rows on the 
other sublattice. In the case of Young tableaux of one column it was found that the ground state is twofold degenerate 
and breaks translational and parity symmetry. Moreover the elementary excitations are massive non relativistic solitons in 
the large ${\cal N}$ limit with a mass of $O({\cal N})$. 

In~\cite{b54} the Lieb-Shultz-Mattis theorem~\cite{b55} was generalized to $SU({\cal N})$ spin chains. The theorem proves 
that a half-integer-S spin chain with essentially any reasonably local Hamiltonian respecting 
translational and rotational symmetry either has zero gap or else has degenerate ground states spontaneously breaking 
translational and parity invariance. In~\cite{b55} it was proved the existence of a unique ground state of the $SU({\cal N})$ 
chains where ${\cal N}$ is even for spins in the antisymmetric ${\cal N}$ tensor representation. Under the assumption of a 
unique ground state  which must be a $SU({\cal N})$ singlet, an infinitesimal energy gap was found for all the representations 
of $SU({\cal N})$ whose Young tableaux contain a number of boxes not divisible by ${\cal N}$. Of course the degeneracy of the ground state 
trivially implyes a zero gap. 

Using the Lieb-Shultz-Mattis I can state that when ${\cal N}$ is even the ground state of the Hamiltonian (\ref{sunn}) is unique and there are gapless 
excitations for the spin representation with a Young tableau of one column and ${\cal N}/2$ rows. 
When ${\cal N}$ is odd the ground state is twofold degenerate, as I illustrated with the simple example of the two site $SU(3)$ chain. 
It is doubtful if there exist gapless excitations in this case and at present time I am investigating this problem.
\chapter{The one-flavor lattice Schwinger model}
\label{1flsm}

In this chapter I study the one-flavor lattice Schwinger model in the strong coupling 
limit using staggered fermions and the hamiltonian approach to lattice gauge theories~\cite{b56}. 
Even though the continuous axial symmetry is broken explicitly by the staggered fermions, the discrete axial symmetry 
remains and appears in the lattice theory as a translation by one site. In the continuum gauge model the chiral symmetry is 
broken by the anomaly. An interesting issue which arises in the lattice regularization of gauge theories is how 
the lattice theory produces the effects of the axial anomaly. Lattice field theory are manifestly gauge invariant by 
construction and therefore, to produce continuum theories, they must find a way to violate axial current conservation. 
Normally, in lattice gauge theories axial anomalies are either cancelled by fermion doubling or 
else the lattice regularization breaks the axial symmetry explicitly. The lattice Schwinger model in the hamiltonian formalism 
with one species of staggered fermions represents the unique example where neither of these occur. The effects of the anomaly 
are not cancelled by doubling since the continuum limit of $(1+1)$-dimensional staggered fermions produce exactly a Dirac fermion which is the matter 
content of the model. Furthermore, even if the continuum axial symmetry is explicitly broken, the remaining discrete axial symmetry must  be spontaneously 
broken. The mass operator $\bar\psi\psi$ is odd under discrete axial transformations, if this symmetry is unbroken then  $<\bar\psi\psi>=0$. 
However, in the continuum theory it is known that \cite{b57},
\begin{equation}
\left< \bar\psi(x)\psi(x)\right>=-\frac{e^\gamma}{2\pi}\frac{e_c}{\sqrt{\pi}}
~~.
\label{chico1}
\end{equation}
where $\gamma=0.577...$ is the Euler constant.
The expectation value in Eq.(\ref{chico1}) is the chiral condensate.
\footnote{It is worth to stress that the sign of the chiral condensate in Eq.(\ref{chico1}) is just a convention. 
The chiral condensate has always the opposite sign of the fermion mass $m$ that appears in the Lagrangian of the theory: it is the order 
parameter for a spontaneous symmetry breaking of the chiral symmetry. The chiral condensate is the analogous of the magnetization for a spin model 
and the mass m plays the role of the magnetic field. So even if one is studying a massless model, in order to 
decide the sign of the chiral condensate, one has to consider a small mass term $m$ that then one sends to zero. 
The minus sign which appears in Eq.(\ref{chico1}) is due to the fact that one considers the $m\longrightarrow 0^{+}$ limit in the Schwinger model 
Lagrangian.}
I shall show that the chiral condensate (\ref{chico1}) is non-zero also on the lattice and the discrete axial symmetry is indeed spontaneously broken. 
In addition to the analysis of the chiral symmetry breaking pattern on the lattice, by performing a strong coupling expansion around 
a gauge invariant ground state I shall derive results converging more rapidly to the continuum than the results provided in~\cite{b58}. 
The one-flavor Schwinger model has also been used as an example of the fact that quantum link models may reproduce the physiscs of 
conventional hamiltonian lattice gauge theories~\cite{b59bis}.

\section{Definition of the model}
\label{dom} 

The continuum Schwinger model \cite{b5,b59} 
 is the prototypical example of a solvable model where the anomaly occurs.
It is 1+1-dimensional electrodynamics with a single charged massless Dirac spinor field.  The action is 
\begin{equation}
S = \int d^{2} x (\overline{\psi}(i\gamma_{\mu}\partial^{\mu}+\gamma_{\mu}
A^{\mu})\psi-\frac{1}{4e^{2}_{c}}F_{\mu\nu}F^{\mu\nu})
\end{equation}
It is invariant under gauge transformations,
\begin{equation}
\psi(x)\longrightarrow e^{i\theta(x)}\psi(x),\psi^{\dag}(x)\longrightarrow \psi^{\dag}(x)
e^{-i\theta(x)}
\end{equation}
\begin{equation}
A_{\mu}\longrightarrow A_{\mu}+\partial_{\mu}\theta(x)
\end{equation}
and, formally, under the axial phase rotation,  
\begin{equation}
\psi(x)\longrightarrow e^{i\gamma_{5} \alpha} \psi(x),
\psi^{\dag}(x)\longrightarrow \psi^{\dag}(x) e^{-i\gamma_{5} \alpha}
\end{equation}
(where $\gamma^{5}=i\gamma^{0}\gamma^{1}$). 

At the classical level the above 
symmetries lead to conservation laws for the vector and axial currents, 
\begin{equation}
j^{\mu}(x)=\overline{\psi}(x)\gamma^{\mu}\psi(x),\quad\quad j_{5}^{\mu}(x)
=\overline{\psi}(x)\gamma^{5}\gamma^{\mu}\psi(x),
\end{equation}
respectively. At the quantum level both currents cannot be simultaneously 
conserved. If the regularization is gauge invariant, so that the vector current 
is conserved, then the axial current obeys the anomaly equation \cite{b60} 
\begin{equation}
\partial_{\mu} j_{5}^{\mu}(x)=\frac{e^{2}_{c}}{2\pi}\epsilon^{\mu \nu}F_{\mu\nu}(x).
\end{equation}
Moreover, the correlation functions of the model do not 
exhibit axial symmetry.

There are only neutral particles in the spectrum of the Schwinger model; notably, the bound state boson 
created by the axial current operator, 
\begin{equation}
\partial^{\mu}\phi(x)=\sqrt{\pi}j_{5}^{\mu}(x)
\label{anomaly1}
\end{equation}
appears in the spectrum as a free pseudoscalar with mass $\frac{e_{c}}{\sqrt{\pi}}$.

In terms of the boson field, the action is 
\begin{equation}
S=\int d^{2}x(\frac{1}{2}\partial_{\mu}\phi
\partial^{\mu}\phi+\frac{e^{2}_{c}}{2\pi}\phi^{2})
\end{equation}
and the correlation functions for the vector and axial currents are obtained 
from the correlation functions of derivatives of the bose fields. 
The coupling constant $e_{c}$ has the dimension of a mass and the model
is super-renormalizable. 

Since the Schwinger model is solvable it has often been used as a field theory 
where methods of lattice gauge theory, particularly the strong coupling expansion can be tested and compared with exact results of the continuum 
model~\cite{b58}. 

In this chapter, I shall revisit the lattice quantization of the Schwinger 
model. My main purpose is to demonstrate two results.  One is to show how the effects of the anomaly appear through spontaneous symmetry breaking.  
The other is to 
demonstrate that a strong coupling expansion around a gauge, parity and charge
conjugation invariant lattice ground state can produce results in 
good agreement with the continuum.   My results turn out to be quite 
different from those of previous strong coupling computations~\cite{b58}.  The difference arises from my careful treatment of the normal 
ordering of the charge density operator, which must be defined so as to 
be compatible with all of the discrete symmetries of the theory.  The 
necessity of careful normal ordering of the charge operator was pointed 
out in~\cite{b21}. My vacuum state, vacuum energy and energies of elementary excitations are
different from those in \cite{b58,b61} and, as I shall show, they give a more accurate extrapolation to the continuum limit.

The Hamiltonian and gauge constraint of the continuum Schwinger model are
\begin{equation}
H=\int dx\left( \frac{e_c^2}{2}E^2(x)+
\psi^{\dagger}(x)\alpha\left(i\partial_x +eA(x)\right)\psi(x)\right)
\end{equation}
\begin{equation}
\partial_x E(x)+\psi^{\dagger}(x)\psi(x)\sim 0
\end{equation}
A lattice Hamiltonian and constraint which reduce to these in the continuum limit are:  
\begin{equation}
H_{S}=\frac{e^{2}_{L}a}{2}\sum_{x}E_{x}^{2}-\frac{it}{2a}\sum_{x}(\psi_{x+1}
^{\dag}e^{iA_{x}}\psi_{x}-\psi_{x}^{\dag}e^{-iA_{x}}\psi_{x+1})
\label{hamilton1}
\end{equation}
\begin{equation}
E_{x}-E_{x-1}+\psi_{x}^{\dag}\psi_{x}-\frac{1}{2}\sim 0\ ,
\label{gauss1}
\end{equation}
where the fermion fields are defined on the sites, $x=-\frac{N}{2},
-\frac{N}{2}+1,...,\frac{N}{2}$, gauge and the electric fields, $ A_{ x}$ and
 $E_{x}$,  on the links $[x; x + 1]$, $N$ is an even integer 
and, when $N$ is finite, I use periodic boundary conditions.  When $N$ is finite, the continuum limit is the Schwinger model on a circle~\cite{b62,b57}.
The coefficient $t$ of the hopping term in (\ref{hamilton1})
plays the role of the lattice light speed. In the naive continuum limit,
$e_L=e_c$ and $t=1$.  However, I shall keep $e_L$ and $t$ as parameters 
which can be adjusted to fit the lattice values of quantities such as the 
mass gap and  the chiral condensate to those which are known in the 
continuum.

The non vanishing (anti-)commutators for the lattice variables are
\begin{equation}
[A_{x},E_{y}]=i\delta_{xy},\quad\quad\{\psi_{x},\psi_{y}^{\dag}\}=\delta_{xy}
\end{equation}

The Hamiltonian and gauge constraint exhibit the discrete symmetries 
\begin{itemize}
\item{}Parity P: 
\begin{equation}
A_{x}\longrightarrow -A_{-x-1},\ E_{x}\longrightarrow -E_{-x-1},\ 
\psi_{x}\longrightarrow (-1)^{x}\psi_{-x},\ \psi_{x}^{\dag}\longrightarrow 
(-1)^{x}\psi_{-x}^{\dag}
\label{par1}
\end{equation}

\item{}Discrete axial symmetry $\Gamma$: 
\begin{equation}
A_{x}\longrightarrow A_{x+1},\ E_{x}\longrightarrow E_{x+1},\ 
\psi_{x}\longrightarrow \psi_{x+1},\ \psi_{x}^{\dag}\longrightarrow 
\psi_{x+1}^{\dag}
\label{chir1}
\end{equation}

\item{}Charge conjugation C:
\begin{equation}
A_{x}\longrightarrow -A_{x+1},\  E_{x}\longrightarrow -E_{x+1},\ 
\psi_{x}\longrightarrow \psi^{\dag}_{x+1},\ \psi_{x}^{\dag}\longrightarrow 
\psi_{x+1}
\label{char1}
\end{equation}
\end{itemize}

When $A_x=0$, the spectrum of the hopping Hamiltonian is $\epsilon(pa)= 
t \sin pa (p\in 
[0,\frac{2 \pi}{a} ])$. In the low energy limit it resembles a massless 
relativistic spectrum for excitations near the Fermi level. 
The two 
intersections of the energy band with the Fermi level provide the continuum 
right and left moving massless fermions. The electron ground state is invariant under charge conjugation only when the Fermi level 
is at $\epsilon(p_F)=0$ (i.e. when exactly half of the fermion states are 
filled and $\sum_x<\rho(x)>=0$).  In the remainder of this paper I consider only this case of half-filling.

One has two possile ways to put the spinors  $\psi=\left(\begin{array}{c}
\psi_{1}\\
\psi_{2}
\end{array}\right)$ on the lattice in the staggered fermion formalism. One can put upper components on even sites and lower components on odd sites or viceversa.
The mass operator, which reduces to $\bar\psi\psi$ in the continuum limit is
the staggered charge density
\begin{equation}
M(x)=\frac{(-1)^x}{a}\left(\psi_x^{\dagger}\psi_x-1/2\right)
\label{sma}
\end{equation}
for the first choice of staggered fermions or
\begin{equation}
M(x)=\frac{(-1)^{x+1}}{a}\left(\psi_x^{\dagger}\psi_x-1/2\right)
\label{smbb}
\end{equation}
in the other case.

These operators (\ref{sma})(\ref{smbb}) are scalars under parity and are even under
charge conjugation but change sign under a discrete axial transformation.

The lattice Schwinger model is equivalent to a one 
dimensional quantum Coulomb gas on the lattice. To see this one can fix the 
gauge, $A_{x} = A$ (Coulomb gauge). Eliminating the non-constant electric 
field and using the gauge constraint, one obtains the effective Hamiltonian
\begin{eqnarray}
H_{S}&=&H_u+H_p
\equiv\left[\frac{e^{2}_{L}}{2 N}E^{2}+\frac{e^{2}_{L}a}{2}
\sum_{x,y}\rho(x) V(x-y)\rho(y)\right]+\nonumber\\
&+&\left[
-\frac{it}{2a}\sum_{x}(\psi_{x+1}^{\dag}e^{iA}\psi_{x}-\psi_{x}^{\dag}e^{-iA}
\psi_{x+1})\right]\ ,
\label{hs1}
\end{eqnarray}
where the charge density is
\begin{equation}
\rho(x)=\psi^{\dag}_x\psi_x-\frac{1}{2}\ \ ,
\end{equation}
and the potential
\begin{equation}
V(x-y)=\frac{1}{N}
\sum^{N-1}_{n=1} e^{i 2\pi n (x-y)/N}\frac{1}{4\sin^2\frac{\pi n}{N}}
\label{cpo}
\end{equation}
is the Fourier transform of the inverse laplacian on the lattice
for non zero momentum.
The constant electric field is normalized so that $[ A, E ] = i$ . 
The constant 
modes of the gauge field decouple in the thermodynamic limit 
$ N \longrightarrow \infty $.
The gauge fixed Hamiltonian (\ref{hs1}) can also be written as a quantum spin model.  Consider the Jordan-Wigner transformation
\begin{equation}
\psi_x=\prod_{y<x}[2iS_3(y)]S^-(x)~~,~\psi^{\dagger}_x=
\prod_{y<x}[-2iS_3(y)]S^+(x)~~,~\psi^{\dagger}_x\psi_x=S^3(x)+1/2
\end{equation}
Then, the Hamiltonian is 
\begin{equation}
H=\frac{t}{2a}\sum_x\left( e^{iA}S^+(x+1)S^-(x)+e^{-iA}S^-(x+1)S^+(x)\right)+
\frac{e_L^2a}{2}\sum_{x,y}S^3(x)V(x-y)S^3(y)+e_L^2E^2/2N
\label{spinham}
\end{equation}
For the moment, let us ignore the constant modes $E$ and $A$. 
The first term in the Hamiltonian 
is the quantum x-y model which has a disordered 
ground state (of course, exact solution of this model is obtained by 
the inverse Jordan-Wigner transformation \cite{b55,b63}).  
The second is a long-ranged Ising interaction.  
I will prove in the next section that, in spite of the low 
dimensionality of this system, the latter term has a N\'eel ordered 
ground state.  This is a result of the infinite range of the Coulomb interaction.  
When both terms are present in the Hamiltonian, they compete, the 
first one favoring disorder and the second one favoring order.  Later, 
when I extrapolate my strong coupling results to the continuum limit, 
I shall assume that the ordered ground state persists for all positive 
values of the dimensionless constant $e_L^2a^2/t$.  The strong coupling 
limit is where this constant is large and the ground state has N\'eel order.  
The continuum limit of the Schwinger model is found where this constant is
 small.  I shall assume that the N\'eel order persists in that limit.  
\footnote{If it did not persist, but there was a second order phase 
transition at some critical value of the constant to a disordered 
state, the continuum limit would be found at that phase transition.}

\section{The low-lying excitation spectrum}
\label{tlles}

I shall examine the Schwinger model in the strong coupling limit. 
To solve the lattice Schwinger model in the strong coupling expansion, it is 
necessary to find states  which are annihilated by the generator of gauge transformations
\begin{equation}
{\cal G}_{x}=E_{x}-E_{x-1}+\psi^{\dag}_{x}\psi_{x}-\frac{1}{2}
\label{gaussl}
\end{equation}
and, at the same time, are eigenstates of the unperturbed Hamiltonian 
\begin{equation}
H_{u}=\frac{e_{L}^{2}}{2}\sum_{x}E_{x}^{2}\ .
\end{equation}
I shall quantize the gauge fields in the functional Schr\"{o}dinger picture where the wave-functions are functions of the 
variables $A_{x}$ and the electric field operators are defined as  
\begin{equation}
E_{x}=\frac{1}{i}\frac{\delta}{\delta A_{x}}\ .
\end{equation}
In~\cite{b21} it was proved that there were two gauge invariant ground states of the system. In fact, when the number, N, 
of lattice sites is even the two states   
\begin{eqnarray}
|\psi>&=&\prod_{x=even}\psi_{x}^{\dag}|0>e^{-i\sum_{x}A_{x}\frac{(-1)^{x}}{4}}\\
|\chi>&=&\prod_{x=odd}\psi_{x}^{\dag}|0>e^{i\sum_{x}A_{x}\frac{(-1)^{x}}{4}}
\end{eqnarray}
are degenerate gauge invariant ground states of $H_u$.
They are gauge invariant since are annihilated by the operator ${\cal G}_{x}$. 
To see that they are indeed ground states, note that when $H_u$ acts on gauge invariant states, it is effectively the sum of two operators, 
the zero mode energy $e_L^2 E^2/2N$ 
where $E=\sum_x E_x/\sqrt{N}$ and the Coulomb energy
\begin{eqnarray}
H_C=\frac{e^{2}_{L}a}{2}
\sum_{x,y}\rho(x) V(x-y)\rho(y)=\frac{e^{2}_{L}a}{2N}\sum_{k\neq
0}\rho(-k)\frac{1}{4\sin^{2}\frac{k}{2}}\rho(k)\ .
\label{huuu}
\end{eqnarray}
where I have used the Fourier transform
\begin{equation}
\psi^{\dag}_{x}\psi_{x}-\frac{1}{2}=
\frac{1}{N}\sum_{p}e^{ipx}\rho(p)\ ,\ p=\frac{2\pi n}{N}\ ,\ n\in \{-\frac{N}{2},...,\frac{N}{2}-1\}
\end{equation}
Furthermore, gauge invariant states have vanishing total charge,
\begin{equation}
\sum_{x}(\psi_{x}^{\dag}\psi_{x}-\frac{1}{2})=0
\end{equation}
The Coulomb energy satisfies the bound
\begin{eqnarray}
\frac{e^{2}_{L}a}{2N}\sum_{k\neq 0}\rho(-k)\frac{1}{4\sin^{2}\frac{k}{2}}\rho(k)\geq \frac{e^{2}_{L}a}{2N}\sum_{k\neq 0}\rho(-k)\frac{1}{4}\rho(k)=\nonumber\\
=\frac{e^{2}_{L}a}{8}\sum_{x}(\psi_{x}^{\dag}\psi_{x}-\frac{1}{2})^{2}
=e_L^2aN/32 
\end{eqnarray} 
(since  
$(\psi_{x}^{\dagger}\psi_{x}-\frac{1}{2})^{2}=\frac{1}{4}$). 
The states $|\psi>$ and $|\chi>$ are eigenstates of the charge density 
with eigenvalues
$\rho(k)=\pm\frac{N}{2}\delta{k,\pi}$.  These charge densities  
saturate the lower bound for the Coulomb energy.  Furthermore, $E|\psi>=0=E|\chi>$ so $|\psi>$ and $|\chi>$ minimize the energy of the zero mode.  Therefore, they are degenerate ground 
states of $H_u$.

The ground states
\label{smb} are invariant under parity and charge conjugation but they 
both break chiral symmetry, i.e. the symmetry under translations by one site.
The ivariance under parity is easy to see
\begin{equation}
P|\psi>=\prod_{x=even}(-1)^{x}\psi_{-x}^{\dagger}e^{-i\sum_{x}-A_{-x-1}\frac{(-1)^{x}}{4}}|0>=|\psi>
\end{equation}
and analogously 
\begin{equation}
P|\chi>=-|\chi>
\end{equation}
For what concerns charge conjugation one has
\begin{equation}
C|\psi>=\prod_{x=even}C \psi_{x}^{\dagger}e^{-i\sum_{x}A_{x}\frac{(-1)^{x}}{4}}C^{-1}C|0>=
\prod_{x=even} \psi_{x+1}e^{-i\sum_{x}A_{x}\frac{(-1)^{x}}{4}}C|0>=|\psi>
\end{equation}
where the last equality is true since, being 
\begin{eqnarray}
\sum_{x}<0|\rho(x)|0>=-\frac{N}{2}\\
\sum_{x}<0|C^{-1}\rho(x)C|0>=\frac{N}{2}
\end{eqnarray}
one has 
\begin{equation}
\prod_{x=even}\psi^{\dagger}_{x}|0>=\prod_{x=even}\psi_{x+1}C|0>
\end{equation}
Analogously one finds
\begin{equation}
C|\chi>=|\chi>
\end{equation}
Under discrete chiral symmetry one gets
\begin{equation}
\Gamma |\psi>=|\chi>\quad;\quad \Gamma |\chi>=|\psi>
\end{equation}
The order parameter of the chiral symmetry, the mass operator $M=\frac{1}{Na}
\sum_xM(x)$, is diagonal in these states and has 
eigenvalue $\frac{1}{2}$. 
\footnote{One has to consider $M(x)$ given by Eq.(\ref{sma}) when applying it to $|\psi>$ and $M(x)$ given by Eq.(\ref{smbb}) when applying it to $|\chi>$.}
In order to have a chiral condensate of negative sign as in the continuum, one has to consider the mass operator $M'(x)=-M(x)$. 
It is interesting that 
the ground states are degenerate and are not invariant under chiral 
symmetry, even in finite volume and in 
one spatial dimension; it is possible due to 
the long-ranged nature of the Coulomb interaction~\cite{b39}. The chiral symmetry is spontaneously broken in the infinite volume limit.
It is essential that the number of sites of the lattice, N, is even. 

Of course once perturbations from the hopping term in the fermion Hamiltonian
are taken into account, when the volume is finite, there are high order terms 
(of order $N$) in the strong coupling perturbation theory which mix the 
ground states that are constructed by perturbing $|\psi>$ and $|\chi>$.
The mixing amplitude is
\begin{equation}
<\chi|H_{p}(\frac{1}{E_{0}-H_{C}}H_{p})^{N-1}|\psi>=-\frac{N^{2}}{2}
\frac{|t|^{N}}{2^{N}}\frac{4^{N-1}}{e_{L}^{2(N-1)}}
\frac{N!}{((\frac{N}{2})!)^{2}} 
\label{mixa}
\end{equation}
Since the amplitude (\ref{mixa}) is negative and real, whatever the diagonal elements are, the state with lower energy is the state
\beq
\frac{1}{\sqrt{2}}(|\psi>+|\chi>)
\eeq
This is an eigenstate of the discrete axial transformation, with eigenvalue 
$1$. Thus the true ground state is chirally symmetric. 

In order to avoid 
this restoration of the symmetry, it is necessary to take the infinite  
N limit so that the two ground states are not mixed to any finite order of 
perturbation theory. Thus, in the perturbative expansion one only has to consider diagonal matrix elements and, consequently, perturbation theory 
for non-degenerate-states. I shall then choose as unperturbed ground 
state $|\psi>$. The results concerning the energies would have been the same 
had I chosen 
$|\chi>$.

Excitations are created by the operators
\begin{equation}
\psi_{j}^{\dag}e^{i\sum_{x=i}^{j}A_{x}}\psi_{i}
\end{equation}
and have energy $\frac{e_{L}^{2}}{2}|i-j|$ greater than the ground state energy. A single electron, in order to have a gauge invariant state, 
must be attached to a line of electric flux which goes from the electron to infinity. This is not allowed at all in a finite volume with 
periodic boundary conditions and on the infinite lattice, the string of electric flux would have infinite energy. Thus electric charges are 
confined and the string tension is given by the electric charge $\frac{e_{L}^{2}}{2}$.

Let us now start the perturbative computation of the excitations mass gap.
I shall work in the Coulomb gauge $\nabla A=0$, so that $A$ is $x$-independent 
and the ground states are
\begin{eqnarray}
|\psi>&=&\prod_{x=even}\psi_{x}^{\dag}|0>\\
|\chi>&=&\prod_{x=odd}\psi_{x}^{\dag}|0>.
\end{eqnarray}   
The Coulomb potential reads 
\begin{equation}
V(x-y)=\frac{1}{N}\sum_{k\neq 0} \frac{1}{4\sin^{2}\frac{k}{2}}e^{ik(x-y)
}=\frac{1}{2N}(x-y)^{2}-\frac{1}{2}|x-y|+\frac{1}{12N}(N^{2}-1).
\end{equation}
The Schwinger Hamiltonian, rescaled by the factor $e_{L}^{2}a$,
 may then be written as
\begin{equation}
H=H_{0}+\epsilon H_{h}=\frac{1}{e_{L}^{2}a}(H_{C}+H_{p})
\end{equation}
with 
\begin{equation}
H_{0}=\sum_{x>y}[\frac{(x-y)^{2}}{2N}-\frac{(x-y)}{2}]\rho(x)\rho(y)
\end{equation}
\begin{equation}
H_{h}=-i\sum_{x}(\psi_{x+1}^{\dag}e^{iA}\psi_{x}-\psi_{x}^{\dag}e^{-iA}
\psi_{x+1})=-i(R-L)
\end{equation}
and
\begin{equation}
\epsilon=\frac{t}{2e^{2}_{L}a^{2}}.
\end{equation}
To the fourth order in $\epsilon$ the perturbative expansion for the energy is 
given by\footnote{
Let us observe that for the ground state 
perturbative expansion the relative minus sign in the hopping 
Hamiltonian is irrelevant whereas it will play an important role in the
calculation of the excited state energy. 
This is easily seen if one notices that at the second order in $\epsilon$  
one gets
\begin{eqnarray}
E_{\psi}^{(2)}&=&<\psi|(L-R)\frac{\Pi_{\psi}}{E_{\psi}^{(0)}-H_{0}}
(R-L)|\psi>=\nonumber\\
&=&<\psi|L\frac{\Pi_{\psi}}{E_{\psi}^{(0)}-H_{0}} 
R|\psi>+<\psi|R\frac{\Pi_{\psi}}{E_{\psi}^{(0)}-H_{0}}L|\psi>\nonumber
\end{eqnarray}}
\begin{equation}
E_{\psi}=E_{\psi}^{(0)}+\epsilon^{2}E_{\psi}^{(2)}+\epsilon^{4}E_{\psi}^{(4)}
\end{equation}
where
\begin{equation}
E_{\psi}^{(0)}=<\psi|H_{0}|\psi>
\end{equation}
\begin{equation}
E_{\psi}^{(2)}=<\psi|H_{h}^{\dag}\frac{\Pi_{\psi}}{E_{\psi}^{(0)}-H_{0}}H_{h}|\psi>
\end{equation}
\begin{eqnarray}
E_{\psi}^{(4)}=<\psi|H_{h}^{\dag}\frac{\Pi_{\psi}}{E_{\psi}^{(0)}-H_{0}}H_{h}^{\dag}
\frac{\Pi_{\psi}}{E_{\psi}^{(0)}-H_{0}}H_{h}\frac{\Pi_{\psi}}{E_{\psi}^{(0)}-H_{0}}H_{h}
|\psi>+\nonumber\\
-<\psi|H_{h}^{\dag}\frac{\Pi_{\psi}}{E_{\psi}^{(0)}-H_{0}}H_{h}|\psi>
<\psi|H_{h}^{\dag}\frac{\Pi_{\psi}}{(E_{\psi}^{(0)}-H_{0})^{2}}H_{h}|\psi>
\end{eqnarray}
and
$1-\Pi_{\psi}$ is a projection operator 
which projects a generic state $|s>$ on $|\psi>$
\begin{equation}
(1-\Pi_{\psi})|s>=|\psi>\ .
\end{equation}

The phases $e^{iA}$ in the hopping Hamiltonian are irrelevant 
until one reaches the $N^{th}$ order in the perturbative expansion. 
As a matter of fact, 
only at this order the transverse electric field becomes important. 
By direct evaluation one gets  
\begin{eqnarray}
E_{\psi}^{(0)}=\frac{N}{32}\\
E_{\psi}^{(2)}=-4N\\
E_{\psi}^{(4)}=192N
\end{eqnarray}
\begin{equation}
E_{\psi}=\frac{N}{32}-4N\epsilon^{2}+192N\epsilon^{4}.
\end{equation}

Next I compute the energies of the low-lying excitations
to the same order in the perturbative expansion.
The lowest-lying excitations are degenerate and are created by the operators 
\begin{eqnarray}
L(x)&=&\psi_{x}^{\dag}e^{iA}\psi_{x+1}\\
R(x)&=&\psi_{x+1}^{\dag}e^{-iA}\psi_{x}
\end{eqnarray}
which move a particle one lattice spacing to the left and right respectively. When they act on the vacuum $|\psi>$ they create the two states 
\begin{eqnarray}
|x,R>&\equiv&R(x)|\psi>=\psi_{x+1}^{\dag}e^{-iA}\psi_{x}|\psi>\\
|x,L>&\equiv&L(x)|\psi>=\psi_{x}^{\dag}e^{iA}\psi_{x+1}|\psi>
\end{eqnarray}
which are degenerate in energy (see Eq.(\ref{bba})).
In order to resolve the degeneracy of these two states, one must find the first nontrivial order of the 
perturbative expansion in which the two states are mixed; this turns out to be the 
second order. It is easy to show that the second order of perturbative expansion is diagonal  in the two states which are created  by the parity odd and 
even operators
\begin{eqnarray}
H_{-}(x)&\equiv&R(x)+L(x)\\ 
H_{+}(x)&\equiv&R(x)-L(x)\ .
\end{eqnarray}
In fact
\begin{equation}
P H_{\pm}(x)=\pm H_{\pm}(-x)\ .
\end{equation}
The parity odd operator creates the state  with lower energy. Thus, it is this the state which 
will have finite energy gap in the continuum limit 
and which will correspond to the pseudoscalar boson of the continuum Schwinger model.
In the continuum one gets this massive excitation by 
applying the vector flux $j^{1}=j_{5}^{0}$ to the vacuum. The excitation created by $H_{+}$ 
is the scalar excitation which in the continuum limit decouples from the spectrum. 
Since one is interested in the masses of these excitations, one has to compute their energies at zero momentum.

On the lattice the construction of the pseudoscalar excitation at zero momentum is 
provided by
\begin{eqnarray}
|\theta>&\equiv&\frac{1}{\sqrt{N}}\sum_{x=1}^N|\theta(x)>=\frac{1}{\sqrt{N}}\sum_{x=1}^{N}H_{-}(x)|\psi>=
\frac{1}{\sqrt{N}}\sum_{x=1}^{N}(\psi_{x}^{\dag}e^{iA}\psi_{x+1}+
\psi^{\dag}_{x+1}e^{-iA}\psi_x)|\psi>=\nonumber\\
&=&\frac{1}{\sqrt{N}}\sum_{x=1}^{N}j^{1}(x)|\psi>=
\frac{1}{\sqrt{N}}\sum_{x=1}^{N}
j_{5}^{0}(x)|\psi>.
\end{eqnarray}
One could also consider the scalar massive excitation 
\begin{eqnarray}
|\varphi>&\equiv&=\frac{1}{\sqrt{N}}\sum_{x=1}^{N}|\varphi(x)>=\frac{1}{\sqrt{N}}\sum_{x=1}^{N}H_{+}(x)|\psi>=\nonumber\\
&=&
\frac{1}{\sqrt{N}}\sum_{x=1}^{N}(\psi_{x}^{\dag}e^{iA}\psi_{x+1}-
\psi^{\dag}_{x+1}e^{-iA}\psi_x)|\psi>=\frac{1}{\sqrt{N}}\sum_{x=1}^{N}j^{5}(x)|\psi>;
\label{pseudo}
\end{eqnarray}
The states $|\theta>$ and $|\varphi>$ are characterized 
by a dimer-antidimer configuration; that is by a pair of particles 
followed or preceded by a pair of antiparticles.

The energy of the state $|\theta>$, 
at the second order in $\epsilon$, is given by 
\begin{equation}
E_{\theta}^{(2)}=\sum_{x,y}\frac{1}{N}<\theta(y)|(L-R)
\Lambda_{\theta}(R-L)|\theta(x)>
\end{equation}
where
\begin{equation}
\Lambda_{\theta}=\frac{\Pi_{\theta}}{E_{\theta}^{(0)}-H_{0}}
\end{equation}
with $E_{\theta}^{(0)}=E_{\psi}^{(0)}+\Delta$, $\Delta=\frac{1}{4}$ and $1-\Pi_{\theta}$ a projection 
operator onto $|\theta>$.
A state obtained by applying the hopping Hamiltonian to the state $|\theta>$ 
has a dimer-antidimer pair; as a consequence its energy will be
raised of $2\Delta$ with respect to $E_{\psi}^{(0)}$. 
\footnote{One must be careful in applying one more time the hopping Hamiltonian.
 Since the
Coulomb Hamiltonian and the hopping Hamiltonian do not commute (see section (\ref{mae}), 
there is a state created 
by appling two times the hopping Hamiltonian to the state $|\theta>$ which has not a gap of $3\Delta$ 
 with respect to the ground
state, but it has a gap of $5\Delta$-characterized by three particles 
followed or preceded by three antiparticles,
that is, it is created by applying $L^{3}$ or $R^{3}$ to the ground state .}
The computation of $E_{\theta}^{(2)}$ (see section (\ref{mae})) leads to
\begin{equation}
E_{\theta}^{(2)}=-
\frac{1}{\Delta}N+\frac{2}{\Delta}=-4N+8
\end{equation}
and 
\begin{equation}
<\theta|H_{h}^{\dag}\Lambda_{\theta}^{2}
H_{h}|\theta>=\frac{N}{\Delta^{2}}-\frac{2}{\Delta^{2}}=16N-32\ .
\end{equation}

Next one should compute 
\begin{eqnarray}
E_{\theta}^{(4)}=\frac{1}{N}\sum_{x,y}<\theta(y)|H_{h}^{\dag}
\Lambda_{\theta}H_{h}^{\dag}\Lambda_{\theta}H_{h}\Lambda_{\theta}H_{h}|\theta(x)>  
+\nonumber\\
-\frac{1}{N^{2}}\sum_{x,y}<\theta(y)|H_{h}^{\dag}\Lambda_{\theta}H_{h}
|\theta(x)>\sum_{x,y}<\theta(y)|H_{h}^{\dag}\Lambda_{\theta}^{2}H_{h}|\theta(x)>
\label{et4}.
\end{eqnarray}
Using the matrix elements given in paragraph (\ref{mae}), one gets
\begin{equation}
E_{\theta}^{(4)}=\frac{3}{\Delta^{3}}N-\frac{9}{\Delta^{3}}=192N-576\ .
\end{equation}
The pseudoscalar excitation energy, to the fourth order in $\epsilon$, 
is then given by
\begin{equation}
E_{\theta}=(\frac{N}{32}+\frac{1}{4})+(-4N+8)\epsilon^{2}+(192N-576)
\epsilon^{4}\ .
\end{equation}

Using again the matrix element given in appendix C, one easily obtains the 
energy $E_{\varphi}$ of the state $|\varphi>$, $i.e.$
\beq
E_{\varphi}=(\frac{N}{32}+\frac{1}{4})+(-4N+24)\epsilon^{2}+(192N-1600)
\epsilon^{4}\ .
\end{equation}

Taking the difference between $E_{\theta}$ ($E_{\varphi}$) 
and $E_{\psi}$, one gets the masses of the  pseudoscalar and scalar 
excitations, 
\beq
\frac{m^P}{e^{2}_{L}a}=\frac{1}{4}+8\epsilon^{2}-576\epsilon^{4}\ ,
\label{mga}
\end{equation}
\begin{equation}
\frac{m^S}{e^{2}_{L}a}=\frac{1}{4}+24\epsilon^{2}-1600\epsilon^{4}\ .
\label{mgb}
\end{equation}

Eqs.(\ref{mga},\ref{mgb}) provide the behavior of $m^P$ and $m^S$ 
for small values of
$z=\epsilon^{2}=\frac{t^{2}}{4e^{4}_{L}a^{4}}$.

As expected from the continuum theory, 
the pseudoscalar particle mass is smaller 
than the scalar particle mass.
Furthermore, since both quantities are intensive, 
there is no $N$ dependence in 
Eqs.(\ref{mga},\ref{mgb}). 

Now let us compute the dispersion relation of the pseudoscalar and scalar excitations to the fourth order in perturbation theory.

At the zero-th order one has no 
momentum dependence in the lattice pseudoscalar excitation since
\begin{equation}
E_{\theta}^{(0)}(p)=\frac{1}{N}\sum_{x,y}<\theta(y)|H_{0}|\theta(x)>e^{ip(x-y)}=\frac{N}{32}+\Delta=\frac{N}{32}+\frac{1}{4}\ .
\label{eth1}
\end{equation}
The momentum dependence starts only at the second order in $\epsilon$, where one gets 
\begin{equation}
E_{\theta}^{(2)}(p)=\frac{1}{N}\sum_{x,y}<\theta(y)|H_{h}^{\dag}\Lambda_{\theta}H_{h}|\theta(x)>e^{ip(x-y)}\ .
\label{eth2}
\end{equation}
Using the matrix elements given in paragraph (\ref{mae}), one has
\begin{equation}
E_{\theta}^{(2)}(p)=-\frac{1}{\Delta}(N-4)-\frac{2}{\Delta}\cos pa=-4N+16-8\cos pa\ .
\end{equation}
The dispersion relation, at the fourth order in perturbation theory, reads
\begin{eqnarray}
E_{\theta}^{(4)}(p)=\frac{1}{N}\sum_{x,y}<\theta(y)|H_{h}^{\dag}\Lambda_{\theta}H_{h}^{\dag}\Lambda_{\theta}H_{h}\Lambda_{\theta}H_{h}|\theta(x)>e^{ip(x-y)
}+\nonumber\\
-\frac{1}{N^{2}}\sum_{x,y}<\theta(y)|H_{h}^{\dag}\Lambda_{\theta}H_{h}|\theta(x)>e^{ip(x-y)}\sum_{x,y}<\theta(y)|H_{h}^{\dag}\Lambda_{\theta}^{2}H_{h}|\theta(x)>e^{ip(x-y)}
\end{eqnarray}
and leads to
\begin{eqnarray}
E_{\theta}^{(4)}(p)&=&\frac{3}{\Delta^{3}}N-\frac{1}{2\Delta^{3}}\cos 2pa+\frac{4}{\Delta^{3}}(2\cos pa-\cos^{2} pa)
-\frac{25}{2}\frac{1}{\Delta^{3}}=\nonumber\\
&=&192N-32\cos 2pa+256(2\cos pa-\cos^{2}pa)-800\ .
\label{eth4}
\end{eqnarray}
Using Eqs.(\ref{eth1}), (\ref{eth2}) and (\ref{eth4}), the dispersion relation of the lattice pseudoscalar excitation is given by 
\begin{eqnarray}
E_{\theta}(p)&=&E_{\theta}^{(0)}(p)+\epsilon^{2}E_{\theta}^{(2)}(p)+\epsilon^{4}E_{\theta}^{(4)}(p)+...=\nonumber\\
&=&E_{\theta}(0)+\frac{2}{\Delta}(1-\cos pa)\epsilon^{2}+[\frac{1}{2\Delta^3}(1-\cos 2pa)-\frac{4}{\Delta^{3}}(1-\cos pa)^{2}]
\epsilon^{4}=\nonumber\\
&=&E_{\theta}(0)+8(1-\cos pa)\epsilon^{2}+[32(1-\cos 2pa)-256(1-\cos pa)^{2}]\epsilon^{4}.
\end{eqnarray}
Performing a small momentum expansion, one gets 
\begin{equation}
E_{\theta}(p)=E_{\theta}(0)+(4\epsilon^{2}+64\epsilon^{4})(pa)^{2}+\dots
\label{mg2}
\end{equation}

The same procedure yields 
\begin{eqnarray}
E_{\varphi}(p)&=&E_{\varphi}^{(0)}(p)+\epsilon^{2}E_{\varphi}^{(2)}(p)+\epsilon^{4}E_{\varphi}^{(4)}(p)+...=\nonumber\\
&=&E_{\theta}(0)-\frac{2}{\Delta}(1-\cos pa)\epsilon^{2}+[\frac{1}{2\Delta^3}(1-\cos 2pa)+\frac{16}{\Delta^{3}}-\frac{4}{\Delta^{3}}(1+\cos pa)^{2}]
\epsilon^{4}=\nonumber\\
&=&E_{\theta}(0)-8(1-\cos pa)\epsilon^{2}+[32(1-\cos 2pa)+1024-256(1+\cos pa)^{2}]\epsilon^{4}.
\end{eqnarray}
which, for small momentum gives
\begin{equation}
E_{\varphi}(p)=E_{\varphi}(0)+(-4\epsilon^{2}+72\epsilon^{4})(pa)^{2}+\dots
\label{mg3}
\end{equation}
The $p^{2}$ terms are consistent with the bosonic nature of the excitations.

Eq.(\ref{mg2}) and Eq.(\ref{mg3}) give us the masses of the pseudoscalar and scalar excitations from the curvature of the energy momentum relation
\begin{equation}
\frac{1}{2m^{P}_{*}e_{L}^{2}a^{3}}=(4\epsilon^2+64\epsilon^4)
\label{mg4}
\end{equation}
\begin{equation}
\frac{1}{2m^{S}_{*}e_{L}^{2}a^{3}}=(-4\epsilon^2+72\epsilon^4)
\label{mg5}
\end{equation}

\subsection{Unperturbed energies of the dimer states}

In this section the energies of the excitations generated 
by applying the hopping Hamiltonian $H_h$ to the ground state $|\psi>$, 
are evaluated.
The action of $H_h$ on the state $|\psi>$ generates a dimer-antidimer pair.
All the states with one dimer-antidimer pair,
and with two dimer-antidimer pairs 
(obtained applying $H_{h}$ twice), have the same energy 
gap with respect to the ground state.
However, not all the states obtained applying $(H_h)^n$, with $n\ge3$,
to $|\psi>$ have the same energy. 
There are, for example, two possible energies, corresponding to a gap of 
$3\Delta=3e^2_L a/4$ and $5\Delta$, for the 
states with three dimer-antidimer pairs. 

The states with a dimer-antidimer pair are given by
\begin{equation}
|S^{1}_{R,L}>=H_{R,L}(y)|\psi>.
\end{equation}
Applying on these states the unperturbed Hamiltonian one gets
\begin{eqnarray}
H_{u}|S^{1}_{R,L}>&=&H_{R,L}(y)    \frac{e^{2}_{L}a}{2}\sum_{x}
(E_{x}\pm\delta_{xy})^{2}|\psi>=\nonumber\\
&=&[E_{\psi}^{0}+\frac{e^{2}_{L}a}{2}\sum_{x}(-\frac{(-1)^{x}}{2}\delta_{xy}+\delta_{xy})]|S^{1}_{R,L}>
\label{bba}
\end{eqnarray}
where $y$ must be even if one applies $R$ and must be odd 
if one applies $L$. One thus sees that all the states $|S_{R,L}^{1}>$ have
a mass gap of $\Delta$.

The states with two dimer-antidimer pairs are given by
\begin{equation}
|S^{2}_{(R,L)(R,L)}>=H_{R,L}(z)H_{R,L}(y)|\psi>
\end{equation}
\begin{equation}
H_{u}|S^{2}_{(R,L)(R,L)}>=\frac{e^{2}_{L}a}{2}H_{R,L}(z)H_{R,L}(y)    \sum_{x}(E_{x}\pm \delta{xz}\pm 
\label{bbb}
\delta_{xy})^{2}|\psi>
\end{equation}
It is easy to show that all these states have an energy gap of $2\Delta$.

Based on Eq.(\ref{bba}) and Eq.(\ref{bbb}) one might expect that the states with three dimer-antidimer pairs have 
an energy gap of $3\Delta$.  This is actually true for all the states of this type
except for the 2 states with 
three particles followed or preceeded by three antiparticles. These states have an 
energy gap of $5\Delta$. If one considers the state
\begin{equation}
|S^{3}_{RRR}>=R(w)R(z)R(y)|\psi>,
\end{equation}
one easily obtains that
\begin{eqnarray}
H_{u}|S_{RRR}^{3}>&=&\frac{e^{2}_{L}a}{2}R(w)R(z)R(y)      \sum_{x}(E_{x}+\delta_{xw}+\delta_{xz}+\delta_{xy})^{2}|  \psi>=\nonumber\\
&=&[E_{\psi}^{0}+\frac{e^{2}_{L}a}{2}\sum_{x}(-\frac{(-1)^{x}}{2}\delta_{xw}-\frac{(-1)^{x}}{2}\delta_{xz}-\frac{(-1)
^{x}}{2}\delta_{xy}+\nonumber\\
&+&2\delta_{xw}\delta_{xz}+2\delta_{xw}\delta_{xy}+2\delta_{xz}\delta_{xy}+
\nonumber\\
&+&\delta_{xw}+\delta_{xz}+\delta_{xy})]
|S^{3}_{RRR}>\ .
\end{eqnarray}
To generate a state with three antiparticles followed by three particles,
one must have
\begin{eqnarray}
z=y-2\\
w=y-1
\end{eqnarray}
The energy gap for this state is $5\Delta$. 
The same result is obtained for the
state $|S^{3}_{LLL}>=L(w)L(z)L(y)|\psi>$
with $z=y+3$ and $w=y+2$, which contains three particles followed by three 
antiparticles.

The reason for  these different behaviours,
lies on the fact that the commutation relation between 
$H_{0}$ and $H_{h}$ is
\begin{eqnarray}
[H_{0},H_{h}]=\frac{e^{2}_{L}a}{4}H_{h}+
\frac{e^{2}_{L}t}{8}(\sum_{j\geq k}-\sum_{k\geq j})
(\psi^{\dag}_{k+1}e^{iA}\psi_{k}-\psi_{k}^{\dag}e^{-iA}\psi_{k+1})\rho_{j}
\label{aaa}
\end{eqnarray}
If the second term in the r.h.s. was absent, all the states 
with $n$ dimer-antidimer pairs would have 
an energy gap of $n\Delta$.

\subsection{Perturbative matrix elements}
\label{mae}

In paragraph I provide the matrix elements 
needed in the strong coupling expansion expression of the mass gap.
The six matrix elements that arise in 
the computation of the second order bosonic excitation energies, are 
\begin{eqnarray}
M_{1}&=&<\psi|R(y)R\Lambda_{\theta}L L(x)|\psi>=\nonumber\\
&=&-\frac{1}{\Delta}(\frac{N}{2}-1)\delta_{xy}-\frac{1}{\Delta}(1-\delta_{xy})\\
M_{2}&=&<\psi|L(y)L\Lambda_{\theta}RR(x)|\psi>=\nonumber\\
&=&-\frac{1}{\Delta}(\frac{N}{2}-1)\delta_{xy}-\frac{1}{\Delta}(1-\delta_{xy})\\
M_{3}&=&<\psi|R(y)L\Lambda_{\theta}RL(x)|\psi>=\nonumber\\
&=&-\frac{1}{\Delta}(\frac{N}{2}-3)\delta_{xy}+\frac{1}{\Delta}(1-\delta_{xy})\\
M_{4}&=&<\psi|L(y)R\Lambda_{\theta}LR(x)|\psi>=\nonumber\\
&=&-\frac{1}{\Delta}(\frac{N}{2}-3)\delta_{xy}+\frac{1}{\Delta}(1-\delta_{xy})\\
M_{5}&=&-<\psi|R(y)L\Lambda_{\theta}LR(x)|\psi>=\nonumber\\
&=&-\frac{1}{\Delta}(\delta_{y,x-1}+\delta_{y,x+1})\\
M_{6}&=&-<\psi|L(y)R\Lambda_{\theta}RL(x)|\psi>=\nonumber\\
&=&-\frac{1}{\Delta}(\delta_{y,x-1}+\delta_{y,x+1})
\label{mi}
\end{eqnarray}
where $M_{1}$ and $M_{3}$ are defined for $x$ and $y$ odd, $M_{2}$ and $M_{4}$ for $x$ and $y$ even, $M_{5}$ for $x$ even and $y$
odd, and $M_{6}$ for $x$ odd and $y$ even.

To compute the fourth order energy gap one needs also 
the matrix elements $M_{i}'$, $i=1,...,6$ obtained by replacing $\Lambda_{\theta}$ with $\Lambda_{\theta}^{2}$ in $M_{i}$
\begin{eqnarray}
M_{1}'=M_{2}'=M_{3}'=M_{4}'&=&\frac{1}{\Delta^{2}}(\frac{N}{2}-1)
\delta_{xy}+\frac{1}{\Delta^{2}}(1-\delta_{xy})\\
M_{5}'=M_{6}'&=&-\frac{1}{\Delta^{2}}(\delta_{y,x-1}
+\delta_{y,x+1})-\frac{2}{\Delta^{2}}(1-\delta_{y,x-1}-\delta_{y,x+1})
\end{eqnarray}
with the same constraints on $x$ and $y$ as before.

One must then  evaluate the twenty matrix elements arising when three $H_R$ 
and three $H_L$ are combined in all possible ways in the expression
 in the first line of Eq.(\ref{et4})
\begin{equation}
<\theta(y)|(L-R)\Lambda_{\theta}(L-R)\Lambda_{\theta}
(R-L)\Lambda_{\theta}(R-L)|\theta(x)>
\end{equation}
Since hermiticity of the Hamiltonian requires that the elements 
obtained changing $R$ with $L$ are equal 
one has only to compute the ten matrix elements, 
\begin{eqnarray}
M_{1}''&=&\frac{1}{N}\sum_{x,y}M''_{1}(x,y)=\frac{1}{N}\sum_{x,y}<\psi|R(y)R\Lambda_{\theta}R\Lambda_{\theta}L\Lambda_{\theta}LL(x)|\psi>\\
M_{2}''&=&\frac{1}{N}\sum_{x,y}M''_{2}(x,y)=\frac{1}{N}\sum_{x,y}<\psi|R(y)L\Lambda_{\theta}L\Lambda_{\theta}R\Lambda_{\theta}RL(x)|\psi>\\
M_{3}''&=&\frac{1}{N}\sum_{x,y}M''_{3}(x,y)=\frac{1}{N}\sum_{x,y}<\psi|R(y)R\Lambda_{\theta}L\Lambda_{\theta}R\Lambda_{\theta}LL(x)|\psi>\\
M_{4}''&=&\frac{1}{N}\sum_{x,y}M''_{4}(x,y)=\frac{1}{N}\sum_{x,y}<\psi|R(y)R\Lambda_{\theta}L\Lambda_{\theta}L\Lambda_{\theta}RL(x)|\psi>\\
M_{5}''&=&\frac{1}{N}\sum_{x,y}M''_{5}(x,y)=\frac{1}{N}\sum_{x,y}<\psi|R(y)L\Lambda_{\theta}R\Lambda_{\theta}R\Lambda_{\theta}LL(x)|\psi>\\
M_{6}''&=&\frac{1}{N}\sum_{x,y}M''_{6}(x,y)=\frac{1}{N}\sum_{x,y}<\psi|R(y)L\Lambda_{\theta}R\Lambda_{\theta}L\Lambda_{\theta}RL(x)|\psi>\\
M_{7}''&=&\frac{1}{N}\sum_{x,y}M''_{7}(x,y)=-\frac{1}{N}\sum_{x,y}<\psi|R(y)R\Lambda_{\theta}L\Lambda_{\theta}L\Lambda_{\theta}LR(x)|\psi>\\
M_{8}''&=&\frac{1}{N}\sum_{x,y}M''_{8}(x,y)=-\frac{1}{N}\sum_{x,y}<\psi|R(y)L\Lambda_{\theta}R\Lambda_{\theta}L\Lambda_{\theta}LR(x)|\psi>\\
M_{9}''&=&\frac{1}{N}\sum_{x,y}M''_{9}(x,y)=-\frac{1}{N}\sum_{x,y}<\psi|R(y)L\Lambda_{\theta}L\Lambda_{\theta}R\Lambda_{\theta}LR(x)|\psi>\\
M_{10}''&=&\frac{1}{N}\sum_{x,y}M''_{10}(x,y)=-\frac{1}{N}\sum_{x,y}<\psi|R(y)L\Lambda_{\theta}L\Lambda_{\theta}L\Lambda_{\theta}RR(x)|\psi>\ .
\end{eqnarray}
where $M_{1}''(x,y),...,M_{6}''(x,y)$ are defined for $x$ and $y$ odd, $M_{7}''(x,y),...,M_{10}''(x,y)$ are defined for $x$ even and $y$ odd.
The results are
\begin{eqnarray}
M_{1}''(x,y)&=&-\frac{1}{2\Delta^{3}}\delta_{x,y}-\frac{1}{4\Delta^{3}}(N-2)(N-4)\delta_{x,y}+\nonumber \\
&-&\frac{1}{4\Delta^{3}}(\delta_{y,x+2}+\delta_{y,x-2})-\frac{1}{\Delta^{3}}(N-4)(1-\delta_{x,y}) \\
M_{2}''(x,y)&=&-\frac{1}{4\Delta^{3}}(N-4)(N-6)\delta_{x,y}\\
M_{3}''(x,y)&=&M_{4}''(x,y)=M_{5}''(x,y)=M_{6}''(x,y)=-\frac{1}{8\Delta^{3}}(N-4)(N-6)\delta_{x,y}+\nonumber \\
&-&\frac{1}{4\Delta^{3}}(N-6)(\delta_{y,x+2}+\delta_{y,y-2})-\frac{1}{4\Delta^{3}}(1-\delta_{x,y}-\delta_{y,x+2}-\delta_{y,x-2})\\
M_{7}''(x,y)&=&M_{8}''(x,y)=M_{9}''(x,y)=M_{10}''(x,y)=\nonumber\\
&=&\frac{1}{2\Delta^{3}}(N-6)(1-\delta_{y,x-1}-\delta_{y,x+1})
\end{eqnarray}
and
\begin{eqnarray}
M_{1}''&=&-\frac{1}{2\Delta^{3}}-\frac{3}{8\Delta^{3}}(N-2)(N-4)\\
M_{2}''&=&M_{3}''=M_{4}''=M_{5}''=M_{6}''=-\frac{1}{8\Delta^{3}}(N-4)(N-6)\\
M_{7}''&=&M_{8}''=M_{9}''=M_{10}''=+\frac{1}{8\Delta^{3}}(N-4)(N-6)
\end{eqnarray}

\section{The chiral condensate}
\label{chirco}

In the continuum Schwinger model, the phenomenon 
of dynamical symmetry breaking of the chiral symmetry, 
is due to the anomaly. 
The order parameter is the mass operator 
$M(x)=\overline{\psi}(x)\psi(x)$ 
which acquires a non zero vacuum expectation value, 
giving rise to the chiral condensate
\cite{b57}
\begin{equation}
\chi_{c}=<\overline{\psi}(x)\psi(x)>
=-\frac{e^{\gamma}}{2\pi}m_{c}=-\frac{e^{\gamma}}{2\pi}\frac{e_{c}}{\sqrt{\pi}}
\label{cond}.
\end{equation}

In this section I compute the lattice chiral condensate 
to the fourth order in perturbation theory.
In the staggered fermion formalism (having put particles on even sites and antiparticles on odd sites), one has 
\begin{equation}
\overline{\psi}(x)\psi(x) \longrightarrow \frac{(-1)^{x}}{a}(\psi^{\dagger}_{x}\psi_{x}-\frac{1}{2})
\end{equation}
The lattice chiral condensate may be obtained by considering
the mass operator
\begin{equation}
M=-\frac{1}{Na}\sum_{x=1}^{N}(-1)^{x}\psi^{\dagger}_{x}\psi_{x}
\label{mmm}
\end{equation}
where the extra minus sign is there to give the same sign to the lattice and continuum chiral condensates (see footnote 1 of this chapter), and evaluating the expectation value of (\ref{mmm}) on the perturbed states $|p_{\psi}>$ generated by applying
$H_{h}$ to $|\psi>$. One has
\begin{equation}
|p_{\psi}>=|\psi>+\epsilon|p_{\psi}^{1}>+\epsilon^{2}|p_{\psi}^{2}>
\end{equation}
where
\begin{eqnarray}
|p_{\psi}^{1}>=-\frac{1}{\Delta}H_{h}|\psi>\\
|p_{\psi}^{2}>=\frac{\Pi_{\psi}}{2\Delta^{2}}H_{h}H_{h}|\psi>.
\end{eqnarray}
The lattice chiral condensate is then given by
\begin{equation}
\chi_{L}=\frac{<p_{\psi}|M|p_{\psi}>}{<p_{\psi}|p_{\psi}>}=\frac{<\psi|M|\psi>+\epsilon^{2}<p_{\psi}^{1}|M|p_{\psi}^{1}>+\epsilon^{4
}<p_{\psi}^{2}|M|p_{\psi}^{2}>}{<\psi|\psi>+\epsilon^{2}<p_{\psi}^{1}|p_{\psi}^{1}>+\epsilon^{4}<p_{\psi}^{2}|p_{\psi}^{2}>}
.
\end{equation}
One gets the following expressions for the wave functions
\begin{eqnarray}
<\psi|\psi>&=&1\\
<p_{\psi}^{1}|p_{\psi}^{1}>&=&\frac{N}{\Delta^{2}}\\
<p_{\psi}^{2}|p_{\psi}^{2}>&=&\frac{N(N-3)}{2\Delta^{4}}
\end{eqnarray}
and for the $M$ operator
\begin{eqnarray}
<\psi|M|\psi>&=&-\frac{1}{2a}\\
<p_{\psi}^{1}|M|p_{\psi}^{1}>&=&-\frac{1}{\Delta^{2}a}(\frac{N}{2}-2)\\
<p_{\psi}^{2}|M|p_{\psi}^{2}>&=&-\frac{1}{4\Delta^{4}a}(\frac{N}{2}-4)\cdot 2\cdot (N-3).
\end{eqnarray}
To the fourth order in $\epsilon$, the lattice chiral condensate is given by 
\begin{equation}
\chi_{L}=-\frac{1}{a}(\frac{1}{2}-
\frac{2}{\Delta^{2}}\epsilon^{2}
+\frac{6}{\Delta^{4}}\epsilon^{4})=-\frac{1}{a}(\frac{1}{2}-32\epsilon^{2}+1536\epsilon^{4}).
\label{lchi}
\end{equation}

\section{Lattice versus continuum}
\label{lvc}

In this section I want to extract 
some physics from the lattice results I obtained; to do this, one should compare the answer 
of the strong coupling analysis of the lattice theory with the exact results of the continuum model.
For this purpose one should extrapolate the strong 
coupling expansion derived under the assumption that the parameter 
$z=\frac{t^{2}}{4e_{L}^{4}a^{4}}\ll1$ to the region in which 
$z\gg1$; this region corresponds to the continuum theory since 
$e_{L}^{4}a^{4}\longrightarrow 0$, $z\longrightarrow \infty$. 
To make this 
extrapolation possible, it is customary~\cite{b26} to make use of Pad\'e approximants, 
which allow to extrapolate a series expansion beyond 
the convergence radius. 
Strong coupling perturbation theory improved with Pad\'e approximants should be compared with the continuum theory.

As we shall see, the gauge invariant strong coupling expansion here proposed,
provides a very accurate estimate of the observables of the continuum theory, 
already at the second order in powers of $z$. This strongly suggests that,
expanding around the gauge, parity and charge conjugation invariant ground state $|\psi>$, leads 
to a perturbative series converging
to the continuum theory faster then the one used in \cite{b58,b61}.  

Let us first compute 
the ratio between the continuum 
value of the meson mass $m_c =e_c/\sqrt{\pi}$ and the lattice coupling constant 
$e_L$, 
by equating the lattice chiral condensate (\ref{lchi}), to its continuum 
counterpart (\ref{cond}) 
\beq
\frac{1}{a} \left(\frac{1}{2}-32 z+1536z^2\right)=\frac{e^{\gamma}}{2\pi}
m_c\ \ .
\label{lmecm}
\eeq
Of course, Eq.(\ref{lmecm}) is true only when Pad\'e approximants are used, 
since - as it stands - the l.h.s. 
holds only for $z\ll 1$, while the r.h.s. provides the value of the chiral 
condensate to be obtained when $z\simeq \infty$.
Using the relation
\begin{equation}
a=(\frac{t^{2}}{4z})^{\frac{1}{4}}\frac{1}{e_{L}}
\end{equation}
one gets from (\ref{lmecm})
\begin{equation}
\frac{m_{c}}{e_{L}}=(\frac{4z}{t^{2}})^{\frac{1}{4}}\frac{2\pi}{e^{\gamma}}(\frac{1}{2}-32z+1536z^2)
\label{fme1}
\end{equation}
As in Ref. \cite{b58}, due to the factor $z^{\frac{1}{4}}$, the fourth power of (\ref{fme1}) should be considered in order to construct a 
non-diagonal Pad\'e approximant. Since the strong coupling expansion has been carried out up to second order in $z$, one is allowed to construct only the 
$[0,1]$ Pad\'e approximant for the polynomial written in (\ref{fme1}). One gets
\begin{equation}
(\frac{m_{c}}{e_{L}})^{4}=\frac{1}{4t^2}(\frac{2\pi}{e^{\gamma}})^{4}\frac{z}{1+256z}
\end{equation}
Taking the continuum limit $z\longrightarrow \infty$ one has
\begin{equation} 
(\frac{m_{c}}{e_{L}})^{4}=\frac{\pi^{4}}{64e^{4\gamma}}\frac{1}{t^{2}}
\label{fme2}
\end{equation}
Next let us compute the same ratio by equating the pseudoscalar mass gap given in (\ref{mga}) to its continuum counterpart $m_c$
\beq
e_L^2 a \left(\frac{1}{4}+8z-576 z^2\right)= m_c\ \ .
\label{mcecc}
\eeq
Again Eq.(\ref{mcecc}) is true only when Pad\'e approximants are used.

Dividing both sides of Eq.(\ref{mcecc}) by $e_{L}$ and taking into account 
that 
\begin{equation}
e_{L}a=(\frac{t^{2}}{4z})^{\frac{1}{4}},
\label{ez}
\end{equation}
one gets 
\begin{equation}
\frac{m_{c}}{e_{L}}=(\frac{t^{2}}{4z})^{\frac{1}{4}}
(\frac{1}{4}+8z-576z^{2})\ ,
\label{moe}
\end{equation}
Taking the fourth power, as I did for the chiral condensate equation, and constructing the $[1,0]$ Pad\'e approximant for the r.h.s. of Eq.(\ref{moe}) 
one gets
\begin{equation}
(\frac{m_{c}}{e_{L}})^{4}=\frac{t^{2}}{4z}(\frac{1}{256}+\frac{z}{2})
\end{equation} 
One may now take the limit  $z\to \infty$, obtaining
\begin{equation}
(\frac{m_{c}}{e_{L}})^{4}=\frac{t^{2}}{8}\ .
\label{lm1}
\end{equation}
Equating Eq.(\ref{fme2}) and Eq.(\ref{lm1}) one gets an equation for $t$ which gives
\begin{equation}
t=\frac{\pi}{8^{\frac{1}{4}}e^{\gamma}}=1.049
\end{equation}
which lies $4.9\%$ above the exact value. It is conforting to see that the lattice theory gives a light velocity very close to 
(but greater than ) $1$. 
Putting this value of $t$ in Eq.(\ref{fme2}) or Eq.(\ref{lm1}) one has
\begin{equation}
\frac{m_{c}}{e_{L}}=0.609
\end{equation}
which lies $7.9\%$ above the exact value $\frac{1}{\sqrt{\pi}}$.

It is worth to stress that I can reproduce very well the continuum results, even if I use just first order (in z) results of the 
strong coupling perturbation theory.

A second way \cite{b58} to extrapolate the polinomial in (\ref{fme1}) is to equate it to the ratio
\begin{equation}
\frac{(1+bz)^{x}}{(1+az)^{x+\frac{1}{4}}}=1-64z+3072z^{2}
\end{equation}
so that one can take the limit $z\longrightarrow \infty$ in (\ref{fme1}). Setting $t=1$ one gets
\begin{equation}
\frac{m_{c}}{e_{L}}=\frac{\sqrt{2}\pi}{e^{\gamma}}\frac{b^x}{a^{x+\frac{1}{4}}}
\label{1m}
\end{equation}
It is worth to say that this extrapolation works for a large range of values of the parameter $x$ between $0$ and $6$ giving an answer that lies between $6.9\%$ below and $10\%$ above the exact answer. In particular for $x=1$ one gets $\frac{m_{c}}{e_{L}}
=0.596$ which lies $5.6\%$ above the exact value. For $x=\frac{3}{4}$ one gets $\frac{m_{c}}{e_{L}}=0.593$ wich differs from the continuum value of the $5.1\%$. For $x=\frac{1}{10}$ one has only $0.7\%$ of error.
In the same way one can extrapolate the polinomial in (\ref{moe}) 
\beq
\frac{(1+ b z)^{x+\frac{1}{4}}}{(1+az)^x}=1+32 z-2304 z^2
\eeq
so that one can then take the limit $z\to\infty$ in (\ref{moe}).
Setting $t=1$ one gets
\beq
\frac{m_c}{e_L}=\frac{1}{4^{\frac{5}{4}}}\left( b \right)^{x+\frac{1}{4}}
\frac{1}{a^{x}}
\label{617}
\eeq
for $x=1$ one gets $\frac{m_{c}}{e_{L}}=0.617$ which lies $9\%$ above the exact value, against the $30 \%$ error of ref.\cite{b58}. For $x=0.5$ one has  
$\frac{m_{c}}{e_{L}}=0.595$ which lies only $5.4\% $ above the exact answer.

In the continuum Schwinger model the ratio between the masses of the scalar
and the pseudoscalar particles, $m^S/m^P$, equals 2, since $m^S$ is the lowest eigenvalue
of the continuum mass spectrum, which starts at $2 m^P$ \cite{b58}.
I shall now evaluate this ratio using the lattice expression of $m^P$ and $m^S$ 
(\ref{mga},\ref{mgb}). $m^S/m^P$ is given by
\beq
\frac{m^S}{m^P}=\frac{\frac{1}{4}+ 24 z - 1600 z^2}
{\frac{1}{4}+ 8 z - 576 z^2}\ \ .
\label{rm}
\eeq
Expanding the r.h.s. of (\ref{rm}) in power series of $z$ 
\beq
\frac{m^S}{m^P}=1+64 z-6144 z^2\ \ ,
\eeq
one may perform the $[1,1]$ Pad\'e approximant
\beq
\frac{m^S}{m^P}=\frac{1+ 160 z}
{1+96 z}\ \ .
\eeq
For $z\to\infty$
\beq
\frac{m^S}{m^P}=1.67
\eeq
which lies $16\%$ below the continuum value. 
My result coincides with the one obtained in \cite{b58}
at the same order in the perturbative expansion.

Taking the $[1,1]$ Pad\'e approximants of Eq.(\ref{mg4}) and Eq.(\ref{mg5}) one can compute in an independent way the ratio between the masses of 
the scalar and pseudoscalar particles
\begin{equation}
\frac{m_{*}^{S}}{m_{*}^{P}}=-\frac{1+18z}{1-16z}
\end{equation}
For $z\longrightarrow \infty$
\begin{equation}
\frac{m_{*}^{S}}{m_{*}^{P}} =1.125
\end{equation}
which lies $43\%$ below the exact value. This means that one should go to the next order in perturbation theory in Eq.(\ref{mg4}) and Eq.(\ref{mg5}) 
to reproduce better the continuum limit, since in Eq.(\ref{mg4}) and Eq.(\ref{mg5}) one has no zeroth order term.
Another test (also suggested in \cite{b58}) to check the validity of the lattice 
computations, may be performed by computing the quantity 
\beq
D=\frac{z^{3/4}}{m^P} \frac{d}{dz}\left(z^{1/4} m^P\right)=
\frac{1}{4} + \frac{z}{m^P}\frac{d m^P}{d z}\ .
\eeq
This should equal $1/4$ if the lattice theory has to reproduce 
the continuum result.

Using (\ref{mga}) one gets
\beq
D=z\frac{8-1152 z}{1/4+8z}\simeq 32 z(1-176 z)\ .
\eeq
Constructing the $[0,1]$ Pad\'e approximant and taking the $z\to \infty$ 
limit, one has
\beq
D=z\frac{32}{1+176 z}\stackrel{z\to\infty}{\longrightarrow}=\frac{2}{11}=0.182
\eeq
which lies $27\%$ below the desired $0.25$.
This coincides with the result obtained in \cite{b58}, 
using a strong coupling expansion up to the order $z^4$.
The agreement of my results with the continuum theory is very encouraging, since 
only terms up to the order $z^2$
have been used in the gauge invariant strong coupling expansion proposed 
in this paper.

\section{Summary}

The investigation, that has been exposed, is aimed at constructing an improved strong coupling 
expansion for the lattice Schwinger model. I showed that, gauge 
invariance together with the discrete symmetries of the model, force 
the ground state to be a N\'eel state.
Moreover gauge invariance 
requires the ground state to have electric fields related to the 
charge density operator by the Gauss's law. 
Thus, for large $e_{L}^{2}$, the ground state energy 
is of order $e_{L}^{2}$ rather 
than $\frac{1}{e_{L}^{2}}$ \cite{b58}; as a consequence,
the strong coupling expansion here constructed 
provides a very accurate extrapolation to the 
continuum theory already at the second order in 
$z=\frac{t^{2}}{4e_{L}^{4}a^{4}}$ . Furthermore, 
the Coulomb energy of elementary excitations is also affected, 
since $-$ besides the Coulomb self 
energy $-$ there is also an interaction energy with the charge of the ground state. 

The strongly coupled  lattice Schwinger model is equivalent to a one-dimensional antiferromagnetic Ising spin chain
with a Coulomb long range interaction, which admits the N\'eel state as 
an exact ground state.
Due to this, in my approach, chiral symmetry 
is broken spontaneously in the lattice model already at 
zero-th order in $z$. I expect that chiral symmetry breaking should persist for all
coupling, since the critical coupling for $D=1$ should be at $e_L^2=0$.
 I conjecture that, for small $e_{L}^{2}$, this behavior 
is a manifestation of a Peierls instability
$-$ the tendency of a one dimensional Fermi gas to form a gap at the Fermi 
surface. This happens with any infinitesimal interaction. 

In the
continuum theory the Schwinger model exhibits 
the Nambu-Goldstone phenomenon: 
global chiral symmetry is spontaneously broken, but no Goldstone boson
appears in the spectrum since the local current is 
anomalous~\cite{b60}. On the lattice there is no anomaly
due to the Nielsen-Ninomiya theorem~\cite{b24}. 

The present analysis shows that in the strong coupling limit, chiral 
symmetry is spontaneously broken at all orders,
since the states $|\psi>$ and $|\chi>$ mix
only at a perturbative order comparable with the volume of the system.
Being chiral
symmetry replaced by the discrete  
symmetry representing invariance under 
tranlations by one lattice site, effects related to the
breaking of the chiral symmetry on the lattice 
should come from the coupling between the two sublattices. 
This is manifest in my strong coupling calculations.
\chapter{The two-flavor lattice Schwinger model}

In this chapter I study the $SU(2)$-flavor lattice Schwinger model in the hamiltonian formalism using staggered fermions. 
The existence of the continuum internal isospin symmetry makes the model much more interesting than the one-flavor case; the 
spectrum is extremely reacher, exhibiting also massless excitations and the chiral symmetry breaking pattern is 
completely different from the one-flavor case. I shall demonstrate~\cite{b36,b36b}
 that the strong coupling limit of the two-flavor 
lattice Schwinger model is mapped onto an interesting quantum spin model $-$ the one-dimensional spin-$1/2$
 quantum Heisenberg antiferromagnet. 
The ground state of the antiferromagnetic chain has been known since many years~\cite{b33} and its energy 
was computed in~\cite{b34}; the complete spectrum as been determined by Faddeev and Takhtadzhyan~\cite{b27} using the algebraic 
Bethe ansatz. 

The two-flavor lattice Schwinger model with non-zero fermion mass $m$ has been analysed in~\cite{b64b} in the limit 
of heavy fermions $m\gg e^{2}$; good agreement with the continuum theory has been found.

There are by now many hints at a correspondence between quantized gauge theories and quantum spin models, aimed at analyzing new phases relevant 
for condensed matter systems~\cite{b16,b17,b17a,b17b,b17c}. Recently Laughlin has argued that there is an analogy between the spectral data of gauge 
theories and strongly correlated electron systems~\cite{b22}. Moreover,
 certain spin ladders have been shown to be related to the two-flavor Schwinger model \cite{b65}. 
 
The correspondence between the SU(2) flavor Schwinger model and the quantum Heisenberg antiferromagnetic chain provides a concrete 
computational scheme in which the issue of the correspondence between quantized gauge theories and quantum spin models may be 
investigated. Because of dimensionality of the coupling constant in (1+1)-dimensions the infrared behavior is governed by the strong coupling limit, 
and it is tempting to conjecture the existence of an exact correspondence between the infrared limits of the Heisenberg and two-flavor Schwinger models. 
I shall derive~\cite{b36,b36b} results which support this conjecture. For example the gapless modes in the spectra have 
identical quantum numbers; within the accurancy of the strong coupling limit, the gapped mode of the two-flavor Schwinger model was also 
identified in the spectrum of the Heisenberg model.

In this chapter I present a complete study~\cite{b36b} of the strong coupling limit of the two-flavor lattice Schwinger model. I firstly compute explicitly the masses 
of the excitations to the second order in the strong coupling expansion; this 
computation needs the knowledge of the spin-spin correlators of the quantum Heisenberg antiferromagnetic chain. 
The continuum massless two-flavor Schwinger model does exhibit neither an isoscalar $\left<\bar\psi\psi\right>$ nor an isovector 
$\left< \bar\psi \sigma^a\psi\right>$ chiral condensate, since this is forbidden by the Coleman theorem~\cite{b39}. 
On the lattice I show that these fermion 
condensates are zero to all the orders in the strong coupling expansion. Moreover I find that the pertinent chiral condensate is 
$<\overline{\psi}_{L}^{(2)}\overline{\psi}_{L}^{(1)}\psi_{R}^{(1)}\psi_{R}^{(2)}>$ and I compute its lattice expression up to the 
second order in the strong coupling expansion. It should be noticed that, in abscence of gauge fields, the chiral condensate is zero, is 
different from zero only when the fermions are coupled to gauge fields. This can be viewed as the manifestation of the 
chiral anomaly in this model. I finally compare the lattice results with the ones of the continuum theory. 

\section{Definition of the model}
\label{2fcl}

The action of the $1+1$-dimensional electrodynamics with two charged 
Dirac spinor fields is
\begin{equation}
S = \int d^{2} x\left[\sum_{a=1}^{2} \overline{\psi}_{a}
(i\gamma_{\mu}\partial^{\mu}+\gamma_{\mu}
A^{\mu})\psi_{a}-\frac{1}{4e^{2}_{c}}F_{\mu\nu}F^{\mu\nu}\right]
\label{action}
\end{equation}
The theory has an internal $SU_L(2)\otimes SU_R(2)$-flavor isospin symmetry; 
the Dirac fields are 
an isodoublet whereas the electromagnetic field is an isosinglet.
It is well known that in $1+1$ dimensions there is no spontaneous breakdown 
of continuous internal symmetries, unless there are anomalies or the Higgs 
phenomenon occurs. Neither mechanism is possible in the two-flavor Schwinger 
model for the $SU_L(2)\otimes SU_R(2)$-symmetry: isovector currents do not 
develop anomalies and there are no gauge 
fields coupled to the isospin currents. The particles belong then to 
isospin multiplets. For what concerns the $U(1)$ gauge symmetry there is an Higgs phenomenon~\cite{b60}.

The action is invariant under the symmetry 
\beq
SU_{L}(2)\otimes SU_{R}(2)\otimes U_{V}(1) \otimes U_{A}(1) \nonumber
\eeq
The group generators act on the fermion isodoublet to give
\begin{eqnarray}
SU_{L}(2) &:& \psi_{a}(x)\longrightarrow (
e^{i\theta_{\alpha}\frac{\sigma^{\alpha}}{2}P_{L}})_{ab}\ \psi_{b}(x)\   , \  
\overline{\psi_{a}}(x)\longrightarrow \overline{\psi_{b}}\ (x)
(e^{-i\theta_{\alpha}\frac{\sigma^{\alpha}}{2}P_{R}})_{ba}\label{s1} \\
SU_{R}(2) &:& \psi_{a}(x)\longrightarrow 
(e^{i\theta_{\alpha}\frac{\sigma^{\alpha}}{2}P_{R}})_{ab}\ \psi_{b}(x)\   ,\   
\overline{\psi_{a}}(x)\longrightarrow \overline{\psi_{b}}(x)\ 
(e^{-i\theta_{\alpha}\frac{\sigma^{\alpha}}{2}P_{L}})_{ba} \\
U_{V}(1) &:& \psi_{a}(x)\ \longrightarrow 
(e^{i\theta(x){\bf 1}})_{ab}\ \psi_{b}(x)\  ,\  
\psi_{a}^{\dagger}(x)\longrightarrow \psi_{b}^{\dagger}(x)\ 
(e^{-i\theta(x){\bf 1}})_{ba} \\
U_{A}(1) &:& \psi_{a}(x)\longrightarrow (e^{i\alpha 
\gamma_{5}{\bf 1}})_{ab}\ \psi_{b}(x)\  ,\  
\psi_{a}^{\dagger}(x)\longrightarrow \psi_{b}^{\dagger}(x)\ 
(e^{-i\alpha \gamma_{5}{\bf 1}})_{ba}
\quad ,
\label{s4} 
\end{eqnarray}
where $\sigma^{\alpha}$ are the Pauli matrices, $\theta_{\alpha}$, $\theta(x)$ and $\alpha$ are real coefficients and
\beq
P_{L}=\frac{1}{2}(1-\gamma_{5})\ ,\ P_{R}=\frac{1}{2}(1+\gamma_{5})\quad .
\eeq
At the classical level the symmetries (\ref{s1}$-$\ref{s4}) lead to 
conservation laws for the isovector, vector and axial currents
\begin{eqnarray}
j_{\alpha}^{\mu}(x)_{R}&=&\overline{\psi}_{a}(x)\gamma^{\mu}P_{R}
(\frac{\sigma_{\alpha}}{2})_{ab}\psi_{b}(x)
\label{ca}\\
j_{\alpha}^{\mu}(x)_{L}&=&\overline{\psi}_{a}(x)\gamma^{\mu}P_{L}
(\frac{\sigma_{\alpha}}{2})_{ab}\psi_{b}(x)
\label{cb}\\
j^{\mu}(x)&=&\overline{\psi}_{a}(x)\gamma^{\mu}{\bf 1}_{ab}\psi_{b}(x)
\label{cc}\\
j^{\mu}_{5}(x)&=&\overline{\psi}_{a}(x)\gamma^{\mu}\gamma ^{5}
{\bf 1}_{ab}\psi_{b}(x)
\end{eqnarray}
It is well known that at the quantum level the vector and axial currents 
cannot be simultaneously conserved, 
due to the anomaly phenomenon \cite{b23}. If the regularization is gauge 
invariant, so that 
the vector current is conserved, then the axial current acquires the 
anomaly which breaks the $U_{A}(1)$-symmetry 
\beq
\partial_{\mu} j_5^{\mu}(x)=2\frac{e_{c}^{2}}{2\pi}\epsilon_{\mu \nu}
F^{\mu \nu}(x)
\label{anomaly}
\eeq
The isoscalar and isovector chiral condensates are zero due to the 
Coleman theorem \cite{b39}; in fact, they would break not only the 
$U_A(1)$ symmetry 
of the action, but also the continuum internal symmetry 
$SU_L(2)\otimes SU_R(2)$ down to $SU_V(2)$. 
There is, however,  a $SU_L(2)\otimes SU_R(2)$ invariant operator, 
which is non-invariant under the $U_A(1)$-symmetry;
it can acquire a non-vanishing VEV without violating 
Coleman's theorem and consequently 
may be regarded as a good order parameter for the 
$U_A(1)$-breaking.
Its expectation value is given by~\cite{b67,b57}
\begin{equation} 
<F>\equiv
<\overline{\psi}_{L}^{(2)}\overline{\psi}_{L}^{(1)}
\psi_{R}^{(1)}\psi_{R}^{(2)}>=(\frac{e^{\gamma}}{4\pi})^{2} 
\frac{2}{\pi} e_{c}^{2}\quad .
\label{chico}
\end{equation}
It describes a process in which two right movers are anihilated
and two left movers are created. 
Note that $F$, being quadrilinear in the fields,
is actually invariant under chiral rotations of $\pi/2$, namely under the 
discrete axial symmetry
\begin{equation}
\psi_a(x)\to \gamma^5\psi_a(x)\quad\bar\psi_a(x)\to-\bar\psi_a(x)\gamma_5\ \ .
\label{disax}
\end{equation}
As a consequence, this part of the chiral symmetry group is not broken by 
the non-vanishing VEV of $F$ (\ref{chico}).

As we shall see in section (\ref{schi}), the lattice theory faithfully reproduces the 
pattern of symmetry breaking of the continuum theory; this happens even if 
on the lattice the $SU(2)$-flavor symmetry is not protected 
by the Coleman theorem. The isoscalar and isovector chiral 
condensates are zero also on the lattice,
whereas the operator $F$ acquires a non-vanishing VEV due to the 
coupling of left and right movers induced by the gauge field.
The continuous axial symmetry is broken explicitly by 
the staggered fermion, but the discrete axial symmetry (\ref{disax}) remains.

The action (\ref{action}) may be presented 
in the usual abelian bosonized form \cite{b68}. Setting
\beq
:\overline{\psi}_{a}\gamma^{\mu}\psi_{a}:=\frac{1}{\sqrt{\pi}}\epsilon^{\mu 
\nu}\partial_{\nu}\Phi_{a}\  ,\  a=1,2\quad ,
\eeq
the electric charge density and the action read
\beq
j_{0}=:\psi^{\dagger}_{1}\psi_{1}+\psi^{\dagger}_{2}\psi_{2}:=
\frac{1}{\sqrt{\pi}}\partial_{x}(\Phi_{1}+\Phi_{2})
\label{chde}
\eeq
\beq
S=\int d^{2}x \left[\frac{1}{2}\partial_{\mu}\Phi_{1}\partial^{\mu}\Phi_{1}
+\frac{1}{2}\partial_{\mu}\Phi_{2}\partial^{\mu}\Phi_{2}-
\frac{e_{c}^{2}}{2\pi}(\Phi_{1}+\Phi_{2})^{2}\right] \quad .
\eeq
By changing the variables to
\begin{eqnarray}
\Phi_{+}&=&\frac{1}{\sqrt{2}}(\Phi_{1}+\Phi_{2})\\
\Phi_{-}&=&\frac{1}{\sqrt{2}}(\Phi_{1}-\Phi_{2})\quad ,
\end{eqnarray}
one has
\beq
S=\int d^{2}x \left(\frac{1}{2}\partial_{\mu}\Phi_{+}
\partial^{\mu}\Phi_{+}+\frac{1}{2}\partial_{\mu}\Phi_{-}
\partial^{\mu}\Phi_{-}-
\frac{e_{c}^{2}}{\pi}\Phi_{+}^{2}\right)\quad .
\label{boa}
\eeq

The theory describes two scalar fields, one massive and one massless. 
$\Phi_{+}$ is an isosinglet as 
evidenced from Eq.(\ref{chde}); its mass $m_{S}=\sqrt{\frac{2}{\pi}}e_{c}$ 
comes from the anomaly Eq.(\ref{anomaly})~\cite{b60}. Local electric charge 
conservation is 
spontaneously broken, but no Goldstone boson appears because the 
Goldstone mode may be gauged away.
$\Phi_{-}$ represents an isotriplet; it has rather involved nonlinear 
transformation properties under a general 
isospin transformation. All three isospin currents can be written in 
terms of $\Phi_{-}$ but only the third component has a simple 
representation in terms of $\Phi_-$; namely
\begin{eqnarray}
j_{\mu} ^{3} (x) = : \overline{\psi} _{a} (x) \gamma_{\mu} 
(\frac{\sigma^{3}}{2})_{ab} \psi_{b} (x):=\quad \quad \quad 
\quad \quad \nonumber\\
   = \frac{1}{2}:\overline{\psi}_{1}(x)\gamma_{\mu}\psi_{1}(x)-
\overline{\psi}_{2}(x)\gamma_{\mu}\psi_{2}(x):=(2\pi)^{\frac{1}{2}}
\epsilon^{\mu \nu}
\partial_{\nu}\Phi_{-}\quad .
\end{eqnarray}
The other two isospin currents $j_{\mu}^{1}(x)$ and $j_{\mu}^{2}(x)$ 
are nonlinear 
and nonlocal functions of $\Phi_{-}$ \cite{b68}; a more symmetrical 
treatment of the bosonized form of the isotriplet currents is available 
within the framework of non abelian bosonization~\cite{b68bis}. For the multiflavor 
Schwinger model this approach has been carried out in~\cite{b68tris}, providing results in 
agreement with~\cite{b68}.

The excitations are most conveniently classified in terms of the 
quantum numbers of $P$-parity and $G$-parity; 
$G$-parity is related to the charge conjugation $C$ by
\beq
G=e^{i\pi \frac{\sigma^{2}}{2}} C\quad .
\eeq
$\Phi_{-}$ is a $G$-even pseudoscalar, while $\Phi_{+}$ is a $G$-odd pseudoscalar
\begin{eqnarray}
\Phi_{-}\  &:&\  I^{PG}=1^{-+}\\
 \Phi_{+}\  &:&\  I^{PG}=0^{--}\ \ .
\end{eqnarray}
The massive meson $\Phi_{+}$ is stable by $G$ conservation since the action 
(\ref{boa}) is invariant under $\Phi_{+}\longrightarrow -\Phi_{+}$.

In the massive $SU(2)$ Schwinger model $-$ when the mass of the fermion 
$m$ is small compared to $e^2$ (strong coupling) $-$ 
Coleman \cite{b68} showed that - in addition to the triplet $\Phi_-$ 
($I^{PG}=1^{-+}$)
the low-energy spectrum exhibits a singlet 
$I^{PG}=0^{++}$  lying on 
top of the triplet $\Phi_-$. In this limit the gauge theory is 
mapped to a sine-Gordon model and  the low lying excitations  
are soliton-antisoliton states. When $m\rightarrow 0$, these
soliton-antisoliton states become massless \cite{b69}; in this 
limit, the analysis of the many body wave functions, carried out in 
ref.\cite{b69}, hints to the existence of  
a whole class of massless states with positive G-parity; 
P-parity however cannot be determined with the procedure developed
in \cite{b69}. 
These are not the only excitations of the model:
way up in mass there is the 
isosinglet $I^{PG}=0^{--}$, ($\Phi_+$), already discussed in 
ref.~\cite{b68}. 
The model exhibits also triplets, 
whose mass $-$ of order $m_{S}$ or greater $-$ stays finite~\cite{b69}; 
among the triplets there is a G-even state
\footnote{K. Harada private communication.}.

The Hamiltonian, gauge constraint and non-vanishing (anti-)commutators
of the continuum two-flavor Schwinger model are
\begin{eqnarray}
H=\int dx&\left[\frac{e^2}{2}E^2(x)+\sum_{a=1}^2
\psi^{\dagger}_a (x)\alpha\left(i\partial_x +eA(x)\right)\psi_a(x)\right]
\label{ham1}\\ 
&\partial_x E(x)\ +\sum_{a=1}^2 \psi^{\dagger}_a
(x)\psi_a (x)\sim 0\label{ga1}\\
&\left[ A(x),E(y)\right]=i\delta(x-y) ~,
 \left\{\psi_a(x),\psi_b^{\dagger}(y)\right\}=\delta_{ab}\delta(x-y)\ \ .
\label{commu1}
\end{eqnarray}
A lattice Hamiltonian, constraint and (anti-) commutators reducing to 
(\ref{ham1},\ref{ga1},\ref{commu1}) 
in the naive continuum limit are 
\begin{eqnarray}
H_{S}=\frac{e^{2}a}{2}\sum_{x=1}^N E_{x}^{2}&-&\frac{it}{2a}\sum_{x=1}^N
\sum_{a=1}^2 \left(\psi_{a,x+1}^{\dag}e^{iA_{x}}\psi_{a,x}
-\psi_{a,x}^{\dag}e^{-iA_{x}}\psi_{a,x+1}\right)\label{hamilton}\nonumber\\
E_{x}-E_{x-1}&+&\psi_{1,x}^{\dag}\psi_{1,x}+\psi_{2,x}^{\dag}\psi_{2,x}-1\sim
0\ ,
\label{gauss}\\
\left[ A_x,E_y\right]=i\delta_{x,y}~&,&
\left\{\psi_{a,x},\psi_{b,y}^{\dagger}\right\}=\delta_{ab}\delta_{xy}\ \ .
\nonumber
\end{eqnarray}
The fermion fields are defined on the sites, $x=1,...,N$, 
the gauge and electric fields, $ A_{ x}$ and
 $E_{x}$,  on the links $[x; x + 1]$, $N$ is an even integer 
and, when $N$ is finite it is convenient to impose periodic boundary conditions.  When $N$ is finite, the continuum limit is the 
two-flavor Schwinger model on a circle \cite{b62}.
The coefficient $t$ of the hopping term in (\ref{hamilton})
plays the role of the lattice light speed. In the naive continuum limit,
$e_L=e_c$ and $t=1$. 

The Hamiltonian and gauge constraint exhibit the discrete symmetries 
\begin{itemize}
\item{}Parity P: 
\begin{equation}
A_{x}\longrightarrow -A_{-x-1},\ E_{x}\longrightarrow -E_{-x-1},\ 
\psi_{a,x}\longrightarrow (-1)^{x}\psi_{a,-x},\ \psi_{a,x}^{\dag}\longrightarrow 
(-1)^{x}\psi_{a,-x}^{\dag}
\label{par}
\end{equation}

\item{}Discrete axial symmetry $\Gamma$: 
\begin{equation}
A_{x}\longrightarrow A_{x+1},\ E_{x}\longrightarrow E_{x+1},\ 
\psi_{a,x}\longrightarrow \psi_{a,x+1},\ \psi_{a,x}^{\dag}\longrightarrow 
\psi_{a,x+1}^{\dag}
\label{chir}
\end{equation}

\item{}Charge conjugation C:
\begin{equation}
A_{x}\longrightarrow -A_{x+1},\  E_{x}\longrightarrow -E_{x+1},\ 
\psi_{a,x}\longrightarrow \psi^{\dag}_{a,x+1},\ \psi_{a,x}^{\dag}\longrightarrow 
\psi_{a,x+1}
\label{char}
\end{equation}

\item{}G-parity:
\begin{eqnarray}
A_{x}\longrightarrow -A_{x+1},\  E_{x}\longrightarrow -E_{x+1}\nonumber\\
\psi_{1,x}\longrightarrow \psi^{\dag}_{2,x+1},\ \psi_{1,x}^{\dag}
\longrightarrow \psi_{2,x+1}\\
\psi_{2,x}\longrightarrow -\psi^{\dag}_{1,x+1},\ \psi_{2,x}^{\dag}
\longrightarrow -\psi_{1,x+1}\ .\nonumber
\end{eqnarray}
\end{itemize}

The lattice two-flavor Schwinger model is equivalent to a one 
dimensional quantum Coulomb gas on the lattice with two kinds of particles. To see this one can fix the 
gauge, $A_{x} = A$ (Coulomb gauge). Eliminating the non-constant electric 
field and using the gauge constraint, one obtains the effective Hamiltonian
\begin{eqnarray}
H_{S}&=&H_u+H_p
\equiv\left[\frac{e^{2}_{L}}{2 N}E^{2}+\frac{e^{2}_{L}a}{2}
\sum_{x,y}\rho(x) V(x-y)\rho(y)\right]+\nonumber\\
&+&\left[
-\frac{it}{2a}\sum_{x}\sum_{a=1}^{2}(\psi_{a,x+1}^{\dag}e^{iA}\psi_{a,x}-\psi_{a,x}^{\dag}e^{-iA}
\psi_{a,x+1})\right]\ ,
\label{hs}
\end{eqnarray}
where the charge density is
\begin{equation}
\rho(x)=\psi^{\dag}_{1,x}\psi_{1,x}+\psi^{\dag}_{2,x}\psi_{2,x}-1\ \ ,
\label{cd1}
\end{equation}
and the potential
\begin{equation}
V(x-y)=\frac{1}{N}
\sum^{N-1}_{n=1} e^{i 2\pi n (x-y)/N}\frac{1}{4\sin^2\frac{\pi n}{N}}
\end{equation}
is the Fourier transform of the inverse laplacian on the lattice
for non zero momentum.
The constant 
modes of the gauge field decouple in the thermodynamic limit 
$ N \longrightarrow \infty $.

\section{The strong coupling limit and the antiferromagnetic Heisenberg Hamiltonian}
\label{2fhm}

In the thermodynamic limit the Schwinger Hamiltonian (\ref{hs}), 
rescaled by the
 factor ${e_{L}^{2}a}/{2}$, reads 
 \begin{equation}
H=H_{0}+\epsilon H_{h}
\end{equation}
with
\begin{eqnarray}
H_{0}&=&\sum_{x>y}\left[\frac{(x-y)^{2}}{N}-(x-y)\right]\rho(x)\rho(y)\quad ,
\label{hu}\\
H_{h}&=&-i(R-L)
\label{hp}
\end{eqnarray}
and $\epsilon=t/e_{L}^{2}a^{2}$.
In Eq.(\ref{hp}) the right $R$ and left $L$ hopping operators are defined ($L=R^{\dagger}$) as
\begin{equation}
R=\sum_{x=1}^{N}R_{x}=\sum_{x=1}^{N}\sum_{a=1}^{2}R_{x}^{(a)}= 
\sum_{x=1}^{N} \sum_{a=1}^2 \psi_{a,x+1}^{\dag}e^{iA}\psi_{a,x}\quad .
\end{equation}
On a periodic chain the commutation relation 
\begin{equation}
[R,L]=0\quad 
\end{equation}
is satisfied.

I shall consider the strong coupling perturbative expansion 
where the Coulomb 
Hamiltonian (\ref{hu}) is the unperturbed Hamiltonian and the hopping 
Hamiltonian 
(\ref{hp}) the perturbation.
Due to Eq.(\ref{cd1}) every configuration with one particle per site 
has zero energy, so that the ground state of the Coulomb 
Hamiltonian (\ref{hu}) is $2^N$ times degenerate. The degeneracy of 
the ground state can be removed only at the second perturbative order 
since the first order is trivially zero. 

At the second order the 
lattice gauge theory is effectively described by the antiferromagnetic 
Heisenberg Hamiltonian. 
The vacuum energy $-$ at order $\epsilon^2$ $-$ reads  
\begin{equation}
E^{(2)}_{0}=<H_{h}^{\dagger}\frac{\Pi}{E_{0}^{(0)}-H_{0}}H_{h}>
\label{secorder}
\end{equation}
where the expectation values are defined on the degenerate subspace and 
$\Pi$ is the operator projecting on a set orthogonal to the states with one 
particle per site. 
Due to the vanishing of the charge density on the ground states of $H_{0}$, 
the commutator
\begin{equation}
[H_0, H_h]=H_h
\label{comm1}
\end{equation}
holds on any linear combination of the degenerate ground states. 
Consequently, from Eq.(\ref{secorder}) one finds
\begin{equation}
E^{(2)}_{0}=-2<RL>\quad .
\label{secorder2}
\end{equation}
On the ground state the combination $R L$ can be written in terms of the 
Heisenberg Hamiltonian.
By introducing the Schwinger spin operators
\begin{equation}
\vec{S}_{x}=\psi_{a,x}^{\dag}\frac{\vec{\sigma}_{ab}}{2}\psi_{b,x}
\end{equation}
the Heisenberg Hamiltonian $H_{J}$ reads
\begin{eqnarray}
H_{J}&=&\sum_{x=1}^{N}\left(\vec{S}_x
\cdot \vec{S}_{x+1}-\frac{1}{4}\right)=\nonumber\\ 
&=&\sum_{x=1}^{N} \left( -\frac{1}{2} 
L_{x}R_{x}-\frac{1}{4}\rho(x) \rho(x+1) \right)
\end{eqnarray}
and, on the degenerate subspace, one has 
\begin{equation}
<H_{J}>=\left<\sum_{x=1}^{N}\left(\vec{S}_x\cdot \vec{S}_{x+1}-\frac{1}{4}
\right)\right>
=\left<\sum_{x=1}^{N}\left(-\frac{1}{2}L_{x}R_{x}\right)\right>\quad .
\label{mainequation}
\end{equation}
Taking into account that products of $L_x$ and $R_y$ at different points 
have vanishing expectation values on the ground states,
and using Eq.(\ref{mainequation}), Eq.(\ref{secorder2}) reads
\begin{equation}
E^{(2)}_{0}=4<H_{J}>\quad .
\label{secorder3}
\end{equation}
The ground state of $H_{J}$ singles out the correct vacuum, on which to 
perform the perturbative expansion. 
In one dimension $H_{J}$ is exactly diagonalizable \cite{b33,b31}. 
In the spin model a flavor 1 particle on a site 
can be represented by a spin up, a flavor 2 particle by a spin down. 
The spectrum of $H_{J}$ exhibits $2^N$ eigenstates; among these, the spin 
singlet with lowest energy is the non degenerate ground state $|g.s.>$. 

I shall construct the strong coupling perturbation theory of the 
two-flavor Schwinger model using $|g.s.>$ as the unperturbed ground state. 
$|g.s.>$ is invariant under translations by one lattice site, which amounts
to invariance under discrete chiral transformations. As a consequence,
At variance with 
the one-flavor model~\cite{b56}, chiral symmetry cannot be 
spontaneously broken even in the infinite coupling limit. 
 
$|g.s.>$ has zero charge density on each site and zero electric flux 
on each link 
\begin{equation}
\rho(x)|g.s.>=0\quad ,\quad E_{x}|g.s.>=0 \quad \quad (x=1,...,N)\quad .
\label{keyequation}
\end{equation}
$|g.s.>$ is a linear combination of all the possible states with 
$\frac{N}{2}$ spins up and $\frac{N}{2}$ spins down. The 
coefficients are not explicitly known for general $N$. In chapter 3 I
 exhibited $|g.s.>$ explicitly
for finite size systems of 4, 6 and 8 sites. 
The Heisenberg energy of $|g.s.>$ is known exactly and, in the 
thermodynamic limit, is \cite{b34,b27}
\begin{equation}
H_{J}|g.s.>=(-N\  \ln\ 2)|g.s.>\quad .
\label{mainenergy}
\end{equation}
Eq.(\ref{mainenergy}) provides the second order correction 
Eq.(\ref{secorder3}) to the vacuum energy, $E_{g.s.}^{(2)}=-4 N\ln 2$.

There exist two kinds of excitations created from 
$|g.s.>$; one kind involves only spin flipping and has lower energy 
since no electric flux is created, the other involves fermion 
transport besides spin flipping and thus has a higher energy. For the 
latter excitations the energy is proportional to the coupling times 
the length of the electric flux: the lowest energy is achieved when the 
fermion is transported by one lattice spacing. 
Of course only the excitations of the first kind can be mapped into 
states of the Heisenberg model. 

In~\cite{b27} the antiferromagnetic Heisenberg model excitations 
have been classified. There it was shown that any excitation may be 
regarded as 
the scattering state of quasiparticles of spin-$1/2$: every 
physical state contains an even number of quasiparticles and 
the spectrum exhibits only integer spin states. 
The two simplest excitations of lowest energy in the thermodynamic 
limit are a triplet and a singlet \cite{b27}; they 
have a dispersion relation depending on the momenta of the two 
quasiparticles. 
For vanishing total momentum (relative to the ground state 
momentum $P_{g.s.} =0$ for $\frac{N}{2}$ even, 
$P_{g.s.} =\pi$ for $\frac{N}{2}$ odd) in the thermodynamic limit
they are degenerate with the ground state. 

In chapter 3 I showed that even for finite size systems, the excited
states can be grouped in families corresponding to the classification
given in~\cite{b27}. I explicitly exhibited all the energy 
eigenstates for $N=4$ and $N=6$. The lowest lying are a triplet and a 
singlet, respectively; they have a well defined relative 
(to the ground state) $P$-parity and $G$-parity
- $1^{-+}$ for the triplet and $0^{++}$ for the singlet.
Since they share the same quantum numbers
these states can be identified, in the limit of vanishing fermion mass, 
with the soliton-antisoliton excitations 
found by Coleman in his analysis of the two-flavor Schwinger model.
A related analysis about the parity of the lowest lying states 
in finite size Heisenberg chains, has been given in~\cite{b70}. 

Moreover in \cite{b27} a whole class $-$ ${\cal M}_{AF}$ $-$ of 
gapless excitations at zero momentum was singled out 
in the thermodynamic limit; these states are eigenstates of the total 
momentum and consequently have positive G-parity at 
zero momentum. The low lying states of the Schwinger model also contain 
\cite{b69} many massless excitations with positive G-parity; they are 
identified \cite{b36,b36b} with the excitations belonging to 
${\cal M}_{AF}$. 
The mass of these states in the Schwinger model can be obtained
from the differences between the excitation
energies at zero momentum 
and the ground state energy. 
The energies of the states $|ex.>$ belonging to the class
${\cal M}_{AF}$ have the same 
perturbative expansion of the ground state. 
Consequently, the states $|ex.>$ 
at zero momentum up to the second order in the strong coupling 
expansion have the same energy 
of the ground state (\ref{secorder}), 
$E^{(2)}_{ex}=-4 N\ln 2$.  To this order the
mass gap is zero. 
Higher order corrections may give a mass gap. 

\subsection{The low lying spectrum of $QCD_2$}

In section (\ref{SC}) I already explained that $QCD_2$ with colour group $SU(2)$ and 1 flavor is also mapped onto the spin-$1/2$ 
quantum Heisenberg antiferromagnetic chain. This result was originally found by Langmann and Semenoff in~\cite{b21}. It is worth stressing 
this fact here, and speculating about the model spectrum.

I claim that the massless excitation spectrum of this non-abelian theory in the strong coupling limit is the same of the massless excitation spectrum 
of the two-flavor massless Schwinger model. Both the models exhibit an exact correspondence with the infrared limit of the Heisenberg antiferromagnetic 
chain. 

The quark-like excitations of $QCD_2$ are gapless spin-$1/2$ kinks or spinons. 
They always appear in even number so that the only physical 
excitations have integer spin and an even number of spinons. This phenomenon is not exactly what one would call confinement. 
These quark-spinons are localizable objects and one can consider their scattering. There are no bound states of spinons in 
the spin-$1/2$ quantum antiferromagnetic chain \cite{b27}. Spinons exist only in even number and this constraint is the spin model counterpart 
of the abscence of coloured states in $QCD_2$. 
One flavor $QCD_2$ exhibits only mesons in its spectrum and it is not possible to generate baryons. I can say nothing, at this level of analysis, 
about the massive excitations of lattice $QCD_2$ which are created by the non abelian currents of the model acting on $|g.s.>$. 
These currents involve charge transport on $|g.s.>$ and create electric flux. The non zero electric flux generates the masses of the excitations, with a 
mechanism similar to that of the two-flavor Schwinger model. 

\section{The meson masses}
\label{2fmm}

In this section  I determine the masses 
for the states obtained by fermion transport
of one site on the Heisenberg model ground state.
My analysis shows that besides the $G$-odd pseudoscalar isosinglet
$0^{--}$ with mass $m_S=e_{L}\sqrt{2/\pi}$, there are also a 
$G$-even pseudoscalar isotriplet $1^{-+}$
and a $G$-odd scalar isotriplet $1^{+-}$ with masses of the order of 
$m_S$ or greater.  
The quantum numbers are relative to those of the ground state
$I_{g.s.}^{PG}=0^{++}$ for $N/2$ even $I_{g.s.}^{PG}=0^{--}$ for $N/2$
odd.

Two states can be created using the spatial
component of the vector $j^{1}(x)$ Eq.(\ref{cc}) and 
isovector $j_{\alpha}^{1}(x)$ Eqs.(\ref{ca},\ref{cb}) 
Schwinger model currents. They are the G-odd pseudoscalar 
isosinglet $I^{PG}=0^{--}$ and the
G-even pseudoscalar isotriplet $I^{PG}=1^{-+}$.  The lattice operators
with the correct quantum numbers creating these states at zero momentum, 
when acting on $|g.s.>$, read 
\begin{eqnarray}
S&=&R+L=\sum_{x=1}^{N}j^{1}(x)\\
T_{+}&=&(T_{-})^{\dagger}=R^{(12)}+L^{(12)}=\sum_{x=1}^{N}j_{+}^{1}(x)
\label{ta}\\
T_{0}&=&\frac{1}{\sqrt{2}}(R^{(11)}+L^{(11)}-R^{(22)}-L^{(22)})=
\sum_{x=1}^{N}j_{3}^{1}(x) \quad .
\label{t0}
\end{eqnarray}
$R^{(ab)}$ and $L^{(ab)}$ in (\ref{ta},\ref{t0}) are the 
right and left flavor changing 
hopping operators ($L^{(ab)}=(R^{(ab)})^{\dagger}$)
$$
R^{(ab)}=\sum_{x=1}^{N}\psi_{a,x+1}^{\dagger}e^{iA}\psi_{b,x}\quad .
$$
The states are given by
\begin{eqnarray}
|S>&=&|0^{--}>=S|g.s.>\\
|T_{\pm}>&=&|1^{-+},\pm1>=T_{\pm}|g.s.>
\label{2tpm}\\
|T_{0}>&=&|1^{-+},0>=T_{0}|g.s.>\ .
\label{to}
\end{eqnarray}
They are normalized as
\begin{eqnarray}
<S|S>&=&<g.s.|S^{\dagger}S|g.s.>=-4<g.s.|H_{J}|g.s.>=4N \ln 2
\label{no1}\\
<T_{+}|T_{+}>&=&\frac{2}{3}(N+<g.s.|H_{J}|g.s.>)=\frac{2}{3}N(1-\ln 2)
\label{no2}
\end{eqnarray}
and 
\begin{equation}
<T_{0}|T_{0}>=<T_{-}|T_{-}>=<T_{+}|T_{+}> \ .
\label{no3}
\end{equation}
In Eqs.(\ref{no1},\ref{no2},\ref{no3}) $<g.s.|g.s.>=1$.

The isosinglet energy, up to the second order in the strong coupling 
expansion, is
$
E_{S}=E_{S}^{(0)}+\epsilon^{2}E_{S}^{(2)}
$ with
\begin{eqnarray}
E_{S}^{(0)}&=&\frac{<S|H_{0}|S>}{<S|S>}=1\quad ,\\
E_{S}^{(2)}&=&\frac{<S|H_{h}^{\dag}\Lambda_{S}H_{h}|S>}{<S|S>}\quad ,
\label{secordex}
\end{eqnarray}
$
\Lambda_{S}=\frac{\Pi_{S}}{E_{S}^{(0)}-H_{0}}
$
and $1-\Pi_{S}$ a projection operator onto $|S>$.
On $|g.s.>$ 
\begin{equation}
[H_{0},(\Pi_{S}H_{h})^n S]=(n+1)(\Pi_{S}H_{h})^n 
,\quad
(n=0,1,\dots),
\end{equation}
holds; Eq.(\ref{secordex}) may then be written in terms of spin correlators as
\begin{equation}
E_{S}^{(2)}=E_{g.s.}^{(2)}+4-\frac{\sum_{x=1}^{N}
<g.s.|\vec{S}_{x}\cdot \vec{S}_{x+2}-\frac{1}{4}|g.s.>}{<g.s.|H_{J}|g.s.>}
\quad .
\label{senes}
\end{equation}
One immediately recognizes that the excitation spectrum is determined 
once $<g.s.|\vec{S}_{x}\cdot \vec{S}_{x+2}|g.s.>$ is known. 
Equations similar to Eq.(\ref{senes}) may be established also at a 
generic order of the strong coupling expansion.    

At the zeroth perturbative order the pseudoscalar triplet is 
degenerate with the isosinglet $E_{T}^{(0)}=E_{S}^{(0)}=1$. 
Following the same procedure as before one may compute the energy of the 
states (\ref{2tpm}) and 
(\ref{to}) to the second order in the strong coupling expansion. To 
this order, the energy is given by
\begin{eqnarray}
&&E_{T}^{(2)}=E_{g.s.}^{(2)}-\Delta_{DS}(T)-\nonumber\\ 
&&\frac{4<g.s.|H_{J}|g.s.>+5\sum_{x=1}^{N}
<g.s.|\vec{S}_{x}\cdot \vec{S}_{x+2}-\frac{1}{4}|g.s.>}{N+<g.s.|H_{J}|g.s.>}
\label{senet}
\end{eqnarray}
where in terms of the vector operator 
$
\vec{V}=\sum_{x=1}^{N}\vec{S}_{x}\wedge \vec{S}_{x+1}
$, one can write $\Delta_{DS}(T)$ as
\begin{eqnarray}
\Delta_{DS}(T_{\pm})&=&12
\frac{<g.s.|(V_{1})^2|g.s.>+<g.s.|(V_{2})^2|g.s.>}
{N+<g.s.|H_{J}|g.s.>}\\
\Delta_{DS}(T_{0})&=&12
\frac{2<g.s.|(V_{3})^2|g.s.>}{N+<g.s.|H_{J}|g.s.>}\ .
\end{eqnarray}
The VEV of each squared component of $\vec{V}$ on 
the rotationally invariant singlet $|g.s.>$ give the same 
contribution $i.e.$ $\Delta_{DS}(T_{\pm})=\Delta_{DS}(T_{0})$: the 
triplet states (as in the continuum theory) have a degenerate mass gap. 
This is easily verified by direct computation on finite size systems; 
when the size of the system is finite one may also show that $\Delta_{DS}$ 
is of zeroth order in N.

The excitation masses are given by 
$m_{S}=\frac{e_{L}^2 a}{2}(E_{S}-E_{g.s.})$ and 
$m_{T}=\frac{e_{L}^2 a}{2}(E_{T}-E_{g.s.})$. 
Consequently, the ($N$-dependent) ground state energy terms 
appearing in $E_{S}^{(2)}$ and $E_{T}^{(2)}$ cancel and what is left 
are only $N$ independent terms.
This is a good check of our computation, being the mass an intensive quantity.
 
In principle one should expect also excitations created acting 
on $|g.s.>$ with the chiral currents, in analogy with the one flavor 
Schwinger model where, as shown in ref.~\cite{b58}, the chiral current 
creates a two-meson bound state. The chiral currents operators for the two 
flavor Schwinger model are given by
\begin{eqnarray}
j^{5}(x)&=&\overline{\psi}(x)\gamma^{5}\psi(x)\\
j^{5}_{\alpha}(x)&=&\overline{\psi}_{a}(x)\gamma^{5}
(\frac{\sigma}{2})_{ab}\psi_{b}(x)\quad .
\end{eqnarray}
The corresponding lattice operators at zero momentum are 
\begin{eqnarray}
S^{5}&=R-L=\sum_{x=1}^{N}j^{5}(x)
\label{s5}\\
T_{+}^{5}&=(T_{-}^{5})^{\dagger}=R^{(12)}-L^{(12)}=
\sum_{x=1}^{N}j_{+}^{5}(x)
\label{ta2}\\
T_{0}^{5}&=\frac{1}{\sqrt{2}}(R^{(11)}-L^{(11)}-R^{(22)}+L^{(22)})=
\sum_{x=1}^{N}j_{3}^{5}(x)\quad .
\label{t02}\\
\end{eqnarray}
The states created by (\ref{s5},\ref{ta2},\ref{t02}) when acting on $|g.s.>$,
are
\begin{eqnarray}
|S^{5}>&=&|0^{++}>=S^{5}|g.s.>\\
|T_{\pm}^{5}>&=&|1^{+-},\pm1>=T_{\pm}^{5}|g.s.>
\label{tpm2}\\
|T_{0}^{5}>&=&|1^{+-},0>=T_{0}^{5}|g.s.>\quad .
\label{to2}
\end{eqnarray}
They are normalized as
\begin{eqnarray}
<S^{5}|S^{5}>&=&<g.s.|S^{5\dagger}S^{5}|g.s.>=-4<g.s.|H_{J}|g.s.>=4N\log 2\\
<T^{5}_{+}|T^{5}_{+}>&=&\frac{2}{3}(N+<g.s.|H_{J}|g.s.>)=\frac{2}{3}
N(1-\log 2)
\end{eqnarray}
and 
\begin{equation}
<T^{5}_{0}|T^{5}_{0}>=<T^{5}_{-}|T^{5}_{-}>=<T^{5}_{+}|T^{5}_{+}>\quad .
\end{equation}

Following the computational scheme used to study $|S>$ and $|T>$, 
one finds that the state $|S^{5}>$ 
is not a particle-like excitation since its energy depends on the 
volume of the system and this would lead to an extensive mass; 
consequently,
there is no $0^{++}$ massive singlet in agreement with \cite{b68}.
For the triplet $|T^{5}>$ one gets
\begin{eqnarray}
E_{T^5}^{(0)}&=&1\\
E_{T^5}^{(2)}&=&E_{g.s.}^{(2)}+\frac{\sum_{x=1}^{N}<g.s.|\vec{S}_{x}
\cdot\vec{S}_{x+2}-
\frac{1}{4}|g.s.>-4<g.s.|H_{J}|g.s.>}{N+<g.s.|H_{J}|g.s.>}\quad .
\label{senet5}
\end{eqnarray} 

Now I can compute the mass spectrum up to the second order in the 
strong coupling expansion. 
Using Eq.(\ref{corrd2n}), the isosinglet mass reads as
\begin{equation}
\frac{m_{S}}{e^2 a}=\frac{1}{2}+1.9509 \quad \epsilon ^2\quad .
\label{ms} 
\end{equation}
For what concerns the isotriplet mass, since the double sum in 
Eq.(\ref{senet}) is given by
\begin{equation}
\Delta_{DS}(T)=8\frac{<g.s.|\vec{V}\cdot \vec{V}|g.s.>}
{N+<g.s.|H_{J}|g.s.>}\quad ,
\end{equation}
using Eq.(\ref{vvapp}), one gets
\begin{equation}
\frac{m_{T}}{e_{L}^2 a}=\frac{1}{2}+0.0972\quad \epsilon ^2\quad .
\label{mtt} 
\end{equation}

The existence of massive isotriplets was already noticed in \cite{b69}, 
and their mass in the continuum theory was numerically computed 
for various values of the fermion mass. 
In particular there is a $G$-parity even isotriplet with mass 
approximately equal to the mass of the isosinglet $0^{--}$.  

The mass of the $|T_{5}>$ isotriplet is
\begin{equation}
\frac{m_{T^5}}{e^2a}=\frac{1}{2}+4.4069\epsilon^2\ . 
\label{mt5}
\end{equation}

Equations (\ref{ms}), (\ref{mtt}) and (\ref{mt5}) 
provide the values of $m_{S}$, $m_{T}$ and 
$m_{T^5}$ for small values of 
$z=\epsilon^2=\frac{t^2}{e_{L}^4a^4}$ up to the second order 
in the strong coupling expansion. 
Whereas (\ref{mtt}) is only approximate (\ref{ms}) and (\ref{mt5}) 
are exact at the second order in the $\epsilon$ expansion.
In section (\ref{lvc2}) I shall 
extrapolate these masses to the continuum limit using the 
standard technique of the Pad\'e approximants.

\section{Chiral condensate}
\label{schi}

In the following I shall first prove that also on the 
lattice either the isoscalar $\left<\bar\psi\psi\right>$ or the isovector 
$\left< \bar\psi \sigma^a\psi\right>$ chiral condensates are 
zero to every order of perturbation theory. 
This should be verified by explicit computation since on the 
lattice the symmetry $SU_{L}(2)\otimes SU_{R}(2)$ 
is already broken by introducing staggered fermions; 
thus, there is no symmetry to prevent the formation of such 
chiral condensates. In the continuum theory, instead, the breaking of the 
$SU_{L}(2)\otimes SU_{R}(2)$ down to $SU_{V}(2)$ is prevented 
by the Coleman theorem \cite{b39}.

In the staggered fermion formalism the isoscalar condensate is given by
\begin{equation}
\sum_{a=1}^2 \overline{\psi_{a}}(x)\psi_{a}(x) \longrightarrow
 \frac{(-1)^{x}}{2a}(\psi^{\dagger}_{1,x}\psi_{1,x}+
 \psi^{\dagger}_{2,x}\psi_{2,x}-1)\quad ;
\end{equation}
it is obtained  by considering the mass operator
\begin{equation}
M=\frac{1}{2Na}\sum_{x=1}^{N}(-1)^{x}(\psi^{\dagger}_{1,x}\psi_{1,x}+ 
\psi^{\dagger}_{2,x}\psi_{2,x})
\label{mmmm}
\end{equation}
and evaluating its expectation value on the perturbed states 
$|p_{g.s.}>$ generated by applying $H_{h}$ to $|g.s.>$.
To the second order in the strong coupling expansion, one has
\begin{equation}
|p_{g.s.}>=|g.s.>+\epsilon|p_{g.s.}^{1}>+\epsilon^{2}|p_{g.s.}^{2}>+\dots 
\end{equation} 
where
\begin{eqnarray}
|p_{g.s.}^{1}>&=&-H_{h}|g.s.> \\
|p_{g.s.}^{2}>&=&\frac{\Pi_{g.s.}}{2}H_{h}H_{h}|g.s.>\quad .
\end{eqnarray}
To the fourth order, (\ref{mmmm}) is given by
\begin{equation}
\chi_{isos.}=\frac{<p_{g.s.}|M|p_{g.s.}>}{<p_{g.s.}|p_{g.s.}>}=
\frac{<g.s.|M|g.s.>+\epsilon^{2}<p_{g.s.}^{1}|M|p_{g.s.}^{1}>+\epsilon^{4
}<p_{g.s.}^{2}|M|p_{g.s.}^{2}>+\ldots}{<g.s.|g.s.>+
\epsilon^{2}<p_{g.s.}^{1}|p_{g.s.}^{1}>+\epsilon^{4}
<p_{g.s.}^{2}|p_{g.s.}^{2}>+\ldots}\quad .
\label{ciso}
\end{equation}
It is very easy to see that $\chi_{isos.}$ is zero to all orders in the 
strong coupling expansion.
Let us introduce the translation operator
\begin{equation}
\hat{T}=e^{i \hat{p} a}\quad ;
\end{equation}
using 
\begin{eqnarray}
\hat{T}M\hat{T}^{-1}&=&-M
\label{t1}\\
\hat{T}H_{h}\hat{T}^{-1}&=&H_{h}
\label{t2}
\end{eqnarray}
and 
\begin{equation}
\hat{T}|g.s.>=\pm|g.s.>
\label{tpm}
\end{equation}
one gets order by order in the strong coupling expansion in Eq.(\ref{ciso})
\begin{equation}
\chi_{isos.}=-\chi_{isos.}\quad .
\end{equation}
In Eq.(\ref{tpm}) the $+$ appears when $N/2$ is even and the $-$ when 
$N/2$ is odd.

The isovector chiral condensate is given by the expectation value of the operator
\begin{equation}
\vec{\Sigma}=\frac{1}{2Na}\sum_{x=1}^{N}(-1)^x\psi_{a,x}^{\dagger}
\vec{\sigma}_{ab}\psi_{b,x}
\end{equation}
on the perturbed states $|p_{g.s.}>$.
Taking into account that
\begin{equation}
\hat{T}\Sigma \hat{T}^{-1}=-\Sigma
\end{equation}
one gets
\begin{equation}
\chi_{isov.}=-\chi_{isov.}\quad ;
\end{equation}
also the isovector chiral condensate is identically zero. 

In the continuum there is, as evidenced in section (\ref{2fcl}), only 
a non-vanishing chiral condensate associated to the anomalous breaking 
of the $U_{A}(1)$ symmetry \cite{b67,b57}. Since 
\begin{equation}
\overline{\psi^{a}}_{L}(x)\psi^{a}_{R}(x)=
\overline{\psi^{a}}(x)\frac{1+\gamma_{5}}{2}\psi^{a}(x)\quad ,
\label{chide}
\end{equation}
the pertinent operator is 
$F=\overline{\psi}_{L}^{(2)}\overline{\psi}_{L}^{(1)}\psi_{R}^{(1)}
\psi_{R}^{(2)}$; 
its expectation value has been computed in 
\cite{b67}\cite{b57} and is given by
\begin{equation}
<\overline{\psi}_{L}^{(2)}\overline{\psi}_{L}^{(1)}\psi_{R}^{(1)}
\psi_{R}^{(2)}>=(\frac{e^{\gamma}}{4\pi})^{2} \frac{2}{\pi} e_{c}^{2}=
(\frac{e^{\gamma}}{4\pi})^{2} m_{S}^{2} \quad.
\label{opua}
\end{equation}

On the lattice one has 
\begin{equation}
\overline{\psi^{a}}_{L}(x)\psi^{a}_{R}(x)\longrightarrow \frac{1}{2a} 
\frac{1}{2}(\psi^{\dagger}_{a,x}\psi_{a,x}-
\psi^{\dagger}_{a,x+1}\psi_{a,x+1}+L_{x}^{(a)}-R_{x}^{(a)})\quad .
\label{chidel}
\end{equation}
The factor $1/2a$ is due to the doubling of the lattice spacing 
in the antiferromagnetic bipartite lattice. 
Upon introducing the occupation number operators $n_{x}^{(a)}=
\psi_{a,x}^{\dagger}\psi_{a,x}$, 
the umklapp operator $F$ is represented on the lattice by
\begin{equation}
F=-\frac{1}{16a^2 N}\sum_{x=1}{N}\left\{
(n_x ^{(1)}-n_{x+1} ^{(1)})(n_x ^{(2)}-n_{x+1} ^{(2)})+
(L_{x}^{(1)}-R_{x}^{(1)})(L_{x}^{(2)}-R_{x}^{(2)})\right\}\quad .
\label{fop}
\end{equation}
The strong coupling expansion carried up to the second order in 
$\epsilon=\frac{t}{e_{L}^{2}a^2}$, 
yields 
\begin{equation}
<F>=\frac{<p_{g.s.}|F|p_{g.s.}>}{<p_{g.s.}|p_{g.s.}>}=
\frac{<g.s.|F|g.s.>+\epsilon^{2}<p_{g.s.}^{1}|F|p_{g.s.}^{1}>}{<g.s.|g.s.>+
\epsilon^{2}<p_{g.s.}^{1}|p_{g.s.}^{1}>}
\label{opuad}
\end{equation}
for the lattice chiral condensate. Since 
\begin{eqnarray}
<g.s.|g.s.>&=&1\\
<p_{g.s.}^{1}|p_{g.s.}^{1}>&=&-4<g.s.|H_{J}|g.s.>\quad ,
\end{eqnarray}
and taking into account that
\begin{eqnarray}
<g.s.|F|g.s.>&=&\frac{1}{8a^{2} N}<g.s.|H_{J}|g.s.>\\
<p^{1}_{g.s.}|F|p^{1}_{g.s.}>&=&
\frac{1}{4a^{2}N}(-2<g.s.|( H_{J})^{2}|g.s.>-\frac{5}{3}
<g.s.|H_{J}|g.s.>+\frac{5}{12}N\nonumber\\
&-&\frac{2}{3}\sum_{x=1}^{N}<g.s.|\vec{S}_{x}\cdot \vec{S}_{x+2}-
\frac{1}{4}|g.s.> )\quad ,
\end{eqnarray}
from Eqs.(\ref{mainenergy}) and (\ref{corrd22}), one gets
\begin{equation}
<F>=\frac{1}{a^{2}}(0.0866-0.4043\epsilon ^{2})\quad .
\label{nopuad}
\end{equation}
A nonvanishing value of the lattice chiral condensate 
is due to the coupling $-$ induced by 
the lattice gauge field $-$ between the right and left fermions. 
This is the relic in the lattice
of the $U_{A}(1)$ anomaly in the continuum theory.

\section{Lattice versus continuum}
\label{lvc2}
I now want to compare our lattice results with the exact 
results of the continuum model; to do this, 
one should extrapolate the strong-coupling expansion 
derived under the assumption that the parameter 
$z=\epsilon ^2=\frac{t^2}{e_{L}^4 a^4} \ll 1$ 
to the region in which $z\gg 1$; this corresponds 
to take the continuum limit since $e_{L}^4 a^4 
\longrightarrow 0$ when $z\longrightarrow \infty$. To make the 
extrapolation possible, it is customary to make use 
of Pad\'e approximants, which allow to 
extrapolate a series expansion beyond the convergence radius. Strong-coupling 
perturbation theory improved by Pad\'e approximants should 
then provide results consistent with the continuum 
theory. As we shall see the strong-coupling expansion derived in this chapter 
provides accurate estimates of the meson masses, already at the first 
order in powers of $z$. 

Let us now evaluate $m_{S}$ and the lattice light velocity $t$.
I first compute the ratio between the continuum value of the meson mass 
$m_{S}=\sqrt{\frac{2}{\pi}}e_{c}$ and the lattice coupling constant $e_{L}$ 
by equating the lattice chiral condensate, Eq.(\ref{nopuad}), 
to its continuum counterpart Eq.(\ref{opua})
\begin{equation}
\frac{1}{a^{2}}(0.0866-0.4043z)=(\frac{e^{\gamma}}{4\pi})^{2} m_{S}^{2}\ .
\label{chima1}
\end{equation}
Eq.(\ref{chima1}) is true only when Pad\'e approximants are used since, 
as it stands, the left hand side holds only 
for $z\ll 1$, while the right-hand side provides the value of the 
chiral condensate to be obtained when $z=\infty$. 
Using 
\begin{equation}
a=\frac{t^{\frac{1}{2}}}{e_{L}z^{\frac{1}{4}}}\quad ,
\end{equation}
one gets from Eq.(\ref{chima1})
\begin{equation}
(\frac{m_{S}}{e_{L}})^{2}=(\frac{4\pi }{e^{\gamma}})^{2}
\frac{z^{\frac{1}{2}}}{t}(0.0866-0.4043z)\quad .
\label{chima2}
\end{equation}
As in Refs.\cite{b58,b56}, due to the factor 
$z^{\frac{1}{2}}$, the second power of Eq.(\ref{chima2})
 should be considered in order to 
construct a non diagonal Pad\'e approximant. 
Since the strong coupling expansion has been carried 
out up to second order in $z$, one is allowed 
to construct only the $[0,1]$ Pad\'e approximant for 
the polynomial written in Eq.(\ref{chima2}). One gets
\begin{equation}
(\frac{m_{S}}{e_{L}})^{4}=(\frac{4\pi}{e^{\gamma}})^{4}
\frac{1}{t^2}\frac{0.0074z}{1+9.3371z}\quad ,
\label{chima3}
\end{equation}
and, taking the continuum limit $z\rightarrow \infty$, one finds
\begin{equation}
(\frac{m_{S}}{e_{L}})^4 =(\frac{4\pi}{e^{\gamma}})^{4} 
\frac{0.0008}{t^{2}}\quad .
\label{chima4}
\end{equation}
 
Next I compute the same mass ratio by equating the 
singlet mass gap given in Eq.(\ref{ms}) to its 
continuum counterpart $m_{S}$
\begin{equation}
e_{L}^{2}a(\frac{1}{2}+1.9509z)=m_{S}\quad .
\label{ma1}
\end{equation}
Again, Eq.(\ref{ma1}) is true only when Pad\'e approximants are used. 
Dividing both sides of Eq.(\ref{ma1}) by $e_{L}$ and taking into account that 
\begin{equation}
e_{L}a=\frac{t^{\frac{1}{2}}}{z^{\frac{1}{4}}}
\end{equation}
one gets
\begin{equation}
\frac{m_{S}}{e_{L}}=\frac{t^{\frac{1}{2}}}
{z^{\frac{1}{4}}}(\frac{1}{2}+1.9509z)\quad .
\label{ma2}
\end{equation}
Taking the fourth power and constructing the $[1,0]$ 
Pad\'e approximant for the right hand side of Eq.(\ref{ma2}) one has
\begin{equation}
(\frac{m_{S}}{e_{L}})^{4}=\frac{t^{2}}{z}(\frac{1}{16}+0.9754z)\quad ;
\label{ma3}
\end{equation}
when $z\rightarrow \infty$, Eq.(\ref{ma3}) gives 
\begin{equation}
(\frac{m_{S}}{e_{L}})^{4}=t^2 0.9754\quad .
\label{ma4}
\end{equation}

The numerical value of the hopping parameter $t$, 
determined if one equates Eq.(\ref{ma4}) and Eq.(\ref{chima4}), is
\begin{equation}
t=\frac{4\pi}{e^{\gamma}}0.1692=1.1940
\label{tequ}
\end{equation}
and lies $19\%$ above the exact value.
Putting this value of $t$ in Eq.(\ref{chima4}) or Eq.(\ref{ma4}) one gets
\begin{equation}
\frac{m_{S}}{e_{L}}=1.0969
\end{equation}
which lies $37\%$ above the exact value $\sqrt{\frac{2}{\pi}}$. 
It is comforting to see that the lattice reproduces in a sensible 
way the continuum 
results even if I use just first order (in $z$) results of the 
strong coupling perturbation theory. 

Using the value of $t$ given in Eq.(\ref{tequ}) one gets 
for the isotriplet mass Eq.(\ref{mtt2})
\begin{equation}
\frac{m_{T}}{e_{L}}=0.5143\quad .
\label{lb}
\end{equation}
By direct computation on an 8 sites chain one gets
\begin{equation}
\frac{m_{T}}{e_{L}}=1.3524\quad .
\label{ub}
\end{equation}
The discrepancy between Eq.(\ref{lb}) and Eq.(\ref{ub}) 
is mainly due to the approximation involved in the computation of 
$<g.s.|\vec{V}^{2}|g.s.>$. 
However, it is safe to believe that my lattice computation 
implies the existence of a massive isotriplet 
$1^{-+}$ with a mass between the lower bound (\ref{lb}) and 
the upper bound (\ref{ub}). This is in agreement 
with the results provided for the continuum theory in \cite{b69}. 
 
Using a similar procedure one may also compute the mass of the 
triplet $1^{+-}$. From Eq.(\ref{mt5}) one gets 
\begin{equation}
\frac{m_{T^{5}}}{e_{L}}=1.3347\quad .
\end{equation}
This triplet, being $G$-odd, is a scattering state of a $0^{--}$ 
singlet with a $1^{-+}$ triplet, which are the fundamental 
excitations of the system. 
The mass of this $1^{+-}$ triplet should be larger than the 
mass of the massive $1^{-+}$ triplet, which should be a 
scattering state of massless 
$1^{-+}$ triplets.

Putting $t=c=1$, $i.e.$ $e_{L}a=\frac{1}{z^{\frac{1}{4}}}$ 
the lattice mass spectrum gets closer to its continuum counterpart;
 for the isosinglet mass, one gets 
\begin{equation}
\frac{m_{S}}{e}=0.9938
\label{t1s}
\end{equation}
while for the isotriplet one gets 
\begin{equation}
\frac{m_{T}}{e_{L}}=0.4695\quad .
\label{mtt2}
\end{equation}
Eq.(\ref{t1s}) provides a value of the isosinglet mass lying 
$24\%$ above the exact answer. 
Again the triplet mass is reproduced with lesser accuracy due to 
the random phase approximation used in the computation of the 
pertinent correlator; 
a better answer is given however by a direct computation on the 
8 sites chain yielding the value $1.2346$ for ${m_{T}}/{e_{L}}$. 

\section{Summary}
In this chapter I used the correspondence between the two-flavor strongly 
coupled lattice Schwinger model and the antiferromagnetic Heisenberg 
Hamiltonian~\cite{b36,b36b} to investigate the spectrum of 
the gauge model. Using the analysis
of the excitations of the finite size chains given in the chapter 3,
I showed the equality of the quantum numbers of the states of the Heisenberg 
model and the low lying excitations of the two-flavor Schwinger model.
I provided also the 
spectrum of the massive excitations of the gauge model; 
in order to extract numerical values for the masses, I explicitly computed  
the pertinent spin-spin correlators of the Heisenberg chain in chapter 3. 
Although the spectrum is determined  only up to the 
second order in the strong coupling expansion
the agreement with the continuum theory is satisfactory. 
 
The massless and the massive excitations of the gauge 
model are created from the 
spin chain ground state with two very different mechanisms: 
massless excitations involve only spin flipping 
while massive excitations are created by fermion transport 
besides spin flipping and do not belong to the spin chain spectrum.
As in the continuum theory, due to the Coleman 
theorem \cite{b68}, the massless excitations are not Goldstone bosons,
but may be regarded as the gapless quantum excitations 
of the spin-$1/2$ antiferromagnetic 
Heisenberg chain \cite{b43}. 

In computing the chiral condensate I showed that, also 
in the lattice theory, 
the expectation value of the umklapp operator $F$ is 
different from zero, while both 
$<\overline{\psi}\psi>$ and $<\overline{\psi}\sigma^{a}\psi>$ 
are zero to every order in the 
strong coupling expansion. 
This implies that both on the lattice and the continuum 
the $SU(2)$ flavor symmetry is preserved whereas the
$U_A(1)$ axial symmetry is broken. The umklapp operator $F$ is 
the order parameter for this symmetry, but being 
quadri-linear in the fermi fields, is invariant, in the continuum, under 
chiral rotation of $\pi/2$ and on the lattice under the corresponding 
discrete axial symmetry (\ref{chir}) (translation by one lattice site). 
This shows that the discrete axial 
symmetry is not broken in both cases.
Our lattice computation enhance this result since the ground state of the 
strongly coupled two-flavor Schwinger model is translationally invariant.

The pattern of symmetry breaking of the continuum is 
exactly reproduced even if the Coleman theorem does not 
apply on the lattice and the anomalous symmetry breaking 
is impossible due to the Nielsen-Ninomiya~\cite{b24} theorem. 
At variance with the strongly coupled one-flavor lattice 
Schwinger model, the anomaly 
is not realized in the lattice theory via the spontaneous breaking of 
a residual chiral symmetry~\cite{b56}, but, rather, 
by explicit breaking of the chiral symmetry due to staggered fermions.
The non-vanishing of $<F>$ may be regarded as
the only relic, in the strongly coupled 
lattice theory, of the anomaly of the 
continuum two-flavor Schwinger model. 
It is due to the coupling induced by the 
gauge field, between the right and left-movers on the lattice.
\chapter{The multiflavor lattice Schwinger models}

In this chapter I study the ${\cal N}$-flavor lattice Schwinger models in the hamiltonian formalism using staggered fermions. I illustrated 
in chapter 5 how the presence of a nontrivial $SU(2)$-flavor symmetry makes the spectrum much richer than the one-flavor 
model spectrum and changes drastically the chiral symmetry breaking pattern. 

The ${\cal N}$-flavor Schwinger models have many features in in common with four dimensional $QCD$: at the classical level they have a 
symmetry group $U_{L}({\cal N})\otimes U_{R}({\cal N})= SU_{L}({\cal N})\otimes SU_{R}({\cal N}) \otimes U_{V}(1)\otimes U_{A}(1)$ that is broken down 
to $SU_{L}({\cal N})\otimes SU_{R}({\cal N}) \otimes U_{V}(1)$ by the axial anomaly exactly like in $QCD$~\cite{b74}. The massless ${\cal N}$-flavor 
Schwinger models describe no real interactions between their particles as one can infer by writing the model action in a bosonized form. 
The model exhibits one massive and ${\cal N}^{2} -1$ massless pseudoscalar ``mesons"~\cite{b75}. 

On the lattice I shall prove that $-$ at the second order in the strong coupling expansion $-$ 
the lattice Schwinger models are effectively described by $SU({\cal N})$ quantum 
antiferromagnetic spin-$1/2$ Heisenberg Hamiltonians with spins in a particular fundamental representation of the $SU({\cal N})$ 
Lie algebra and with nearest neighbours couplings. The features of the model are very different depending on if ${\cal N}$ is odd or even.
 When ${\cal N}$ is odd, the ground state energy in the strong coupling limit 
is proportional to $e_{L}^2$, the square of the electromagnetic coupling constant. In contrast, when ${\cal N}$ is even the ground state energy 
in the strong coupling limit is of order 1. This difference arises from the proper definition of the charge density
\begin{equation}
\rho(x)=\sum_{a=1}^{\cal N} \psi_{a,x}^{\dagger}\psi_{a,x}-\frac{\cal N}{2}
\label{ron}
\end{equation}
where the constant ${\cal N}/2$ has been subtracted from the charge density operator in order to make it odd under 
the charge conjugation transformation. 
As a consequence, when ${\cal N}$ is even, $\rho(x)$ admits zero eigenvalues and the ground state does 
not support any electric flux, while when ${\cal N}$ is odd the ground 
state exhibits a staggered configuration of the charge density and electromagnetic fluxes. 

In the continuum the Coleman theorem~\cite{b39}  prevents the formation of either an isoscalar chiral condensate 
$<\overline{\psi}\psi>$ or an isovector chiral condensate $<\overline{\psi}T^{a}\psi>$ $-$ where $T^{a}$ is an $SU({\cal N})$ generator $-$
 for every model with an internal $SU({\cal N})$-flavor symmetry. We shall see that this feature should be reproduced on the lattice also 
 for this class of models.

\section{The continuum  ${\cal N}$-flavor Schwinger models}
\label{cnsm}

The continuum $SU({\cal N})$-flavor Schwinger models are defined by the action 
\begin{equation}
S = \int d^{2} x(\sum_{a=1}^{\cal N} \overline{\psi}_{a}(i\gamma_{\mu}\partial^{\mu}+\gamma_{\mu}
A^{\mu})\psi_{a}-\frac{1}{4e^{2}_{c}}F_{\mu\nu}F^{\mu\nu})
\label{na1n}
\end{equation}
where the ${\cal N}$ fermions have been introduced in a completely symmetric way. Although the theory described by (\ref{na1n}) strictly parallels 
what has been shown in chapter 5 for the $SU(2)$ model, I shall now report a detailed analysis both for the sake of clarity and 
to show that some difference appears between ${\cal N}$ even and odd. 

The Dirac fields are an ${\cal N}$-plet, $i.e.$ transform according to the fundamental representation of 
the flavor group while the 
electromagnetic field is an $SU({\cal N})$ singlet. The flavor symmetry of the theory cannot be spontaneously broken for the same reasons as 
in the $SU(2)$ case. 
The particles of the theory belong to $SU({\cal N})$ multiplets. The action is invariant under the symmetry 
\beq
SU_{L}({\cal N})\otimes SU_{R}({\cal N})\otimes U_{V}(1) \otimes U_{A}(1) \nonumber
\eeq
The symmetry generators act as follows
\begin{eqnarray}
SU_{L}({\cal N}) &:& \psi_{a}(x)\longrightarrow (e^{i\theta_{\alpha}T^{\alpha}P_{L}})_{ab}\ \psi_{b}(x)\   , \  
\overline{\psi_{a}}(x)\longrightarrow \overline{\psi_{b}}\ (x)(e^{-i\theta_{\alpha}T^{\alpha}P_{R}})_{ba} \\
SU_{R}({\cal N}) &:& \psi_{a}(x)\longrightarrow (e^{i\theta_{\alpha}T^{\alpha}P_{R}})_{ab}\ \psi_{b}(x)\   ,\   
\overline{\psi_{a}}(x)\longrightarrow \overline{\psi_{b}}(x)\ (e^{-i\theta_{\alpha}T^{\alpha}P_{L}})_{ba} \\
U_{V}(1) &:& \psi_{a}(x)\ \longrightarrow (e^{i\theta(x){\bf 1}})_{ab}\ \psi_{b}(x)\  ,\  
\psi_{a}^{\dagger}(x)\longrightarrow \psi_{b}^{\dagger}(x)\ (e^{-i\theta(x){\bf 1}})_{ba} \\
U_{A}(1) &:& \psi_{a}(x)\longrightarrow (e^{i\alpha \gamma_{5}{\bf 1}})_{ab}\ \psi_{b}(x)\  ,\  
\psi_{a}^{\dagger}(x)\longrightarrow \psi_{b}^{\dagger}(x)\ (e^{-i\alpha \gamma_{5}{\bf 1}})_{ba} 
\end{eqnarray}
where $T^{\alpha}$ are the generators of the $SU({\cal N})$ group, $\theta_{\alpha}$, $\theta(x)$ and $\alpha$ are real coefficients and
\begin{equation}
P_{L}=\frac{1}{2}(1-\gamma_{5})\ ,\ P_{R}=\frac{1}{2}(1+\gamma_{5})\quad .
\end{equation}
At the classical level the above symmetries lead to conservation laws for the isovector, vector and axial currents
\begin{eqnarray}
j_{\alpha}^{\mu}(x)_{R}&=&\overline{\psi}_{a}(x)\gamma^{\mu}P_{R}(T_{\alpha})_{ab}\psi_{b}(x)\quad ,
\label{can}\\
j_{\alpha}^{\mu}(x)_{L}&=&\overline{\psi}_{a}(x)\gamma^{\mu}P_{L}(T_{\alpha})_{ab}\psi_{b}(x)\quad ,
\label{cbn}\\
j^{\mu}(x)&=&\overline{\psi}_{a}(x)\gamma^{\mu}{\bf 1}_{ab}\psi_{b}(x)\quad ,
\label{ccn}\\
j^{\mu}_{5}(x)&=&\overline{\psi}_{a}(x)\gamma^{\mu}\gamma ^{5}{\bf 1}_{ab}\psi_{b}(x)\quad .
\end{eqnarray}
At the quantum level the vector and axial currents cannot be simultaneously conserved. If the regularization is gauge invariant, so that 
the vector current is conserved, then the axial current acquires the anomaly which breaks the symmetry $U_{A}(1)$~\cite{b74} 
\beq
\partial_{\mu} j_5^{\mu}(x)={\cal N}\frac{e_{c}^{2}}{2\pi}\epsilon_{\mu \nu}F^{\mu \nu}(x)\quad .
\label{anomalyn}
\eeq
The isoscalar $<\overline{\psi}\psi>$ and isovector 
$<\overline{\psi}T^{\alpha}\psi>$ chiral condensates are zero due to the Coleman theorem~\cite{b39}, in fact they would break not only the $U_A(1)$ symmetry 
of the action but also the continuum internal symmetry $SU_L({\cal N})\otimes SU_R({\cal N})$ down to $SU_V({\cal N})$. 
There is an order parameter just for the breaking of the $U_A(1)$ symmetry~\cite{b57,b67}, the operator
\begin{equation} 
<\overline{\psi}_{L}^{({\cal N})}\ldots \overline{\psi}_{L}^{(1)}\psi_{R}^{(1)}\ldots 
\psi_{R}^{({\cal N})}>=(\frac{e^{\gamma}}{4\pi})^{\cal N} (\sqrt{\frac{\cal N}{\pi}} e_{c})^{\cal N}\quad .
\label{chicu}
\end{equation}
Under a discrete chiral rotation
\begin{equation}
\psi_{L}\rightarrow \gamma_{5} \psi_{L}=-\psi_{L}\quad ,\quad \psi_{R}\rightarrow \gamma_{5} \psi_{R}=\psi_{R}
\label{dga5}
\end{equation}
the operator (\ref{chicu}) of course transforms as
\begin{equation}
 <\overline{\psi}_{L}^{({\cal N})}\ldots \overline{\psi}_{L}^{(1)}\psi_{R}^{(1)}\ldots 
\psi_{R}^{({\cal N})}>\rightarrow (-1)^{{\cal N}} <\overline{\psi}_{L}^{({\cal N})}\ldots \overline{\psi}_{L}^{(1)}\psi_{R}^{(1)}\ldots 
\psi_{R}^{({\cal N})}>
\end{equation}
The umklapp operator is even under (\ref{dga5}) when ${\cal N}$ is even and this implies that notwithstanding the fact that the continuous chiral 
rotations $U_{A}(1)$ are broken by the non-zero VEV (\ref{chicu}), the discrete chiral symmetry (\ref{dga5}) is unbroken. 
When ${\cal N}$ is odd also the discrete chiral symmetry (\ref{dga5}) is broken by the non-zero VEV (\ref{chicu}). 

The usual abelian bosonization procedure may again be applied provided that ${\cal N}$ Bose fields are introduced~\cite{b74,b75,b76}
\beq
:\overline{\psi}_{a}\gamma^{\mu}\psi_{a}:=\frac{1}{\sqrt{\pi}}\epsilon^{\mu \nu}\partial_{\nu}\Phi_{a}\  ,\  a=1,\ldots {\cal N}
\eeq
The electric charge density and the action read
\beq
j_{0}=:\sum_{a=1}^{\cal N}\psi^{\dagger}_{a}\psi_{a}:=\frac{1}{\sqrt{\pi}}\partial_{x}(\sum_{a=1}^{\cal N}\Phi_{a})\quad ,
\label{chden}
\eeq
\beq
S=\int d^{2}x (\frac{1}{2}\sum_{a=1}^{\cal N}\partial_{\mu}\Phi_{a}\partial^{\mu}\Phi_{a}+
\frac{e_{c}^{2}}{2\pi}(\sum_{a=1}^{\cal N}\Phi_{a})^{2})
\label{na2n}
\eeq
The mass matrix is determined by the last term in Eq.(\ref{na2n}) and must be diagonalized. 
The field degrees of freedom span the vector space on which the mass matrix is defined. 
The action must be expressed in terms of an orthonormal basis of field vectors, in order to have a properly normalised kinetic energy term. 
The original $\Phi^{a}$ in Eq.(\ref{na2n})  are orthonormal basis vectors, but they are not eigenvectors of the mass matrix. The mass matrix has one 
non-zero eigenvalue $\frac{e_{c}^{2}}{{\cal N}\pi}$  with associated eigenvector $\frac{1}{\sqrt{\cal N}}\sum_{a=1}^{\cal N}\Phi^{a}$ 
and all the other eigenvalues are zero. The remaining eigenvectors can be made orthonormal by the following change of variables
\begin{equation}
\tilde{\Phi}^a=O^{a}_{b}\Phi^{b}
\end{equation}
where the orthogonal matrices $O^{a}_{b}$ are~\cite{b76}
\begin{eqnarray}
O_{b}^{1}&=&\frac{1}{\sqrt{\cal N}}(1,1,\ldots,1)\quad ,\\
O_{b}^{2}&=&\frac{1}{\sqrt{{\cal N}({\cal N}-1)}}(1,1,\ldots, -{\cal N}+1)\quad ,\\
O_{b}^{3}&=&\frac{1}{\sqrt{({\cal N}-1)({\cal N}-2)}}(1,1,\ldots,-{\cal N}+2,0)\quad ,\\
&\vdots&\nonumber\\
O_{b}^{\cal N}&=&\frac{1}{\sqrt{2}}(1,-1,0,\ldots,0)\quad .
\end{eqnarray}
In terms of these new fields $\tilde{\Phi}^{a}$ the action (\ref{na2n}) reads
\begin{equation}
S=\int d^2x(\frac{1}{2}\sum_{a=1}^{\cal N} \partial_{\mu}\tilde{\Phi}_{a}\partial^{\mu}\tilde{\Phi}^{a}-\frac{1}{2}\mu^2 (\tilde{\Phi}^{1})^2)
\label{na3n}
\end{equation}
where $\mu^{2}={\cal N}\frac{e_{c}^2}{\pi}$. 
The action (\ref{na3n}) describes ${\cal N}$ non interacting fields, one massive and ${\cal N}-1$ massless. 
The multiflavor Schwinger model can also be studied in the framework of non abelian bosonization~\cite{b68}, where the relationship 
between isovector currents and bosonic excitations appears in a more symmetrical form~\cite{b68tris}.  

The Hamiltonian, gauge constraint and non-vanishing (anti-)commutators
of the continuum ${\cal N}$-flavor Schwinger models are
\begin{eqnarray}
H=\int dx[\frac{e^2}{2}E^2(x)+\sum_{a=1}^{\cal N}&
\psi^{\dagger}_a (x)\alpha\left(i\partial_x +eA(x)\right)\psi_a
(x)]\label{ham1n}\\ \partial_x E(x)\ +&\sum_{a=1}^{\cal N} \psi^{\dagger}_a
(x)\psi_a (x)\sim 0\label{ga1n}\\
\left[ A(x),E(y)\right]=i\delta(x-y) ~,&
 \left\{\psi_a(x),\psi_b^{\dagger}(y)\right\}=\delta_{ab}\delta(x-y)
 \label{commu1n}
\end{eqnarray}

\section{The lattice ${\cal N}$-flavor Schwinger models}

On the lattice the Hamiltonian, constraint and (anti-) commutators reducing to (\ref{ham1n},\ref{ga1n},\ref{commu1n}) 
in the naive continuum limit are 
\begin{eqnarray}
H_{S}=\frac{e_{L}^{2}a}{2}\sum_{x=1}^N E_{x}^{2}&-&\frac{it}{2a}\sum_{x=1}^N
\sum_{a=1}^{\cal N} (\psi_{a,x+1}^{\dag}e^{iA_{x}}\psi_{a,x}
-\psi_{a,x}^{\dag}e^{-iA_{x}}\psi_{a,x+1})\label{hamiltonn}\nonumber\\
E_{x}-E_{x-1}&+&\sum_{a=1}^{\cal N}\psi_{a,x}^{\dag}\psi_{a,x}-\frac{\cal N}{2}\sim
0\ ,
\label{gaussn}\\
\left[ A_x,E_y\right]=i\delta_{x,y}~&,&
\left\{\psi_{a,x},\psi_{b,y}^{\dagger}\right\}=\delta_{ab}\delta_{xy}
\nonumber
\end{eqnarray}
The fermion fields are defined on the sites, $x=1,\ldots, N$, gauge and the electric fields, $ A_{ x}$ and
 $E_{x}$,  on the links $[x; x + 1]$, $N$ is an even integer 
and, when $N$ is finite it is convenient to impose periodic boundary conditions.  When $N$ is finite, the continuum limit is the 
${\cal N}$-flavor Schwinger model on a circle~\cite{b62}.
The coefficient $t$ of the hopping term in (\ref{hamiltonn})
plays the role of the lattice light speed. In the naive continuum limit,
$e_L=e_c$ and $t=1$. 
 
The lattice ${\cal N}$-flavor Schwinger model is equivalent to a one 
dimensional quantum Coulomb gas on the lattice with ${\cal N}$ kinds of particles. To see this one can fix the 
gauge, $A_{x} = A$ (Coulomb gauge). Eliminating the non-constant electric 
field and using the gauge constraint, one obtains the effective Hamiltonian
\begin{eqnarray}
H_{S}&=&H_u+H_p
\equiv\left[\frac{e^{2}_{L}}{2 N}E^{2}+\frac{e^{2}_{L}a}{2}
\sum_{x,y}\rho(x) V(x-y)\rho(y)\right]+\nonumber\\
&+&\left[
-\frac{it}{2a}\sum_{x}\sum_{a=1}^{\cal N}(\psi_{a,x+1}^{\dag}e^{iA}\psi_{a,x}-\psi_{a,x}^{\dag}e^{-iA}
\psi_{a,x+1})\right]\ ,
\label{hsnn}
\end{eqnarray}
where $\rho(x)$ is given in Eq.(\ref{ron}) and the Coulomb potential $V(x-y)$ is given in Eq.(\ref{cpo}). 
The constant electric field is normalized so that $[ A, E ] = i$ . 
The constant 
modes of the gauge field decouple in the thermodynamic limit 
$ N \rightarrow \infty$.
In the thermodynamic limit the Schwinger Hamiltonian (\ref{hsnn}), rescaled by the
 factor ${e_{L}^{2}a}/{2}$, reads 
 \begin{equation}
H=H_{0}+\epsilon H_{h}
\end{equation}
with
\begin{eqnarray}
H_{0}&=&\sum_{x>y}[\frac{(x-y)^{2}}{N}-(x-y)]\rho(x)\rho(y)\quad ,
\label{hunn}\\
H_{h}&=&-i(R-L)
\label{hpn}
\end{eqnarray}
and $\epsilon=t/e_{L}^{2}a^{2}$.
In Eq.(\ref{hpn}) the right $R$ and left $L$ hopping operators are defined ($L=R^{\dagger}$) as
\begin{equation}
R=\sum_{x=1}^{N}R_{x}=\sum_{x=1}^{N}\sum_{a=1}^{\cal N}R_{x}^{(a)}= 
\sum_{x=1}^{N} \sum_{a=1}^{\cal N} \psi_{a,x+1}^{\dag}e^{iA}\psi_{a,x}\quad .
\end{equation}
On a periodic chain the commutation relation 
\begin{equation}
[R,L]=0\quad 
\end{equation}
is satisfied.

\begin{figure}[htb]
\begin{center}
\setlength{\unitlength}{0.00041700in}%
\begingroup\makeatletter\ifx\SetFigFont\undefined
\def\x#1#2#3#4#5#6#7\relax{\def\x{#1#2#3#4#5#6}}%
\expandafter\x\fmtname xxxxxx\relax \def\y{splain}%
\ifx\x\y   
\gdef\SetFigFont#1#2#3{%
  \ifnum #1<17\tiny\else \ifnum #1<20\small\else
  \ifnum #1<24\normalsize\else \ifnum #1<29\large\else
  \ifnum #1<34\Large\else \ifnum #1<41\LARGE\else
     \huge\fi\fi\fi\fi\fi\fi
  \csname #3\endcsname}%
\else
\gdef\SetFigFont#1#2#3{\begingroup
  \count@#1\relax \ifnum 25<\count@\count@25\fi
  \def\x{\endgroup\@setsize\SetFigFont{#2pt}}%
  \expandafter\x
    \csname \romannumeral\the\count@ pt\expandafter\endcsname
    \csname @\romannumeral\the\count@ pt\endcsname
  \csname #3\endcsname}%
\fi
\fi\endgroup
\begin{picture}(2424,7224)(4789,-6973)
\thicklines
\put(4801,-961){\line( 1, 0){1200}}
\put(4801,-2161){\line( 1, 0){1200}}
\put(4801,-3361){\line( 1, 0){1200}}
\put(4801,-4561){\line( 1, 0){1200}}
\put(4801,-5761){\line( 1, 0){1200}}
\put(4801,239){\line( 1, 0){1200}}
\put(4801,-6961){\line( 1, 0){1200}}
\put(7051,-3361){\makebox(0,0)[lb]{\smash{\SetFigFont{10}{12.0}{rm}$\frac{\cal N}{2}$}}}
\put(4801,239){\line( 0,-1){7200}}
\put(6001,239){\line( 0,-1){7200}}
\put(7201,-2761){\vector( 0, 1){3000}}
\put(7201,-3961){\vector( 0,-1){3000}}
\end{picture}
\end{center}
\caption{The representation of $SU({\cal N})$ at each site when ${\cal N}$ is even}
\label{ne}
\end{figure}
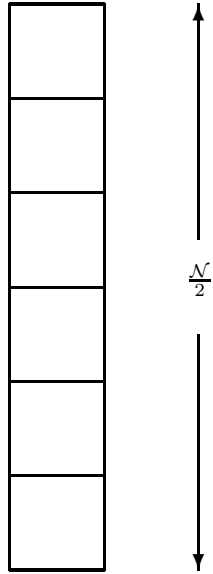

When ${\cal N}$ is even the ground state of the Hamiltonian (\ref{hunn}) is the state $|g.s.>$ with $\rho(x)=0$ on every site, $i.e.$
 with every site half-filled
\begin{equation}
\prod_{a=1}^{\cal N} \psi_{ax}^{\dagger}\psi_{ax}|g.s.>=\frac{\cal N}{2}|g.s.>\quad .
\end{equation}
It is easy to understand that $\rho(x)=0$ on every site in the ground state by observing that the Coulomb Hamiltonian 
(\ref{hunn}) is a non-negative operator and that the states with zero charge density are zero eigenvalues of (\ref{hunn}).  
$|g.s.>$ is an highly degenerate state; in fact at each site $x$ the quantum configuration is 
\begin{equation}
\prod_{a=1}^{\frac{\cal N}{2}}\psi_{ax}^{\dagger}|0>\quad .
\label{qcnp}
\end{equation}
The state (\ref{qcnp}) is antisymmetric in the indices $a=1,\ldots,\frac{\cal N}{2}$; $i.e.$ it takes on any 
orientation of the vector in the representation of the flavor symmetry group $SU({\cal N})$ with Young tableau given in fig.(\ref{ne}).  
The energy of $|g.s.>$ is of order 1, since it is non zero only at the second order in the strong coupling 
expansion.

\begin{figure}[htb]
\begin{center}
\setlength{\unitlength}{0.00041700in}%
\begingroup\makeatletter\ifx\SetFigFont\undefined
\def\x#1#2#3#4#5#6#7\relax{\def\x{#1#2#3#4#5#6}}%
\expandafter\x\fmtname xxxxxx\relax \def\y{splain}%
\ifx\x\y   
\gdef\SetFigFont#1#2#3{%
  \ifnum #1<17\tiny\else \ifnum #1<20\small\else
  \ifnum #1<24\normalsize\else \ifnum #1<29\large\else
  \ifnum #1<34\Large\else \ifnum #1<41\LARGE\else
     \huge\fi\fi\fi\fi\fi\fi
  \csname #3\endcsname}%
\else
\gdef\SetFigFont#1#2#3{\begingroup
  \count@#1\relax \ifnum 25<\count@\count@25\fi
  \def\x{\endgroup\@setsize\SetFigFont{#2pt}}%
  \expandafter\x
    \csname \romannumeral\the\count@ pt\expandafter\endcsname
    \csname @\romannumeral\the\count@ pt\endcsname
  \csname #3\endcsname}%
\fi
\fi\endgroup
\begin{picture}(6024,6024)(3589,-5773)
\thicklines
\put(3601,-961){\line( 1, 0){1200}}
\put(3601,-2161){\line( 1, 0){1200}}
\put(3601,-3361){\line( 1, 0){1200}}
\put(3601,-4561){\line( 1, 0){1200}}
\put(3601,-5761){\line( 1, 0){1200}}
\put(3601,239){\line( 0,-1){6000}}
\put(4801,239){\line( 0,-1){6000}}
\put(6001,-2161){\vector( 0, 1){2400}}
\put(6001,-3361){\vector( 0,-1){2400}}
\put(3601,239){\line( 1, 0){1200}}
\put(7201,239){\line( 1, 0){1200}}
\put(9301,-2161){\makebox(0,0)[lb]{\smash{\SetFigFont{10}{12.0}{rm}$\frac{{\cal N}-1}{2}$}}}
\put(7201,-961){\line( 1, 0){1200}}
\put(7201,-2161){\line( 1, 0){1200}}
\put(7201,-3361){\line( 1, 0){1200}}
\put(7201,-4561){\line( 1, 0){1200}}
\put(7201,239){\line( 0,-1){4800}}
\put(8401,239){\line( 0,-1){4800}}
\put(9601,-1561){\vector( 0, 1){1800}}
\put(9601,-2761){\vector( 0,-1){1800}}
\put(5701,-2761){\makebox(0,0)[lb]{\smash{\SetFigFont{10}{12.0}{rm}$\frac{{\cal N}+1}{2}$}}}
\end{picture}
\end{center}
\caption{The representation of $SU({\cal N})$ at each site of the even sublattice and odd sublattice when ${\cal N}$ is odd}
\label{no}
\end{figure}
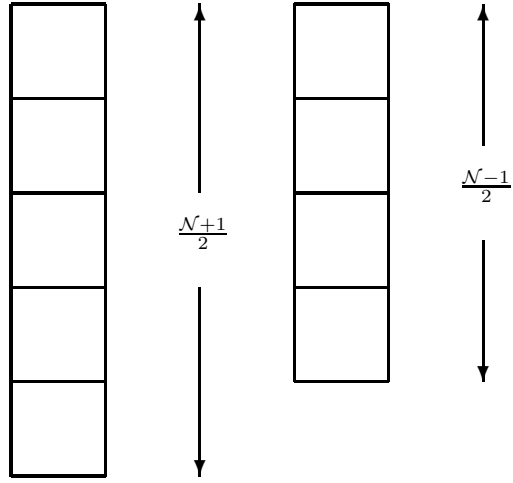

When ${\cal N}$ is odd the ground states of the Hamiltonian (\ref{hunn}) are characterized by the staggered charge distribution
\begin{equation}
\rho(x)=\pm \frac{1}{2}(-1)^{x}
\label{gsron}
\end{equation}
since (\ref{gsron}) minimizes the Coulomb Hamiltonian (\ref{hunn}); one can have $\rho(x)=+1/2$ on the even sublattice and $\rho(x)=-1/2$ on the odd sublattce 
or viceversa. 
The electric fields generated by the charge distribution (\ref{gsron}) are
 \begin{equation}
E_{x}=\pm \frac{1}{4}  (-1)^{x}
\label{exn}
\end{equation}    
Since now
\begin{equation}
H_{0}|g.s.>=\frac{1}{16}|g.s.>
\end{equation}
the ground state energy is of order $e_{L}^{2}$. 
The states $|g.s.>$ are highly degenerate since they can take up any orientation in the vector space which carries the representation of the 
$SU({\cal N})$ group with the Young tableaux given in fig.(\ref{no}).   

Either when ${\cal N}$ is even or when ${\cal N}$ is odd the ground state degeneracy is resolved at the second order in the strong coupling 
expansion. First order perturbations to the vacuum energy vanish. The vacuum energy at order $\epsilon ^{2}$ reads
\begin{equation}
E_{0}^{(2)}=<H_{h}^{\dagger}\frac{\Pi}{E_{0}^{(0)}-H_{0}}H_{h}>
\label{2vev}
\end{equation}
where the expectation values are defined on the degenerate subspace of ground states and $\Pi$ is a projection operator projecting 
orthogonal to the states of the degenerate subspace. 
Due to the commutation relation
\begin{equation}
\left[H_{0},H_{h}\right]=\frac{N-1}{N}H_{h}-2\sum_{x,y}\left[V(x-y)-V(x-y-1)\right](L_{y}+R_{y})\rho(x)
\end{equation}
Eq.(\ref{2vev}) can be rewritten as
\begin{equation}
E_{0}^{(2)}=-2<RL>\quad .
\label{22vev}
\end{equation}
On the ground state the combination $RL$ can be written in terms of the Heisenberg Hamiltonian  of a generalized $SU({\cal N})$ 
antiferromagnet. By introducing as in chapter 5 the Schwinger spin operators
\begin{equation}
\vec{S}_{x}=\psi_{ax}^{\dagger}T_{ab}^{\alpha}\psi_{bx}
\end{equation}
where $T^{\alpha}$ are now the generators of the $SU({\cal N})$ group,  the $SU({\cal N})$ Heisenberg Hamiltonian reads
\begin{equation}
H_{J}=\sum_{x=1}^{N}(~\vec{S}_{x}\cdot \vec{S}_{x+1}-\frac{\cal N}{8}+\frac{1}{2\cal N}\rho(x)\rho(x+1)~)=-\frac{1}{2}\sum_{x=1}^{N}L_{x}R_{x}
\label{nhj}
\end{equation}
When ${\cal N}$ is even, on the degenerate ground states one has
\begin{equation}
<H_{J}>=<\sum_{x=1}^{N}(\vec{S}_{x}\cdot \vec{S}_{x+1}-\frac{\cal N}{4})>=<-\frac{1}{2}\sum_{x=1}^{N}L_{x}R_{x}>
\label{nhje}
\end{equation}
while when ${\cal N}$ is odd one has
\begin{equation}
<H_{J}>=<\sum_{x=1}^{N}(\vec{S}_{x}\cdot \vec{S}_{x+1}-\frac{{\cal N}^{2}+1}{8\cal N})>=<-\frac{1}{2}\sum_{x=1}^{N}L_{x}R_{x}>\quad .
\label{nhjo}
\end{equation}
Taking into account that the products of $L_{x}$ and $R_{y}$ at different points have vanishing expectation values on the ground states and using 
Eq.(\ref{nhje}) or Eq.(\ref{nhjo}), Eq.(\ref{22vev}) reads
\begin{equation}
 E_{0}^{(2)}=4<H_{J}>\quad .
\end{equation}
The problem of determining the correct ground state, on which to perform the perturbative expansion, is then reduced again to the diagonalization of the 
$SU({\cal  N})$ Heisenberg spin-$1/2$ Hamiltonian  (\ref{nhj}). As I already pointed out in chapter 3, generalized $SU({\cal N})$ antiferromagnetic 
chains have not been yet analysed in the literature in such a detailed way as the $SU(2)$ chains. Consequentely, 
the study  of the lattice $SU({\cal N})$ flavor lattice Schwinger models become extremely complicated for ${\cal N}>2$. 
Nonetheless, the computational scheme that I developed for the $U(1)$- and $SU(2)$-flavor models in chapters 4 and 5, should 
work for a generic $SU({\cal N})$-flavor model. 

The ground state of the gauge models is very different when ${\cal N}$ is even or ${\cal N}$ is odd. When ${\cal N}$ is even, the ground state 
$|G.S.>$ of 
the spin Hamiltonian (\ref{nhj}) is non-degenerate and translationally invariant, and since it is the ground state of the gauge model in the infinite coupling 
limit, there is 
no spontaneous breaking of the chiral symmetry for any $SU(2{\cal N})$-flavor lattice Schwinger model. 
In contrast, when ${\cal N}$ is odd, the ground state $|G.S.>$ of 
the spin Hamiltonian (\ref{nhj}) is degenerate of order two and is not translationally invariant and consequently any $SU(2{\cal N}+1)$-flavor lattice Schwinger model 
exhibits spontaneous symmetry breaking of the discrete axial symmetry. By translating of one lattice spacing $|G.S.>$ one gets the other one. 
The ${\cal N}$-flavor lattice Schwinger models excitations are also generated from 
$|G.S.>$ by two very different mechanisms, that I already described for the two-flavor model in chapter 5. There are excitations 
involving only flavor changes of the fermions without changing the charge density $\rho(x)$ which corrispond to spin flips in the $SU(2)$ invariant 
model. These excitations are massless. At variance massive excitations involve fermion transport besides flavor changes and are created 
by applying to $|G.S.>$ the latticized currents of the Schwinger models which vary the on site value of $\rho(x)$.      

Very different is the case of the massive multiflavor Schwinger models~\cite{b77}. When ${\cal N}$ is odd, the presence of a non-zero fermionic mass $m$ 
removes the degeneracy and selects one of the two $|G.S.>$ as the non degenerate ground state. When ${\cal N}$ is even the ground state 
remains translationally invariant in the strong coupling limit $e_{L}^{2}\gg m$. In the weak coupling limit $m\gg e_{L}^{2}$ the 
discrete chiral symmetry is broken for every ${\cal N}$. 
\newpage
\addcontentsline{toc}{chapter}{Summary and conclusions}
\vspace*{90pt}
\textbf{\Huge Summary and conclusions}\par
\vspace{50pt}
\noindent 
In this thesis I showed explicitly that the strong coupling limit of the multiflavor lattice Schwinger models is effectively 
described by quantum spin-$1/2$ generalized $SU({\cal N})$ antiferromagnetic Heisenberg chains. Exploiting the mapping existing between 
these gauge and spin models, exact results about antiferromagnetic chains have been used to analyse the 
chiral symmetry breaking on the lattice. In particular an interesting question has been how the effects of the axial anomaly 
appear in the lattice regularization. 
Since~\cite{b24} fermion theories on a lattice have an equal number of species of left- and right-handed 
Weyl particles in the continuum limit, there is no axial anomaly in a lattice theory.  

I studied in detail  the $U(1)$-flavor and $SU(2)$-flavor models, archetypes for all $SU(2{\cal N}+1)$ and $SU(2{\cal N})$ models. 
The results are very different depending on if the number of flavors is even or odd. The $SU(2{\cal N})$-flavor Schwinger 
models exhibit non-degenerate and translationally invariant ground states in the strong coupling limit and their energy is of order 1. 
At variance, the $SU(2{\cal N}+1)$ models have two degenerate ground states, break the symmetry of translation by one lattice spacing 
and their energy is of order $e_{L}^{2}$. 
The difference between ${\cal N}$ even and odd arises, since, also on the lattice, the charge density operator must 
be odd under charge conjugation; therefore the constant ${\cal N}/2$ should be subtracted from the charge  density operator (\ref{ron}). 
As a consequence, when the ${\cal N}$ is odd, the ground state supports electric fluxes while this becomes impossible when ${\cal N}$ is even. 

In the one-flavor model I proved that the discrete axial symmetry is spontaneously broken on the lattice, reproducing the effects due to 
the axial anomaly in the continuum. In the infinite coupling limit the ground state of the gauge model is the ground state of an antiferromagnetic 
Ising chain with long range spin-spin Coulomb interactions. The Schwinger model excitations are created by acting on the ground state with the latticized 
currents. The excitation masses are determined in the strong coupling expansion by computing the excitation energies 
up to the fourth order in the strong coupling expansion. Using suitable Pad\'e approximants, I extrapolated to the continuum 
the lattice results and I found that the lattice answers are in very good agreement with the continuum theory. 

In the two-flavor Schwinger model, I demonstrated that in the strong coupling limit the gauge model is effectively described by the spin-$1/2$ 
antiferromagnetic Heisenberg Hamiltonian and the ground state of the gauge model in the infinite coupling limit is the ground 
state of the spin chain. 
Since the antiferromagnetic Heisenberg model is exactly solvable in one-dimension, it is very interesting to exhibit a gauge theory 
with the same low lying excitation spectrum. 

The antiferromagnetic spin chain admits spin-$1/2$ quarklike spinon excitations 
but has physical states with integer spin and an even number of spinons. Spinons behave like $U(1)$ quarks and they can be identified in conventional 
materials and models by high-energy spectroscopy and inconsistences in sum rules~\cite{b22}. The two-flavor Schwinger model strictly parallels 
this scenario. The fundamental particles are spin-$1/2$ fermions but the spectrum exhibits only integer spin bosonic excitations~\cite{b5}. I proved that the 
spinons have the same quantum numbers of the gauge model excitations. While spinons are the gapless excitations created from the spin 
chain ground state by spin flipping, the massive excitations of the gauge model are created from the spin chain ground state by fermion transport besides 
spin flipping and do not belong to the antiferromagnetic Heisenberg spectrum. The lattice mass spectrum, properly extrapolated to the continuum by using 
Pad\'e approximants, reproduces in a satisfactory way the continuum results. The pattern of chiral symmetry breaking of the continuum model is 
exactly reproduced on the lattice, even if the Coleman theorem does not apply and the anomalous symmetry breaking is impossible~\cite{b24}. 
In fact, both the isoscalar $<\overline{\psi}\psi>$ and the isovector $<\overline{\psi}\sigma^{a}\psi>$ chiral condensates are 
zero to every order in the strong coupling expansion, while the expectation value of the umklapp operator $F$ is different from zero. 
  
The multiflavor Schwinger models are the gauge theories associated to the quantum spin-$1/2$ antiferromagnetic Heisenberg chains. 
It would be interesting to extend my analysis to $QCD_{2}$ with different flavor and color groups. The number of colors ${\cal N}_{C}$ determines the 
representation of the antiferromagnetic chain, since the spin is $S={\cal N}_{C}/2$. Varying ${\cal N}_{C}$ one changes the spin $S$. 
In particular, by considering the $SU(2)$-flavor case, $i.e.$ $QCD_{2}$ with two kinds of quarks, one finds that in the strong coupling 
limit the gauge model is effectively described by the antiferromagnetic Heisenberg chain with spin-$S$ representation. Since 
half-integer spin chains are expected to exhibit gapless excitations, while integer spin chains have a gap~\cite{b43}, it would be an interesting problem to 
investigate if also $QCD_{2}$ exhibits a gapless or gapped spectrum depending on if ${\cal N}_{C}$ is odd or even. 
Two dimensional $QCD$ affords an excellent opportunity to study 
various dynamical questions of gauge theories, since many of its qualitative features are also valid in four dimensions. $QCD_{2}$ 
still resists analytic solutions for general $SU({\cal N}_{C})$ colour groups, except in the planar limit ${\cal N}_{C}\rightarrow \infty$~\cite{b78,b79}. 
In fact, `t Hooft~\cite{b78} analysed the $QCD_{2}$ spectrum in the continuum exploiting a perturbation expansion with respect to $1/{\cal N}_{C}$ and showed 
that the ${\cal N}_{C}\rightarrow \infty$ limit, keeping $g{\cal N}_{C}$ constant ($g$ is the strong interactions coupling constant), 
corresponds to taking only the planar Feynman diagrams with no fermion loops~\cite{b79}. The ${\cal N}_{C}\rightarrow \infty$ limit of two flavor 
$QCD_{2}$ should be compared with the $S\rightarrow \infty$ of the antiferromagnetic Heisenberg chain. The study of the spin chain in the large $S$ 
limit can be performed using the spin-wave theory and it would be fascinating to single out ``gluon" and ``hadron" degrees of freedom 
in the antiferromagnetic chain.
\newpage 
\addcontentsline{toc}{chapter}{\protect Pad\'e approximants}
\vspace*{90pt}
\textbf{\Huge Pad\'e approximants}\par
\vspace{50pt}
\noindent 
For the sake of clarity and to make the thesis self-contained, I define here what are Pad\'e approximants~\cite{b26,b80}. 
Although the method of Pad\'e approximant has a long history, dating back more than one century, it is only in the 1960s that the procedure 
has become a part of the ``bag of tricks" of the working physicist~\cite{b26}. 
When a power series representation of a function diverges, it indicates the presence of singularities. The divergence of the series 
reflects the inability of a polynomial to approximate a function adeguately near a singularity. The basic idea of summation theory is to 
represent $f(z)$, the function in question, by a convergent expression. In Euler summation this expression is the limit of 
a convergent series, while in Borel summation this expression is the limit of a convergent integral. 
The difficulty with Euler and Borel summation is that all of the terms of the divergent series must be known exactly before 
the ``sum" can be found even approximately. In realistic perturbation problems only a few terms of a perturbative series can be calculated 
as we have seen in chapters 4 and 5. Therefore, it is needed a summation algorithm, which requires as input only a finite number of terms
 of the divergent series. The Pad\'e summation is a method having such a property. 

The idea of Pad\'e summation is to replace a power series such as $\sum_{n=0}^{\infty}a_{n}z^{n}$ by a sequence of rational functions of the form
\begin{equation}
P_{M}^{N}(z)=\frac{\sum_{n=0}^{N}A_{n}z^{n}}{\sum_{n=0}^{M}B_{n}z^{n}}
\label{pad1}
\end{equation}
where one can choose $B_{0}=1$ without loss of generality. As it stands from Eq.(\ref{pad1}) $P_{M}^{N}(z)$ is the ratio of 
a polynomial of degree $N$ and one of degree $M$. The $M+N+1$ coefficients $A_{0},A_{1},\ldots,A_{n},B_{1},B_{2},\ldots,B_{M}$ are determined 
in such a way that the first $M+N+1$ terms in the Taylor series expansion of $P_{M}^{N}(z)$ match the first $M+N+1$ terms of the power series 
$\sum_{n=0}^{\infty}a_{n}z^{n}$. The rational functions $P_{M}^{N}(z)$ are called Pad\'e approximants. Constructing Pad\'e 
approximants $P_{M}^{N}(z)$ is very useful; in fact, if $\sum_{n=0}^{\infty}a_{n}z^{n}$ is a power series representation of a function $f(z)$, 
then in many instances $P_{M}^{N}(z)\rightarrow f(z)$ as $M,N\rightarrow \infty$, even if $\sum_{n=0}^{\infty}a_{n}z^{n}$ is a 
divergent series. The sequence of Pad\'e approximants (\ref{pad1}) with $M=N$ is called the diagonal sequence. Let us sketch 
the computation of $P_{1}^{0}(z)$. One has to expand $P_{1}^{0}(z)$ in a Taylor series
\begin{equation}   
P_{1}^{0}(z)=\frac{A_{0}}{1+B_{1}z}=A_{0}-A_{0}B_{1}z+O(z^{2})\quad (z\rightarrow 0)
\label{pad2}
\end{equation}
and comparing the series (\ref{pad2}) with the first two terms in the power series representation of $f(z)=\sum_{n=0}^{\infty}a_{n}z^{n}$ 
one gets two equations: $a_{0}=A_{0}$, $a_{1}=-A_{0}B_{1}$ from which
\begin{equation}
P_{1}^{0}=\frac{a_{0}}{1-\frac{a_{1}}{a_{0}}z}
\end{equation}
As it clear from this very simple example, to construct a Pad\'e approximant $P_{M}^{N}(z)$ one does not need the full power series representation 
of a function, but just the first $M+N+1$ terms. Since Pad\'e approximant involve only algebraic operations, they are more convenient for computational 
purposes than Borel summation, which requires one to integrate on an infinite range the analytic continuation of a 
function defined by a power series. Pad\'e approximants often work quite well, even beyond their proven range of applicability. 
For an analysis of the convergence theory of Pad\'e approximants, see Ref.~\cite{b80}.

\listoffigures
\listoftables


\newpage 
\addcontentsline{toc}{chapter}{Bibliography}
\begin{thebibliography}{99}
\bibitem{b0} See for instance P. W. Anderson, {\it Basic Notions of Condensed Matter Physics}, Frontiers in Physics, Addison Wesley Pub. Company 1983; 
C. Itzykson and J.-B. Zuber, {\it  Quantum Field Theory}, McGraw-Hill, Inc. 1985, and references therein. 
\bibitem{b1} K. G. Wilson, Phys. Rev. {\bf D2}, 1438 (1970); Phys. Rev. {\bf B4}, 3174 (1971); Phys. Rev. Lett. {\bf 28}, 548 (1972); 
K. G. Wilson and M. E. Fisher, Phys. Rev. Lett. {\bf 28}, 240 (1972); K. G. Wilson and J. Kogut, Phys. Rep. {\bf 12 C}, 75 (1974).
\bibitem{b2} K. G. Wilson, Phys. Rev. {\bf D10}, 2445 (1974).
\bibitem{b3}L. D. Landau and E. M. Lifshitz, {\it Statistical Physics}, Pergamon Press, Oxford, 1969; 
A. Z. Patashinskii and V. L. Pokrovskii, {\it Fluctuation Theory of Phase Transitions}, Pergamon Press, Oxford, 1979.
\bibitem{b4} G. `t Hooft, unpublished (1972); H. D. Politzer, Phys. Rev. Lett. {\bf 30}, 1346 (1973); D. J. Gross and F. Wilczek, Phys. Rev. Lett. 
{\bf 30}, 1343 (1973); Phys. Rev. {\bf D8}, 3633 (1973).
\bibitem{b6} See for instance B. W. Lee, {\it Chyral Dynamics}, Gordon and Breach Science Pub. 1972.
\bibitem{b7b}P. Fomin, V. Gusynin, V. Miransky and Yu. Sitenko, Riv. Nuov. Cim. {\bf 6},  1 (1983); V. Miransky, Nuovo Cimento {\bf 90A}, 149 (1985).
\bibitem{b7}See for instance J. B. Kogut, Rev. Mod. Phys. {\bf 55}, 775 (1983) and references therein.          
\bibitem{b8} K. G. Wilson,{\it ``New Phenomena in Subnuclear Physics"}, Erice, ed. A. Zichichi (New York, Plenum, 1975).
\bibitem{b9} T. Banks, J. Kogut and L. Susskind, Phys. Rev. {\bf D13}, 1043 (1976); 
L. Susskind, Phys. Rev. {\bf D16}, 3031 (1977).
\bibitem{b10} T. Banks, S. Raby, L. Susskind, J. Kogut, D. R. T. Jones, P. N. Scharbach and D. K. Sinclair, Phys. Rev. {\bf D15}, 1111 (1977). 
\bibitem{b11} C. Borgs, Comm. Math. Phys. {\bf 116}, 343 (1988); E. Seiler, {\it Gauge theories as a problem of constructive quantum field theory 
and statistical mechanics}, Lecture Notes in Physics Vol. {\bf 159}, Berlin, Heidelberg, New York: Springer 1982.
\bibitem{b13} J. Kogut and L. Susskind, Phys. Rev. {\bf D11}, 395 (1975). 
\bibitem{b14} J. Smit, Nucl. Phys. {\bf B175}, 307 (1980).
\bibitem{b15} See for a review E. Fradkin, {\it Field Theories and Condensed Matter Systems}, Addison Wesley Pub. Company, 1991.
\bibitem{b16} I. Affleck and J. B. Marston, Phys. Rev. {\bf B37}, 3773 (1988); J. B. Marston, Phys. Rev. Lett. {\bf 61}, 1914 (1988). 
\bibitem{b17}I. Affleck, Z. Zou, T. Hsu and P. W. Anderson, Phys. Rev. {\bf B38}, 745 (1988). 
\bibitem{b17a}P. Wiegmann, Prog. Theor. Phys. Suppl. {\bf 107}, 243 (1992). 
\bibitem{b17b}C. Mudry and E. Fradkin, Phys. Rev. {\bf B49}, 5200 (1994); Phys. Rev. {\bf B50}, 11409 (1994). 
\bibitem{b17c}D. H. Kim and P. A. Lee, cond-mat/9810130.
\bibitem{b18} D. Hofstaeder, Phys. Rev. {\bf B14}, 2239 (1976).
\bibitem{b19} G. W. Semenoff and L. C. Wijewardhana, Phys. Rev. {\bf D45}, 1342 (1992). 
\bibitem{b20} D. Schmeltzer and A. R. Bishop, Phys. Rev. {\bf B41}, 9603 (1990); M. C. Diamantini and P. Sodano, Phys. Rev. {\bf B45}, 5737 (1992). 
\bibitem{b21} G. W. Semenoff, Mod. Phys. Lett. {\bf A7}, 2811 (1992); E. Langmann and G. W. Semenoff, Phys. Lett. {\bf B297}, 175 (1992); 
M. C. Diamantini, E. Langmann, G. W. Semenoff and P. Sodano, Nucl. Phys. {\bf B405}, 595 (1993). 
\bibitem{b22} R. B. Laughlin, cond-mat/9802180; Phys. Rev. Lett. {\bf 79}, 1726 (1997).
\bibitem{b23} S. L. Adler and W. A. Bardeen, Phys. Rev. {\bf 182}, 1517 (1969); 
J. S. Bell and R. Jackiw, Nuovo Cimento {\bf A60}, 47 (1969). 
\bibitem{b24} H. B. Nielsen and M. Ninomiya, Nucl. Phys. {\bf B185}, 20 (1981); {\bf B193}, 173 (1981); Phys. Lett. {\bf B105}, 219 (1981); {\bf B130}, 
389 (1983).
\bibitem{b5} J. Schwinger, Phys. Rev. {\bf 125}, 397 (1962); Phys. Rev. {\bf 128}, 2425 (1962).     
\bibitem{b26} G. A. Baker, Jr., {\it Essential of Pad\'e approximants} (Academic, New York, 1975).
\bibitem{b27}L. D. Faddeev and L. A. Takhtadzhyan, Phys. Lett. {\bf A85},  375 (1981); 
L. D. Faddeev and L. A. Takhtadzhyan, Zapiski Nauchnych Seminarov LOMI, {\bf 109}, 134 (1981), english translation in J. Sov. Math. {\bf 24}, 241 (1984).
\bibitem{b28}J. B. Kogut, Rev. Mod. Phys. {\bf 51}, 659 (1979); M. Creutz, {\it  Quarks , Gluons and Lattices}, Cambridge University press, 1983; 
I. Montvay and G. M\"{u}nster, {\it Quantum Fields on a Lattice}, Cambridge University press, 1994; H. J. Rothe, {\it  Lattice Gauge Theories: An Introduction}, 
World Scientific, 1997; R. Gupta, {\it  Introduction to Lattice QCD}, Les Houches Lectures 1998, hep-lat/9807028; G. Grignani and G. W. Semenoff, 
{\it Introduction to some common topics in gauge theory and spin systems}, in {\it Field Theories for Low Dimensional Condensed Matter Systems: 
Spin Systems and Strongly Correlated Electrons}, ed. by 
R. B. Laughlin, G. Morandi, P. Sodano, A. Tagliacozzo, V. Tognetti, Spinger Verlag in press.   
\bibitem{b29} F. Wegner, J. Math. Phys. {\bf 12}, 2259 (1971).
\bibitem{b30} H. Kluberg-Stern, A. Morel, O. Napoly and R. Petersson, Nucl. Phys. {\bf B220}, 447 (1983).
\bibitem{b31} E. H. Lieb and D. C. Mattis, {\it Mathematical Physics in one dimension}, New York Academic Press (1961); D. C. Mattis, 
{\it The Theory of Magnetism}, Harper \& Row 1965; W. J. Caspers, {\it Spin Systems}, World Scientific 1989; I. Affleck, 
J. Phys. Cond. Mat. {\bf 1}, 3047 (1989); I. Affleck, {\it Field Theory Methods and Quantum Critical Phenomena}, in {\it Fields, Strings and 
Critical Phenomena}, ed. by E. Brezin and J. Zinn-Justin, North Holland (1989); D. C. Mattis, {\it The Many-Body Problem}, World Scientific 1993; 
V. E. Korepin, N. M. Bogoliubov and A. G. Izergin, {\it Quantum Inverse Scattering Method and Correlation Functions}, Cambridge University press 1993; 
A. Auerbach, Interacting Electrons and Quantum Magnetism, Springer Verlag 1994; A. M. Tsvelik, {\it Quantum Field Theory in Condensed Matter Physics}, 
Cambridge University press 1995; L. D. Faddeev, Int. J. Mod. Phys. {\bf A10}, 1845 (1995), hep-th/9605187; 
R. B. Laughlin, D. Giuliano, R. Caracciolo and Olivia L. White, {\it Quantum Number Fractionalization in 
Antiferromagnets}, in {\it Field Theories for Low Dimensional Condensed Matter Systems: Spin Systems and Strongly Correlated Electrons}, ed. by 
R. B. Laughlin, G. Morandi, P. Sodano, A. Tagliacozzo, V. Tognetti, Spinger Verlag in press;  
A. Auerbach, F. Berruto and L. Capriotti, {\it Quantum Magnetism Approach to Strongly Correlated Electrons}, 
in {\it Field Theories for Low Dimensional Condensed Matter Systems: Spin Systems and Strongly Correlated Electrons}, ed. by 
R. B. Laughlin, G. Morandi, P. Sodano, A. Tagliacozzo, V. Tognetti, Spinger Verlag in press.
\bibitem{b32} P. W. Anderson, Phys. Rev. {\bf B86}, 694 (1952); R. Kubo, Phys. Rev. {\bf 87}, 568 (1952).
\bibitem{b33} H.Bethe, Z. Physik {\bf 71}, 205 (1931).
\bibitem{b34} L. Hulth\'en, Arkiv. Mat. Astron. Fysik {\bf 26A} (11), 1 (1938).
\bibitem{b35}R. B. Laughlin, unpublished (1995).
\bibitem{b36}F. Berruto, G. Grignani, G. W. Semenoff and P. Sodano, Phys. Rev. {\bf D59}, 034504 (1999).  
\bibitem{b36b}F. Berruto, G. Grignani, G. W. Semenoff and P. Sodano, hep-th/9901142, submitted to Annals of Physics.
\bibitem{b37}F. D. M. Haldane, Phys. Rev. Lett. {\bf 60}, 635  (1988); 
ibid. {\bf 66}, 1529 (1991); B. S. Shastry, Phys. Rev. Lett. {\bf 60}, 639  (1988).
\bibitem{b38} M. Takahashi, J. Phys. {\bf C10}, 1289 (1977); cond-mat/9708087.
\bibitem{b39}N. D. Mermin, H. Wagner, Phys. Rev. Lett. {\bf 22}, 1133 (1966); 
S. Coleman, Commun. Math. Phys. {\bf 31}, 259 (1973).
\bibitem{b40} F. Calogero, J. Math. Phys. {\bf 10}, 2197 (1969); 
B. Sutherland, Phys. Rev. {\bf A4}, 2019 (1971); ibid. {\bf 5}, 1537 (1972).
\bibitem{b41} See for example J. C. Talstra, {\it Integrability and Applications of the Exactly-Solvable Haldane-Shastry Quantum Spin Chain}, 
Ph. D. Thesis, Princeton University, cond-mat/9509178 and references therein.
\bibitem{b42} V. Kalmeyer and R. B. Laughlin, Phys. Rev. Lett. {\bf 59}, 2095 (1987); Phys. Rev. {\bf B39}, 163 (1989).
\bibitem{b42bis}R. B. Laughlin, Ann. Phys. (N. Y.) {\bf 159}, 220 (1985).
\bibitem{b43} F. D. M. Haldane, Phys. Rev. Lett. {\bf 50}, 1153 (1983).
\bibitem{b44}See for example {\it The Hubbard Model}, ed. by A. Montorsi, World Scientific 1992.
\bibitem{b46} S. Coleman, Ann. Phys. (N.Y.) {\bf 101}, 239 (1976).
\bibitem{b45}J. des Cloizeaux and J. J. Pearson, Phys. Rev. {\bf 128}, 2131 (1962).
\bibitem{b47} V. E. Korepin, A. G. Izergin, F. H. L. Essler and D. B. Uglov, Phys. Lett. {\bf A190}, 182 (1994).
\bibitem{b48} S. Lukyanov, cond-mat/9712314.
\bibitem{b49}I. Affleck, J. Phys. {\bf A31}, 4573 (1998).  
\bibitem{b50} S. V. Tyablikov, Ukrain. Math. Zh. {\bf 11}, 287 (1959); 
D. N. Zubarev, Soviet Physics-Uspekhi {\bf 3}, 320 (1960).
\bibitem{b51} H. Q. Lin and D. K. Campbell, J. Appl. Phys. {\bf 69}, 5947 (1991).
\bibitem{b52} B. Sutherland, Phys. Rev. {\bf B12}, 3795 (1975); P. P. Kulish and Yu. Reshetikhin, Sov. Phys. JETP {\bf 53}, 108 (1981); 
A. Doikou and R. I. Nepomechie, hep-th/9803118.
\bibitem{b53} I. Affleck, Phys. Rev. Lett. {\bf 54}, 966 (1985); 
N. Read and S. Sachdev, Phys. Rev. Lett. {\bf 62}, 1694 (1989);
N. Read and S. Sachdev, Nucl. Phys. {\bf B316}, 609 (1989); 
N. Read and S. Sachdev, Phys. Rev. {\bf B42}, 4568 (1990).
\bibitem{b54}I. Affleck and E. H. Lieb, Lett. Math. Phys. {\bf 12}, 57 (1986).
\bibitem{b55} E. Lieb, T. Schultz and D. Mattis, Ann. Phys. {\bf 16}, 407 (1961).   
\bibitem{b56}F. Berruto, G. Grignani, G. W. Semenoff and P. Sodano, Phys. Rev. {\bf D57}, 5070 (1998).
\bibitem{b57}J. E. Hetrick and Y. Hosotani, Phys. Rev. {\bf D38}, 2621 (1988); J. E. Hetrick, Y. Hosotani  and S. Iso, 
Phys. Lett. {\bf B350}, 92 (1995); Y. Hosotani, R. Rodriguez, J. E. Hetrick and S. Iso, hep-th/9606129; Y. Hosotani, hep-th/9606167.
\bibitem{b58}T. Banks, L. Susskind and J. Kogut, Phys. Rev. {\bf D13}, 1043 (1976); A. Carrol, J. Kogut, D. K. Sinclair and L. Susskind, Phys. Rev. 
{\bf D13}, 2270 (1976).
\bibitem{b59}J. Lowenstein and J. A. Swieca, Ann. Phys. (N.Y.) {\bf  68}, 172 (1971).
\bibitem{b59bis}S. Chandrasekharan,  hep-lat/9809084.
\bibitem{b60}S. Coleman, R. Jackiw and L. Susskind, Ann. Phys. (N.Y.) {\bf 93}, 167 (1975).
\bibitem{b61}C. J. Hamer, Z. Weihong and J. Oitmaa, Phys. Rev. {\bf D56}, 55 (1997).
\bibitem{b62}N. S. Manton, Ann. Phys. (N. Y.) {\bf 159}, 220 (1985).
\bibitem{b63}Y. Nambu, Prog. Theor. Phys. {\bf 116}, 1474 (1950). 
\bibitem{b64b}J. P. Steinhardt, Phys. Rev. {\bf D16}, 1782 (1977).
\bibitem{b65} Y. Hosotani, J. Phys. {\bf A30}, L757 (1997); hep-th/9809066.
\bibitem{b67} C. Gattringer and E. Seiler, Ann. Phys. {\bf 233}, 97 (1994).
\bibitem{b68}S. Coleman, Ann. Phys. (N.Y.) {\bf 101}, 239 (1976).
\bibitem{b68bis}E. Witten, Comm. Math. Phys. {\bf 92}, 455 (1984).
\bibitem{b68tris}D. Gepner, Nucl. Phys. {\bf B252}, 481 (1985).
\bibitem{b69}K. Harada, T. Sugihara, M. Taniguchi and M. Yahiro, Phys. Rev. {\bf D49}, 4226 (1994).
\bibitem{b70}S. Eggert and I. Affleck, Phys. Rev. {\bf B46}, 10866 (1992).
\bibitem{b74} See for example E. Abdalla, M. C. B. Abdalla and K. D. Rothe, {\it Nonperturbative Methods in 2Dimensional Quantum Field Theory}, 
World Scientific 1991. 
\bibitem{b75}C. Gattringer, {\it $QED_{2}$ and the $U(1)$ Problem}, Dissertation University of Graz 1995, hep-th/9503137.
\bibitem{b76}M. Sadzikowski and P. Wegrzyn, Mod. Phys. Lett. {\bf A11}, 1947 (1996).
\bibitem{b77}J. P. Steinhardt, {\it Lattice Theory of $SU(N)$ Flavor Quantum Electrodynamics in $(1+1)$-dimensions}, Ph.D. Thesis, 
Harvard University (1978).  
\bibitem{b78}G. `t Hooft, Nucl. Phys. {\bf B75}, 461 (1974).
\bibitem{b79}G. `t Hooft, Nucl. Phys. {\bf B72}, 461 (1974).
\bibitem{b80}C. M. Bender and S. A. Orszag, {\it Advanced mathematical methods for scientists and engineers}, McGraw-Hill Publ. Comp. 1978.

\end{thebibliography}
\end{document}